\documentclass[manuscript]{aastex6}

\shorttitle{Optical spectral variations of blazars}
\shortauthors{Zhang et al.}

\watermark{DRAFT}

\begin{document}


\title{Optical Spectral Variations of a Large Sample of Fermi Blazars }


\author{Bing-Kai Zhang, Xiao-Yun Zhao and Qi Wu}
\affil{Department of Physics, Fuyang Normal University, Fuyang 236037, China}
\affil{Key Laboratory of Functional Materials and Devices for Informatics of Anhui Higher Education Institutes, Fuyang
Normal University, Fuyang 236037, China}


\email{zhangbk$_$ynu@163.com}









\begin{abstract}

We have investigated the optical spectral behavior of a large sample of Fermi blazars (40 FSRQs and 13 BL Lacs), and found two new universal optical spectral behaviors. In the low state the optical spectrum gradually becomes softer (steeper) or harder (flatter) but more and more slowly when the brightness increases, and then tends to stable in the high state, which are briefly named the redder-stable-when-brighter (RSWB) and bluer-stable-when-brighter (BSWB) behaviors, respectively.
34 FSRQs and 7 BL Lacs exhibit clear RSWB behavior, and 2 FSRQs and 5 BL Lacs show distinct BSWB behavior, which mean that FSRQs favor more RSWB than BSWB behavior, while BL Lacs have no clear preference among both behaviors.
We have put forward a unified nonlinear formula to quantitatively characterize the optical spectral behaviors of FSRQs and BL Lacs, which can fit both kinds of behaviors very well.
We argue that the RSWB and BSWB behaviors originate from the same mechanism, and they are the universal optical spectral behaviors for blazars. The frequently observed redder-when-brighter (RWB) and bluer-when-brighter (BWB) trends can be considered to be the approximations of the behaviors of RSWB and BSWB, respectively. The rarely observed stable-when-brighter (SWB) trend can also be viewed as an approximation or a special case of the RSWB or BSWB behavior.
We have developed a model with two constant-spectral-index components which can not only explain well both two kinds of optical spectral behaviors, but also successfully interpret the differential behaviors between FSRQs and BL Lacs.

\end{abstract}





\section{Introduction}
Blazars are an extreme subclass of active galactic nuclei (AGNs), also a dominant class of extragalactic high energy $\gamma-$ray source in the universe.
Across the entire electromagnetic spectrum from radio to very high energy gamma rays, blazars are observed to exhibit strong emission which is dominated by the non-thermal emission.

The spectral energy distributions (SEDs) of blazars are characterized by two peaks locating at low (infrared-optical/UV, even to soft X-ray) and high (hard X-ray to MeV-GeV $\gamma$-ray) energies, respectively \citep{nieppola06,falomo14}. The low energy component is generally believed to originate from the synchrotron emission of the relativistic electrons in the magnetic field of the jet.
There exist some sources of emission external to the jet, such as broad line region (BLR), dusty torus and accretion disk, which has significant contributions to the low energy component \citep[e.g.][]{ghisellini19}.
The high energy one maybe arise from inverse Compton scattering of the low energy photons by the same relativistic electrons in the jet \citep[e.g.][]{dermer93,sokolov04,bottcher07}. Alternatively, there is also a possibility that the high energy $\gamma$-ray emission is due to the synchrotron radiation of ultra-relativistic protons or the photo-pion production followed by pion decay \citep[e.g.][]{mucke03,murase12,bottcher13}.

According to the optical emission/absorption line features, blazars are generally grouped into two categories, i.e., BL Lacertae objects (BL Lacs) and Flat Spectrum Radio Quasars (FSRQs) \citep{urry95}. BL Lacs show featureless optical spectra with weak/narrow or no spectral lines whereas FSRQs  exhibit flat-spectrum radio spectra with strong and typical broad emission lines.
On the basis of the location of synchrotron peak frequency, blazars have been divided into three subclasses, i.e., high synchrotron peaked (HSP), intermediate synchrotron peaked (ISP) and low synchrotron peaked (LSP) blazars \citep{abdo10,fan16,ghisellini17}.

Blazars have been observed to show violent and rapid variability on different timescales from years to hours, event to minutes. Generally, spectral variations and flux variations always occur together. The relationship between the spectrum and brightness, which is usually represented graphically in the color index vs. magnitude or spectral index vs. flux diagram, has been explored by many investigations. Some individual sources have been investigated, and exhibit BWB trend, such as
OJ 287 \citep{takalo89,dai11,gupta19},
S5 0716+714 \citep{ghisellini97,man16,yuan17,li17,xiong20},
BL Lacertae \citep{villata02,villata04,papadakis07,weaver20},
3C 273 \citep{dai09},
1ES 1011+496 \citep{bottcher10},
PG 1553+113 \citep{raiteri15},
1ES 0806+524 \citep{pandey20},
AO 0235+16 \citep{raiteri01}.
Some objects show RWB trend, such as
PKS 0736+017 \citep{ramirez04},
3C 454.3 \citep{villata06,sarkar19},
PKS 0208-512 \citep{chatterjee13},
CTA 102 \citep{raiteri17}.

In addition to individual sources, the spectral behavior for large samples is also investigated as a whole \citep[e.g.][]{vagnetti03,gu06,rani10,ikejiri11,bonning12,gaur12,wierzcholska15,zhang15,mao16,li18,meng18,gaur19,safna20}. The previous results show that most BL Lac objects exhibit a bluer-when-brighter (BWB) chromatism, while most FSRQs display a redder-when-brighter behavior. However, there are some exceptions. For example, 14 FSRQs display BWB trend and only one FSRQ is RWB behavior in a sample of 29 SDSS FSRQs \citep{gu11}, and all six BL Lacs show no BWB trends \citep{sandrinelli14}.
For individual cases, there are also some special cases, such as BL Lac object PG 1553+113 in 2013 showed a general bluer-when-brighter trend \citep{raiteri15}. However, the source did not show a definite trend when analysing the color behaviour \citep{wierzcholska15,gupta16}. In addition, no significant correlation between $V-R$ color index and $R$ magnitude was found on any observing night in 2016 \citep{pandey19}. For FSRQ 3C 345 the bluer-when-brighter and redder-when-brighter chromatisms were simultaneously observed in this object when using different pairs of pass-bands to compute the colors \citep{wu11}. For 3C 454.3 and PKS 0537-441, RWB to BWB trend was detected \citep{raiteri08,zhang13,rajput19}.

Previous observational studies of blazars at optical frequencies have revealed complex flux-color correlation patterns. In general, there are five types of color (spectral) behaviors: BWB in whole data sets; RWB in whole data sets; cycles or loop-like patterns; RWB in the low state while BWB in the high state (RWB to BWB); stable-when-brighter (SWB) or no correlation with brightness in whole data sets \citep[see][and references therein]{zhang14}. However, there seems to be lack of a universal form to describe the optical spectral behavior of blazars.

As the optical data is accumulated more and more, we will investigate the optical spectral behaviors with a large sample of blazars and try to explore the universal properties of optical spectrum with almost simultaneous optical $BVR$ passband measurements. In the second section of this paper, we will introduce the sample and their optical data, and then calculate the optical spectral indices and describe their behaviors in Section 3. After this, we will discuss the optical spectral properties in great detail in Section 4 and give some brief conclusions in Section 5.

\begin{figure}
\epsscale{.60}
\plotone{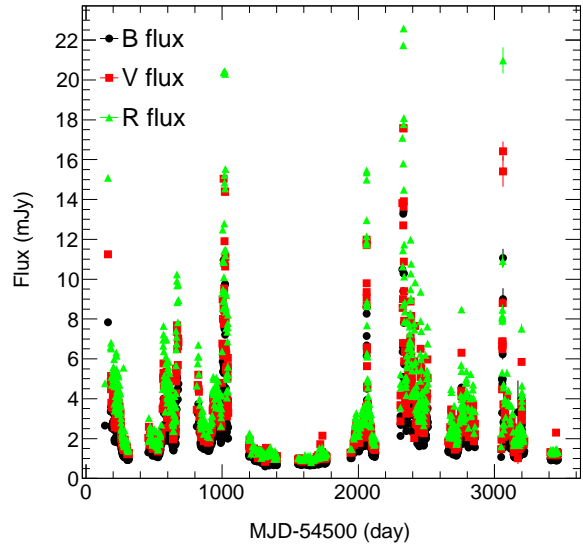}
\caption{The optical light curve of 3C 454.3.\label{3C454lc}}
\end{figure}

\section{Blazar sample and optical data}
\begin{figure}
\epsscale{.60}
\plotone{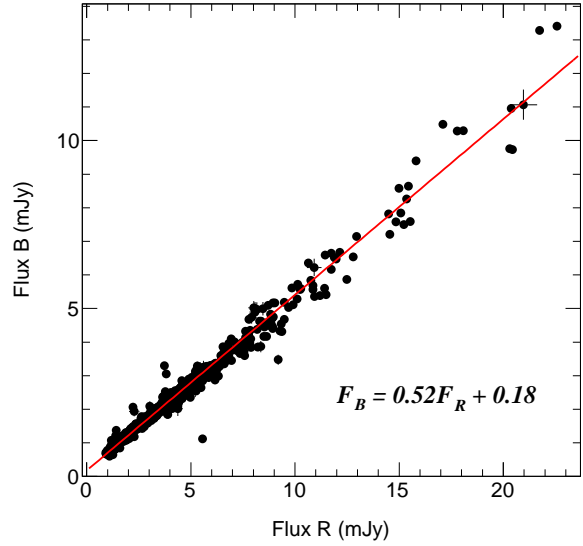}
\caption{The relation between optical $B$ and $R$ bands of 3C 454.3.\label{3C454b-r}}
\end{figure}

The meter-class telescopes of the Small and Moderate Aperture Research Telescope System (SMARTS) \citep[see][]{bonning12} provide a great opportunity to better understand the spectral behaviour of blazars. Since 2008, they have been used to carry out photometric monitoring of numerous blazars on a regular cadence at both optical and near-infrared wavelengths. So far, a large number of $Fermi$ sources (about 110 blazars) have been monitored over a span of 1 decade, and large amount of near simultaneous $BVRJK$ data has been accumulated and released through the web site \footnote{http://www.astro.yale.edu/smarts/glast/home.php}.
However,
the number of monitoring data varies greatly between different sources.
To obtain high-quality optical spectral behavior of each source, the more observations, the better the result. However, the balance between the number of observed data and the sample size must be taken into account. In this analysis, we collect the blazars with more than 10 simultaneous observation data pairs in optical $BVR$ bands. Therefore, 53 sources have survived, which include 40 flat spectrum radio quasars (FSRQs) and 13 BL Lac objects. Among them, there are 7 TeV sources.
These 53 $Fermi$ blazars have been listed in Table 1, which present the source name as well as its redshift, optical type, SED type, and position of synchrotron peak in Columns 1 to 5 in order.
Most of the sources have been monitored simultaneously in optical $BVR$ passbands, and a few sources have been only observed in two passbands. The Galactic interstellar reddening has been corrected. The values of Galactic extinction, $A_{\lambda}$, of $V$ band for each source have been listed in Column 6 of Table 1, which are taken from the NASA Extragalactic Database (NED\footnote{https://ned.ipac.caltech.edu/}). Then the calibrated magnitudes are converted into flux densities. Serving as an example, the optical $BVR$ light curves for 3C 454.3 have been exhibited in Figure~\ref{3C454lc}. It displays large amplitude of variations and very similar trends in three bands. Figure~\ref{3C454b-r} plots the relation between the fluxes of optical $B$ and $R$ bands. It presents a very good linear correlation. The fractional variability amplitude ($F_{var}$) has been employed to quantitatively describe the variations, which is defined as $F_{var} = \sqrt{(S^{2}-<\sigma^{2}>)/<f>^{2}}$,  where $f$ is the flux, $\sigma$ is the error and $S^{2}$ is the sample variance. For 3C 454.3, the $F_{var}$ of $R$ band is 0.79, which demonstrates the proportion of the amplitude of change to the average. The $F_{var}$s of $R$ bands for the other sources have all been investigated and listed in Column 7 of Table 1. The vast majority of sources exhibit significant variations. Only four objects (PKS 0301-243, PKS 0454-46, PKS 0637-75 and PKS 1004-217) show mild and weak variations.

\section{Spectral behavior}
In general, the flux $F_{\nu}$ and the frequency $\nu$ follow a power law, with the following equation
\begin{equation}
  F_{\nu}\propto\nu^{\alpha},
  \label{eq:index}
\end{equation}
where $\alpha$ is the spectral index.

In order to improve the quality of the optical spectral index, we make full use of the measurements of all three $BVR$ pass-bands. We fit the quasi-simultaneous $BVR$ measurements with Equation~\ref{eq:index} and obtain the spectral index $\alpha$. If there are only two pass-band measurements, we get the spectral index by $\alpha = (logF_{\nu_{2}}-logF_{\nu_{1}})/(log\nu_{2} - log\nu_{1})$. The number of $\alpha$ has been filled in Column 8 of Table 1. For each blazar, the mean value of spectral index $\alpha$ has been calculated and listed in Column 9 of Table 1.

To reveal the relation between the optical spectral index $\alpha$ and the flux $F_{R}$ for each source,
a linear regression has always been applied to analyze the spectral behavior (color behavior) of blazars. However, we have noticed that the variation trends of spectral index with flux do not always follow linear relations well. Therefore, the linear regression analysis is clearly inappropriate. Thus, we have employed a continuous \textbf{segmented} linear function to fit the curve of $\alpha$ versus $F_{R}$,
 \begin{equation}
\alpha=\left\{
\begin{array}{lcl}
 k_{1}F_{R}+b_{1}, & & if \quad F_{R} \leq F_{cri}\\
 k_{2}F_{R}+b_{2}, & & if \quad F_{R} > F_{cri}
\end{array}
 \right.
 \label{cplfit}
\end{equation}
in which $k_{1}$, $k_{2}$, $b_{1}$, $b_{2}$ and $F_{cri}$ are all free parameters. During the fitting procedure, we set a restriction of $k_{1}F_{cri}+b_{1} = k_{2}F_{cri}+b_{2}$, to ensure that the \textbf{segmented} function is continuous.

\begin{figure}
\epsscale{1.0}
\plottwo{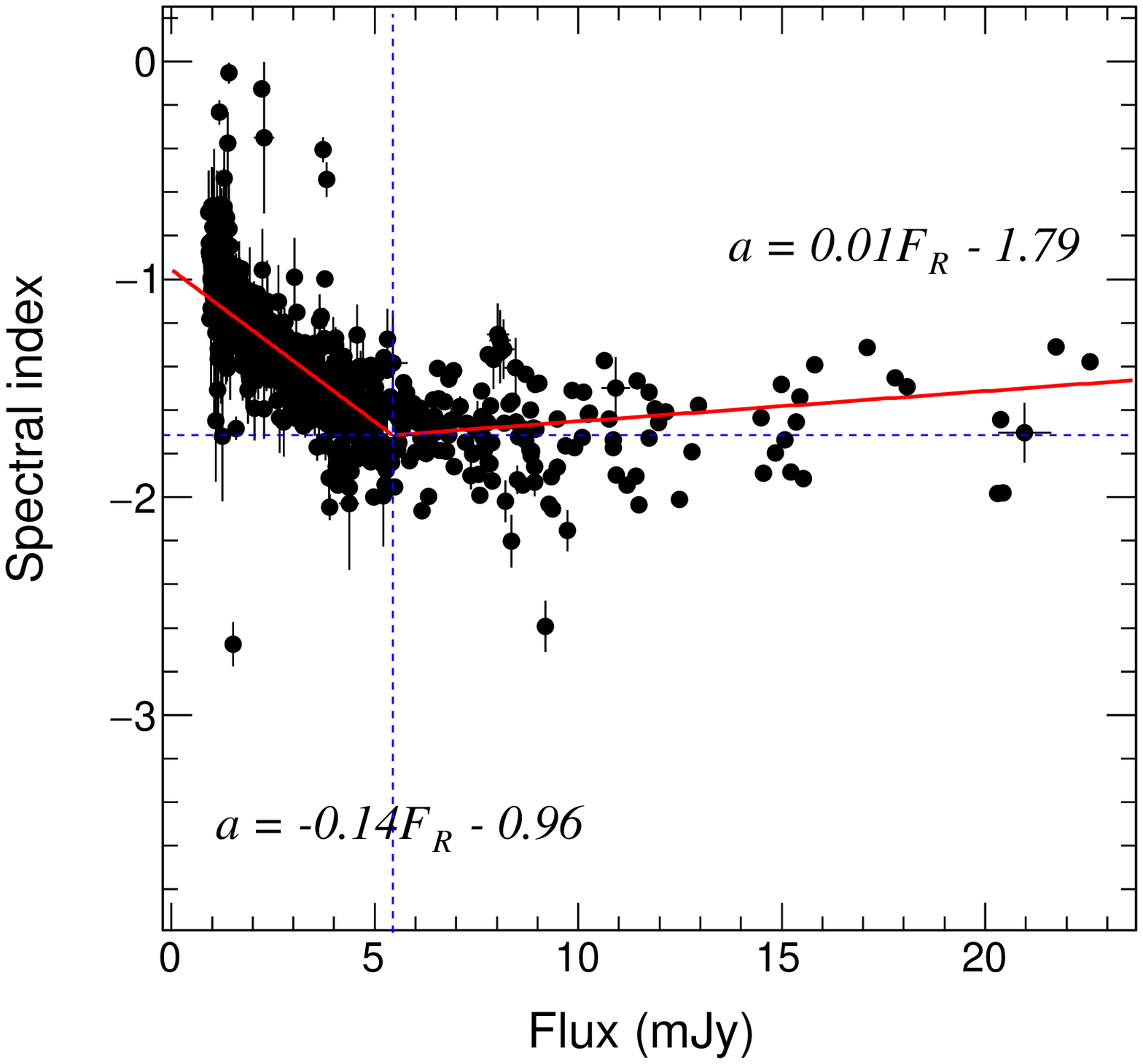}{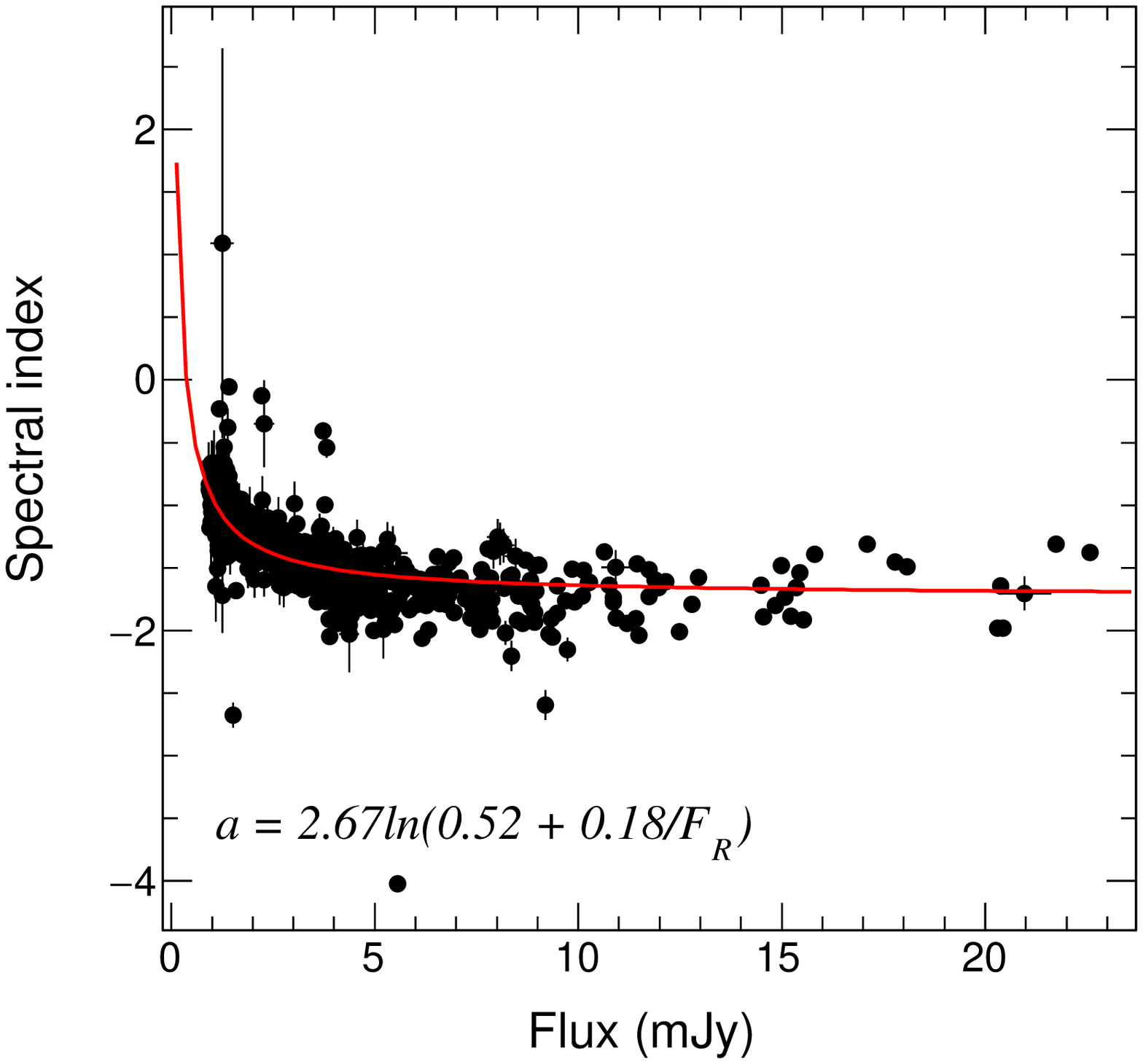}
\caption{The spectral index vs optical $R$ band flux for 3C 454.3. The red solid lines in the left and right panels represent the fitting results with Equations~\ref{cplfit} and~\ref{lnfit}, respectively. \label{3C454alphaflux}}
\end{figure}

For 3C 454.3, the fitting result is
 \begin{equation}
\alpha=\left\{
\begin{array}{llll}
 -0.14F_{R} - 0.96, &if &F_{R} \leq 5.44 &mJy\\
 +0.01F_{R} - 1.79, &if &F_{R} > 5.44 &mJy.
\end{array}\right.
 \label{fit3C454}
\end{equation}
They have been superimposed in the left panel of Figure~\ref{3C454alphaflux}. One can see that a \textbf{segmented} function can generally describe the curve well. All $\alpha-F_{R}$ curves for other blazars have also been fitted with Equation~\ref{cplfit}, and the corresponding results have been illustrated in the left panels of Figures~\ref{0017-0512} -~\ref{OJ287} in the appendix. In most cases, with the brightness increasing, the source spectra become steeper in the low state, however, when the flux is above a certain value (critical flux, $F_{cri}$), the sources show a different behavior, and they exhibit an approximately stable behavior, i.e. the spectral indices do not obviously vary with the brightness in the high state. Two parts mean that there should exist two different mechanisms or emission components in the source. The critical flux point can be viewed as a watershed or transition of two different mechanisms or emission components.

Now, we try to come up with a new formula to describe the relation between optical spectral index and flux.
\begin{equation}
\alpha = 2.67ln(a+b/F_{R}),
\label{lnfit}
\end{equation}
where $\alpha$ and $F_{R}$ are optical spectral index and $R$ band flux, respectively. $a$ and $b$ are two free parameters. This equation represents a continuous and smooth curve.  For 3C 454.3, we have fitted the variations of $\alpha$ with $F_{R}$ by this equation, and obtained
\begin{equation}
\alpha = 2.67ln(0.52+0.18/F_{R}).
\label{lnfit3C454}
\end{equation}
The corresponding fitting curve has been plotted in the right panel of Figure~\ref{3C454alphaflux}. One can see that Equation~\ref{lnfit} fits the $\alpha-F_{R}$ curve very well.
With the increase of flux, the spectral index declines gradually, but more and more slowly, until the spectral index tends to be stable.
The fitting results of other sources have been plotted in the right panels of Figures~\ref{0017-0512} -~\ref{OJ287} in the appendix. In the vast majority of cases (48 out of 53 sources), Equation~\ref{lnfit} reflects the optical spectral behavior very well.
As the source brightness increases, the spectrum gradually and smoothly steepens, but changes more and more slowly and eventually tends to stabilize. In other words, the color becomes redder and redder, and then stable when the source brightens. We refer to this kind of spectral behavior as the \emph{redder-stable-when-brighter (RSWB)} trend in brief.
There are a few sources showing another opposite kind of spectral behavior which is briefly designated as the \emph{bluer-stable-when-brighter (BSWB)} trend here. In these sources, their colors turn bluer and bluer, and then gradually stable with the source brightening from the low state to the high state.
In addition, five blazars (e.g., PKS 1127-14, PKS 2354-021 and 3C 446) can not be successfully fitted with Equation~\ref{lnfit}, since their diagrams of $\alpha$ against $F_{R}$ look very scattered and seem a little weird.
The spectrum characteristics have been given in Column 10 of Table 1 for each source.

\begin{deluxetable}{lllllcrrlc}
\tablenum{1}
\tablecaption{The properties of blazars\label{tab:sample}}
\tablewidth{0pt}
\tablehead{
\colhead{Object} & \colhead{z}  &\colhead{Class}
&\colhead{SED} &\colhead{log($\nu$)} &\colhead{A$_{V}$}& \colhead{F$_{var}$} &\colhead{N}
&\colhead{$\overline{\alpha}$}& \colhead{Behavior}
}
\decimalcolnumbers
\tabletypesize{\tiny}
\startdata
 PMN J0017-0512 &0.227  &FSRQ   &LSP &13.689 &0.084 &0.71$\pm$0.09 &32  &-0.13$\pm$0.28 &RSWB\\
 4C +01.02      &2.099  &FSRQ   &LSP &13.181 &0.066 &0.39$\pm$0.03 &96  &-0.74$\pm$0.42 &RSWB\\
 PKS 0208-512   &1.033  &FSRQ   &LSP &13.422 &0.056 &1.04$\pm$0.03 &624 &-1.31$\pm$0.51 &RSWB\\
 PKS 0235+164   &0.94   &BL Lac &LSP &13.313 &0.219 &1.34$\pm$0.06 &223 &-3.21$\pm$0.54 &RSWB\\
 PKS 0250-225   &1.427  &FSRQ   &LSP &13.25  &0.081 &0.42$\pm$0.04 &53  &-1.93$\pm$1.40 &RSWB\\
 PKS 0301-243\tablenotemark{*}   &0.260  &BL Lac &HSP &15.433 &0.061 &0.09$\pm$0.01 &45  &-0.93$\pm$0.10 &BSWB\\
 PKS 0336-01    &0.850  &FSRQ   &LSP &13.107 &0.253 &0.72$\pm$0.09 &35  &-2.60$\pm$0.49 &RSWB\\
 PKS 0402-362   &1.417  &FSRQ   &LSP &13.023 &0.015 &0.46$\pm$0.02 &149 &-0.75$\pm$0.30 &RSWB\\
 PKS 0426-380   &1.105  &BL Lac &LSP &13.137 &0.066 &0.66$\pm$0.03 &312 &-1.68$\pm$0.38 &RSWB\\
 PKS 0440-00    &0.844  &FSRQ   &LSP &13.306 &0.140 &0.80$\pm$0.12 &21  &-0.41$\pm$0.63 &RSWB\\
 PKS 0454-234   &1.003  &FSRQ   &LSP &13.167 &0.126 &0.68$\pm$0.03 &342 &-1.57$\pm$0.26 &---\\
 PKS 0454-46    &0.858  &FSRQ   &LSP &12.764 &0.047 &0.11$\pm$0.01 &40  &-0.59$\pm$0.22 &---\\
 S3 0458-02     &2.291  &FSRQ   &LSP &13.522 &0.203 &0.38$\pm$0.04 &53  &-1.37$\pm$0.35 &RSWB\\
 PKS 0502+049   &0.954  &FSRQ   &LSP &13.471 &0.219 &1.64$\pm$0.10 &129 &-0.79$\pm$0.33 &RSWB\\
 PKS 0528+134   &2.07   &FSRQ   &LSP &12.344 &2.357 &0.29$\pm$0.02 &136 &-0.71$\pm$1.03 &RSWB\\
 PMN J0531-4827 &0.8116 &BL Lac &LSP &13.11  &0.085 &1.07$\pm$0.05 &191 &-1.43$\pm$0.70 &RSWB\\
 PKS 0537-441   &0.892  &BL Lac &LSP &13.352 &0.102 &1.08$\pm$0.08 &89  &-0.68$\pm$0.27 &RSWB\\
 PKS 0637-75    &0.653  &FSRQ   &LSP &13.048 &0.263 &0.06$\pm$0.01 &43  &-0.05$\pm$0.21 &RSWB\\
 PKS 0736+01\tablenotemark{*}    &0.189  &FSRQ   &ISP &14.125 &0.375 &0.45$\pm$0.05 &49  &-1.07$\pm$0.18 &RSWB\\
 PKS 0805-07    &1.837  &FSRQ   &LSP &13.533 &0.388 &0.96$\pm$0.10 &42  &-2.18$\pm$0.33 &RSWB\\
 PMN J0850-1213 &0.566  &FSRQ   &LSP &13.401 &0.116 &0.67$\pm$0.06 &71  &-0.84$\pm$0.62 &RSWB \\
 PKS 1004-217   &0.33   &FSRQ   &ISP &14.559 &0.162 &0.07$\pm$0.01 &52  &-0.27$\pm$0.31 &RSWB\\
 4C +01.28      &0.89   &BL Lac &LSP &13.176 &0.073 &0.73$\pm$0.09 &32  &-2.14$\pm$0.15 &RSWB\\
 PKS B1056-113  &-      &BL Lac &LSP &13.56  &0.075 &0.30$\pm$0.03 &51  &-2.97$\pm$0.12 &BSWB\\
 PKS 1127-14    &1.184  &FSRQ   &LSP &12.989 &0.100 &0.23$\pm$0.03 &31  &-0.02$\pm$0.49 &---\\
 PKS 1144-379   &1.048  &BL Lac &LSP &13.141 &0.266 &0.66$\pm$0.04 &176 &-1.32$\pm$0.60 &RSWB\\
 PKS 1244-255   &0.635  &FSRQ   &LSP &13.454 &0.236 &0.30$\pm$0.03 &54  &-0.60$\pm$0.29 &RSWB\\
 PKS 1329-049   &2.15   &FSRQ   &LSP &13.098 &0.08  &0.82$\pm$0.12 &24  &-4.59$\pm$0.63 &RSWB\\
 PKS B1406-076  &1.494  &FSRQ   &LSP &13.192 &0.095 &0.50$\pm$0.02 &221 &-1.21$\pm$0.52 &RSWB\\
 PKS 1424-41    &1.522  &FSRQ   &LSP &13.517 &0.335 &0.97$\pm$0.03 &560 &-1.55$\pm$0.27 &RSWB\\
 PKS 1510-089\tablenotemark{*}   &0.36   &FSRQ   &LSP &13.699 &0.275 &0.74$\pm$0.02 &648 &-0.77$\pm$0.47 &RSWB\\
 AP Lib\tablenotemark{*}         &0.048  &BL Lac &ISP &14.185 &0.378 &0.24$\pm$0.02 &114 &-1.91$\pm$0.32 &BSWB\\
 PKS 1550-242   &0.332  &FSRQ   &LSP &13.25  &0.478 &0.54$\pm$0.07 &32  &-1.28$\pm$0.35 &BSWB\\
 PKS B1622-297  &0.815  &FSRQ   &LSP &13.299 &1.245 &0.35$\pm$0.02 &253 &-0.52$\pm$0.42 &RSWB\\
 PKS 1730-13    &0.902  &FSRQ   &LSP &13.074 &1.424 &0.38$\pm$0.02 &244 &-1.12$\pm$0.69 &RSWB\\
 PKS 1824-582   &1.531  &FSRQ   &-   &-      &0.273 &0.32$\pm$0.06 &13  &-1.79$\pm$0.34 &RSWB\\
 PKS 1954-388   &0.63   &FSRQ   &LSP &13.012 &0.224 &0.42$\pm$0.04 &45  &-0.97$\pm$0.43 &RSWB\\
 PKS 2023-07    &1.388  &FSRQ   &LSP &13.418 &0.104 &0.91$\pm$0.07 &89  &-1.23$\pm$0.43 &RSWB\\
 PKS 2032+107   &0.601  &FSRQ   &ISP &14.032 &0.402 &0.57$\pm$0.08 &24  &-1.38$\pm$0.39 &RSWB\\
 PKS 2052-47    &1.491  &FSRQ   &LSP &13.191 &0.098 &0.78$\pm$0.04 &207 &-1.05$\pm$0.45 &RSWB\\
 PKS 2142-75    &1.138  &FSRQ   &LSP &13.335 &0.304 &0.28$\pm$0.02 &164 &-1.16$\pm$0.27 &RSWB\\
 PKS 2155-304\tablenotemark{*}   &0.116  &BL Lac &HSP &15.968 &0.06  &0.36$\pm$0.01 &558 &-0.80$\pm$0.11 &BSWB\\
 PKS 2233-148   &0.33   &BL Lac &LSP &13.08  &0.121 &0.62$\pm$0.11 &17  &-0.58$\pm$0.20 &RSWB\\
 PKS 2255-282   &0.926  &FSRQ   &LSP &13.036 &0.093 &0.17$\pm$0.04 &12  &-0.92$\pm$0.50 &RSWB\\
 PKS 2326-502   &0.518  &FSRQ   &LSP &12.906 &0.036 &0.99$\pm$0.06 &158 &-1.28$\pm$0.38 &RSWB\\
 PMN J2345-1555 &0.621  &FSRQ   &LSP &13.522 &0.067 &0.58$\pm$0.04 &88  &-0.42$\pm$0.16 &RSWB\\
 PKS 2345-16    &0.576  &FSRQ   &LSP &13.168 &0.071 &0.78$\pm$0.10 &30  &-0.95$\pm$0.50 &RSWB\\
 PKS 2354-021   &0.812  &BL Lac &LSP &13.378 &0.114 &0.41$\pm$0.09 &12  &-0.61$\pm$0.68 &---\\
 3C 279\tablenotemark{*}         &0.536  &FSRQ   &LSP &13.106 &0.078 &0.85$\pm$0.02 &761 &-1.77$\pm$0.26 &BSWB\\
 3C 446         &1.404  &FSRQ   &LSP &13.316 &0.207 &0.37$\pm$0.06 &22  &-5.20$\pm$3.52 &---\\
 3C 454.3       &0.859  &FSRQ   &LSP &13.344 &0.289 &0.79$\pm$0.02 &846 &-1.39$\pm0.33$ &RSWB\\
 CTA 102        &1.037  &FSRQ   &LSP &12.719 &0.198 &0.27$\pm$0.03 &32  &-2.53$\pm$0.16 &RSWB\\
 OJ 287\tablenotemark{*}         &0.3056 &BL Lac &LSP &13.732 &0.077 &0.44$\pm$0.01 &503 &-1.61$\pm$0.22 &BSWB\\
\enddata
\tablenotetext{*}{TeV blazar}
\end{deluxetable}

\section{Discussion}

The optical spectral behaviors of 53 blazars have been investigated. Generally, the spectral indices of blazars vary clearly with the source brightness. 77\% of the blazars (41 out of 53 sources) follow the RSWB trend.
They share a common feature, that is, during their variation processes from low to high states, the spectrum becomes steeper and steeper, but more and more slowly, until it tends to be stable. That is to say, the sources become redder gradually and then stable when they brighten.

Although RWB color tendencies have previously been reported for most of FSRQs \cite[e.g.][]{bonning12,zhang15,safna20},
the redder-stable-when-brighter (RSWB) trend has not been explicitly mentioned before.
However,
in previous studies, there are also some hints of the RSWB behavior. For example, 3C 454.3 seems to exist a kind of ``saturation", and RWB trend is not visible any longer in the brightest part \citep{villata06,fan18}. During the campaign of the Whole Earth Blazar Telescope (WEBT), CTA 102 exhibits a RWB trend followed by a very slight BWB trend when it brightens \citep{raiteri17}.
The signs of RWB then to SWB trend in $B-J$ vs $J$ diagrams can also be recognized for several blazars, such as PMN J0850-1213, PKS 2023-07, PKS 2052-47, PKS 1510-089 and 3C 454.3 \citep{bonning12,safna20}. With the observations of optical $R$ and infrared $J$ bands, \cite{zhang15} mentioned that six sources, e.g. PKS 1510-089 and PKS B1622-297, show RWB behavior in fainter states, and then keep to stabilize (SWB) in brighter states.
Although there are some signs of RSWB behaviors in previous studies, this type of spectral behaviors has not been clearly defined before. Therefore, this is the first time that the RSWB behavior has been found and proposed in such a large sample of Fermi blazars.

As a counterpart of RSWB behavior, another spectral phenomenon of bluer-stable-when-brighter (BSWB) has also been found in this work, which exhibits a feature of becoming bluer in fainter state and then stable trend in brighter state as the sources brighten.
This phenomena has also been noticed by \cite{xiong20} for S5 0716+714 and \cite{zhang21} for 3C 279.
In total, only 7 of 53 blazars are found to show the BSWB behavior.

If we examine in detail different subclasses of blazars, we can find that 34 out of 40 FSRQs (i.e., 85\%) and 7 out of 13 BL Lacs (i.e., 54\%) present the RSWB behavior.  In addition, only 2 (i.e., 5\%) FSRQs and 5 (i.e., 38\%) BL Lacs (e.g. 3C 279, AP Lib, PKS 2155-304) have been found to show the BSWB behavior.
In other words, the overwhelming majority of FSRQs exhibit the RSWB behavior and only a few FSRQs show the BSWB behavior. The number of BL Lacs displaying the RSWB behavior has only a slight advantage over the number of those exhibiting the BSWB behavior.
One can also view that
all of 4 TeV BL Lacs and 1 of 3 TeV FSRQs render the BSWB property.
Of course, there are individual blazars showing no clear variations of spectral index, such as PKS 2032+107 and PKS 2233-148. Their spectral indices almost remains a stable trend throughout the entire data sequence.

Customarily, a single linear function is employed to fit the spectral (or color) behavior. However, a straight line does not always succeed to describe the relation of $\alpha$ versus $Flux$.
In this work, we propose a continuous \textbf{segmented} linear function (Equation~\ref{cplfit}) and a nonlinear function (Equation~\ref{lnfit}) to represent the relation between optical spectral index $\alpha$ and the flux $F_{R}$. On closer inspection of the fitting results, one can conclude that, in general, both Equations~\ref{cplfit} and~\ref{lnfit} can fit the great majority of the plots well.
 It is obvious that the $\alpha$ and $F_{R}$ do not follow a linear relationship. In fact, even in the low state, it seems not to be a linear relation between $\alpha$ and $flux$. In Equation~\ref{lnfit}, a single nonlinear function with two free parameters can fit the curves very well, which means it make the transition even smoother.
 \textbf{Bayesian information criterions (BIC) are also calculated for each source. The results show Equation~\ref{lnfit} has smaller BICs than Equation~\ref{cplfit} for all sources.}
 So, Equation~\ref{lnfit} is significantly better than Equation~\ref{cplfit} in describing the curves of $\alpha - flux$.

Generally, two main kinds of behaviors of BWB and RWB are mostly reported, and different mechanisms have been proposed to explain them.
The BWB behavior may be due to the presence of two components contributing to the overall emission \citep{fiorucci04,ikejiri11},  one-component synchrotron model \citep{fiorucci04}, the injection of fresh electrons \citep{kirk98}, or the time delay between the different optical bands \citep{wu07}.
The RWB behavior is generally interpreted as due to a strong contribution of a blue thermal emission from the accretion disk
 \citep{villata06,rani10,bonning12,isler17}, or changes in the Doppler factor \citep{raiteri17}.
However, there is a lack of a universal model to interpret the two behaviors. From this study, one can clearly view that the optical spectral index changing with the flux is not a sudden transformation process, but a continuous, smooth, non-linear, and progressive (gradual) process. Most of them exhibit the RSWB behavior, only a small fraction show the BSWB behavior. These two behaviors are clearly distinct in the low state, but they have a common feather that they both tend to be stable in the high state.

The fact that two behaviors can well be fitted by the same formula in this study suggest that there should arise from the same mechanism.
In order to explain these two behaviors, we have proposed a new model with two constant-spectral-index components,
where the observed optical emission of blazars (labelled as $F_{obs}$) consists of a stable or less variable thermal emission component (labelled as $F_{ther}$) primarily coming from the accretion disc, and a highly variable non-thermal emission component (labelled as $F_{syn}$) coming from the relativistic jet,
while the spectral indices of both components are invariant,
which are labelled as $\alpha_{ther}$ and $\alpha_{syn}$, respectively.
Thus, the observed optical flux
$F_{obs}$ = $F_{ther}$ + $F_{syn}$.
We define
$k_{ther} = F_{ther,B}/F_{ther,R}$, $k_{syn}$ = $F_{syn,B}/F_{syn,R}$, then we can obtain that the spectral indices of thermal and non-thermal components are $\alpha_{ther}$ = $2.67 ln(k_{ther})$ and
$\alpha_{syn} = 2.67 ln(k_{syn})$, respectively.
Then we can derive the expression of the observed spectral index $\alpha_{obs}$ changing with the $R$ band flux $F_{obs,R}$,
\begin{equation}
\alpha_{obs} = 2.67 ln[ k_{syn} + (k_{ther} - k_{syn})\frac{F_{ther,R}}{F_{obs,R}} ],
\label{la}
\end{equation}
which is identical to the Equation~\ref{lnfit}. From this equation, we can see that the optical spectral index is determined jointly by four physical quantities, namely $\alpha_{syn}$, $\alpha_{ther}$, $F_{ther,R}$ and $F_{obs,R}$.
However, for a specific source, the change of the spectral index $\alpha_{obs}$ depends only on the flux $F_{obs,R}$, due to the other 3 quantities are constant.

In the low state, if no non-thermal contribution, i.e. $F_{obs,R}$ = $F_{ther,R}$, then $\alpha$ = $\alpha_{ther}$. Whereafter, with the increase of non-thermal contributions, $\alpha$ begins to vary nonlinearly towards the value of the $\alpha_{syn}$. When $F_{obs,R}$ $\gg$ $F_{ther, R}$, $\alpha$ is close to $\alpha_{syn}$ and tend to be stable in the high state.
The spectral index $\alpha$ varies between $\alpha_{ther}$ and $\alpha_{syn}$, and along a specific direction. So the spectral behavior lies on the quantitative relationship between  $\alpha_{ther}$ and $\alpha_{syn}$.

\begin{figure}
\epsscale{.60}
\plotone{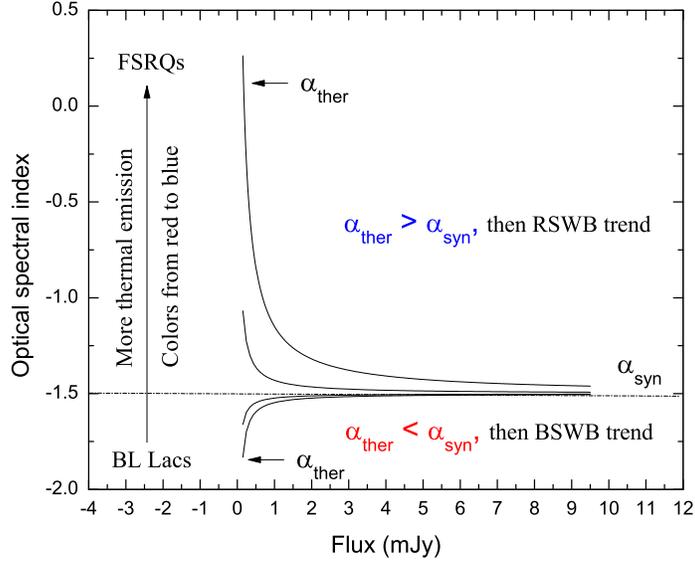}
\caption{The sketch of optical spectral behavior of blazars.\label{sketch}}
\end{figure}

Blazars have strong relativistic jets in which the contribution of non-thermal emission dominates over that of the thermal accretion disc emission. The optical spectral indices $\alpha_{syn}$ s of the non-thermal component distribute around $-1.5$
\citep{chiang02,fiorucci04}.
 As to the thermal emission component,  the stronger it is, the bluer the color will be; otherwise, the weaker it is, the redder the color will be.
FSRQs are believed to have a luminous accretion disk, while BL Lacs have a weaker one.
\cite{ghisellini10} derived that the accretion disk luminosities of FSRQs are in the range of 10$^{45}$ $-$ $10^{47}$ erg s$^{-1}$, and those of BL Lacs are always below $10^{45}$ erg s$^{-1}$.
Luminous accretion disk of FSRQs means the thermal emission component from accretion disk is much bluer
than the non-thermal component, namely, $\alpha_{ther}$ $>$ $\alpha_{syn}$. Thus, almost all FSRQs are expected to display redder-stable-when-brighter (RSWB) behaviors, which is very consistent with the behaviors of FSRQs studied in this work. Only two FSRQs here (PKS 1550-242, 3C 279) show the BSWB behavior, which imply that these two FSRQs have weak thermal emission. The accretion disk of 3C 279 has been derived to have a relatively low luminosity with a value of 3$\times$10$^{45}$ erg s$^{-1}$ \citep{ghisellini10} among the FSRQ group.
The schematic drawing of the spectral behavior is illustrated in Figure~\ref{sketch}. FSRQs are mainly distributed in its upper part.

In contrast to FSRQs, BL Lacs generally have weaker accretion disk and therefore have less contributions of thermal emission, which means the color of their thermal emission should be much redder than that of FSRQs. That is, the spectral index of the thermal component of BL Lacs, $\alpha_{ther}$, is small. So that BL Lac objects locate in the lower part of Figure~\ref{sketch}.
By comparison with the non-thermal emission component, the thermal emission component of part of BL Lac objects are also redder, i.e., $\alpha_{ther}$ $<$ $\alpha_{syn}$. Thus these BL Lacs are predicted to exhibit bluer-stable-when-brighter (BSWB) behavior. 5 out 13 BL Lacs here exhibit this behavior. However, if a BL Lac object has bluer thermal component than non-thermal one (i.e., $\alpha_{ther}$ $>$ $\alpha_{syn}$), therefore this BL Lac, similar to FSRQs, is predicted to present the RSWB behavior. In this work, 7 out 13 BL Lacs display the RSWB behavior, which suggest these 7 objects should have brighter accretion disks than other BL Lacs. According to the result derived by \cite{ghisellini10}, the accretion disks of these objects are more luminous than those of other BL Lacs. For example, PKS 0537-441 and AO 0235+164 have accretion disks with luminosity values of 1.2$\times$10$^{46}$ and 4.5$\times$10$^{45}$ erg s$^{-1}$, respectively, which are significantly brighter than other BL Lacs and can be comparable to FSRQs. In addition, the seven BL Lacs showing the RSWB behavior almost belong to the HBLs or IBLs.

If a FSRQ is observed only in the low state,  one will only see the left part of the RSWB trend which can be approximately considered as the RWB trend frequently seen. Similarly, for BL Lacs, the spectral behavior in the low state will only display a RWB or BWB trend. If observed only in the high state,  both FSRQs and BL Lacs
will exhibit a stable-when-brighter trend which are illustrated in the right part of RSWB and BSWB trends.
In addition, we can derive that blazars, whether FSRQs or BL Lacs, will keep stable-when-brighter behaviors at any state as long as their thermal spectral index is near to the non-thermal one.
It should be noted that if a $\alpha$-$Magnitude$ or $Color$ $index$-$Magnitude$ diagram is adopted, the behavior of RSWB or BSWB is easily obscured, and becomes inconspicuous, therefore it looks more like RWB or BWB trend, especially when the source is in a low state.

\section{Conclusions}

Through the study on a large sample of Fermi blazars (53 blazars), we have found two new phenomena of the optical spectral behavior for blazars, and named them the redder-stable-when-brighter (RSWB) behavior and bluer-stable-when-brighter (BSWB) behavior, respectively. The RSWB behavior represents that a source displays redder-when-brighter in low states, and then keep stable in high states. In other words, the optical spectrum becomes gradually steeper when the source brightens in low states, and then remains invariable in high states. 85\% of FSRQs and 54\% of BL Lacs studied in this work exhibit the RSWB behavior. The BSWB behavior represents that a source exhibits bluer-when-brighter trend in low states, and then keep stable in high states. Only 2 FSRQs (5\%) and 5 BL Lacs  (38.5\%)  studied here show the RSWB phenomenon.

We have constructed a new nonlinear formula to fit the relationship between the optical spectral index and the flux. The formula works very well almost at any case. Whether for FSRQs or BL lacs, whether for RSWB or BSWB behaviors, whether in high states or in low states, the formula can well describe the relation between the spectral index and the flux.

We have developed a model with two constant-spectral-index components to quantitatively explain the new phenomena. In this model, the optical emission of blazars is composed of two components, a stable thermal emission and a variable non-thermal emission. Both of these two components are supposed to have invariant spectral indices. The variations of spectral indices arise from the changes of the contribution of non-thermal component. In the low state, the optical spectrum is mainly affected by the thermal component, and then gradually dominated by the non-thermal component when the source brightens. Whether the spectral behavior is RSWB or BSWB, depends only on the contrast of the spectral indices of thermal and non-thermal components. If the non-thermal component is redder than the thermal component, the spectral behavior will be RSWB which occurs in almost all FSRQs and partial BL Lacs. If the opposite is true, then the spectral behavior will be BSWB which happens mainly in partial BL Lacs.

We suggest that the RSWB behavior as well as the BSWB behavior may be universal trends of optical spectrum (color) variations, since the trends of RWB, BWB and SWB can all be naturally seen as special cases of the RSWB and BSWB trends.


\acknowledgments
\acknowledgments

We are grateful to the referee for his/her valuable suggestions, and the $SMARTS$
group for their data.
This work has been supported by National Natural Science Foundation
 of China (Grant No. U1831124).

\underline{}

\clearpage
\appendix

\renewcommand\thefigure{\Alph{section}\arabic{figure}}
\section{The figures of the spectral index vs optical $R$ band flux for 52 blazars}
\setcounter{figure}{0}
\clearpage

\begin{figure}
\epsscale{0.7}
\plottwo{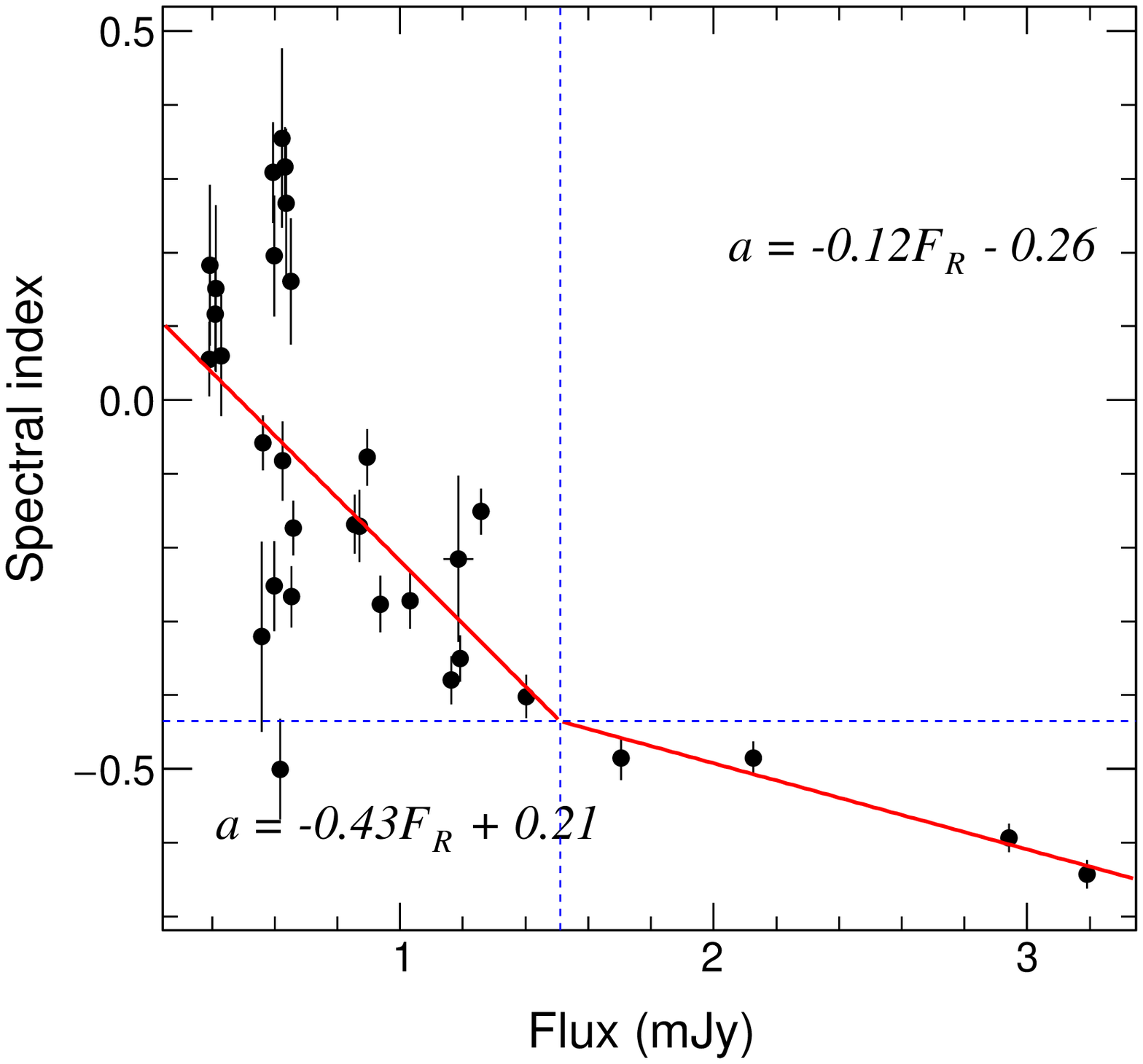}{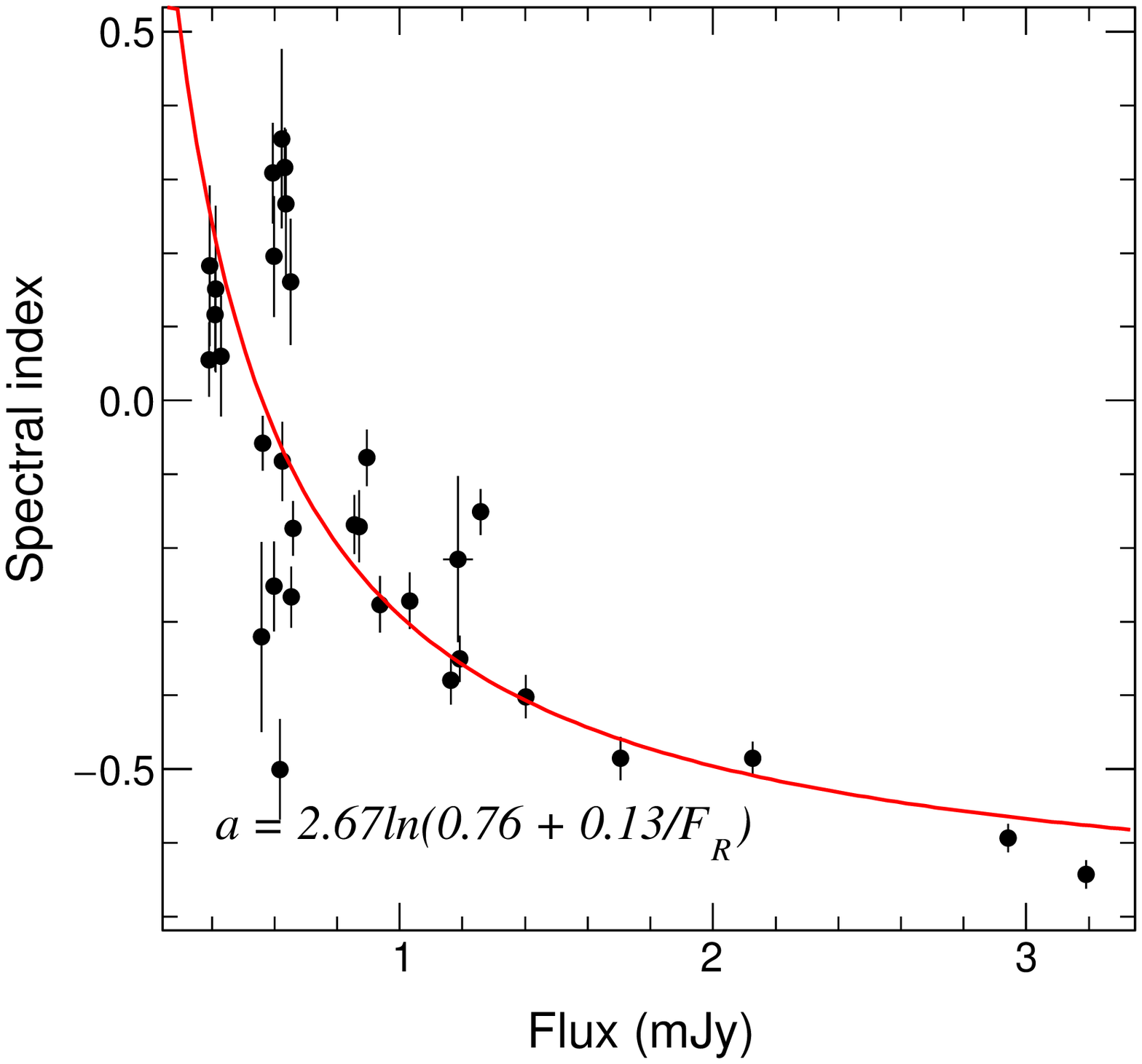}
\caption{Same as Figure~\ref{3C454alphaflux} in the main text, for PMN J0017-0512.
\label{0017-0512}}
\end{figure}

\begin{figure}
\epsscale{0.7}
\plottwo{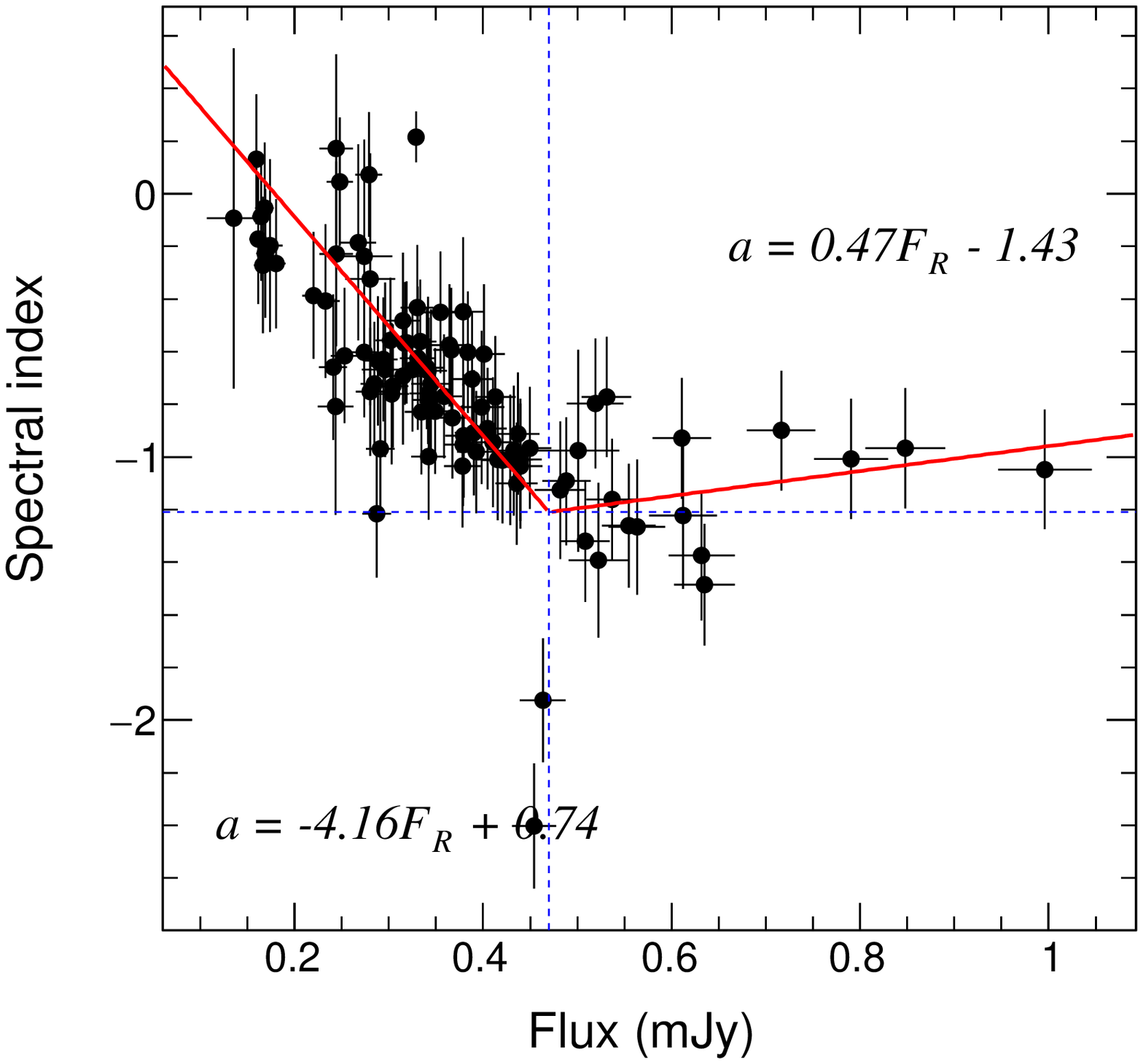}{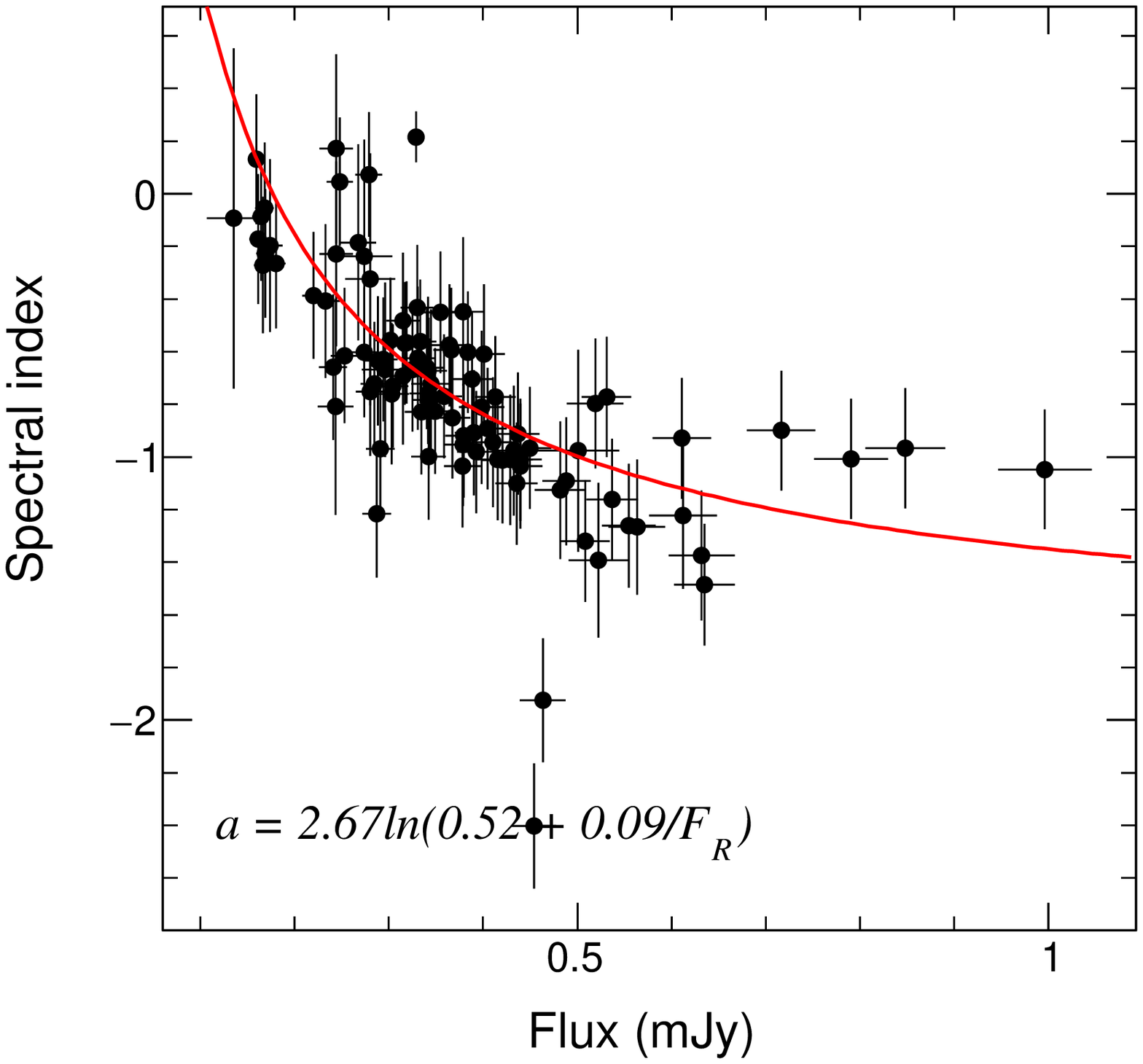}
\caption{Same as Figure~\ref{3C454alphaflux} in the main text, for 4C +01.02. \label{0106+0119}}
\end{figure}

\begin{figure}
\epsscale{0.7}
\plottwo{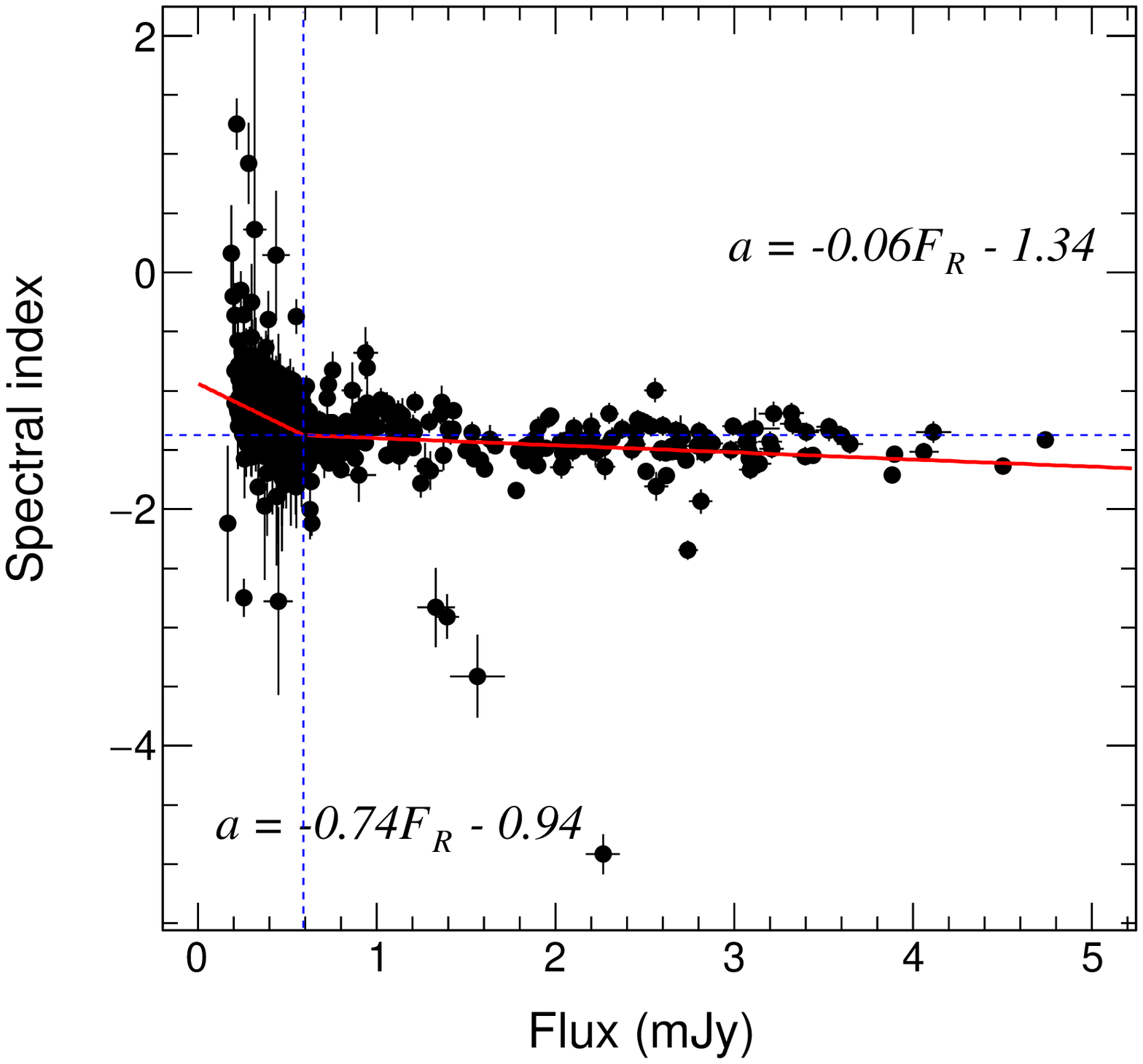}{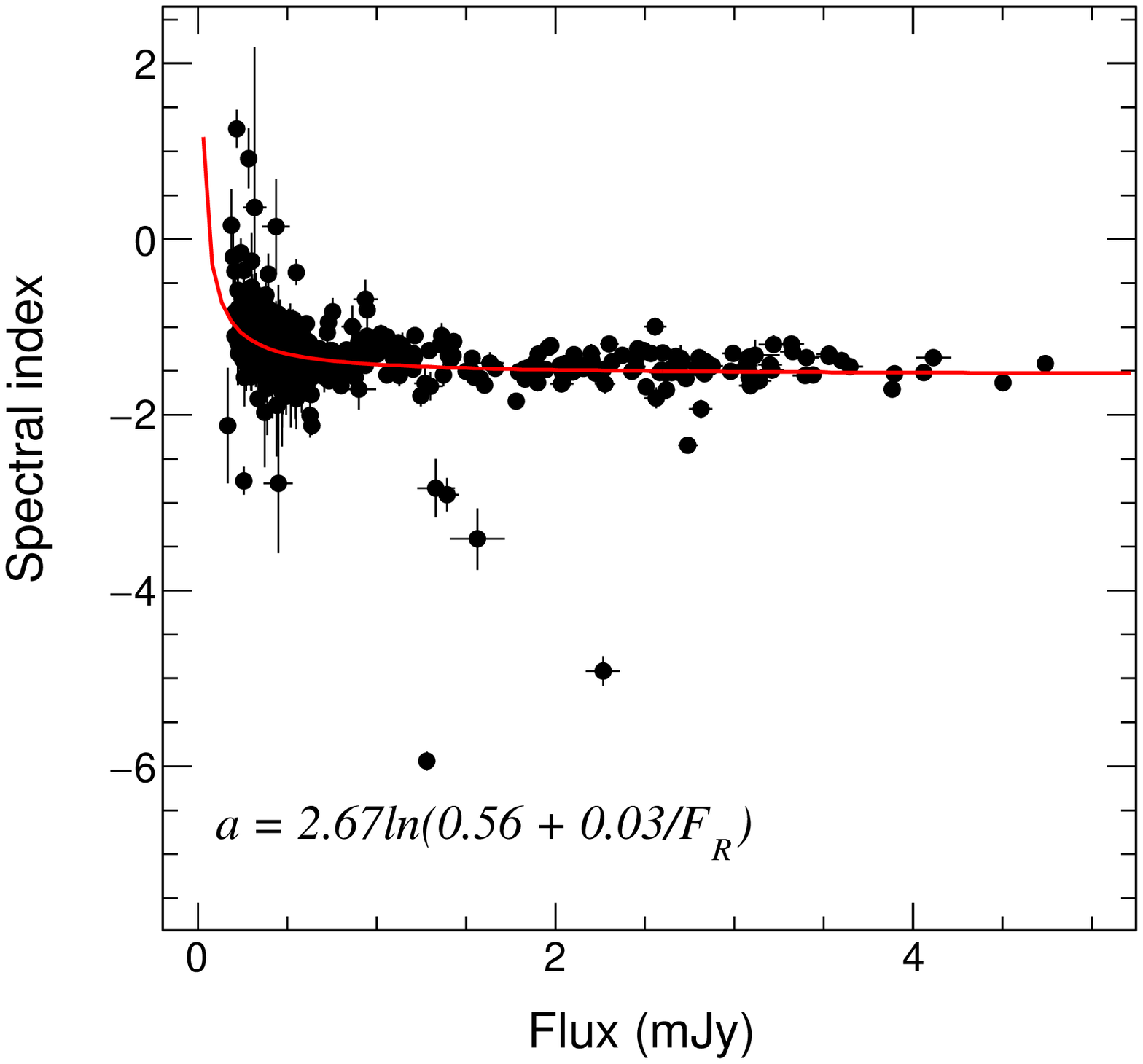}
\caption{Same as Figure~\ref{3C454alphaflux} in the main text, for PKS 0208-512. \label{0208-512}}
\end{figure}

\clearpage

\begin{figure}
\epsscale{0.7}
\plottwo{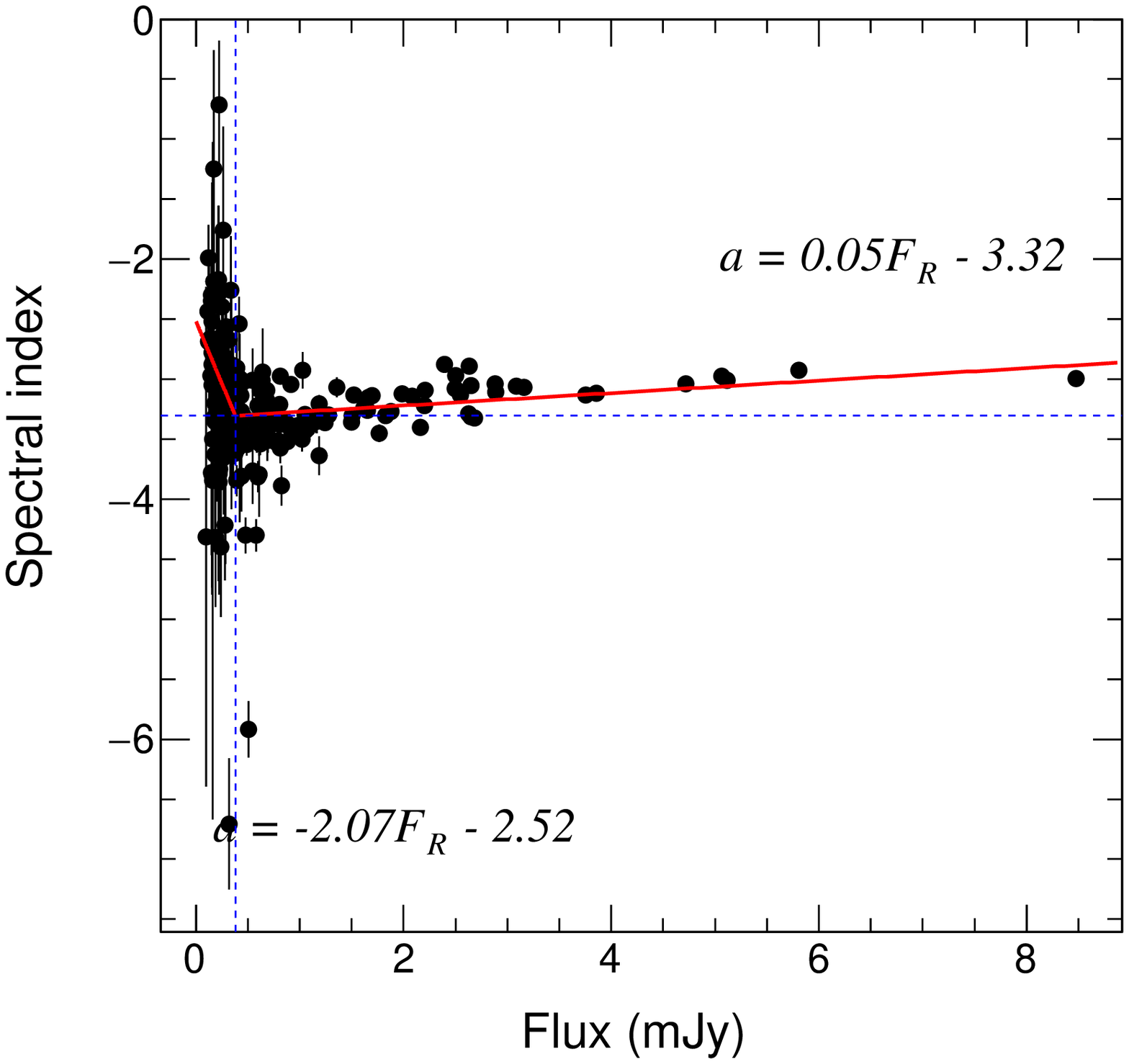}{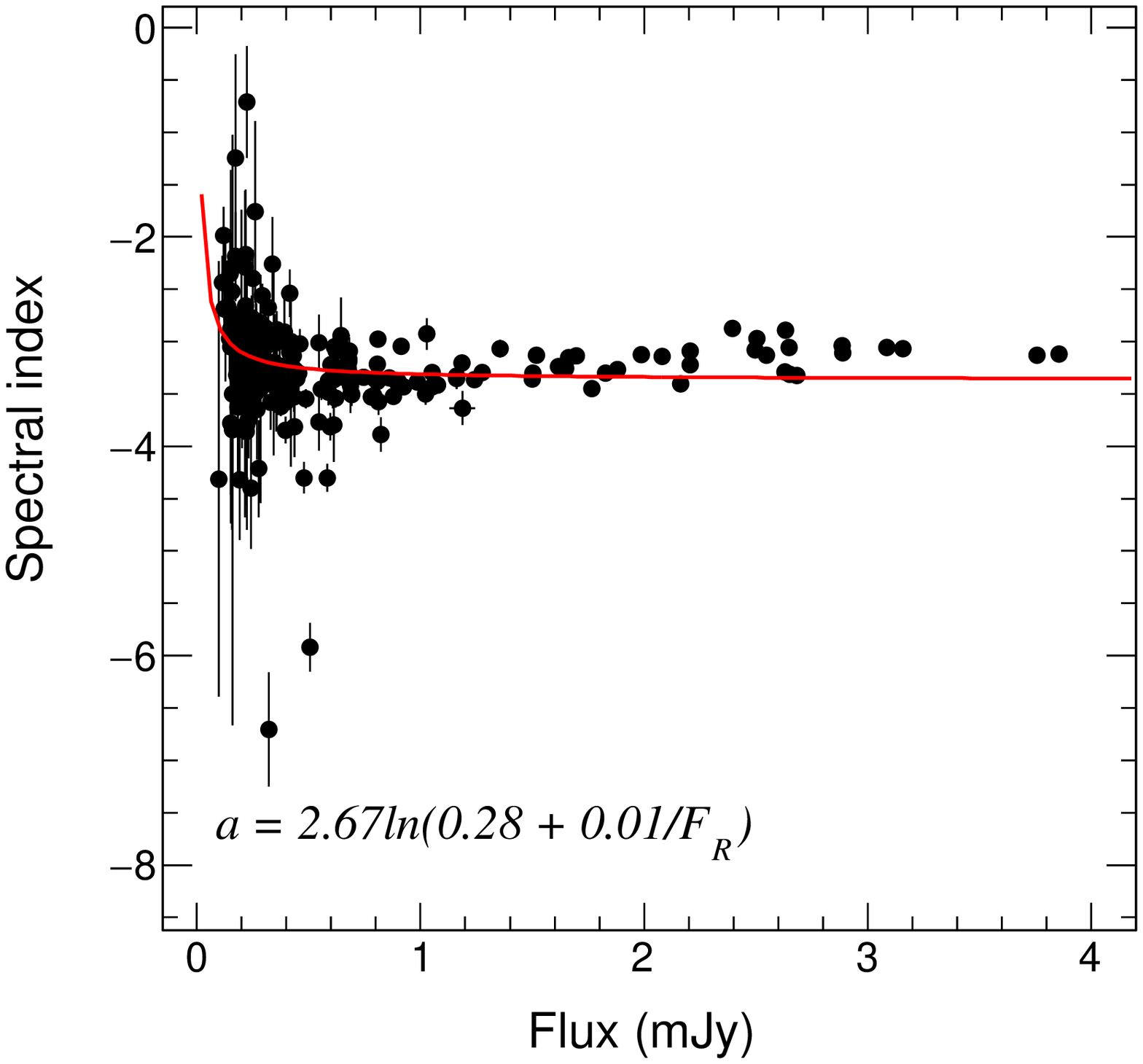}
\caption{Same as Figure~\ref{3C454alphaflux} in the main text, for PKS 0235+164. \label{0235+164}}
\end{figure}

\begin{figure}
\epsscale{0.7}
\plottwo{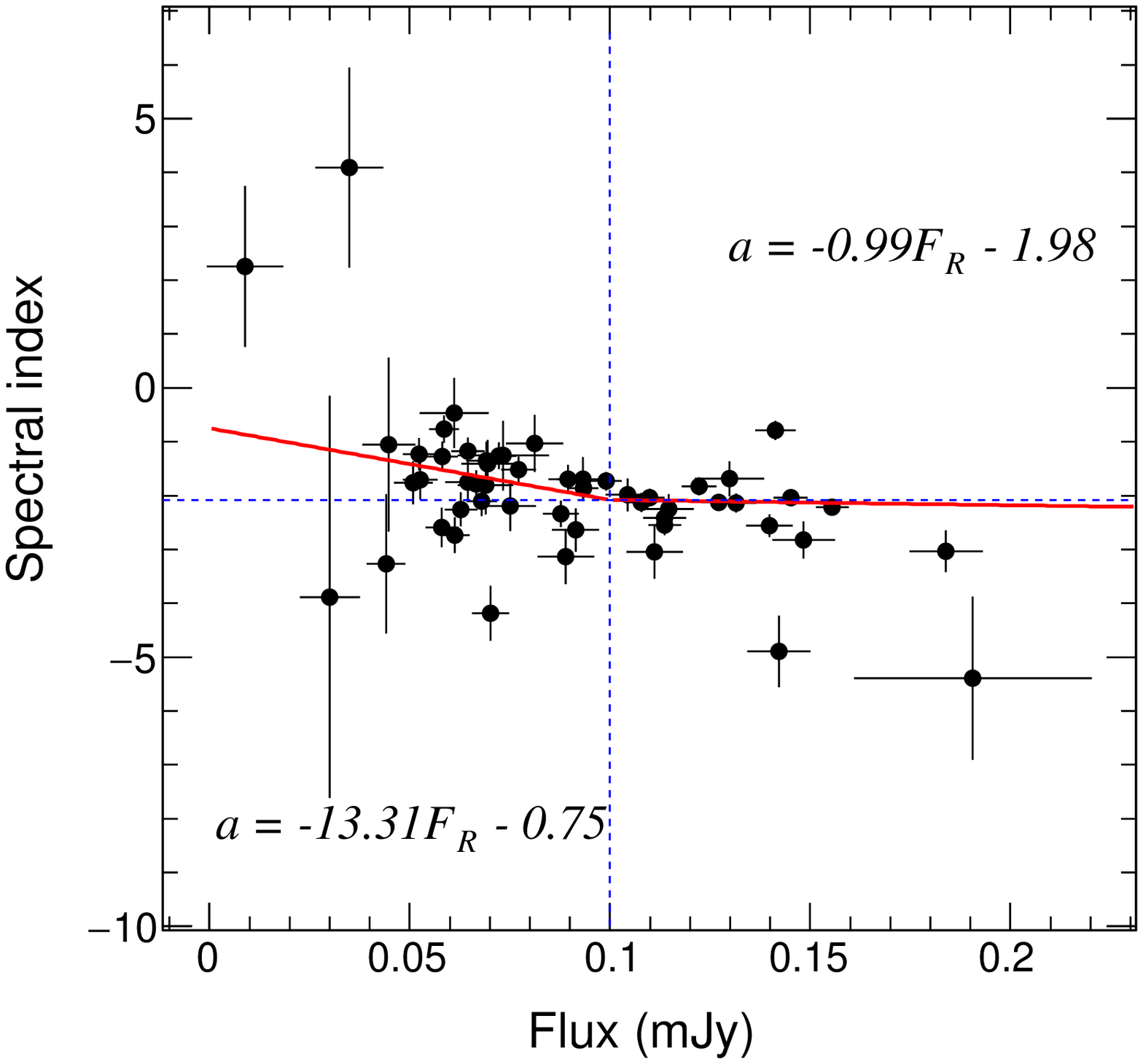}{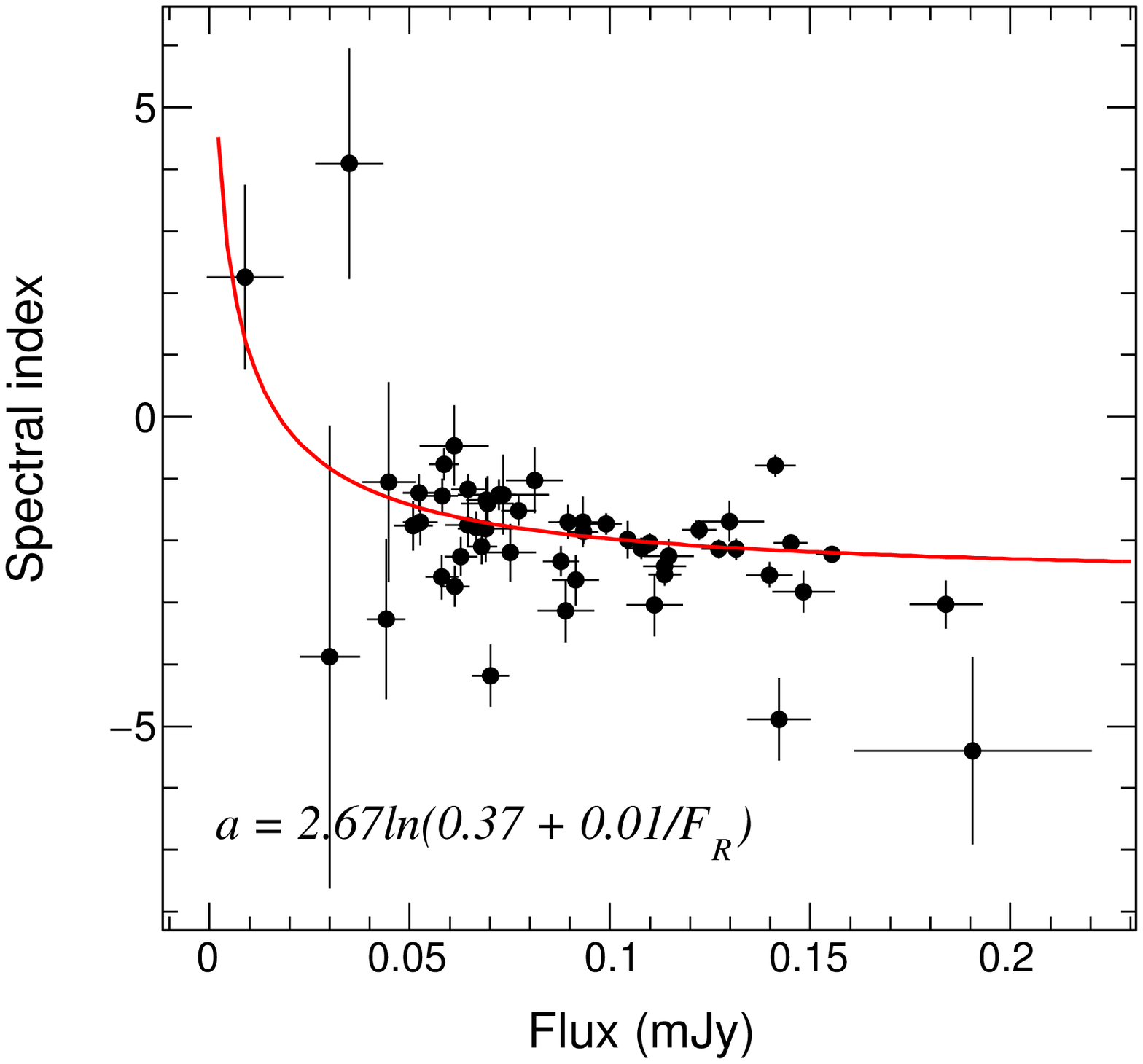}
\caption{Same as Figure~\ref{3C454alphaflux} in the main text, for PKS 0250-225. \label{0250-225}}
\end{figure}

\begin{figure}
\epsscale{0.7}
\plottwo{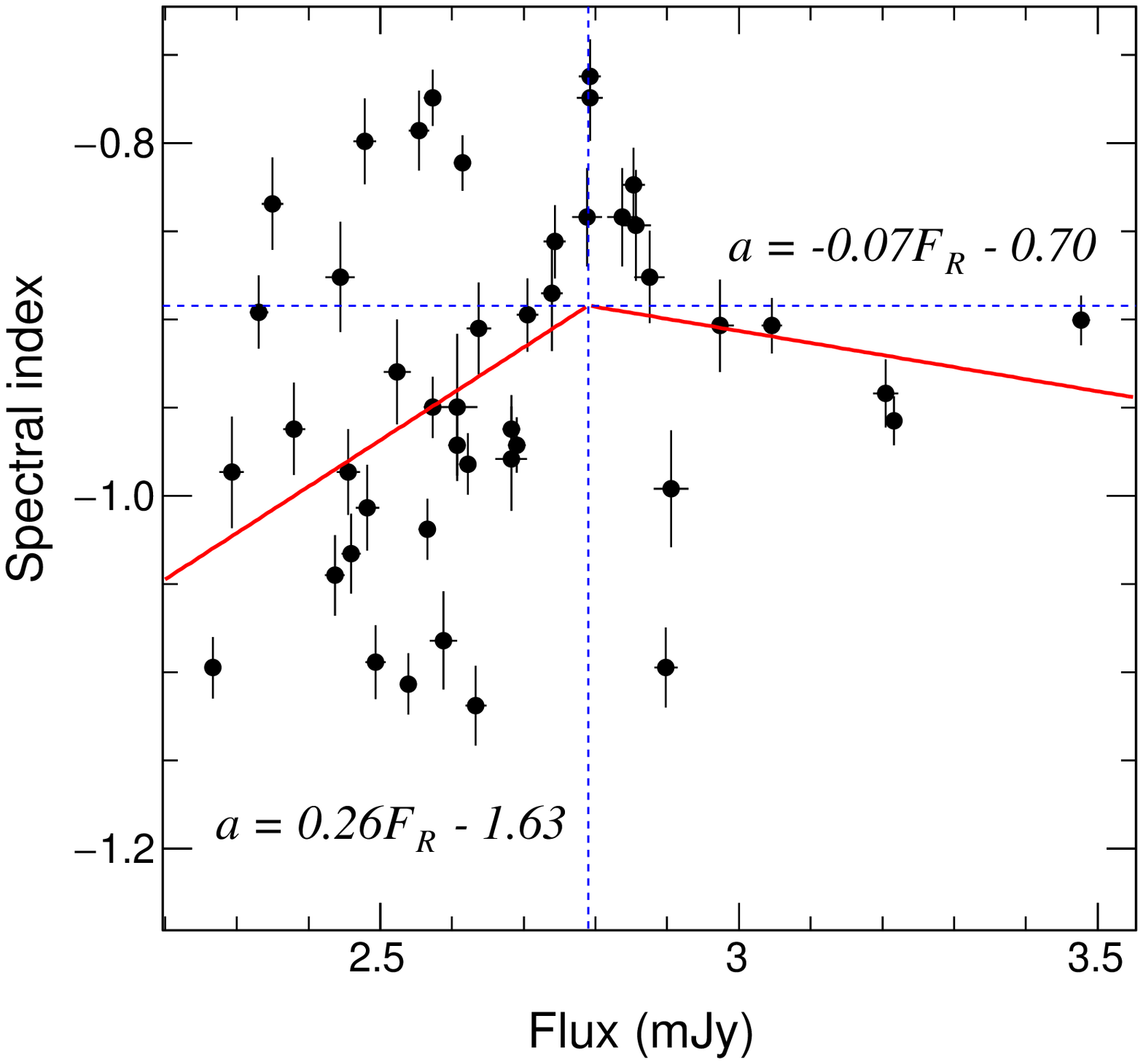}{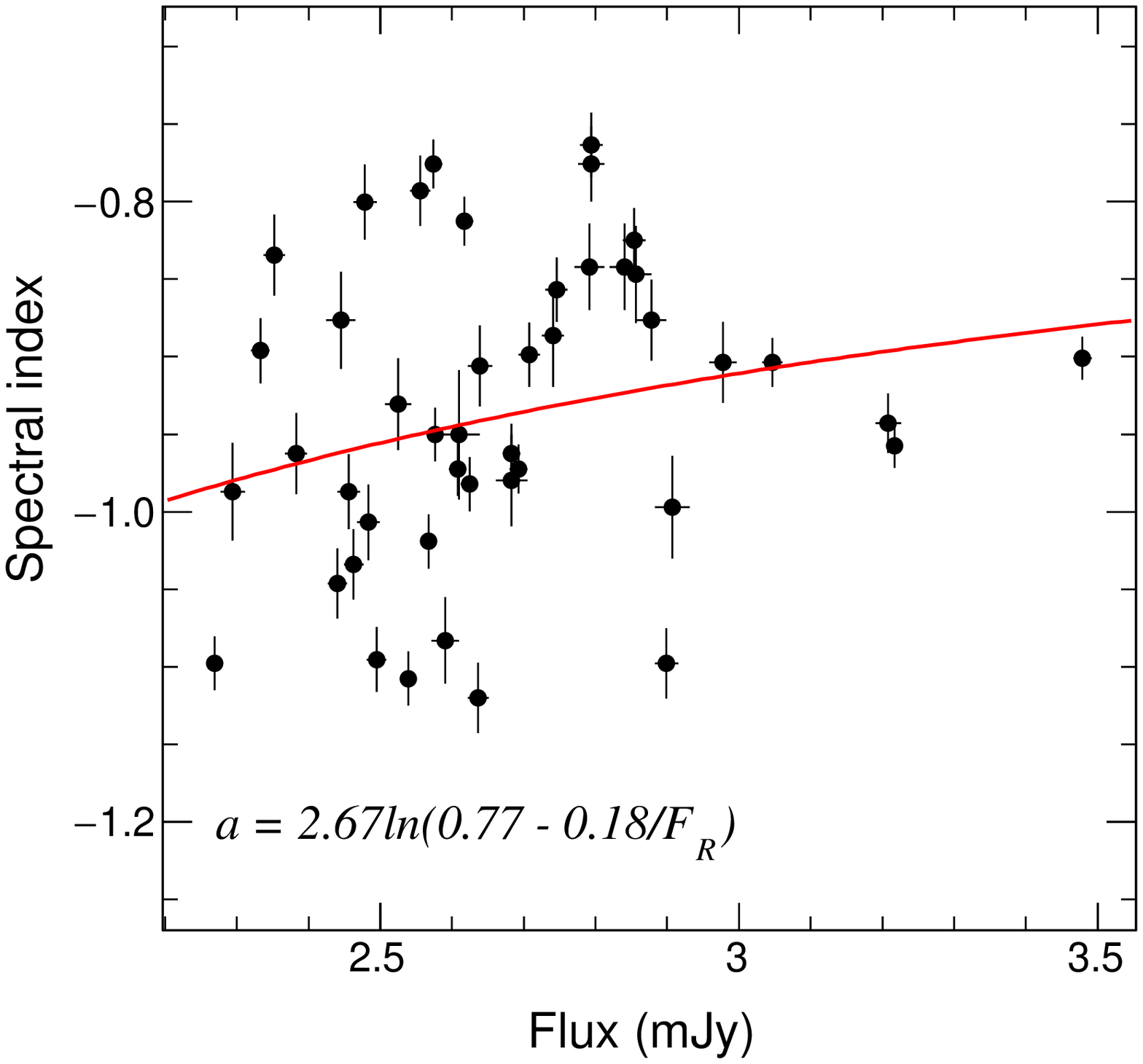}
\caption{Same as Figure~\ref{3C454alphaflux} in the main text, for PKS 0301-243. \label{0301-243}}
\end{figure}

\begin{figure}
\epsscale{0.7}
\plottwo{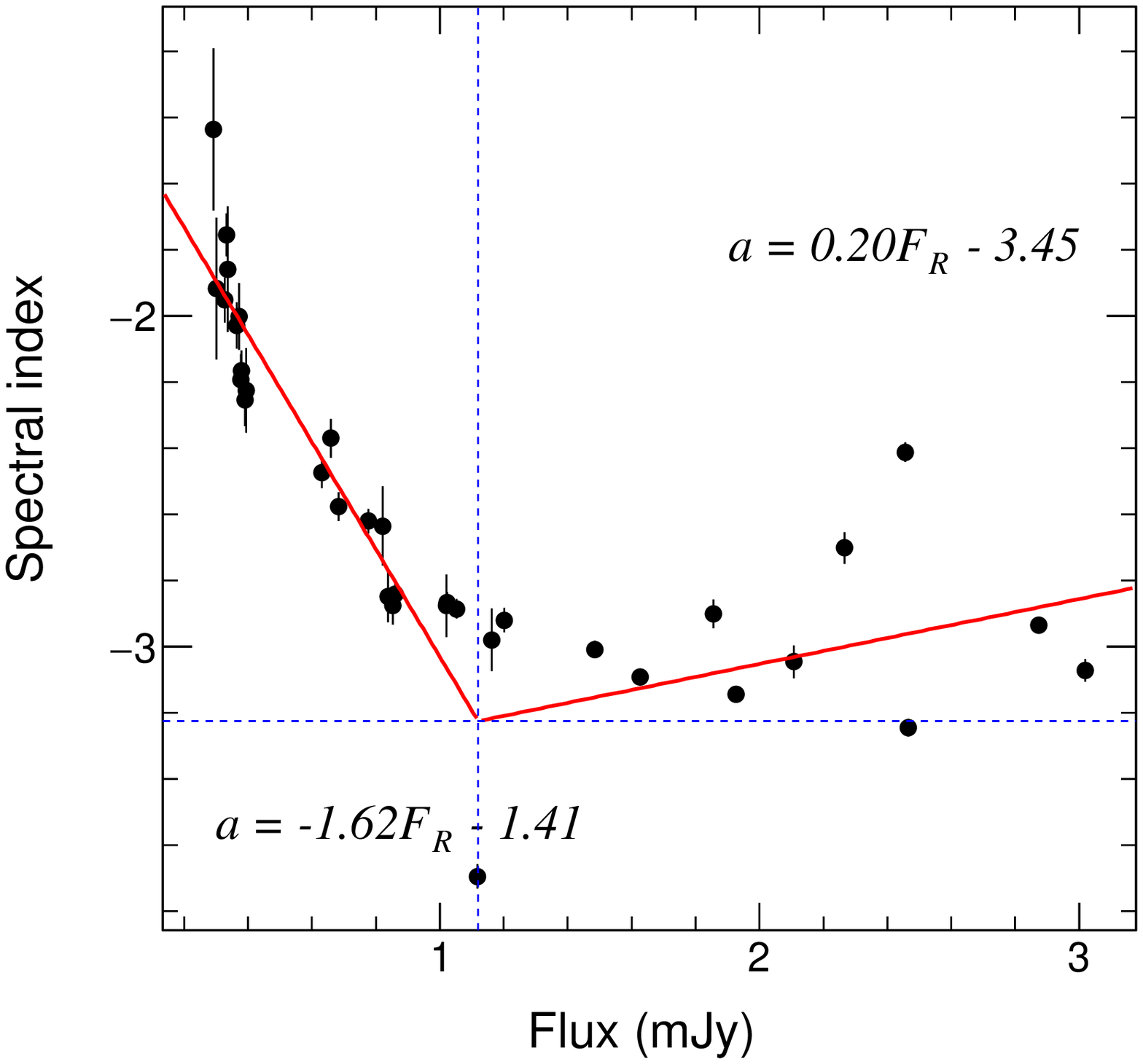}{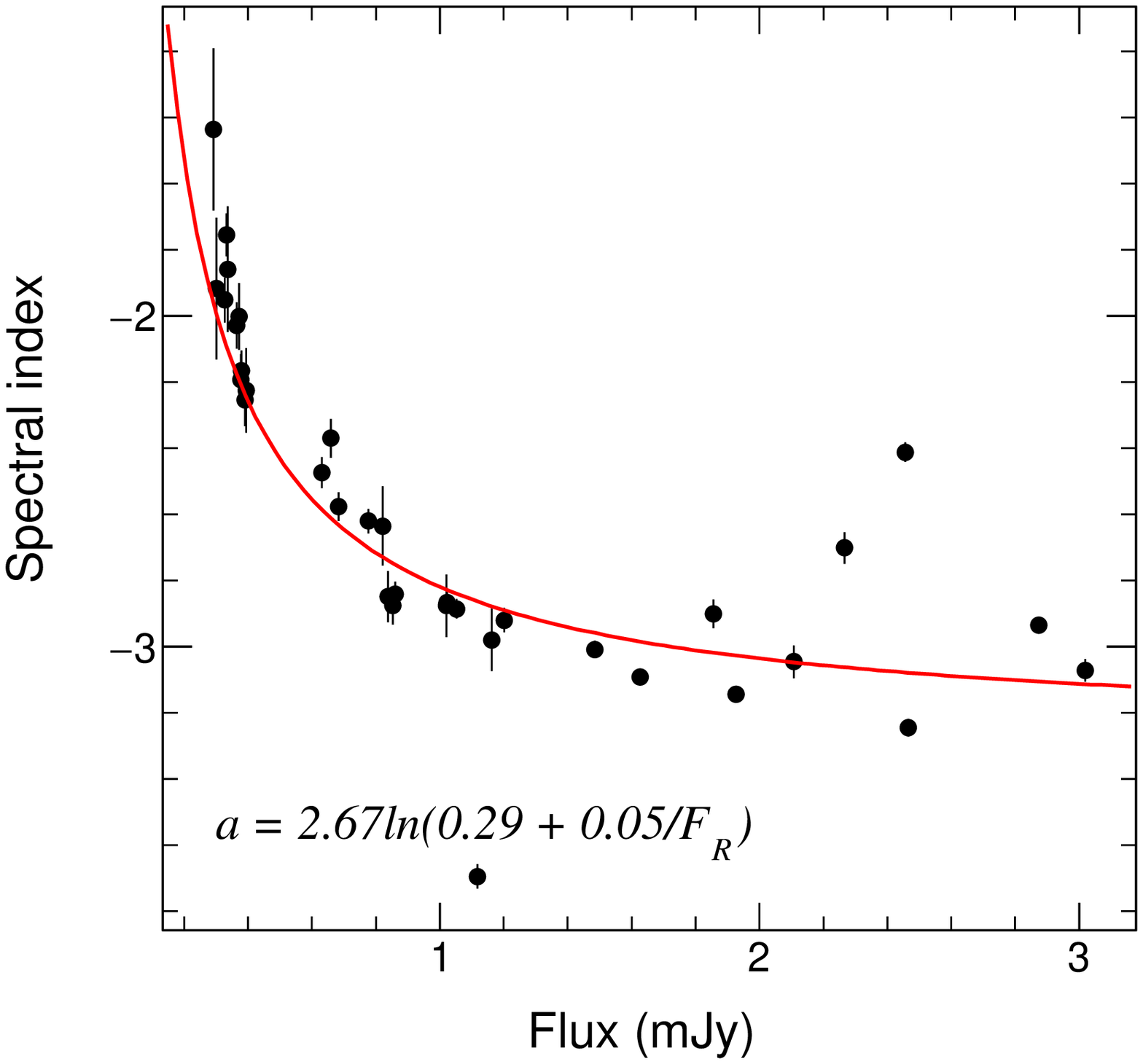}
\caption{Same as Figure~\ref{3C454alphaflux} in the main text, for PKS 0336-01. \label{0336-01}}
\end{figure}

\begin{figure}
\epsscale{0.7}
\plottwo{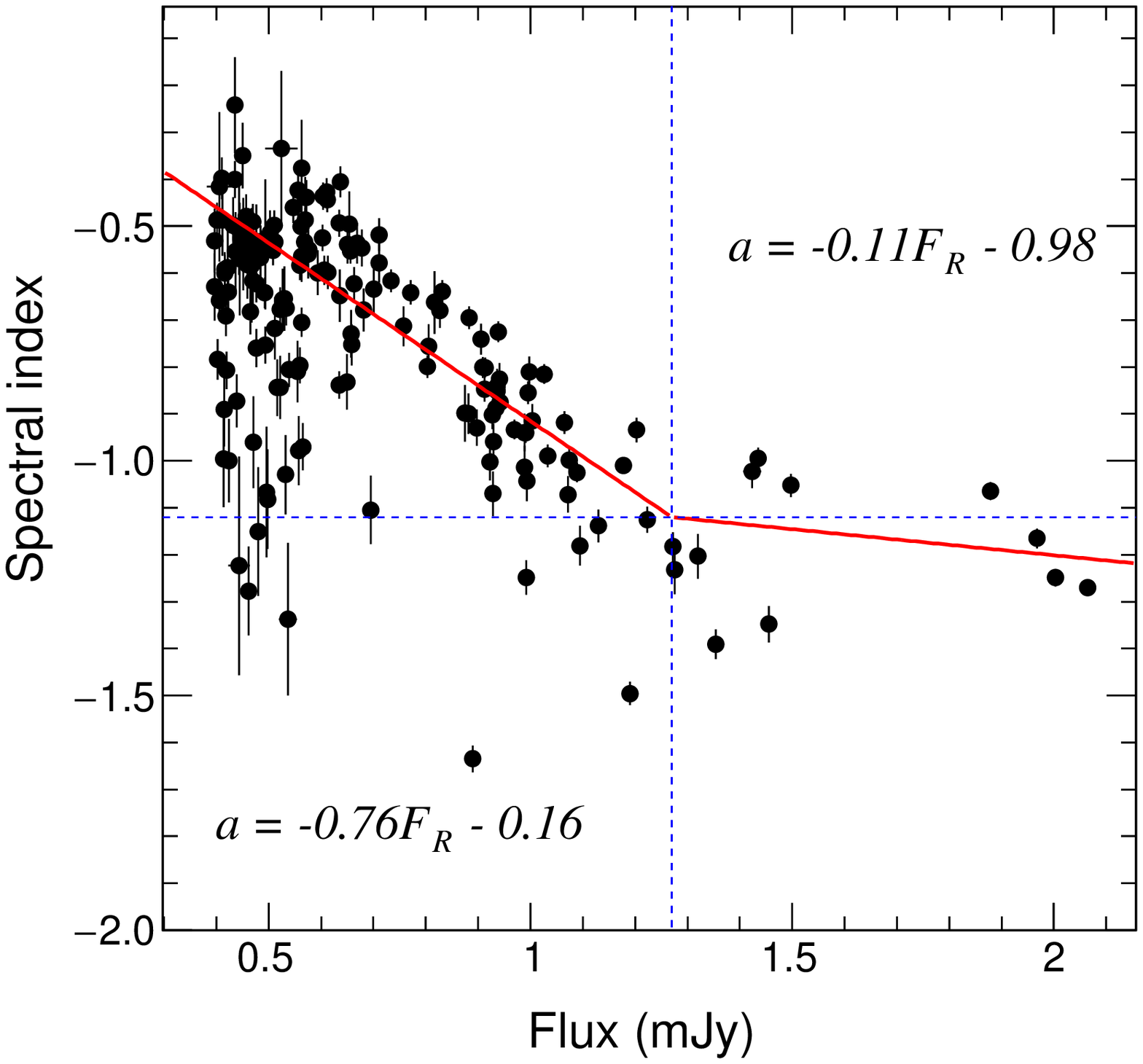}{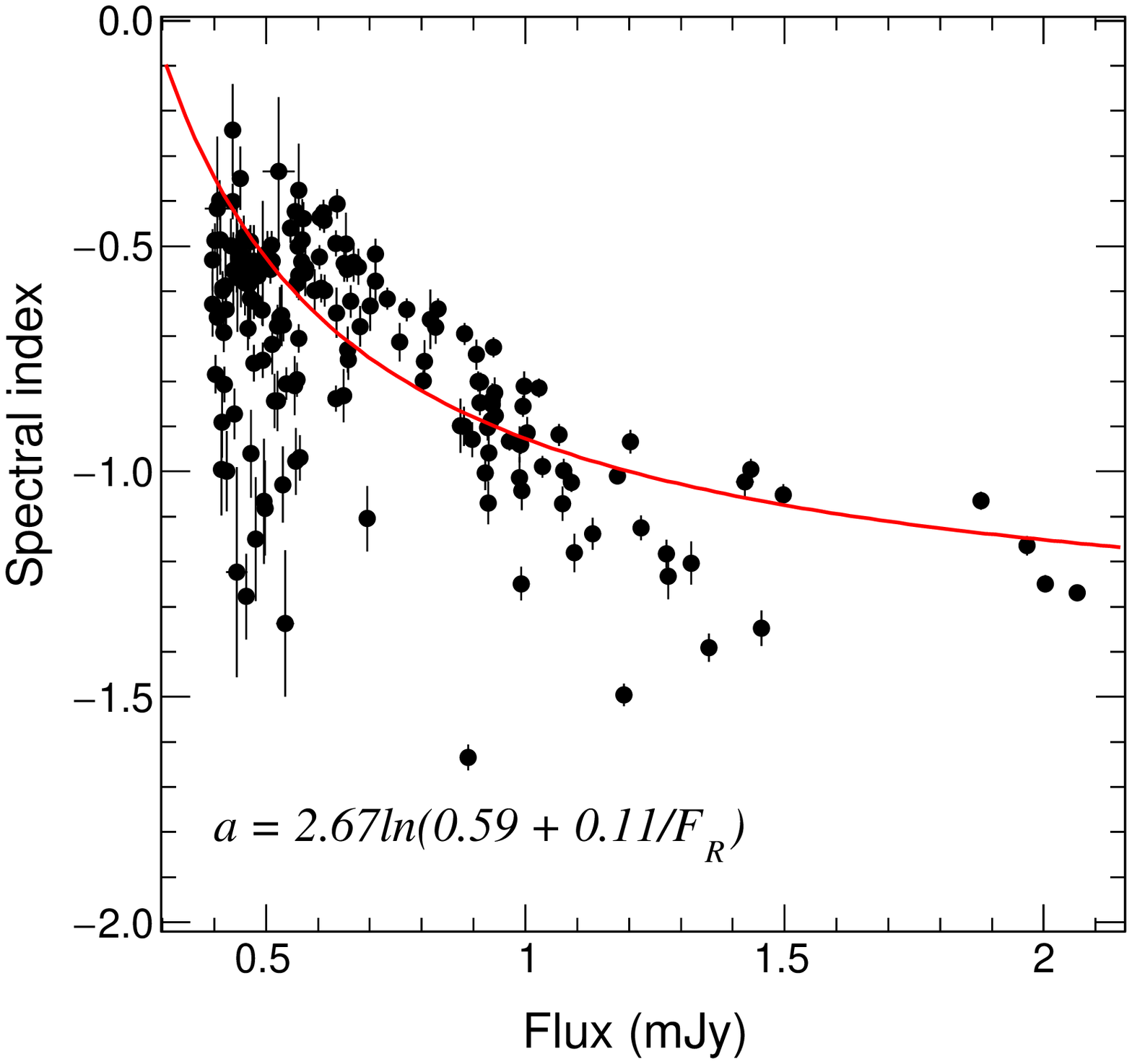}
\caption{Same as Figure~\ref{3C454alphaflux} in the main text, for PKS 0402-362. \label{0402-362}}
\end{figure}

\begin{figure}
\epsscale{0.7}
\plottwo{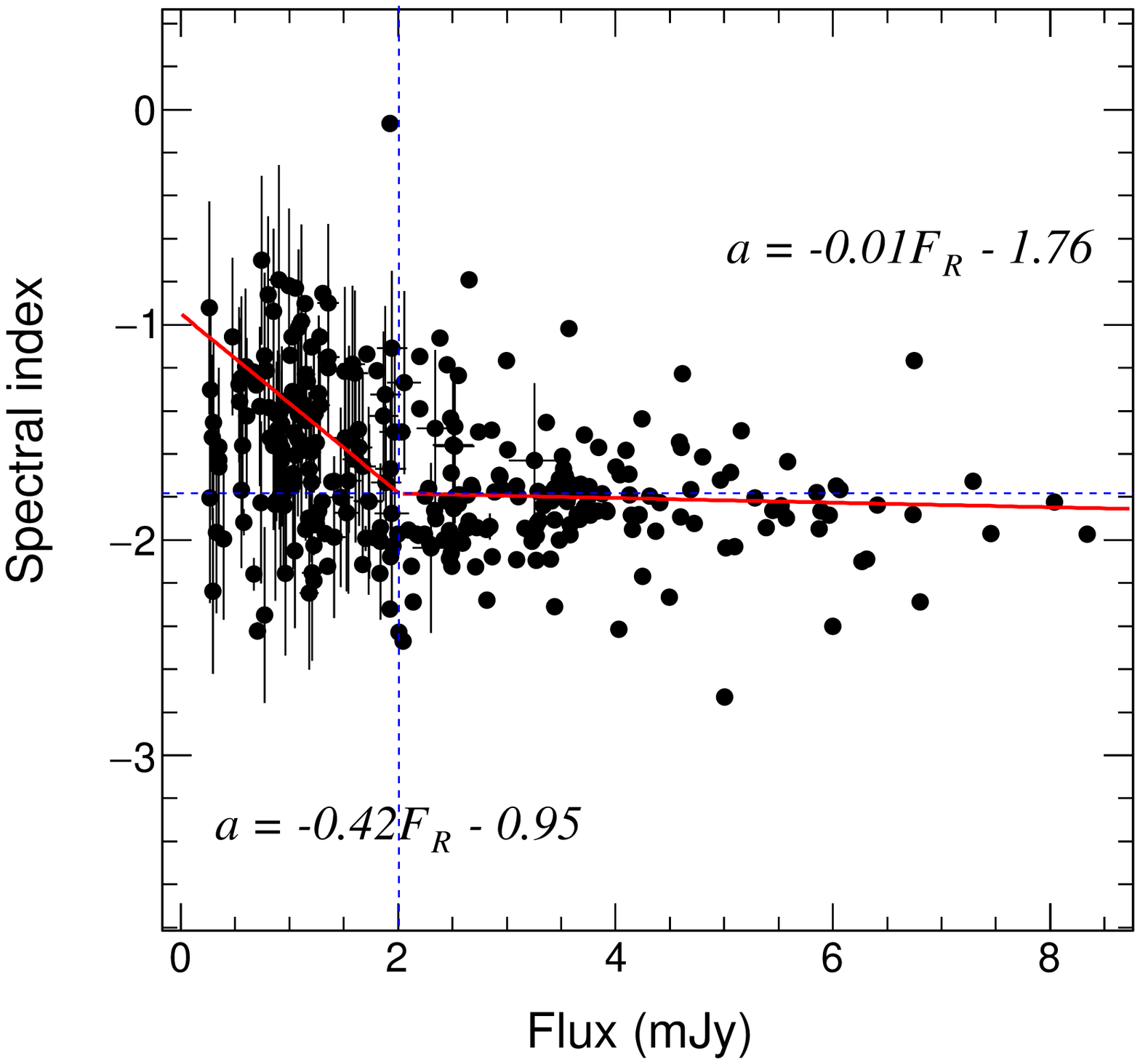}{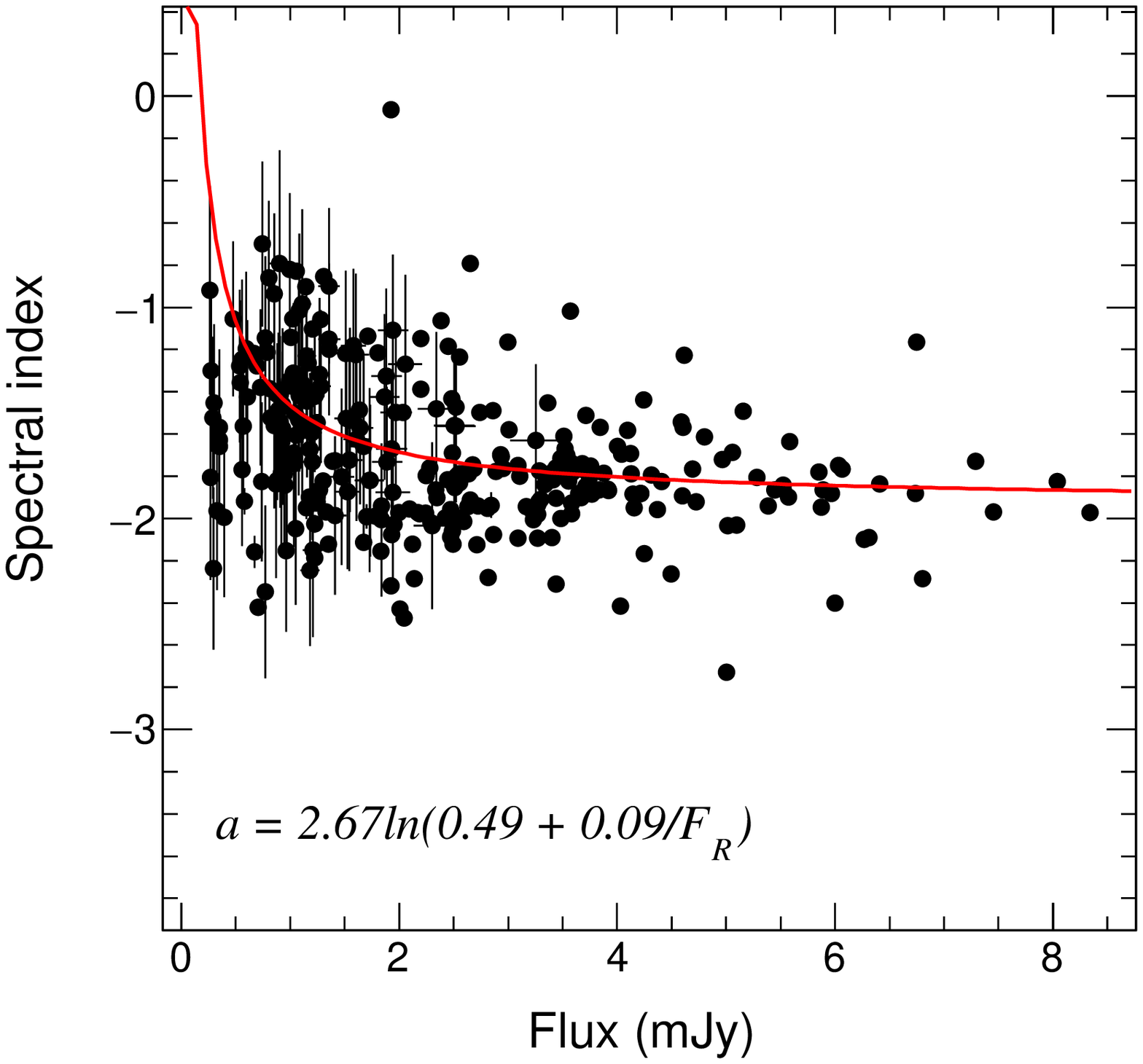}
\caption{Same as Figure~\ref{3C454alphaflux} in the main text, for PKS 0426-380. \label{0426-380}}
\end{figure}

\begin{figure}
\epsscale{0.7}
\plottwo{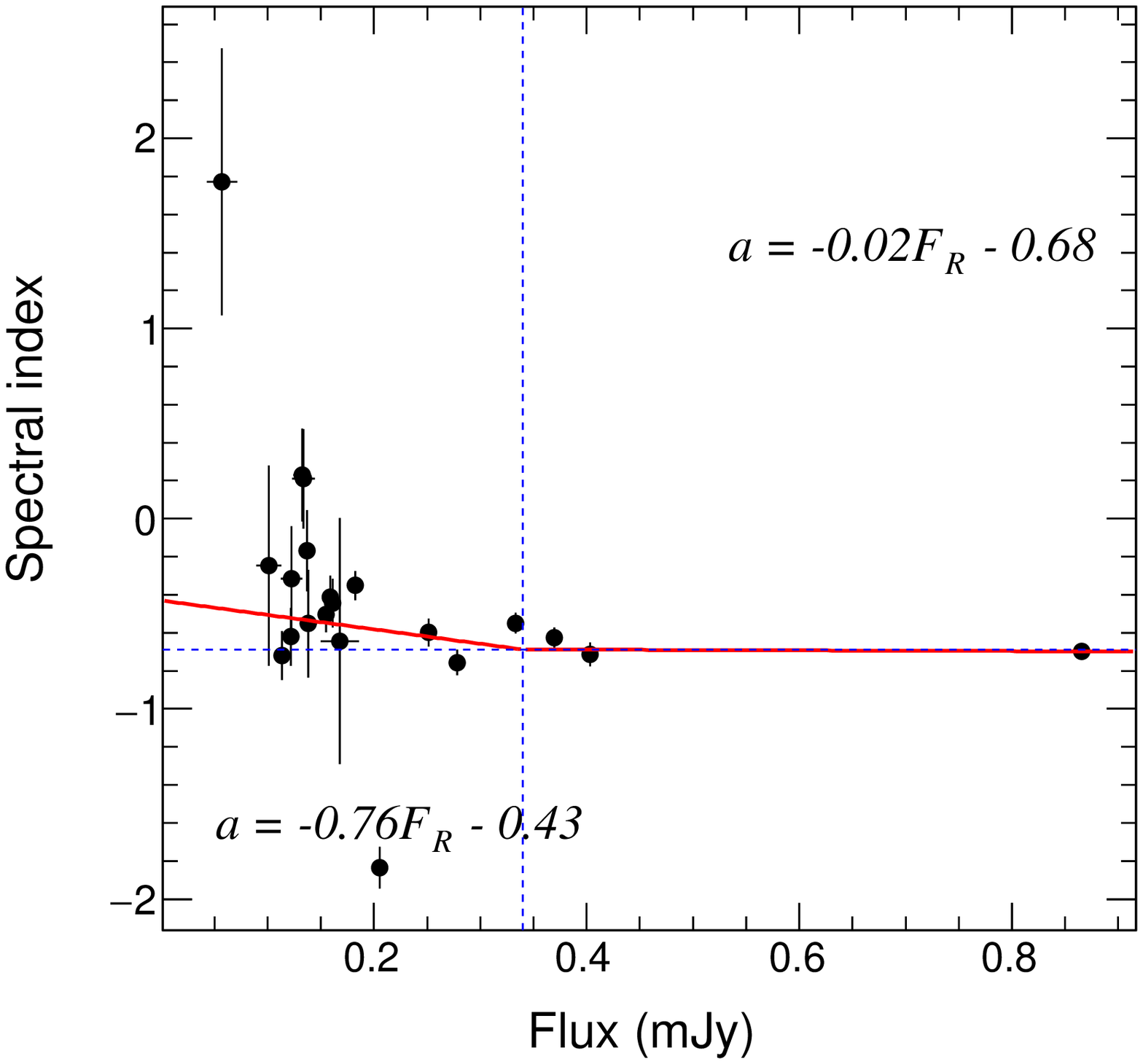}{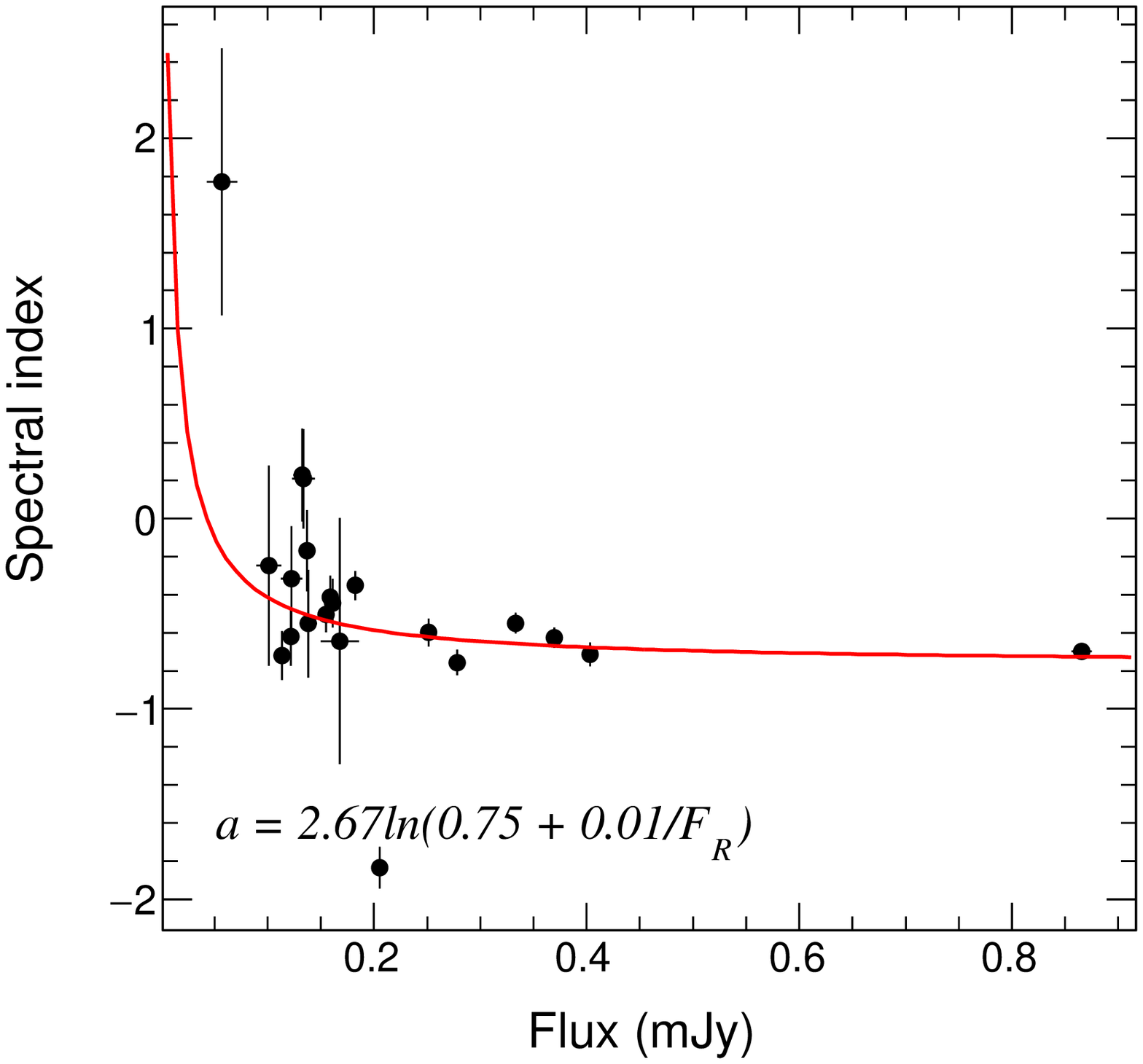}
\caption{Same as Figure~\ref{3C454alphaflux} in the main text, for PKS 0440-00. \label{0440-00}}
\end{figure}

\begin{figure}
\epsscale{0.7}
\plottwo{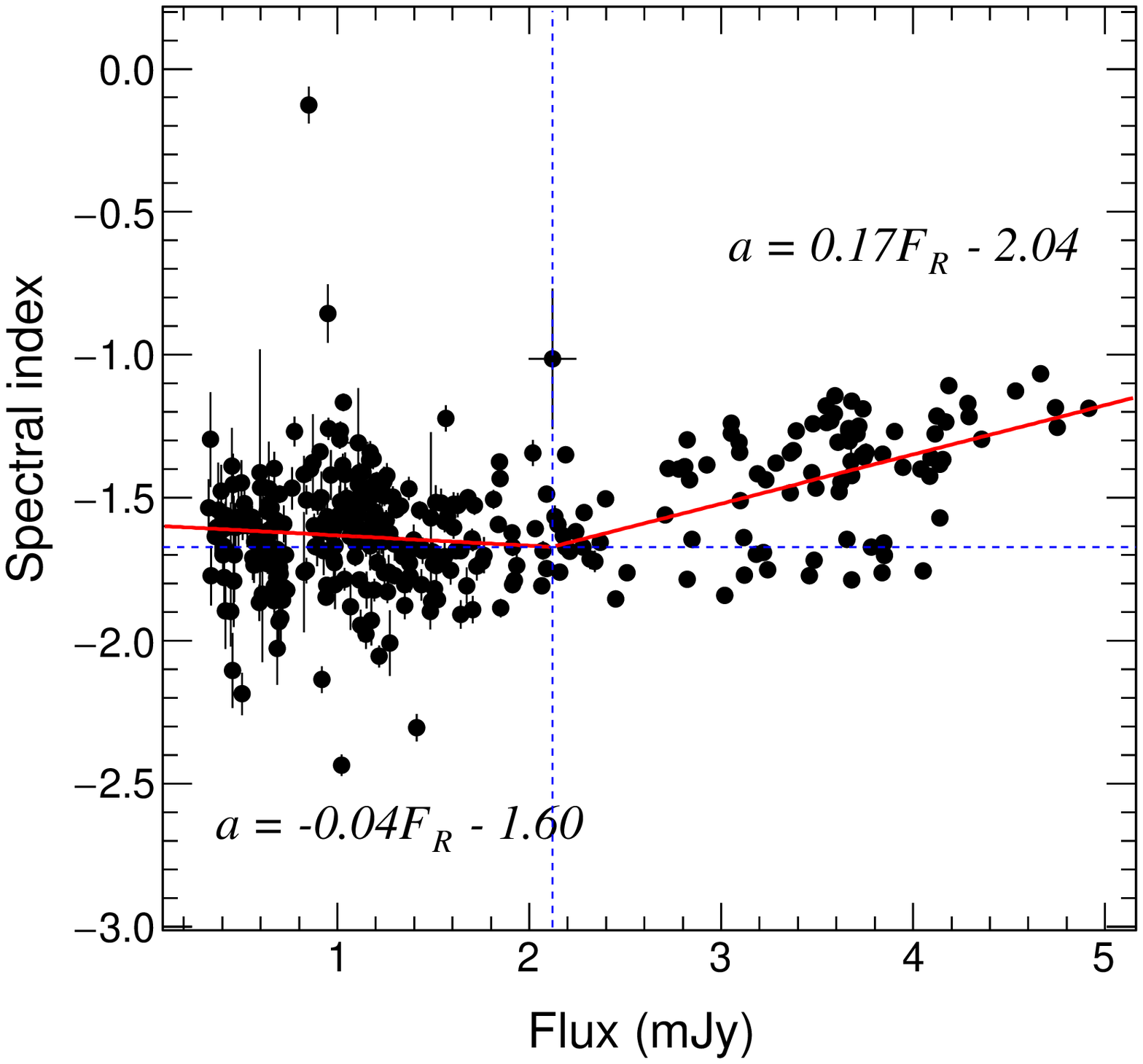}{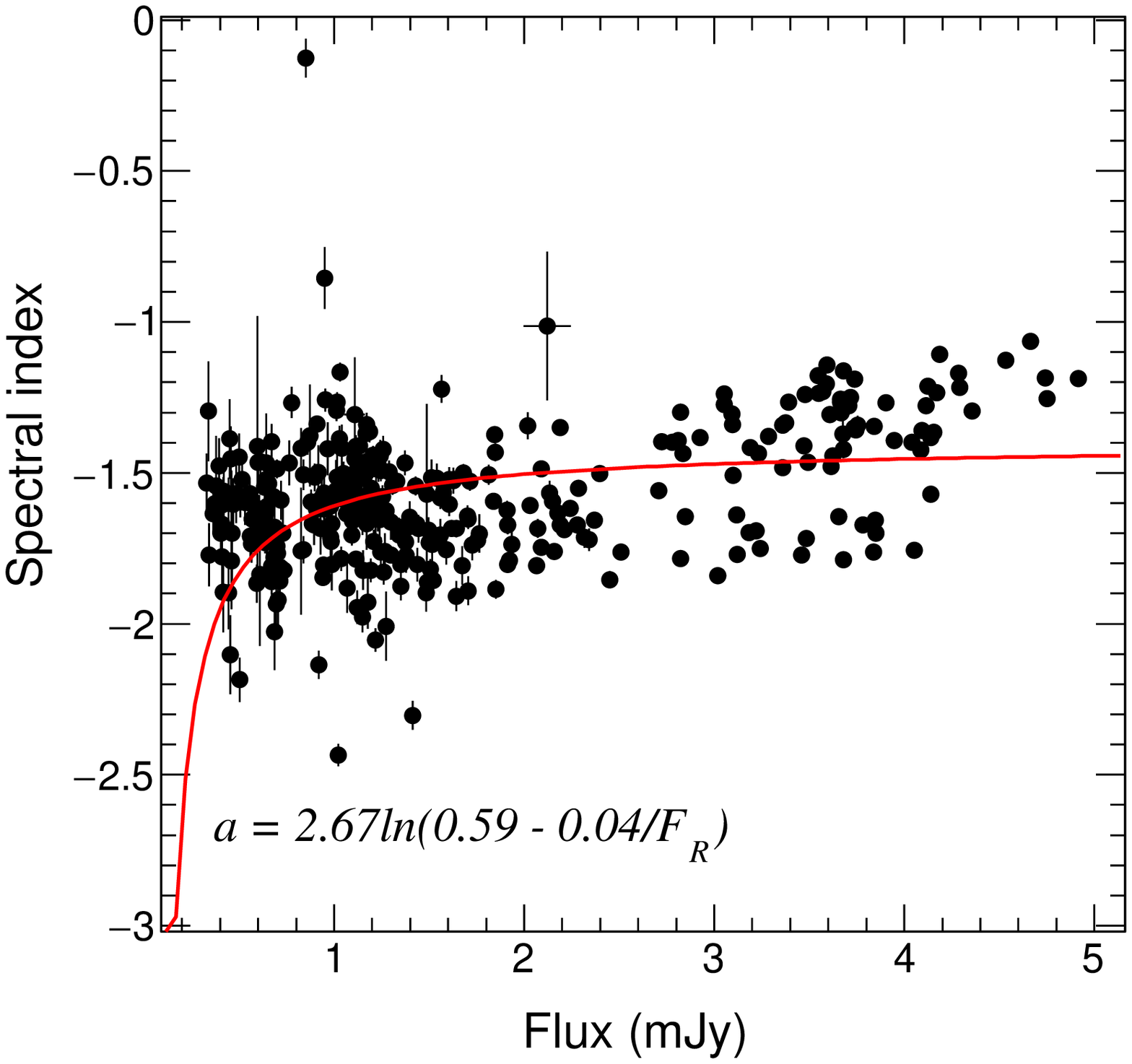}
\caption{Same as Figure~\ref{3C454alphaflux} in the main text, for PKS 0454-234. \label{0454-234}}
\end{figure}

\begin{figure}
\epsscale{0.7}
\plottwo{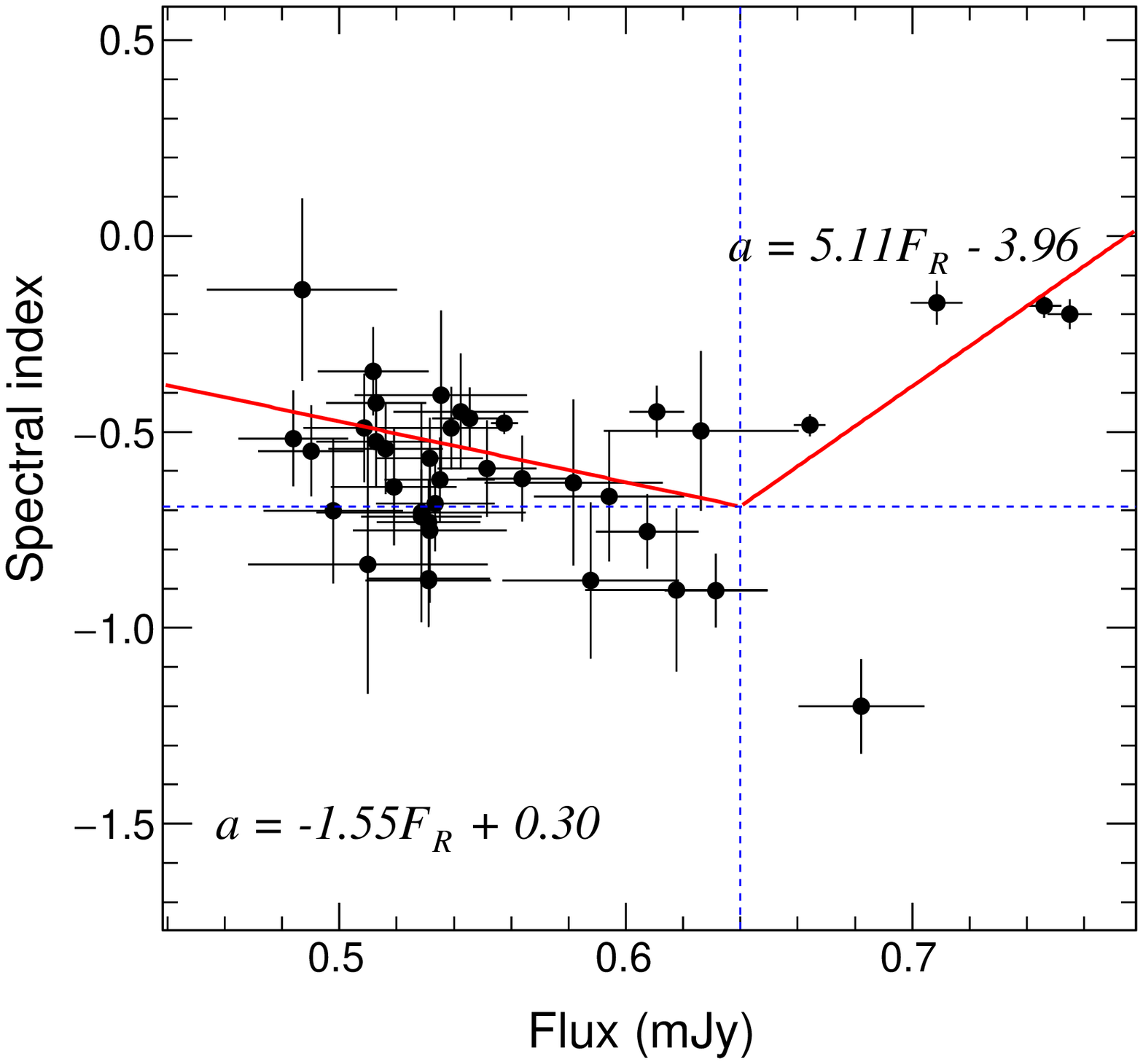}{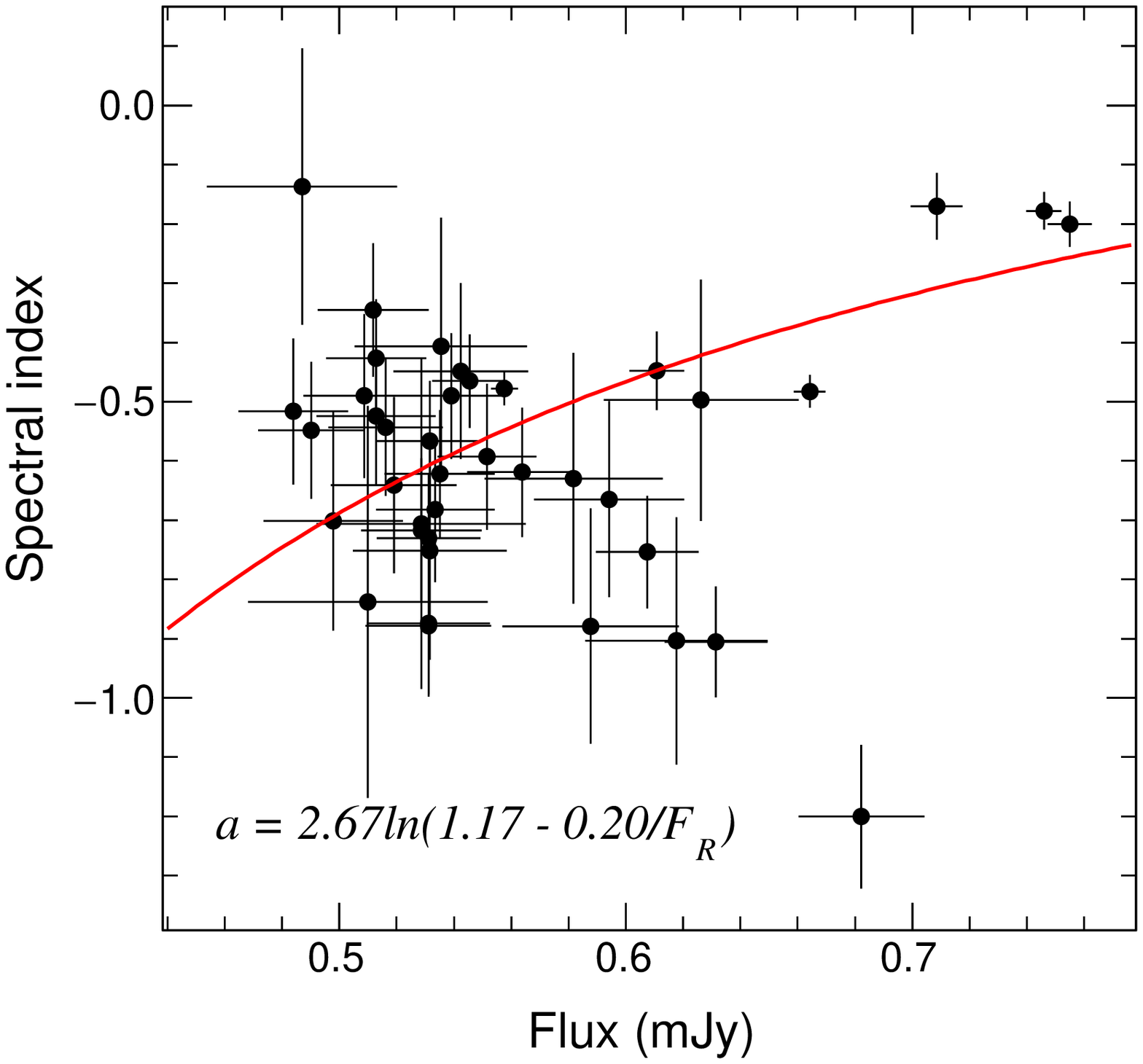}
\caption{Same as Figure~\ref{3C454alphaflux} in the main text, for PKS 0454-46. \label{0454-46}}
\end{figure}

\begin{figure}
\epsscale{0.7}
\plottwo{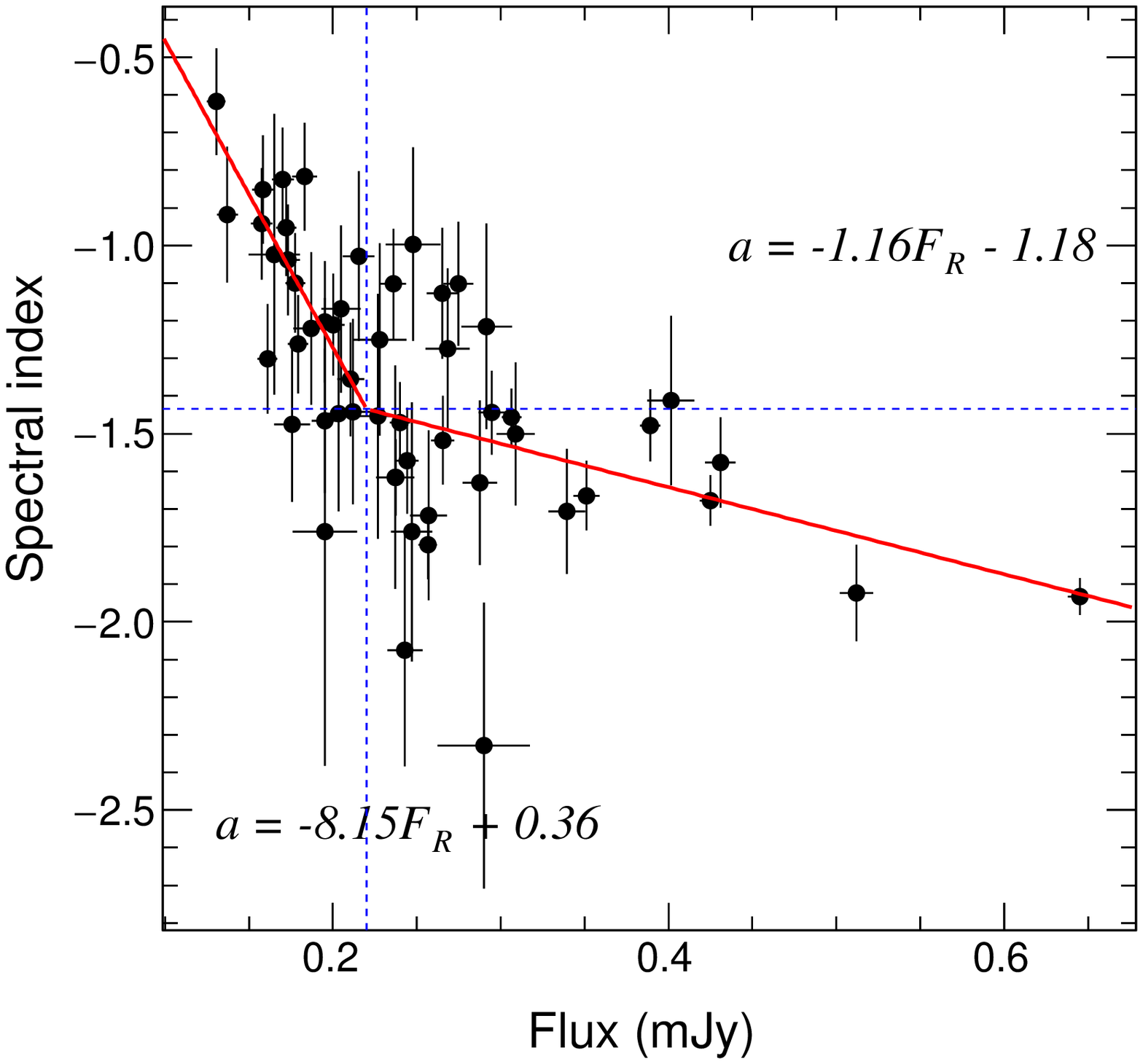}{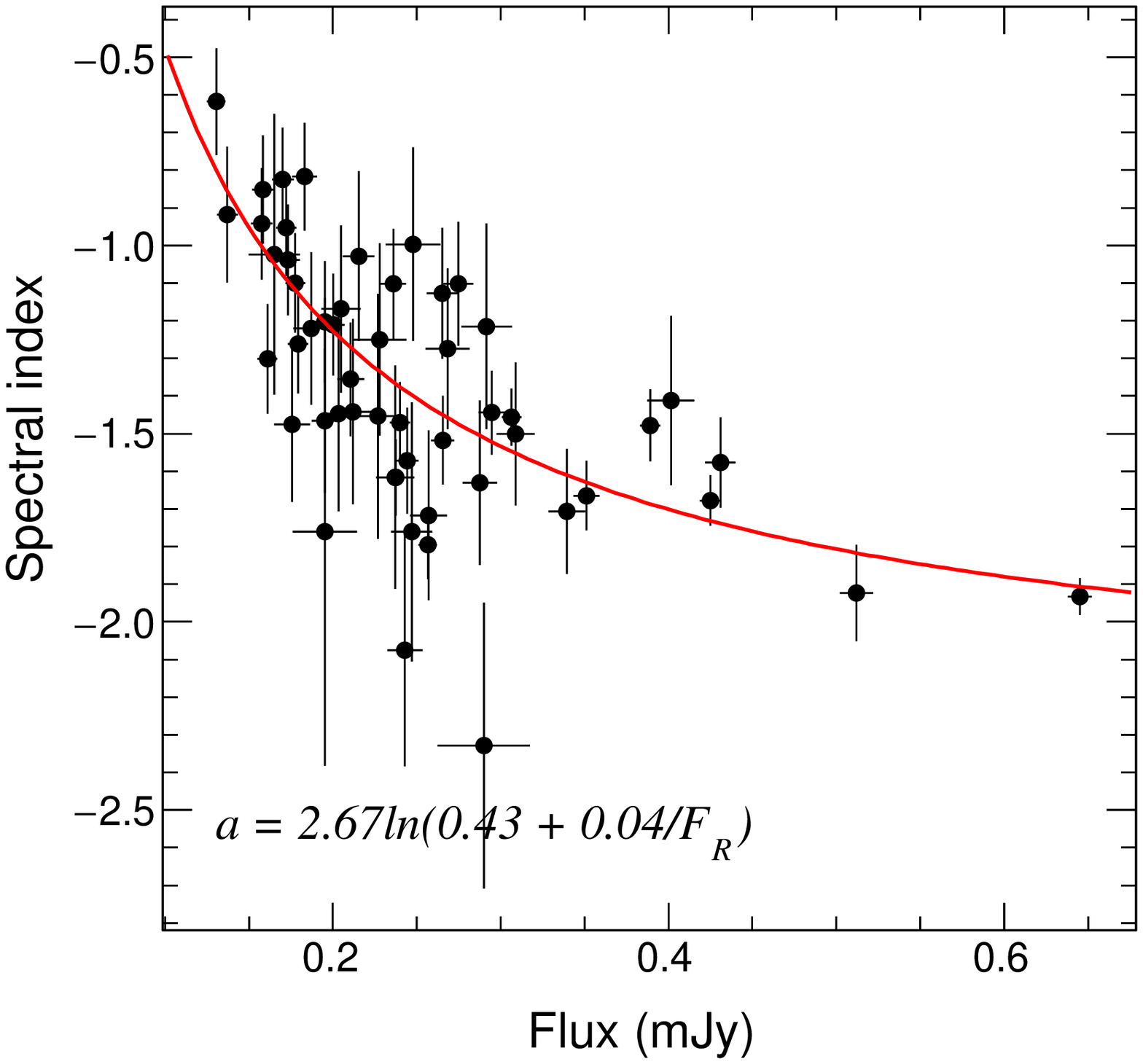}
\caption{Same as Figure~\ref{3C454alphaflux} in the main text, for S3 0458-02. \label{0458-02}}
\end{figure}

\begin{figure}
\epsscale{0.7}
\plottwo{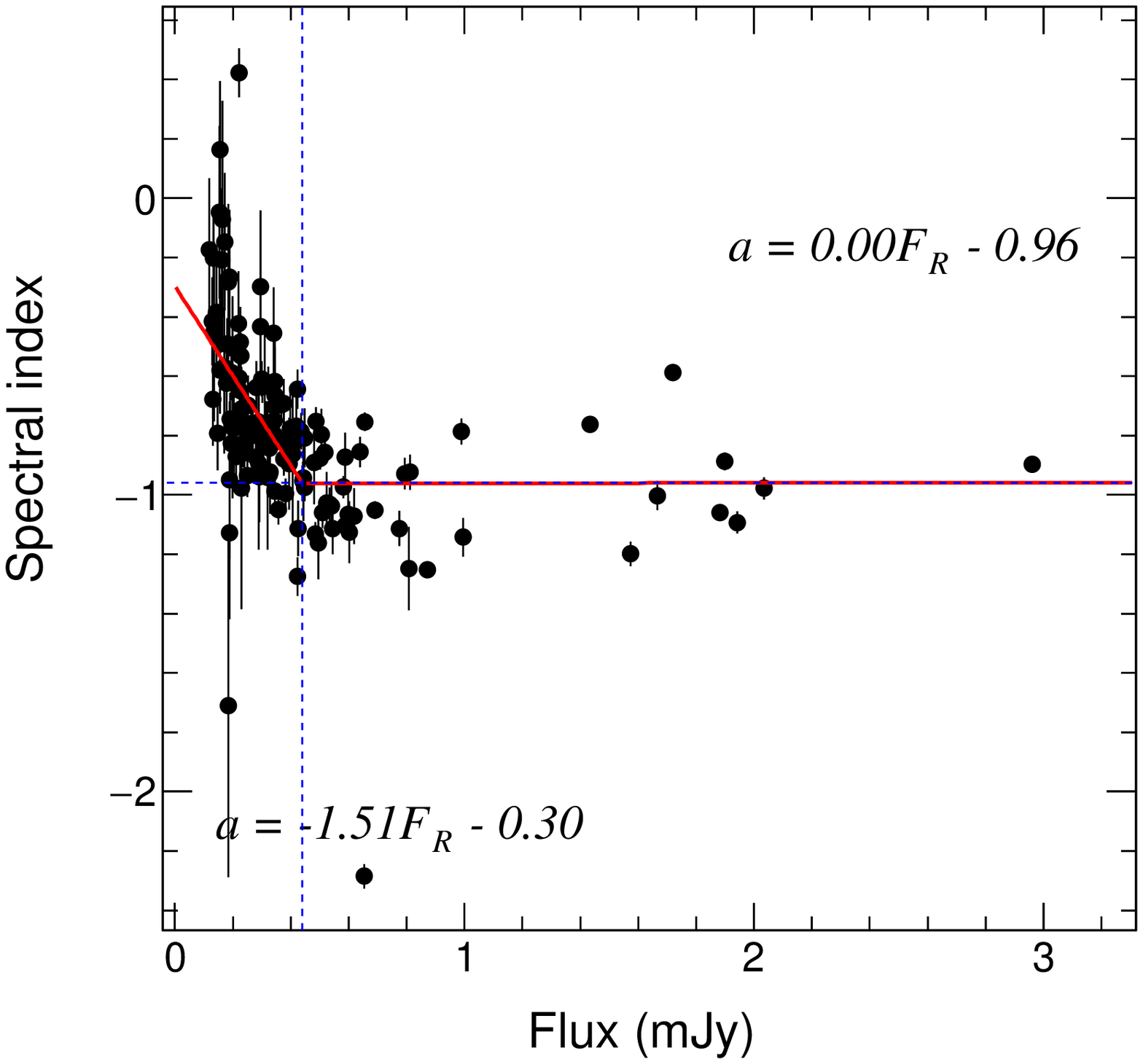}{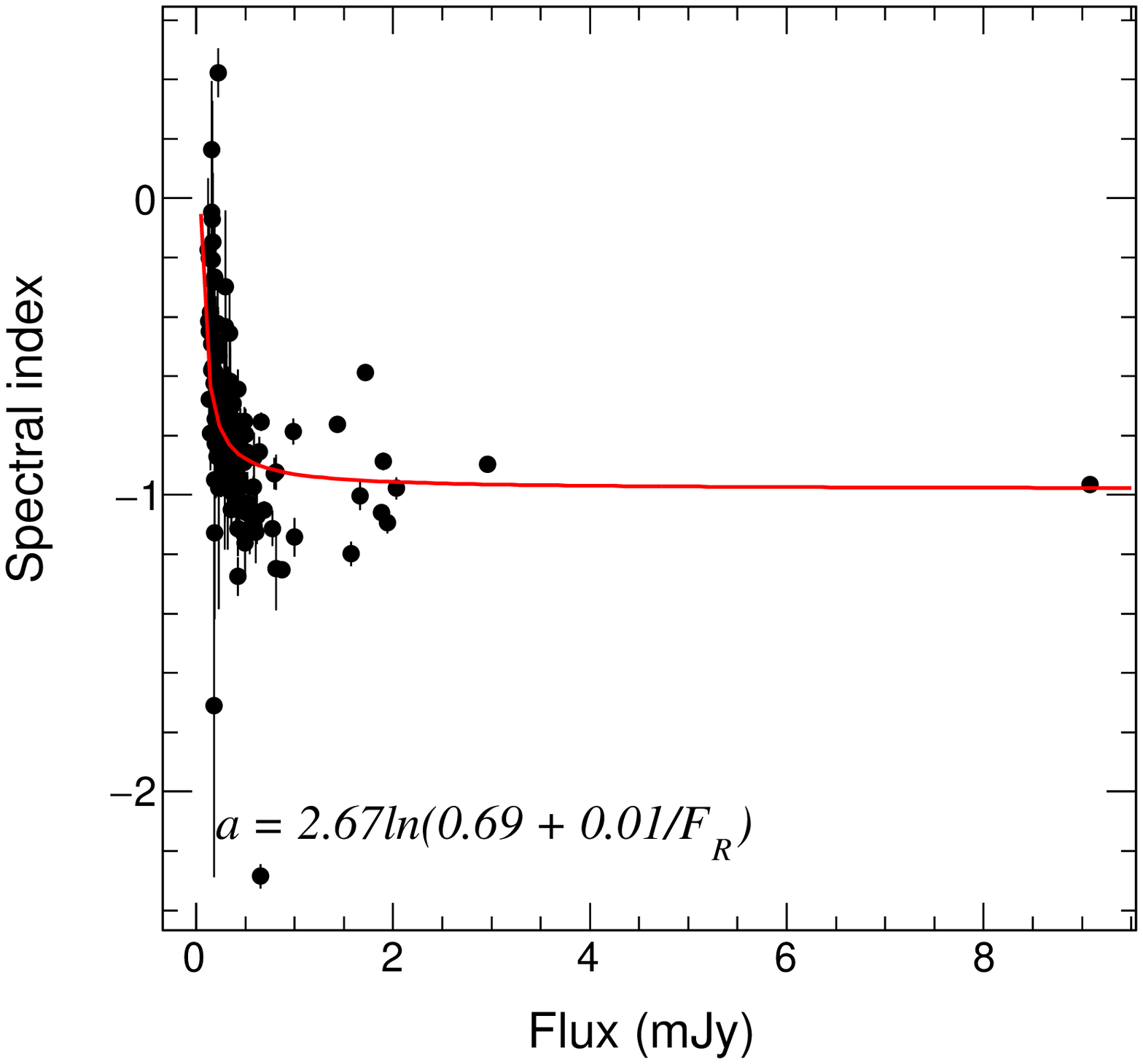}
\caption{Same as Figure~\ref{3C454alphaflux} in the main text, for PKS 0502+049. \label{0502+049}}
\end{figure}

\begin{figure}
\epsscale{0.7}
\plottwo{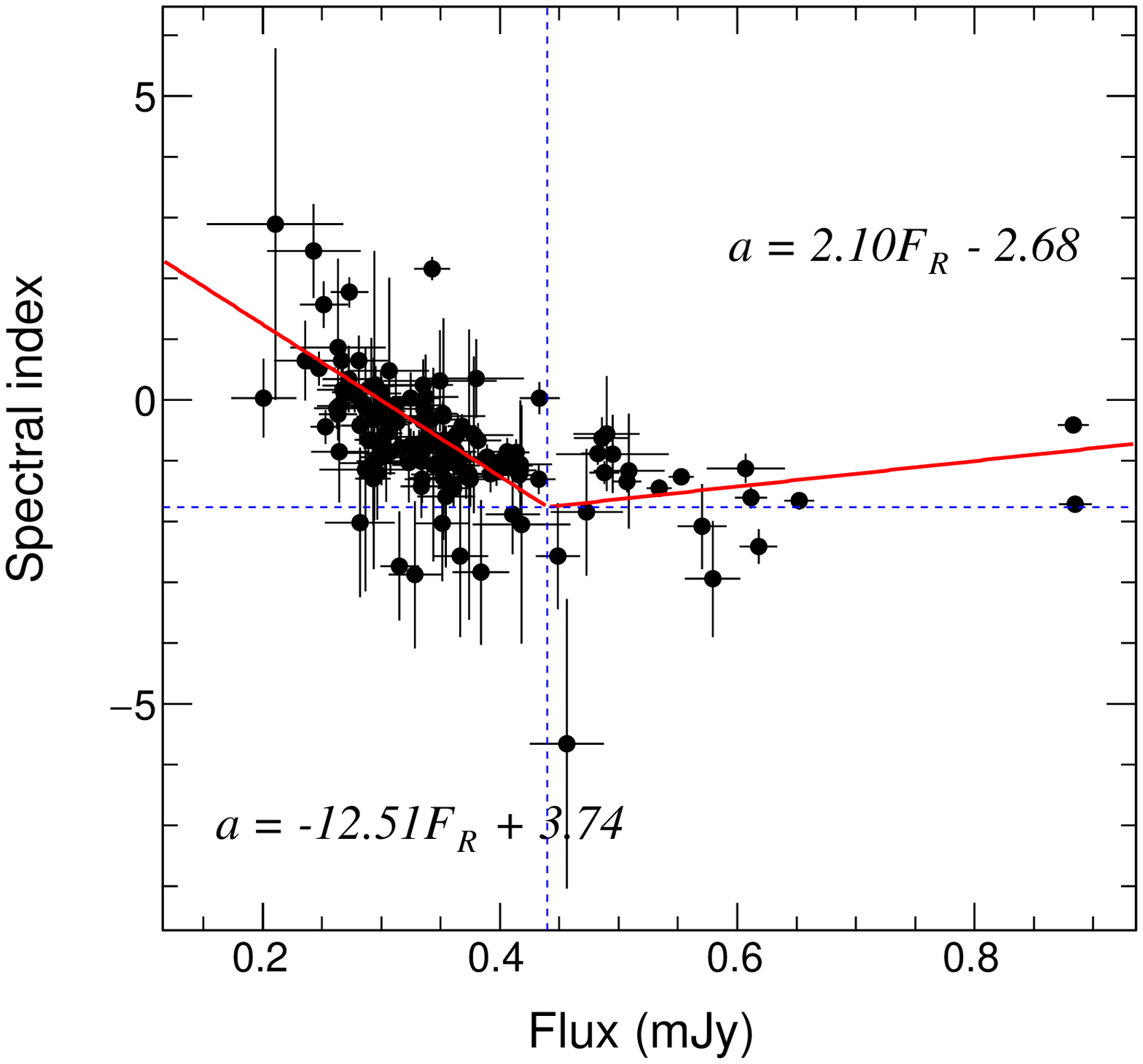}{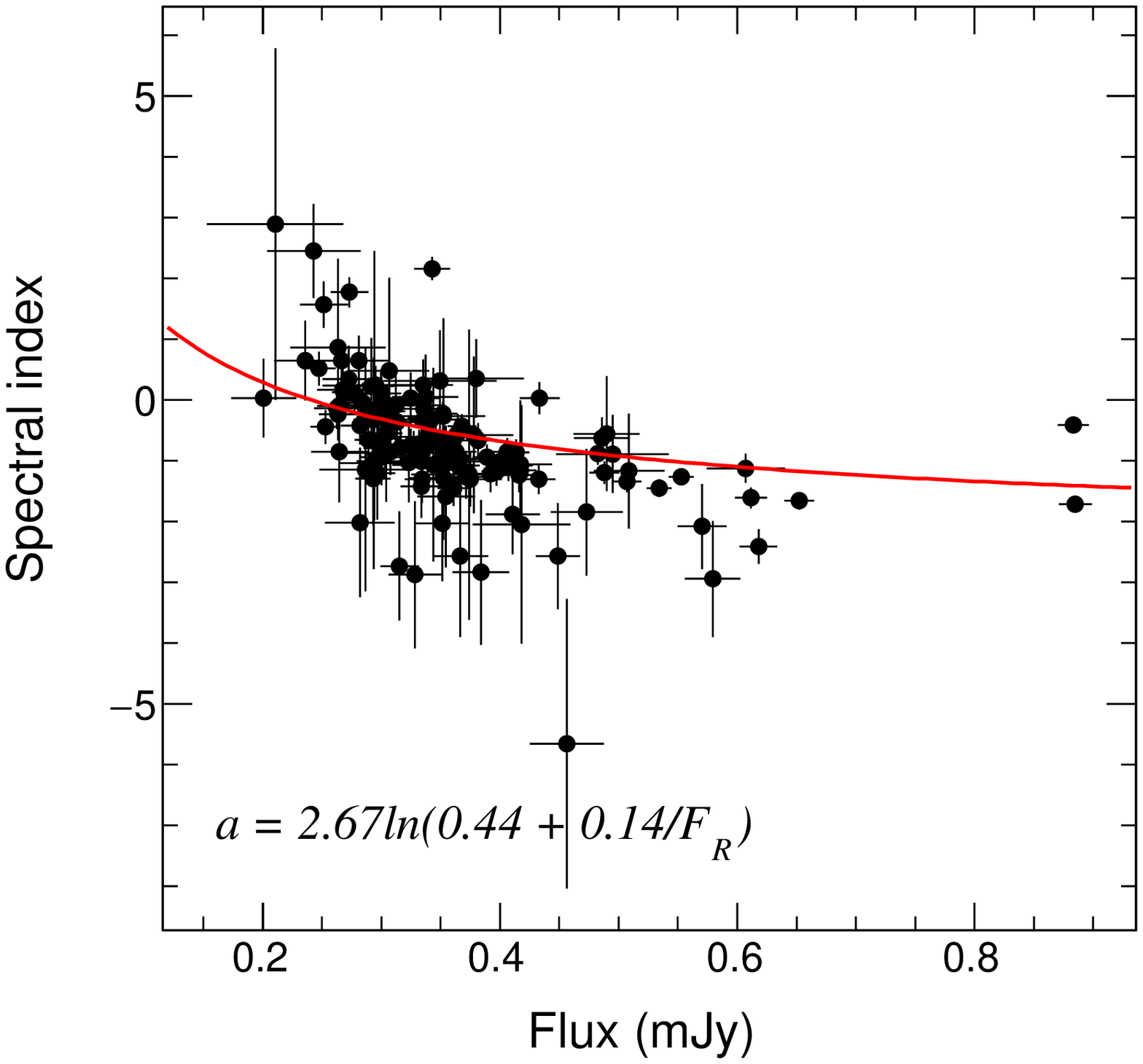}
\caption{Same as Figure~\ref{3C454alphaflux} in the main text, for PKS 0528+134. \label{0528+134}}
\end{figure}

\clearpage

\begin{figure}
\epsscale{0.7}
\plottwo{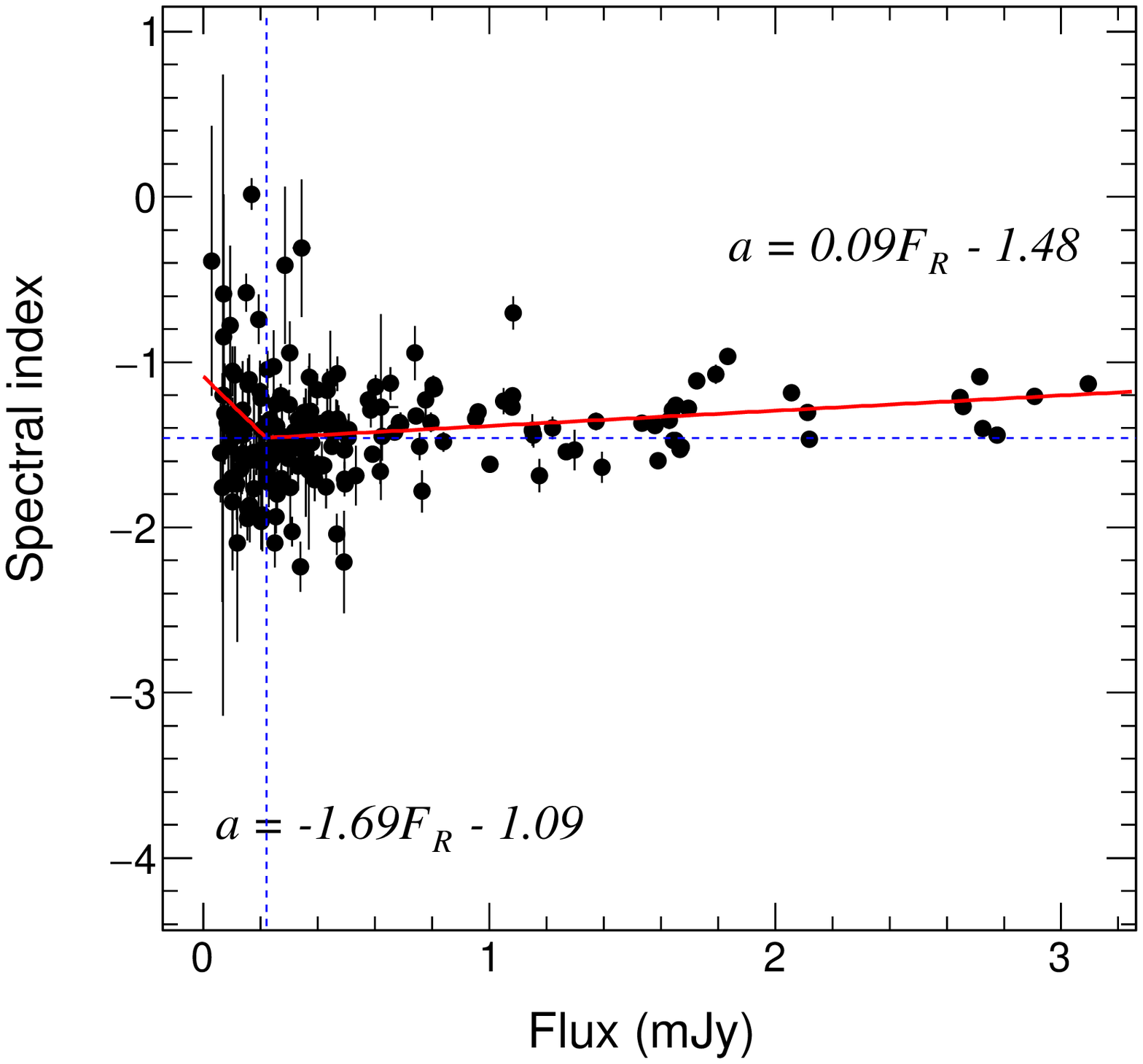}{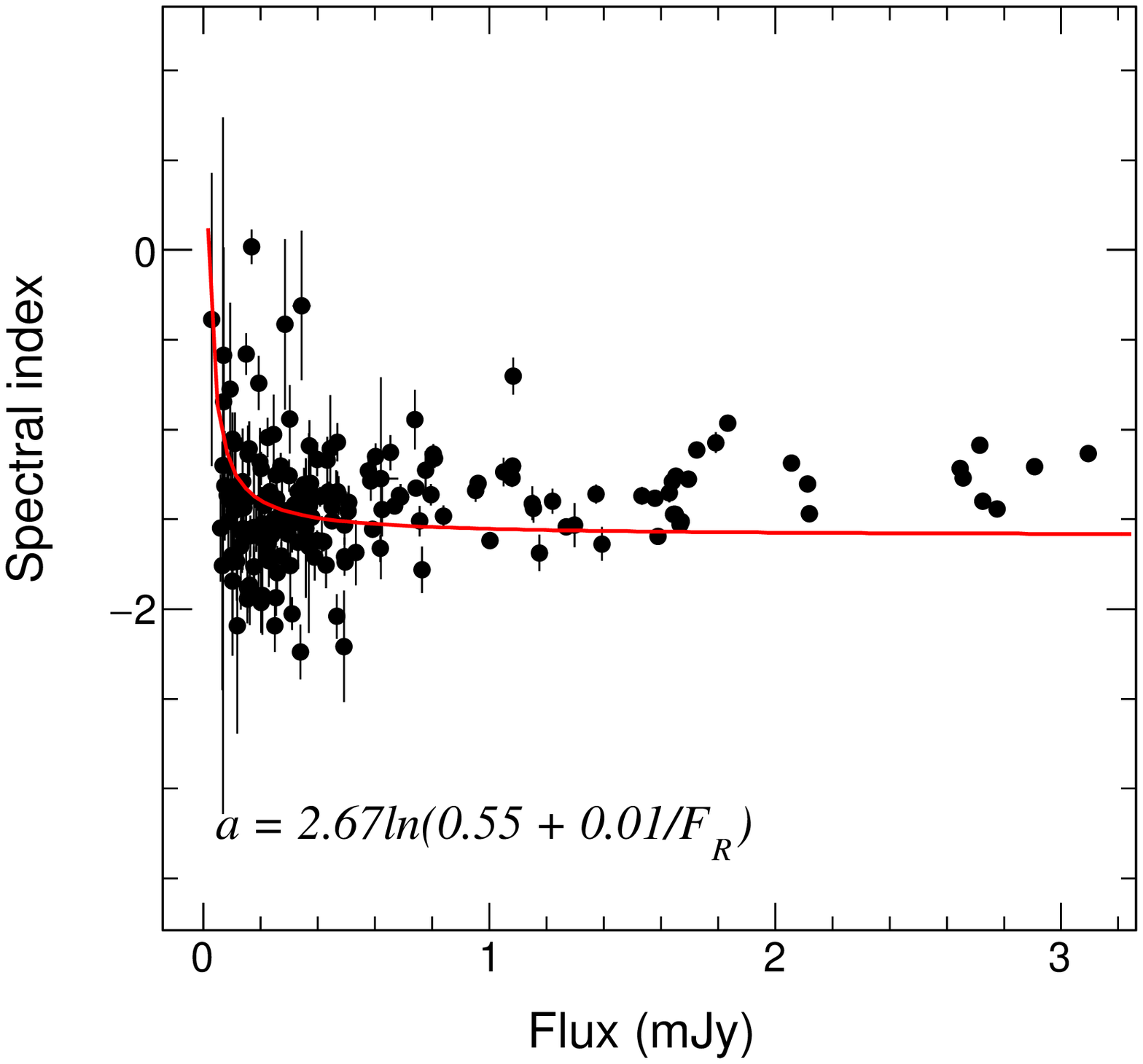}
\caption{Same as Figure~\ref{3C454alphaflux} in the main text, for PMN J0531-4827. \label{0531-4827}}
\end{figure}

\begin{figure}
\epsscale{0.7}
\plottwo{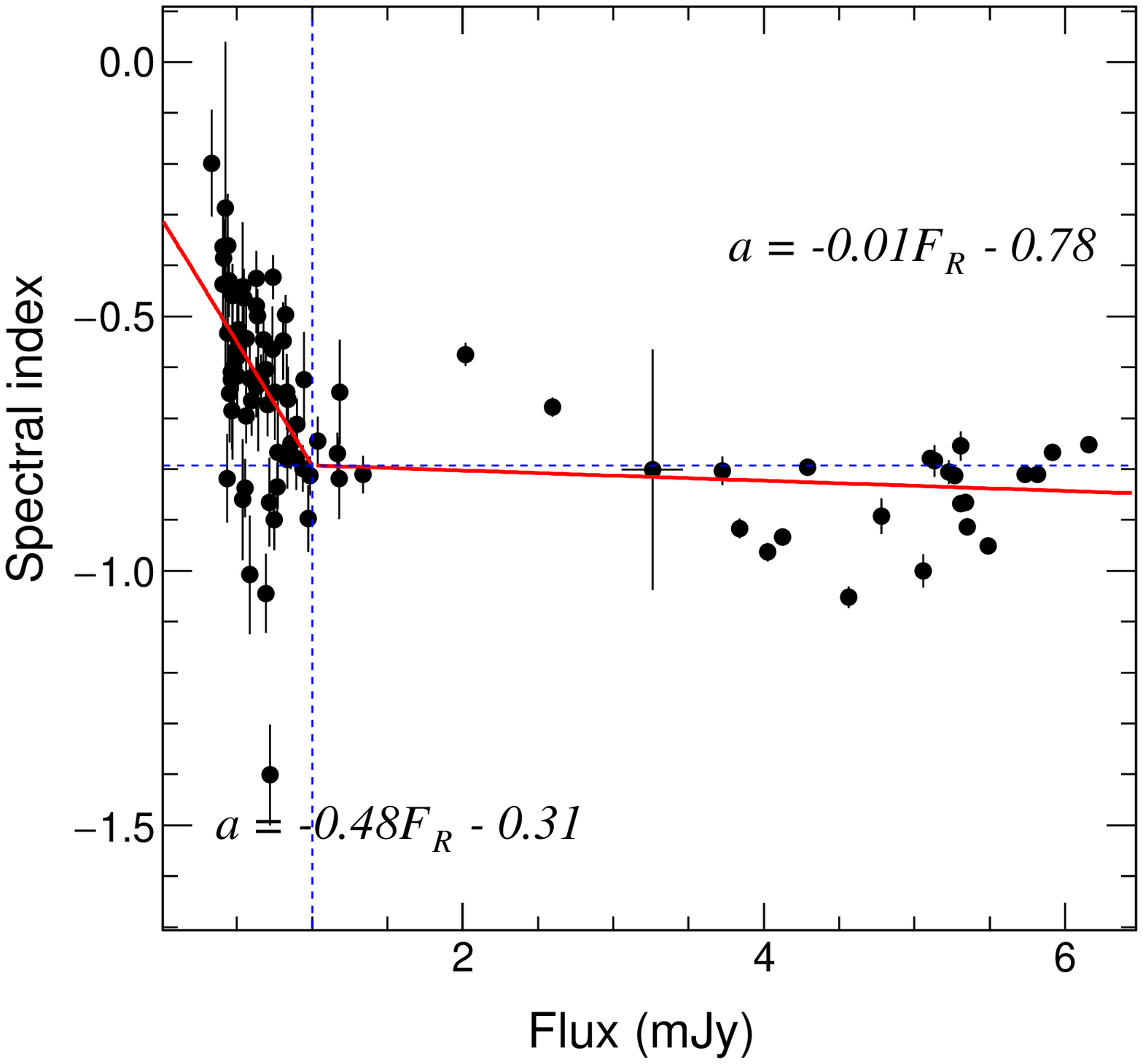}{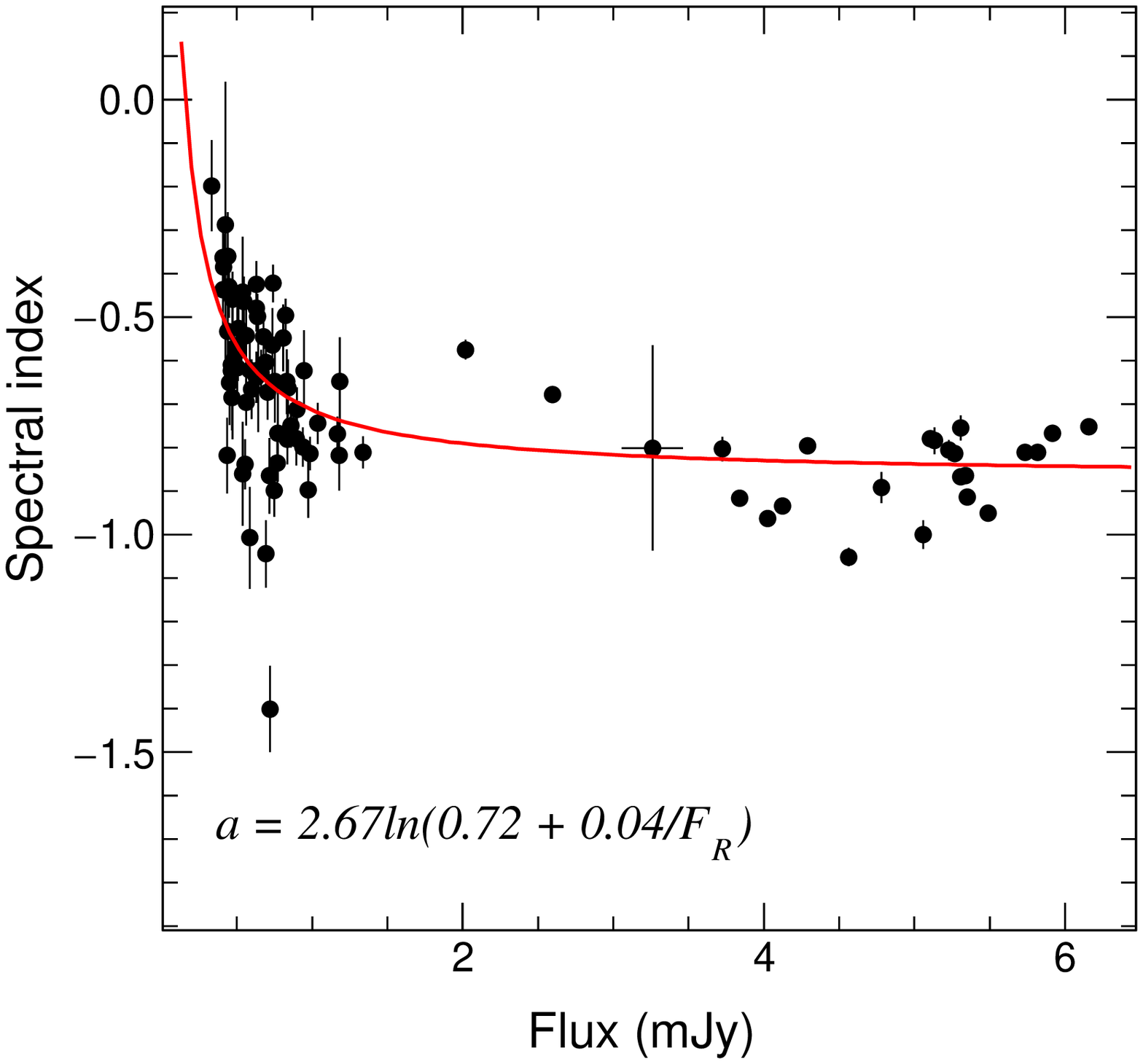}
\caption{Same as Figure~\ref{3C454alphaflux} in the main text, for PKS 0537-441. \label{0537-441}}
\end{figure}

\begin{figure}
\epsscale{0.7}
\plottwo{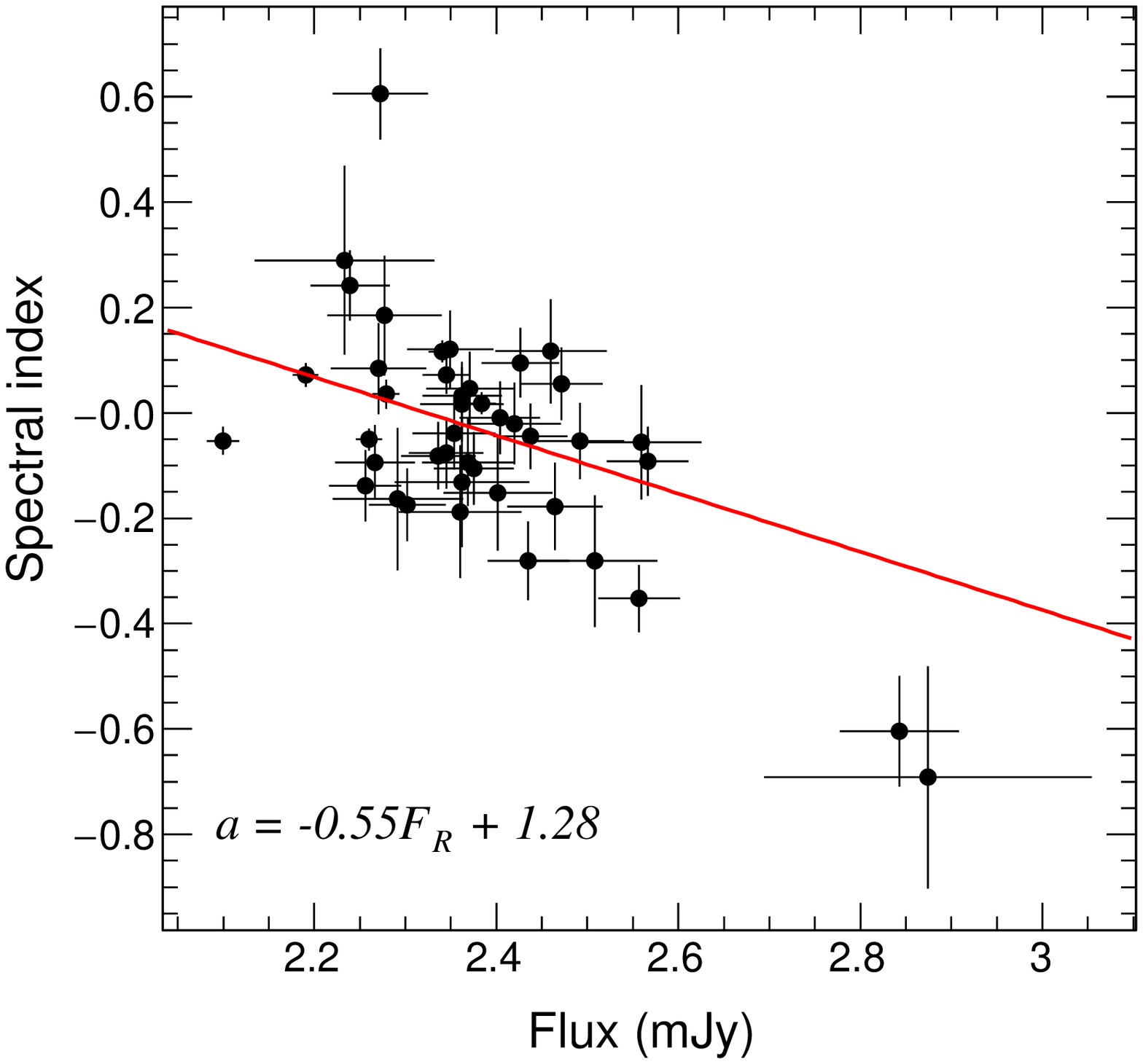}{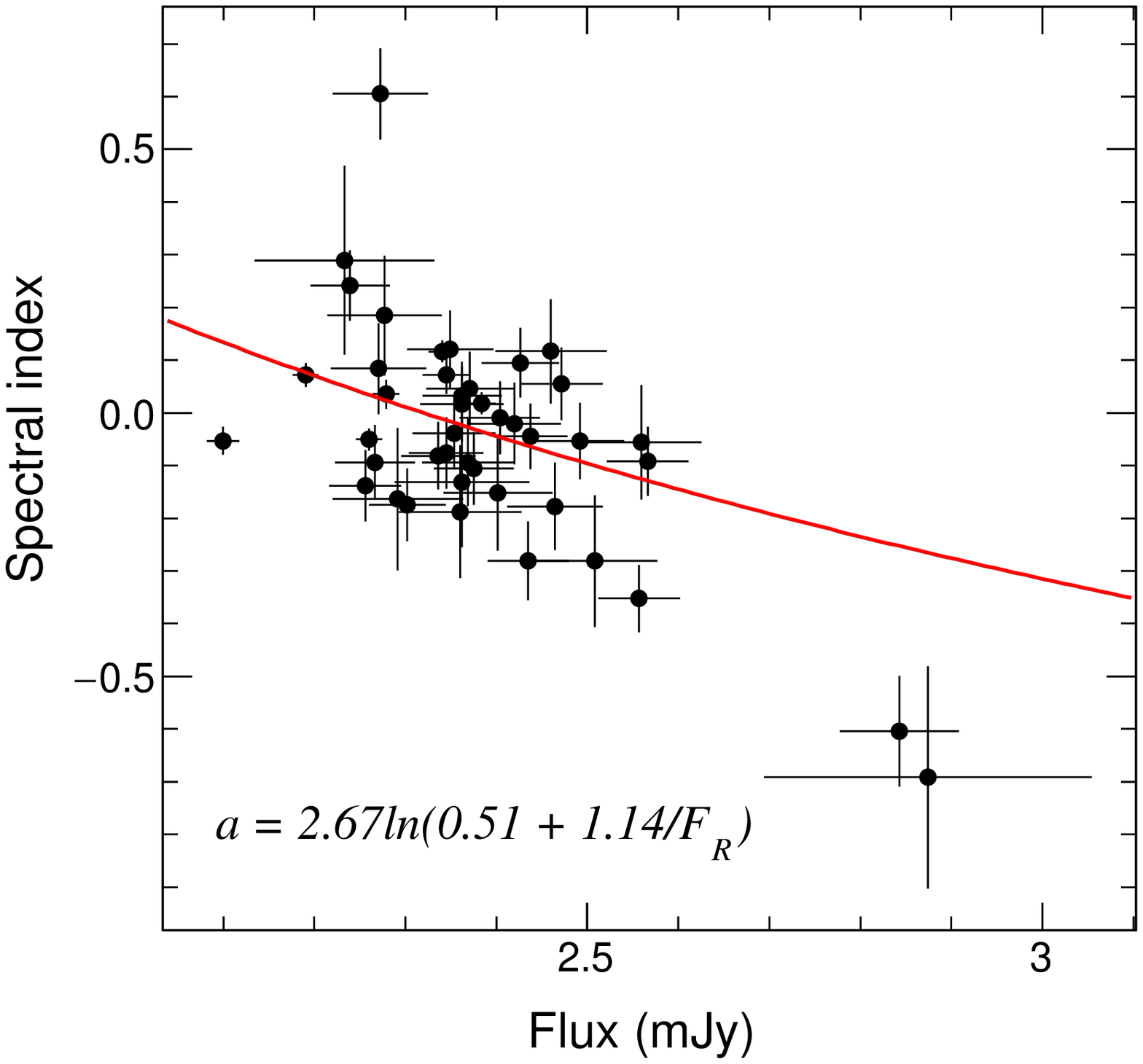}
\caption{Same as Figure~\ref{3C454alphaflux} in the main text, for PKS 0637-75. \label{0637-75}}
\end{figure}

\clearpage

\begin{figure}
\epsscale{0.7}
\plottwo{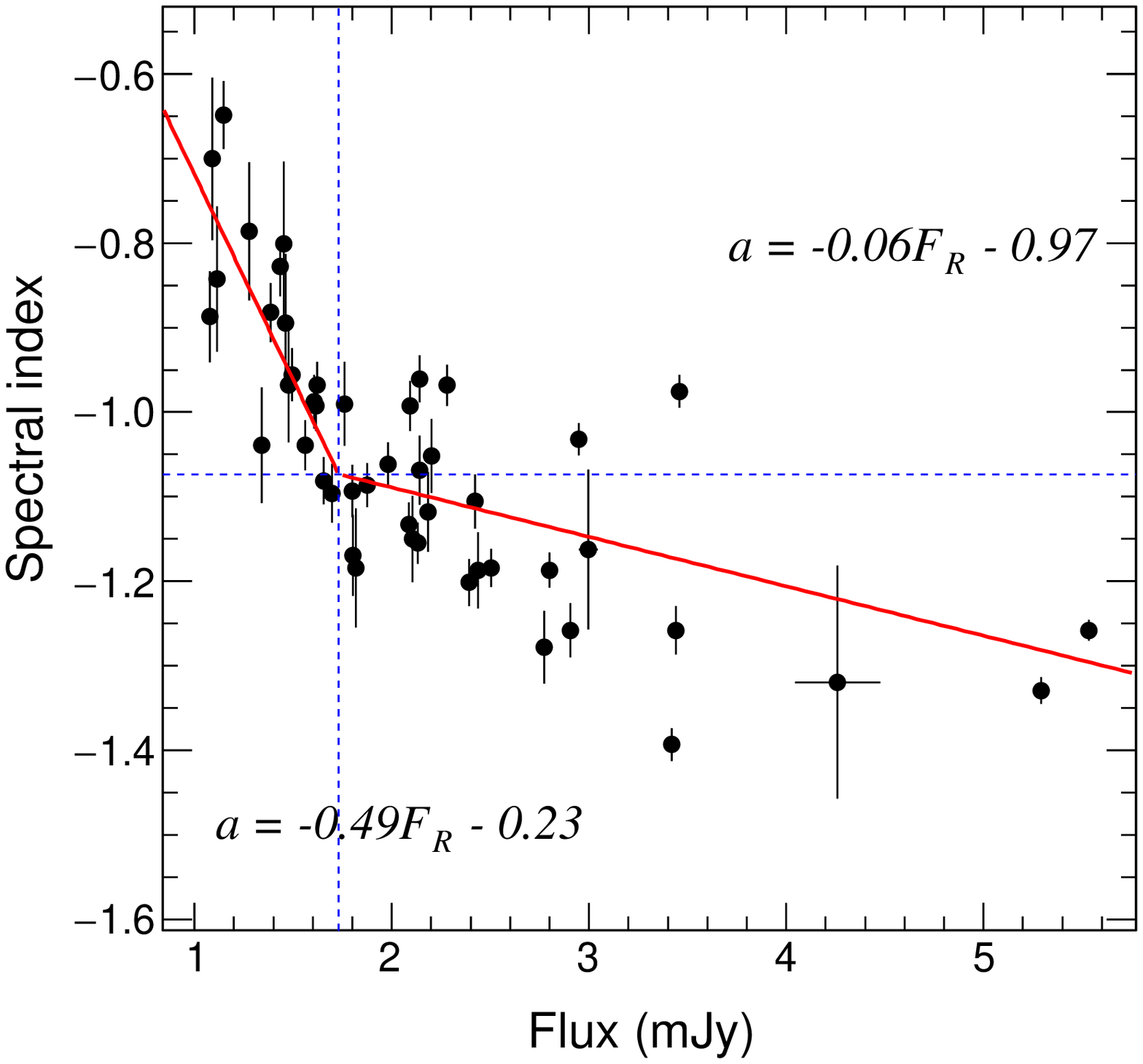}{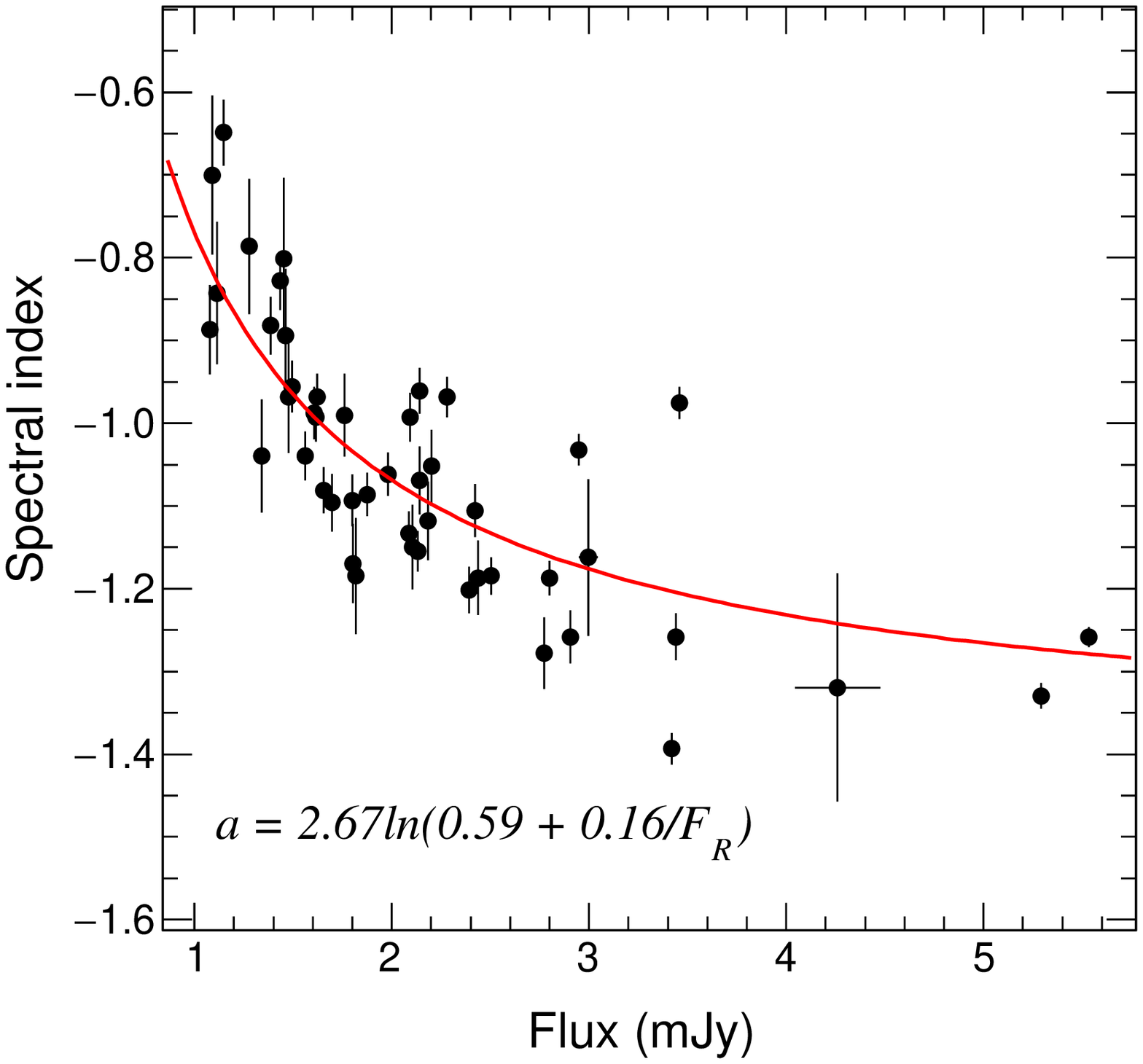}
\caption{Same as Figure~\ref{3C454alphaflux} in the main text, for PKS 0736+01. \label{0736+01}}
\end{figure}

\begin{figure}
\epsscale{0.7}
\plottwo{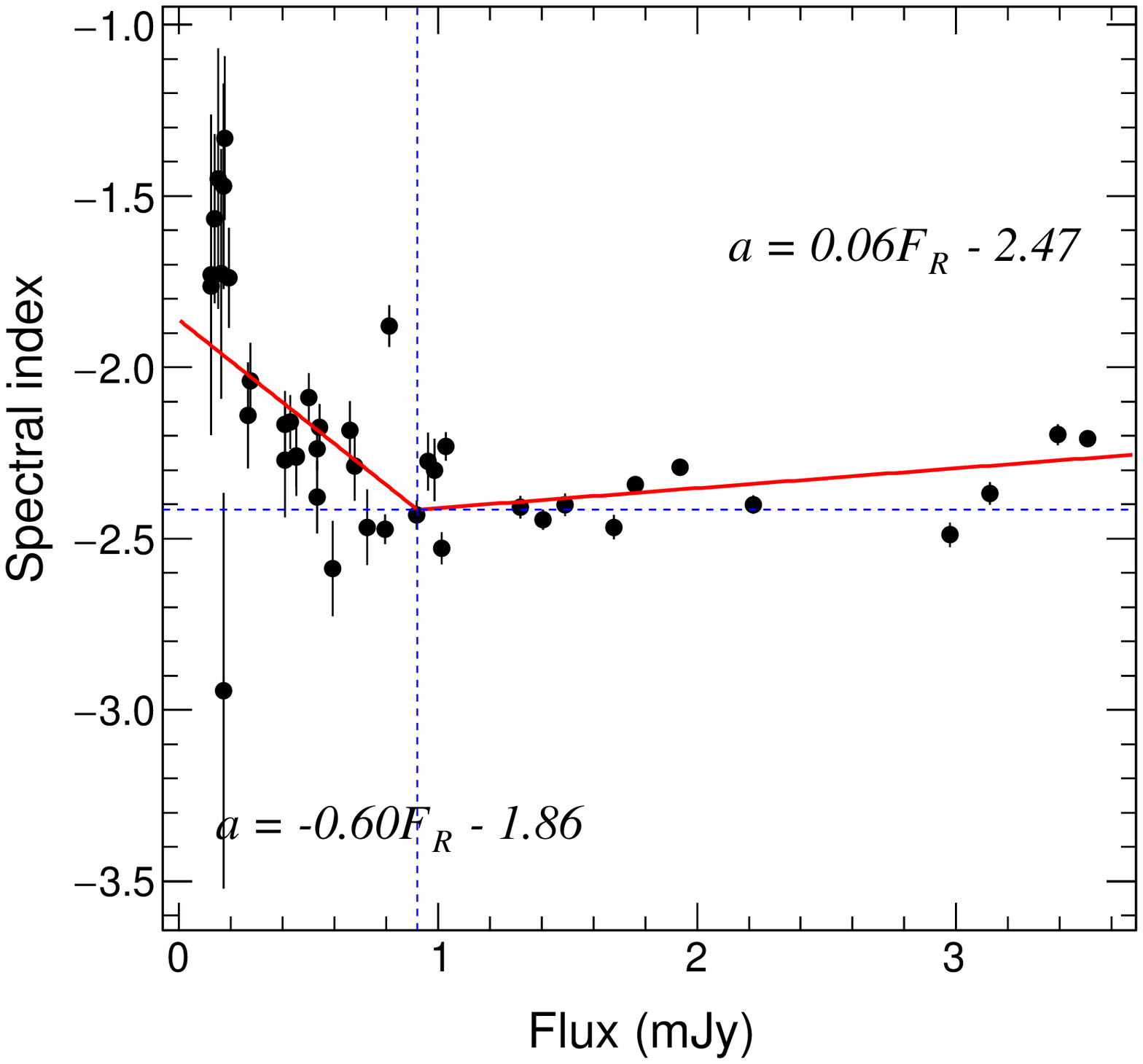}{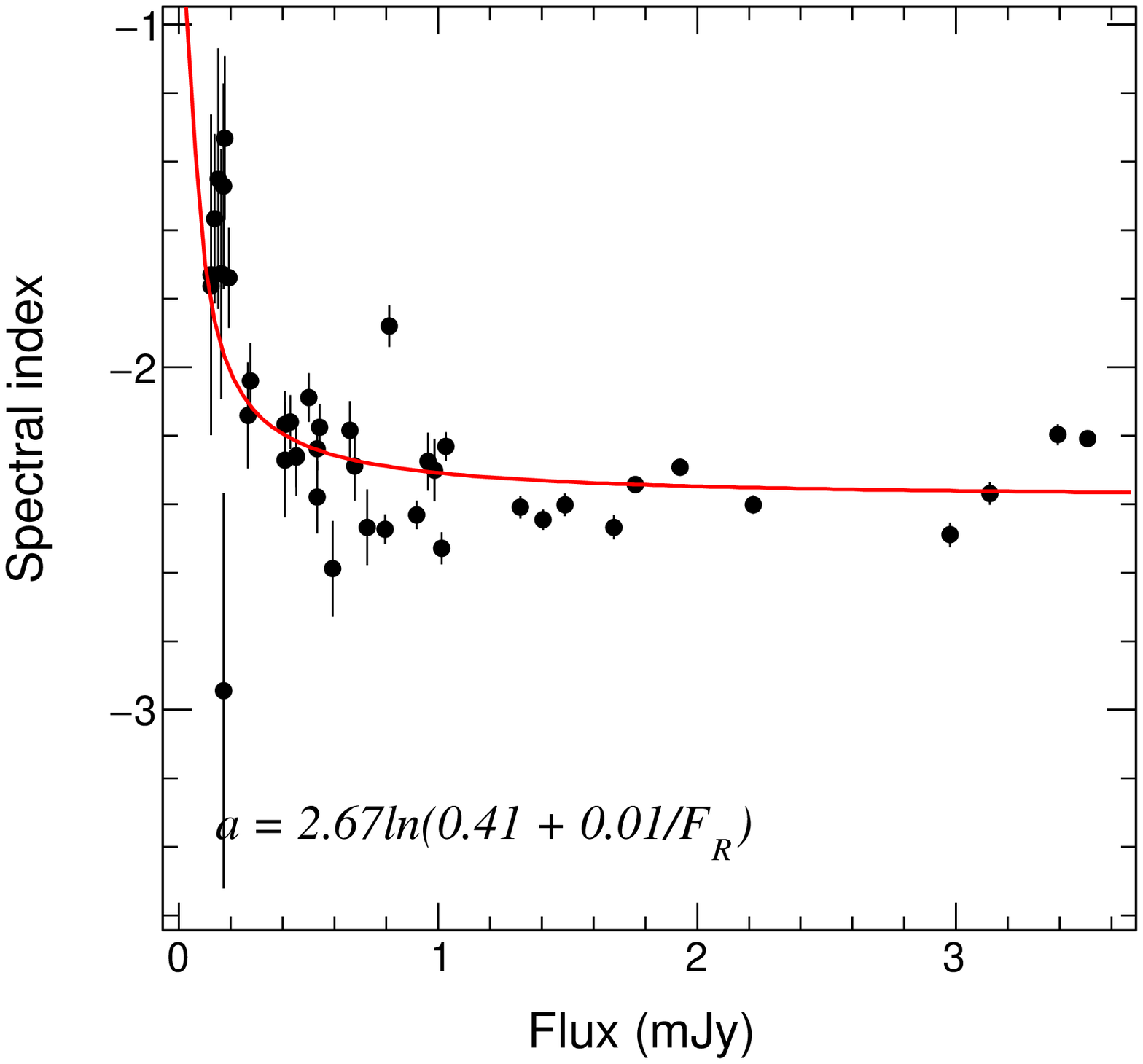}
\caption{Same as Figure~\ref{3C454alphaflux} in the main text, for PKS 0805-07. \label{0805-077}}
\end{figure}

\begin{figure}
\epsscale{0.7}
\plottwo{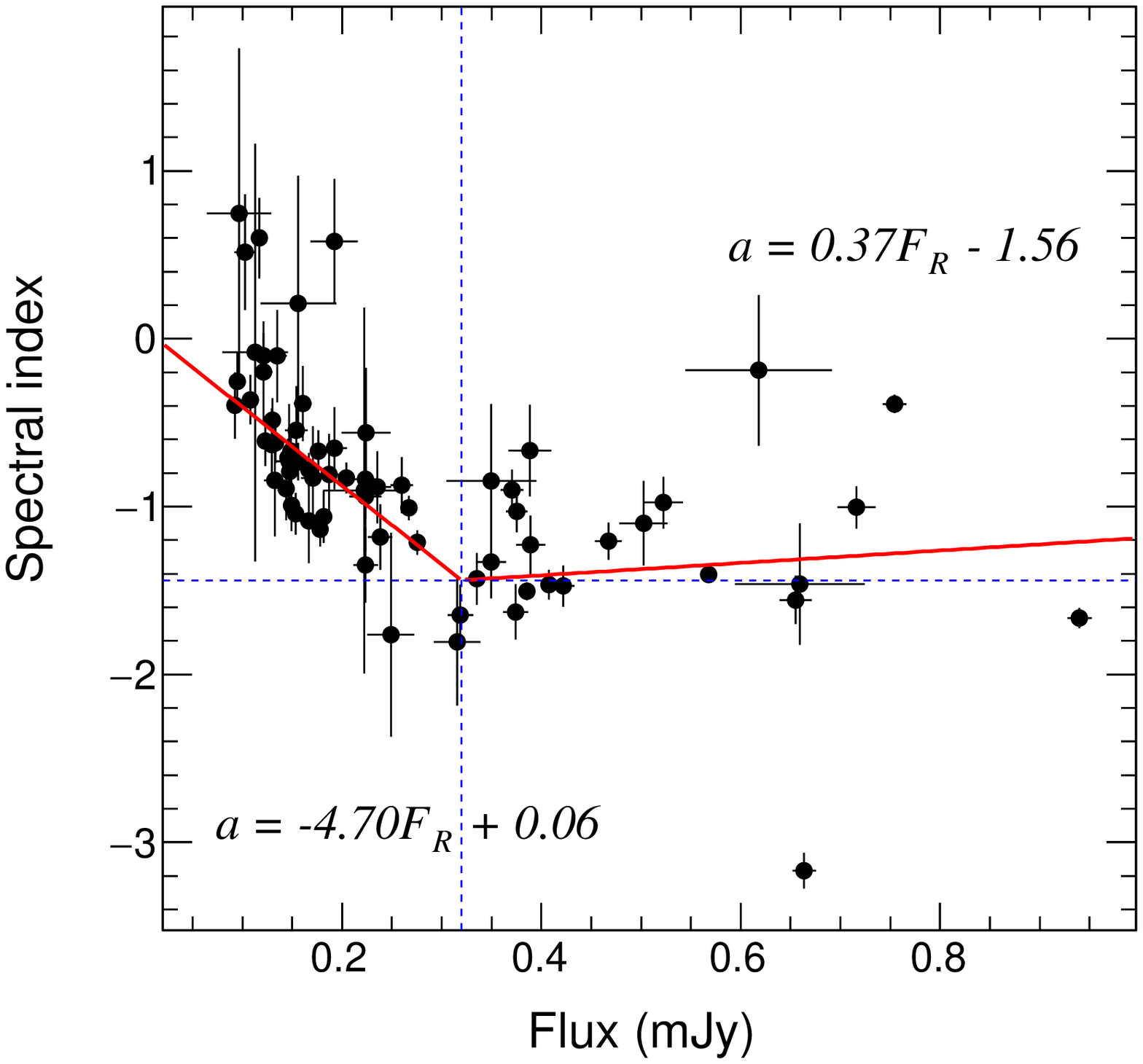}{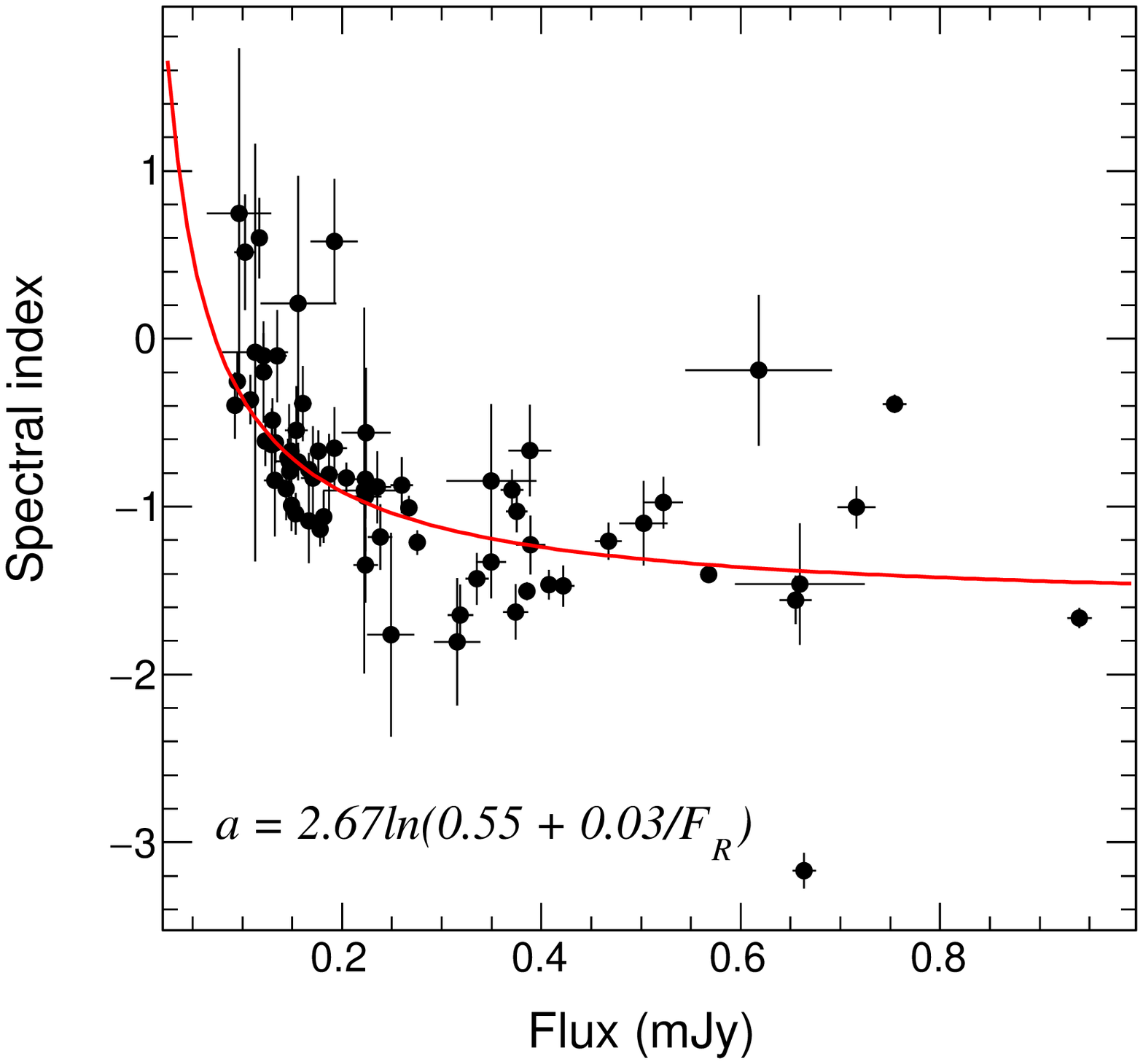}
\caption{Same as Figure~\ref{3C454alphaflux} in the main text, for PMN J0850-1213. \label{0850-1213}}
\end{figure}

\begin{figure}
\epsscale{0.7}
\plottwo{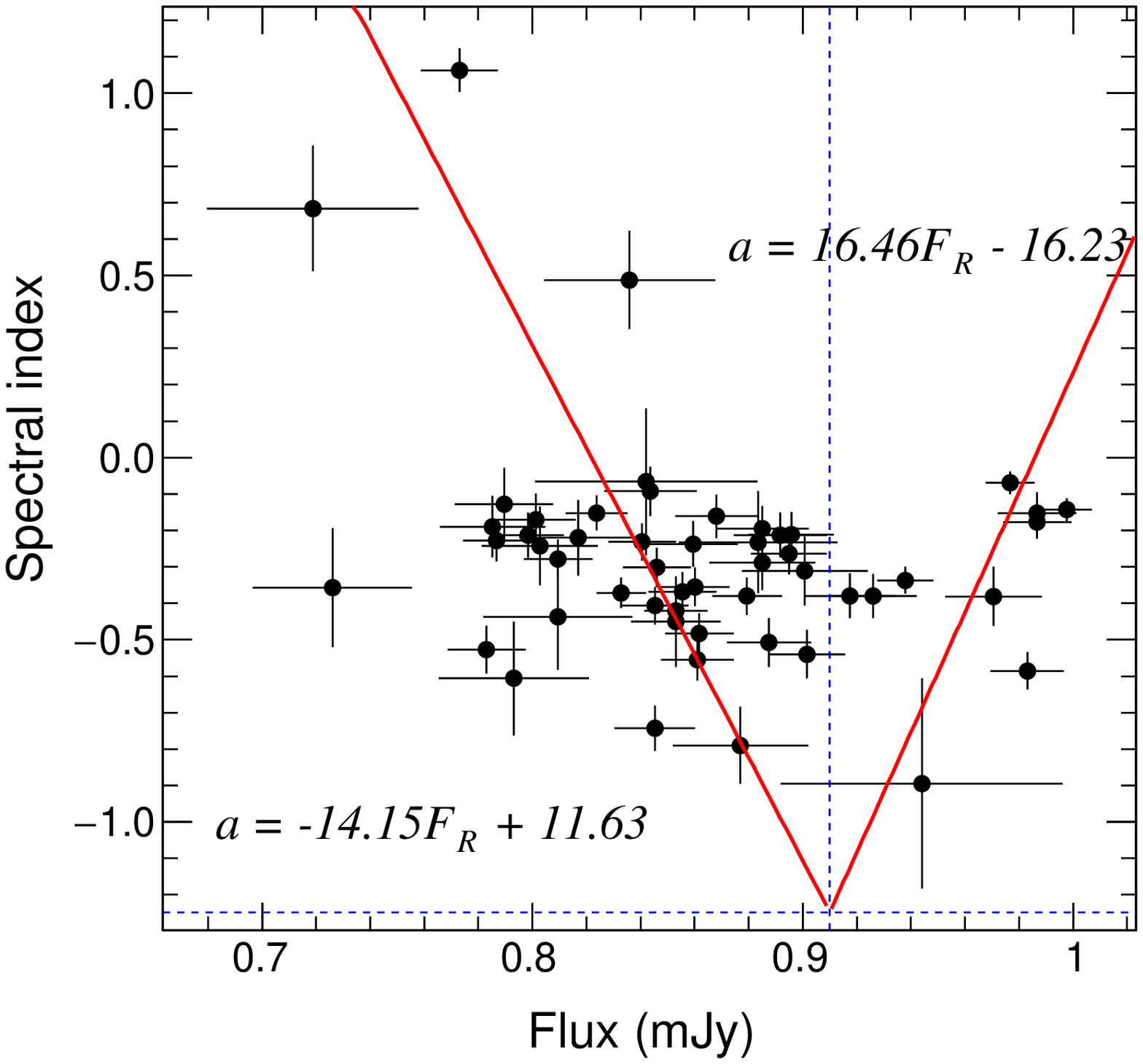}{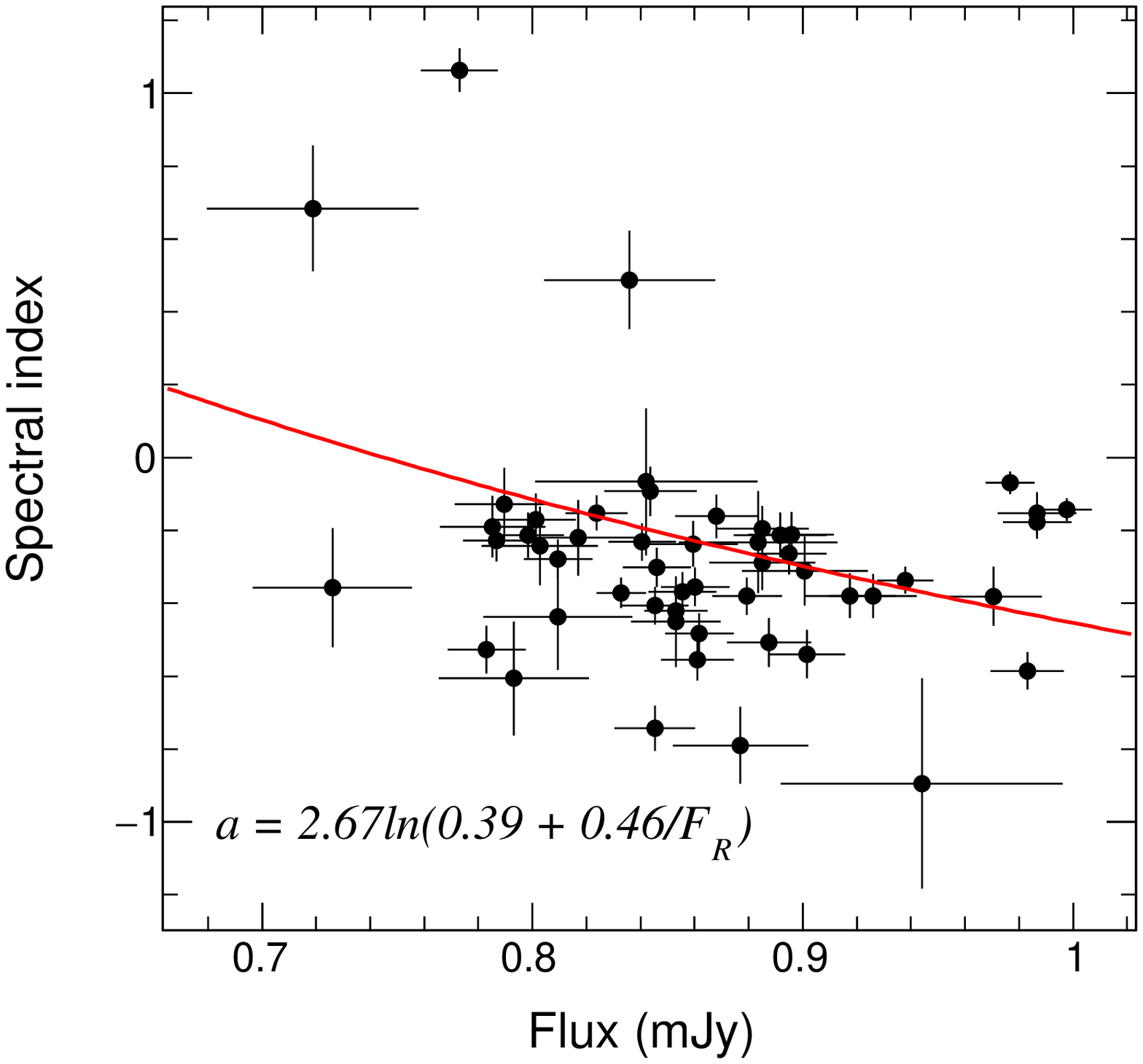}
\caption{Same as Figure~\ref{3C454alphaflux} in the main text, for PKS 1004-217. \label{1004-217}}
\end{figure}

\begin{figure}
\epsscale{0.7}
\plottwo{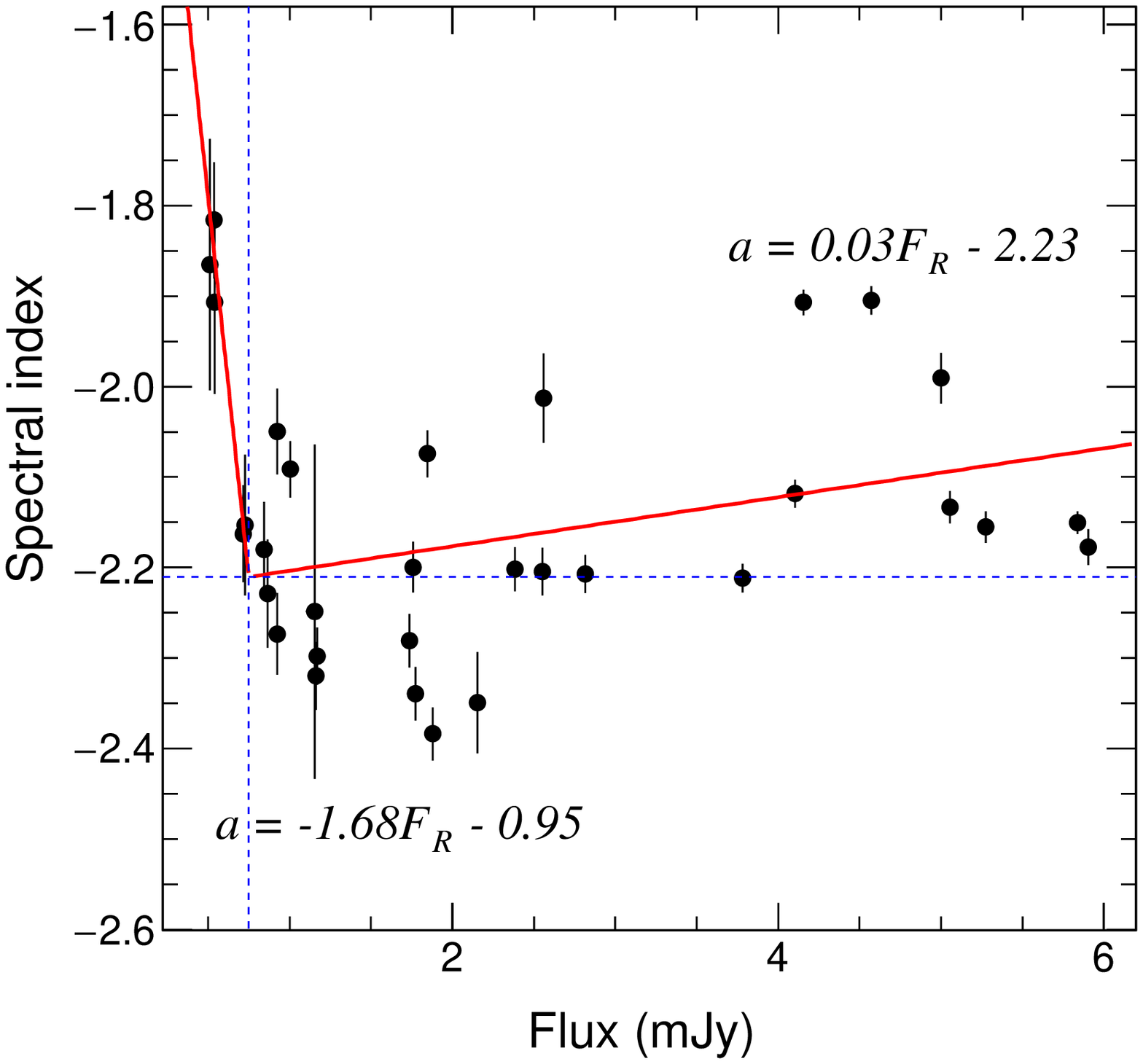}{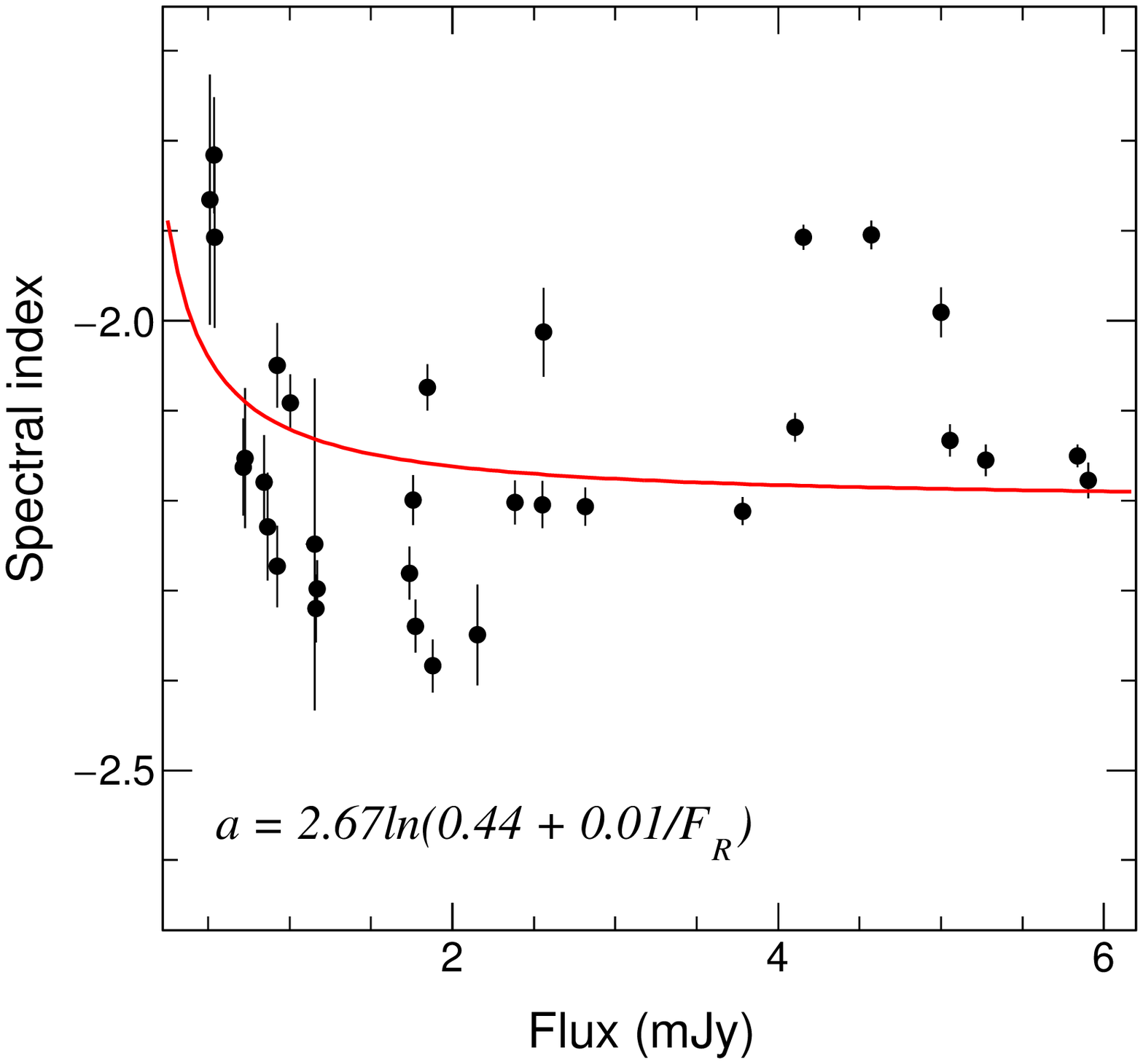}
\caption{Same as Figure~\ref{3C454alphaflux} in the main text, for 4C +01.28. \label{1055+018}}
\end{figure}

\begin{figure}
\epsscale{0.7}
\plottwo{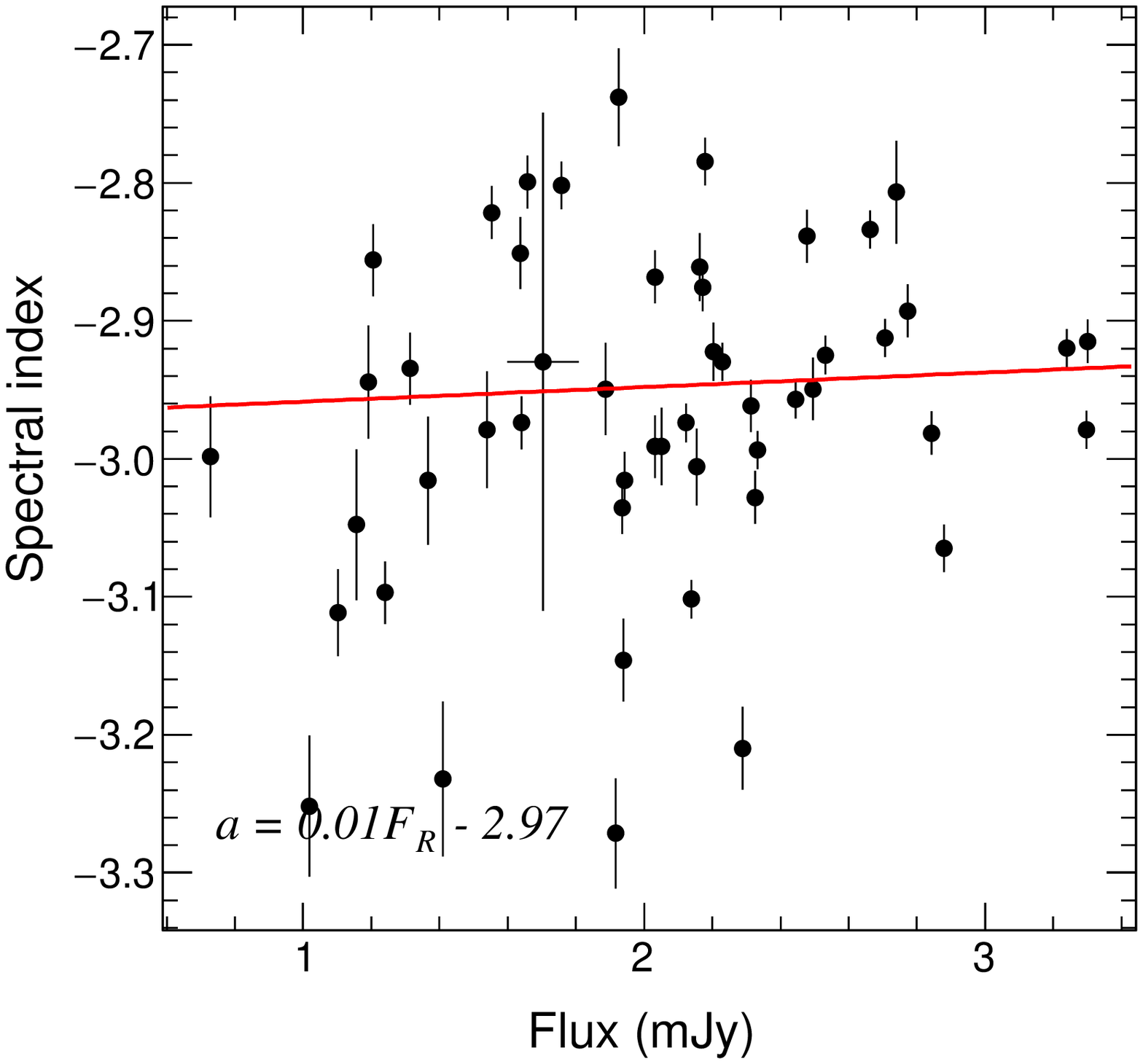}{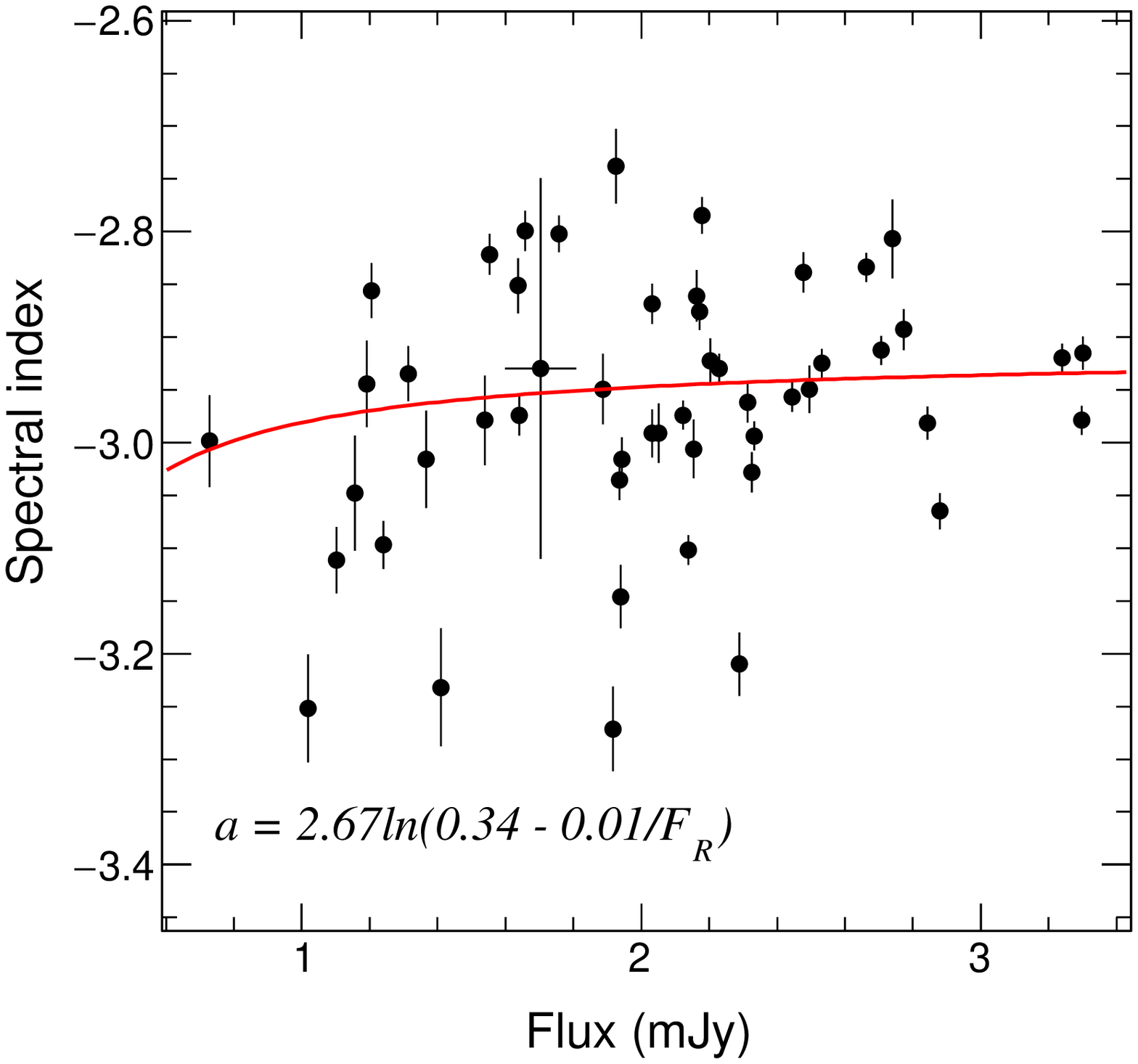}
\caption{Same as Figure~\ref{3C454alphaflux} in the main text, for PKS B1056-113. \label{1059-1134}}
\end{figure}

\clearpage
\begin{figure}
\epsscale{0.7}
\plottwo{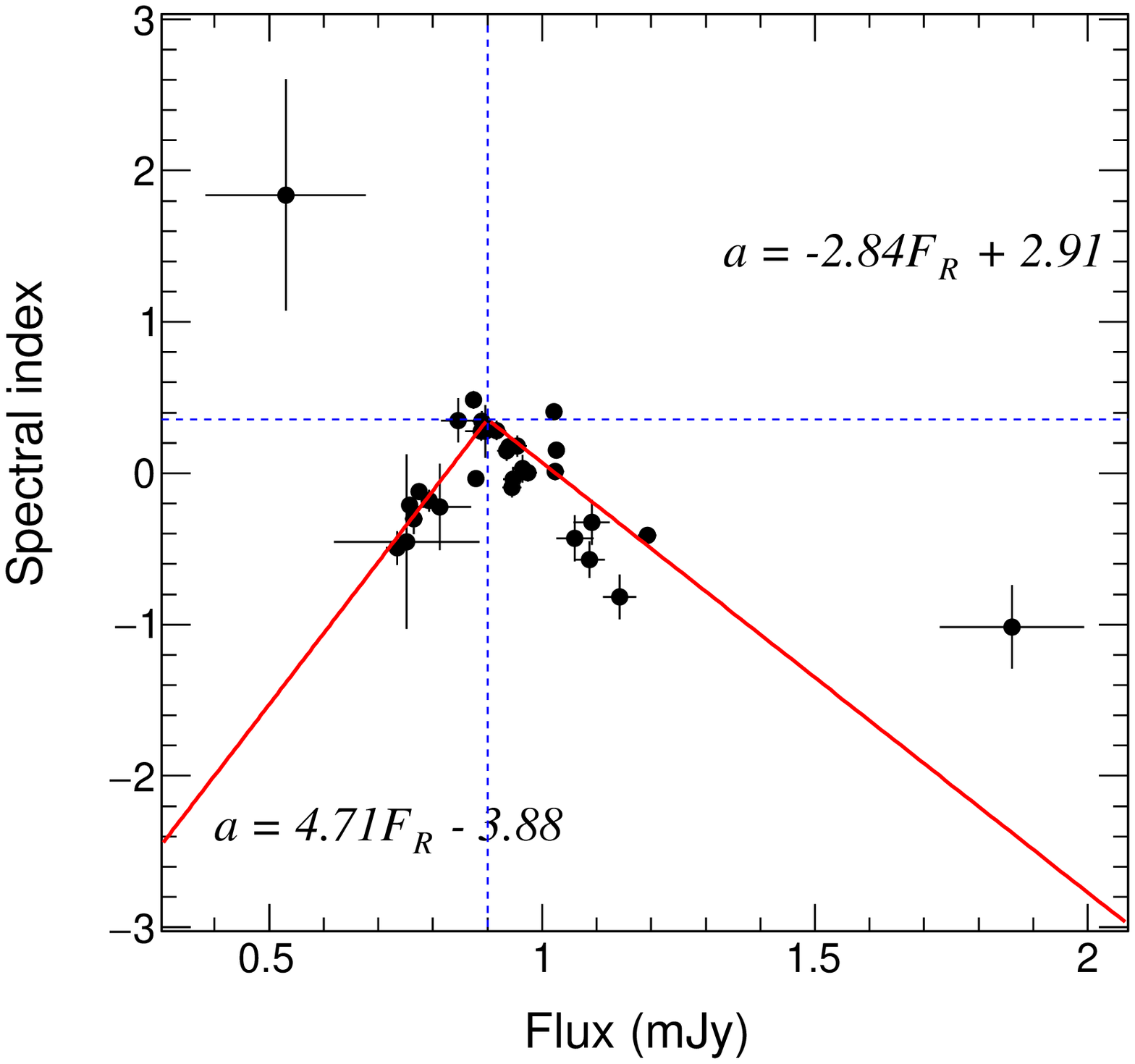}{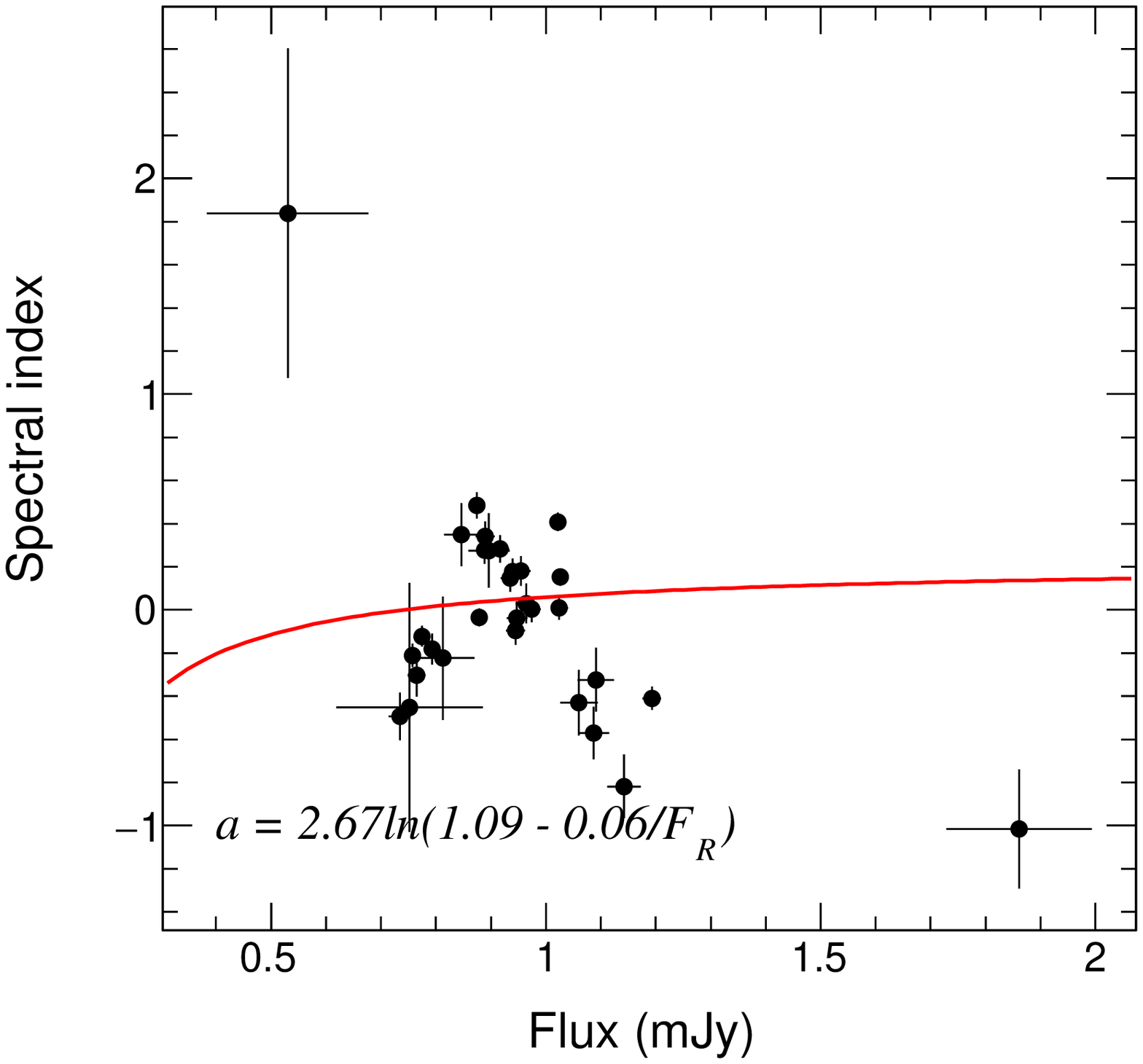}
\caption{Same as Figure~\ref{3C454alphaflux} in the main text, for PKS 1127-14. \label{1127-14}}
\end{figure}

\begin{figure}
\epsscale{0.7}
\plottwo{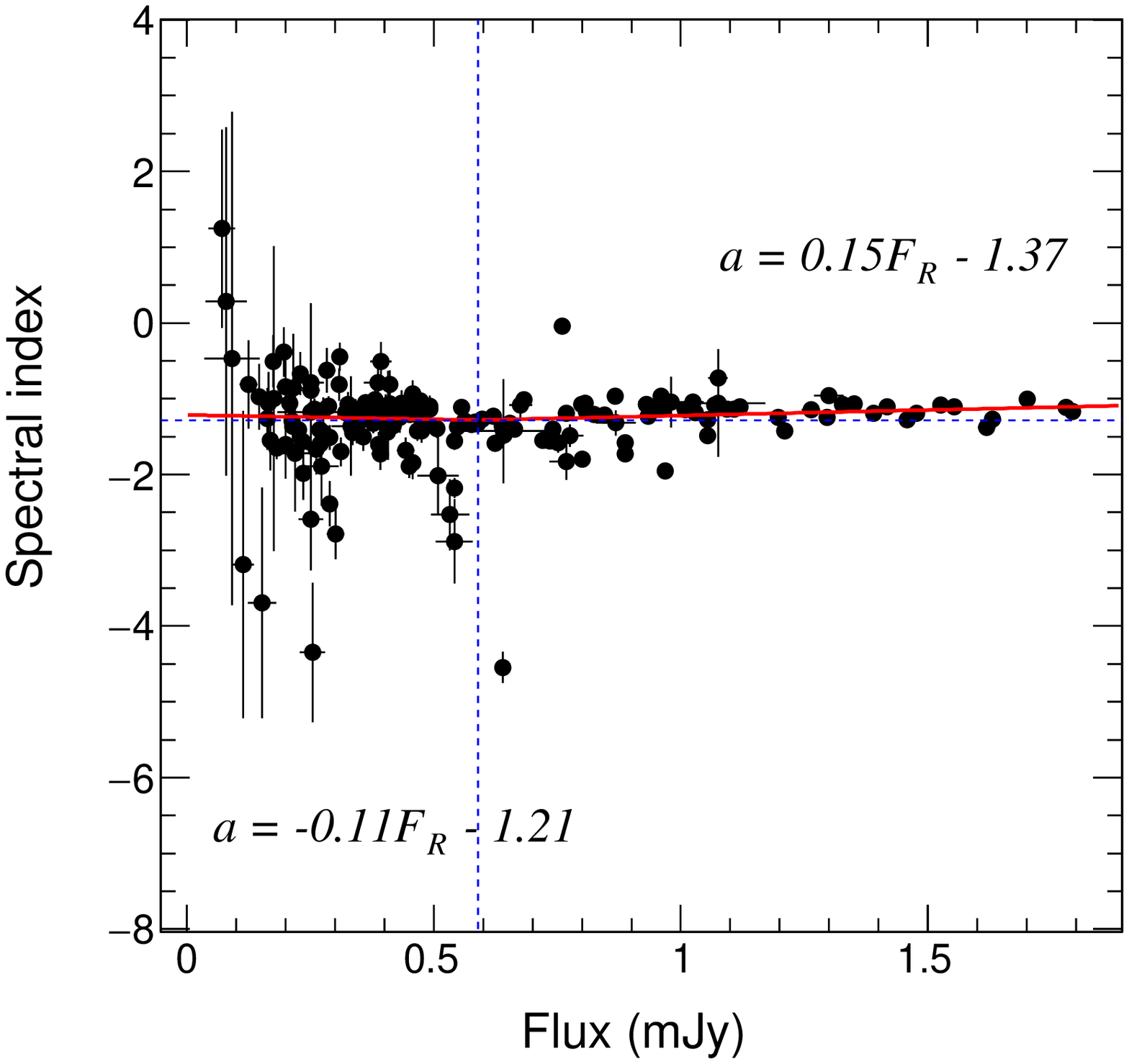}{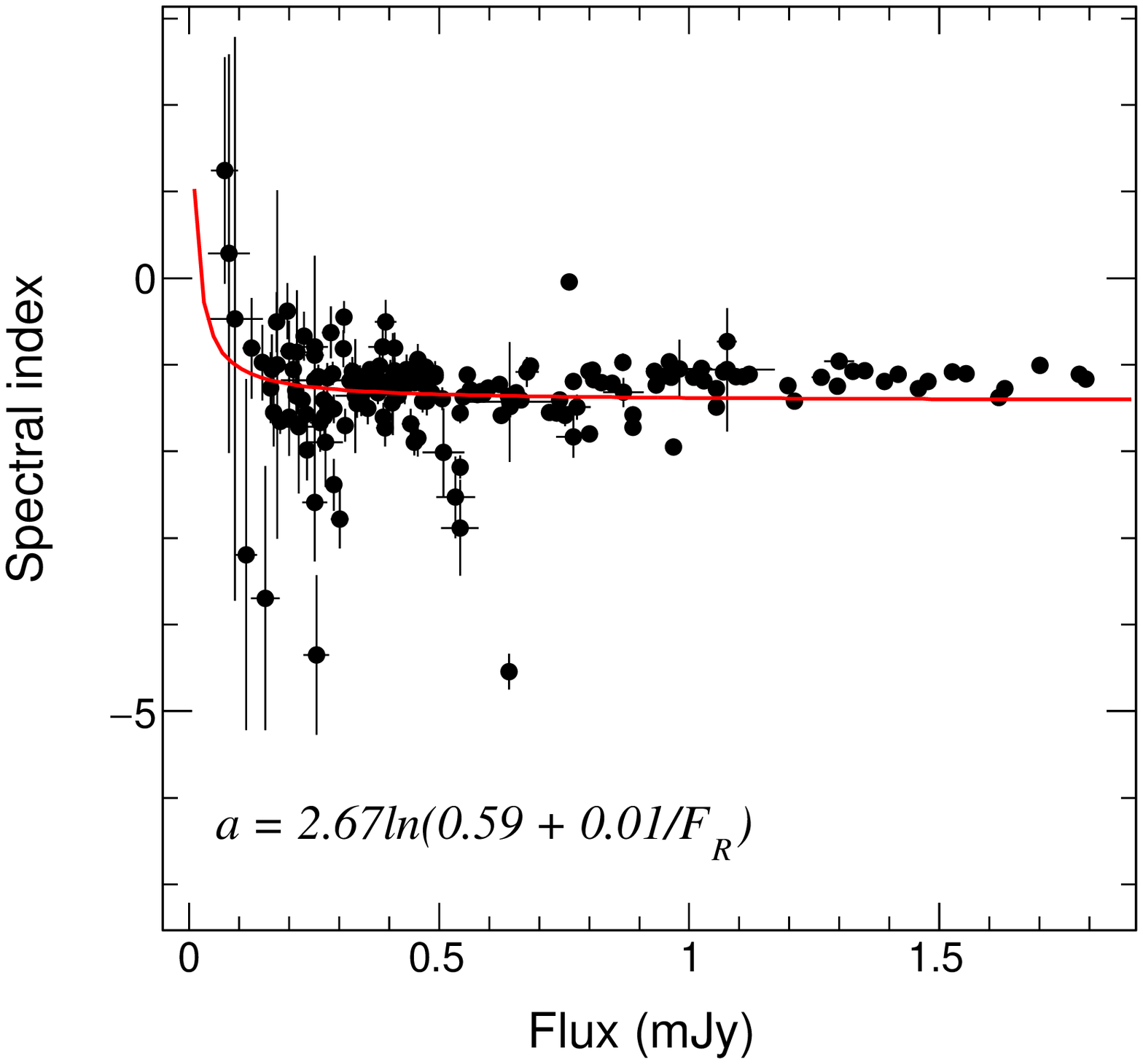}
\caption{Same as Figure~\ref{3C454alphaflux} in the main text, for PKS 1144-379. \label{1144-379}}
\end{figure}

\begin{figure}
\epsscale{0.7}
\plottwo{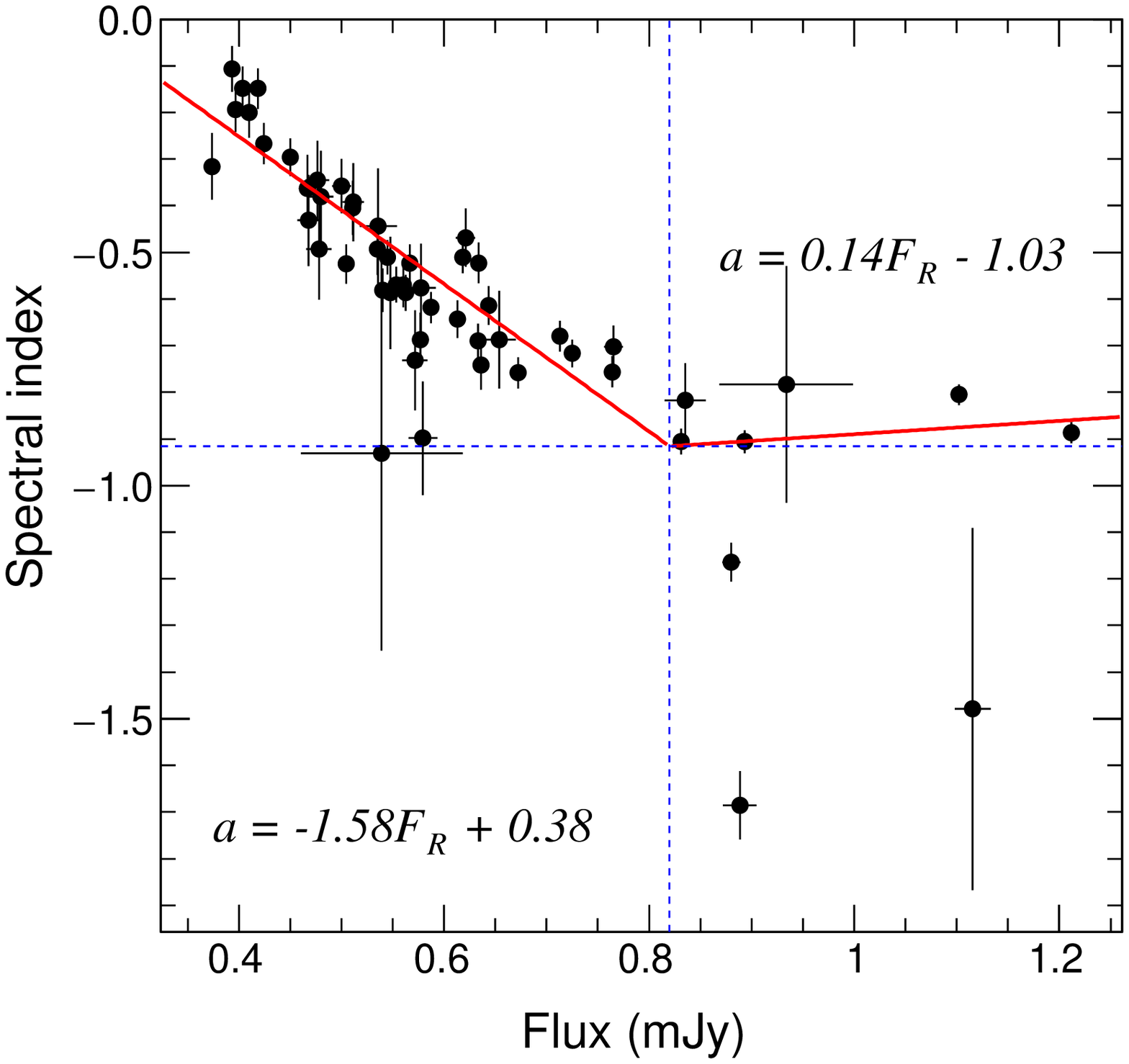}{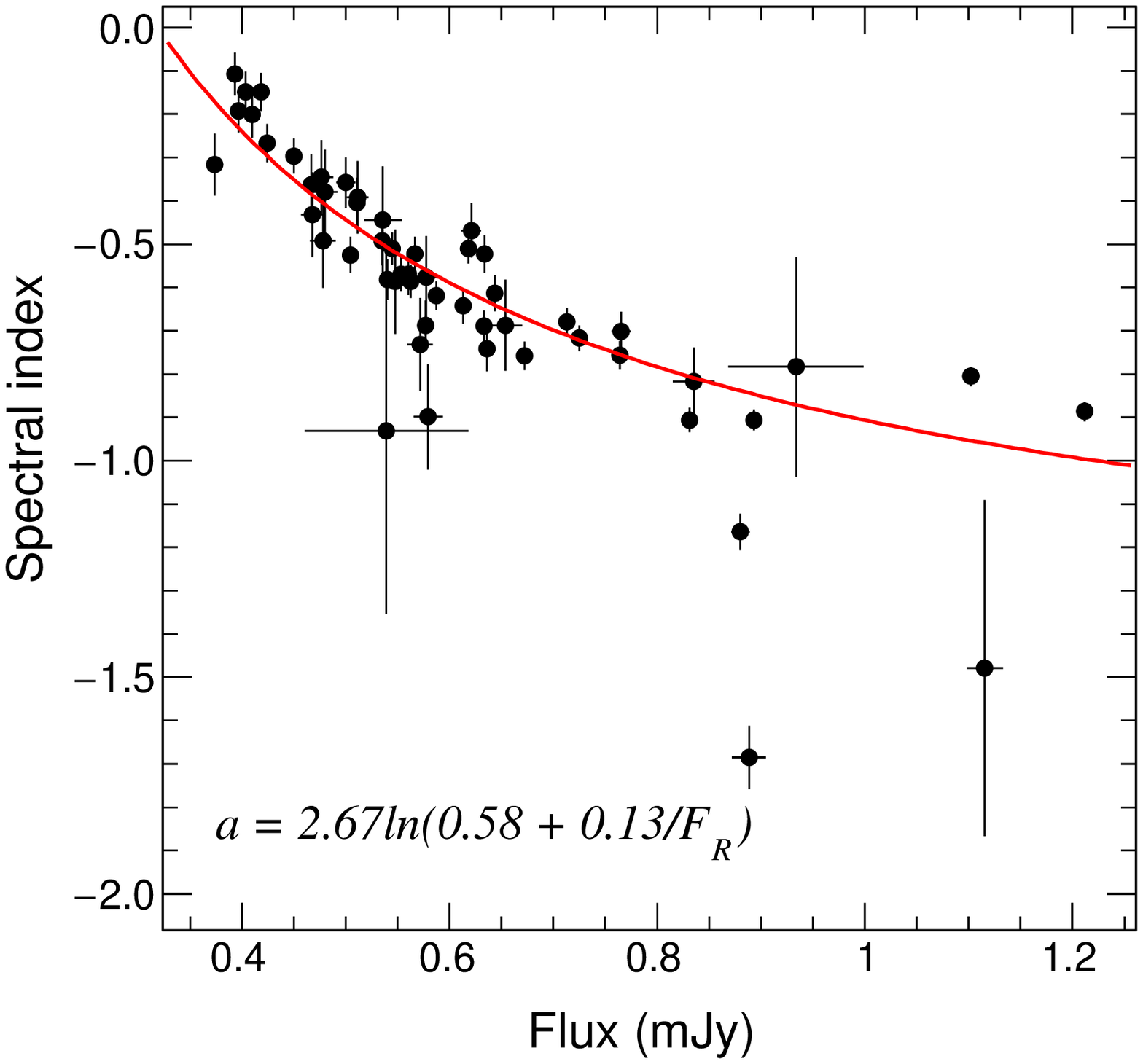}
\caption{Same as Figure~\ref{3C454alphaflux} in the main text, for PKS 1244-255. \label{1244-255}}
\end{figure}

\begin{figure}
\epsscale{0.7}
\plottwo{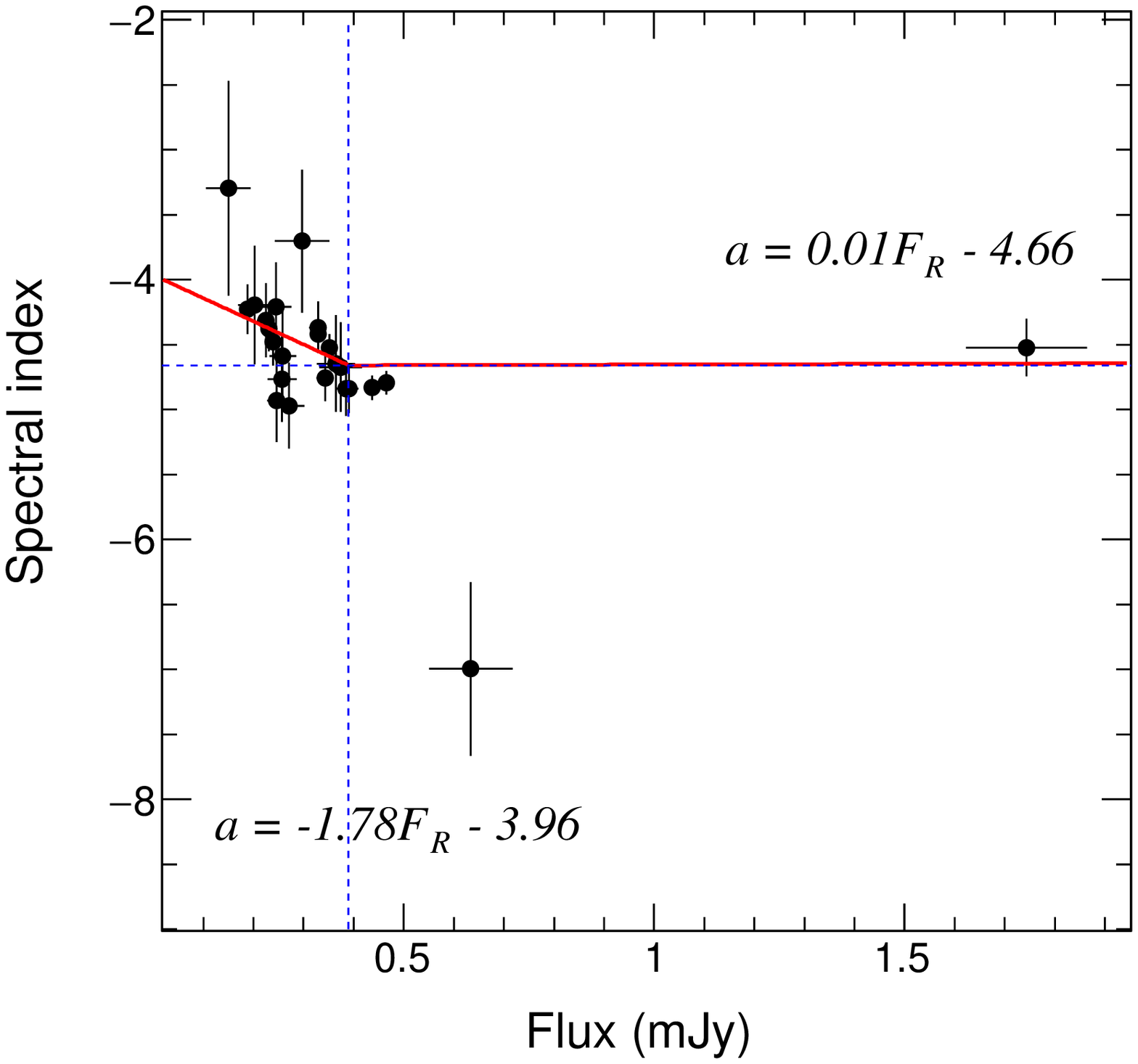}{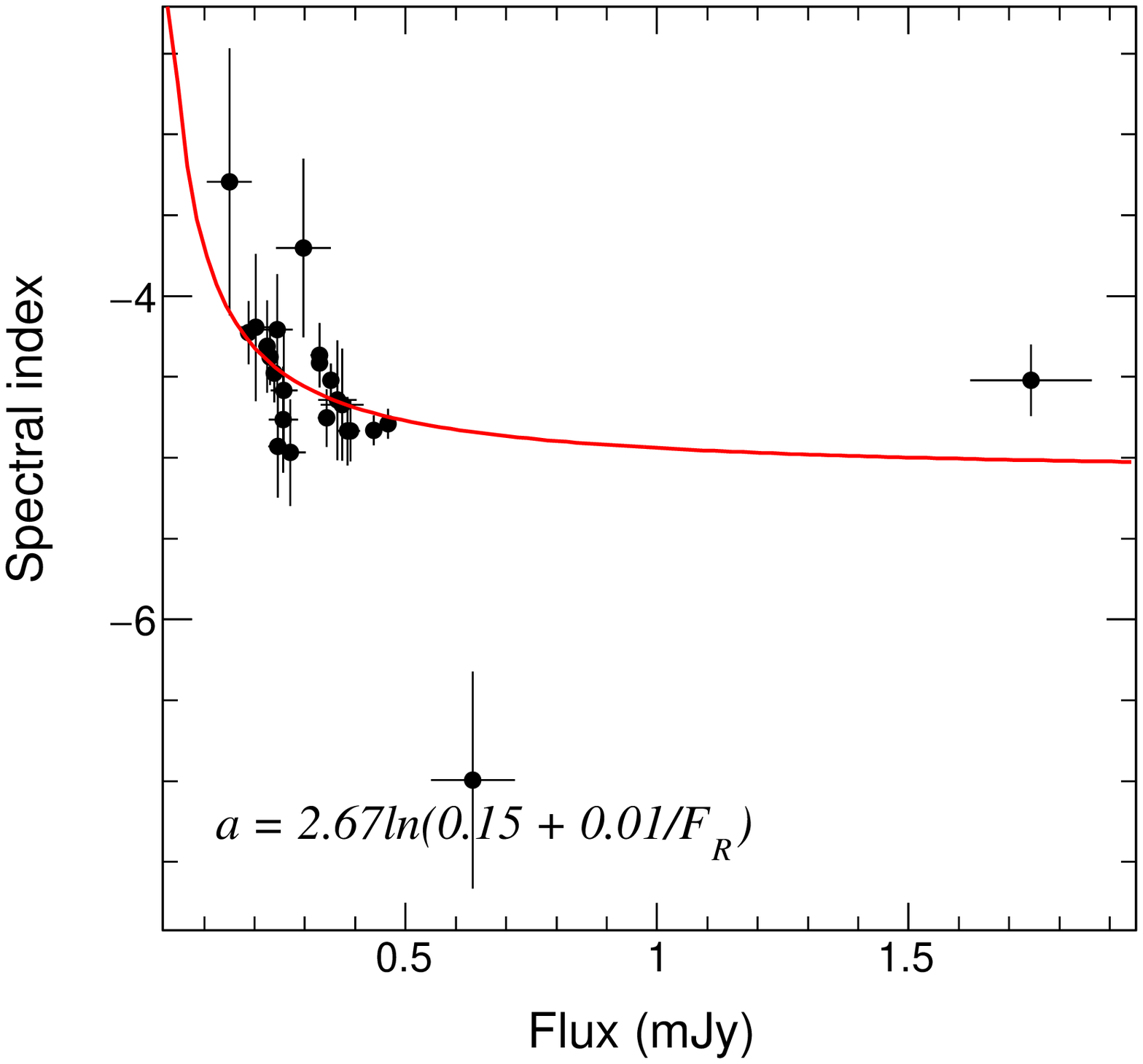}
\caption{Same as Figure~\ref{3C454alphaflux} in the main text, for PKS 1329-049. \label{1329-049}}
\end{figure}

\begin{figure}
\epsscale{0.7}
\plottwo{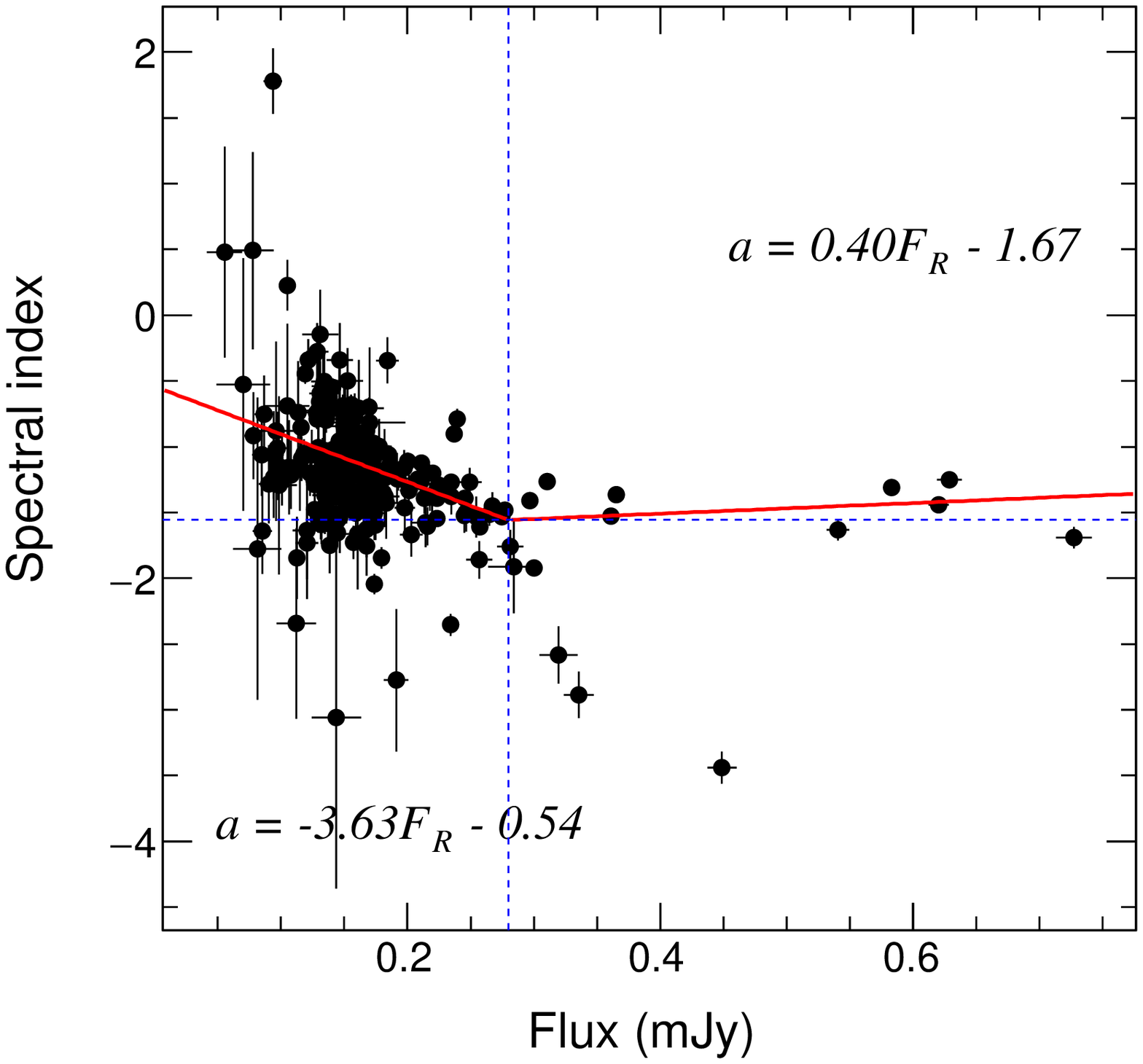}{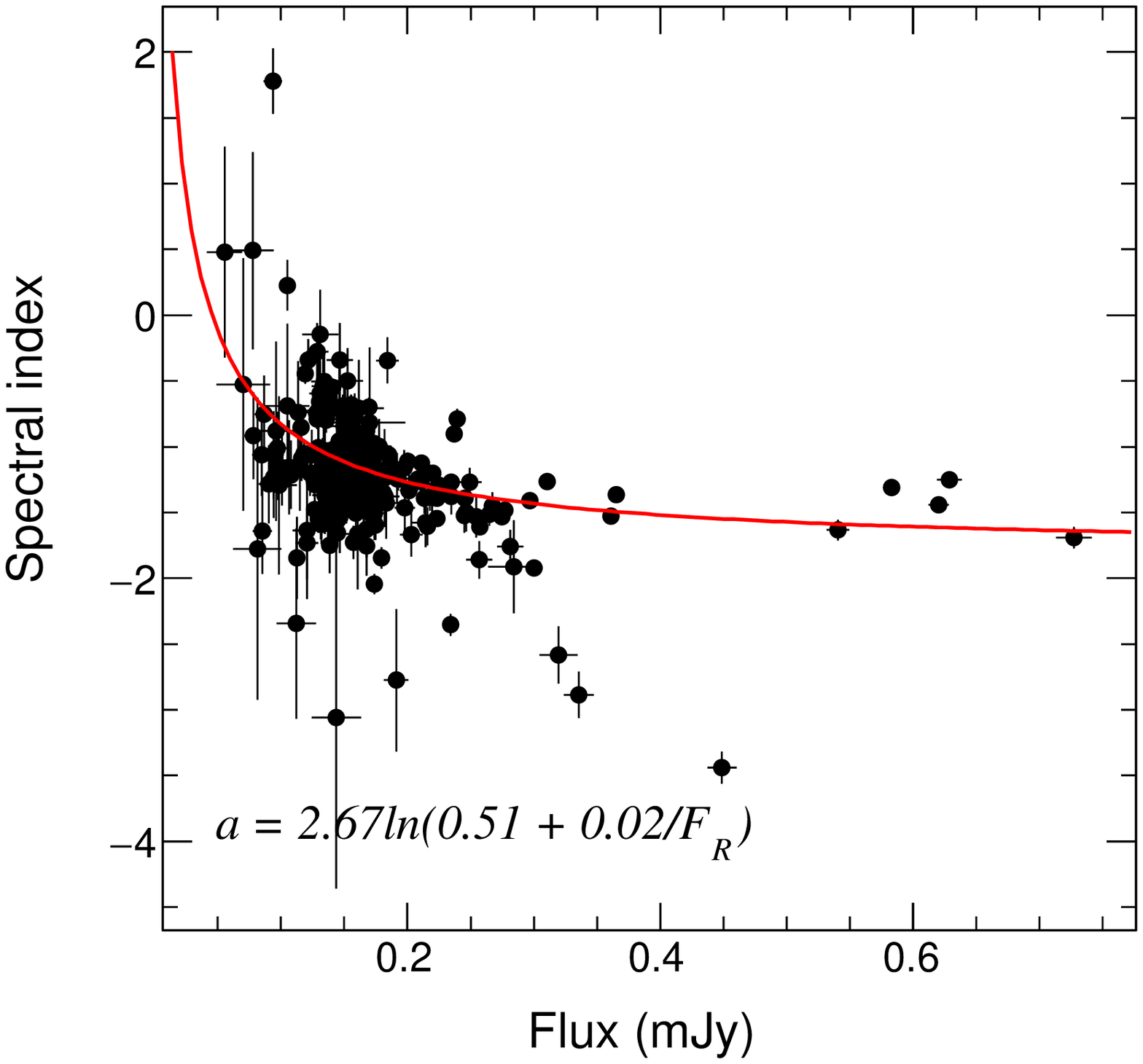}
\caption{Same as Figure~\ref{3C454alphaflux} in the main text, for PKS B1406-076. \label{1406-076}}
\end{figure}

\begin{figure}
\epsscale{0.7}
\plottwo{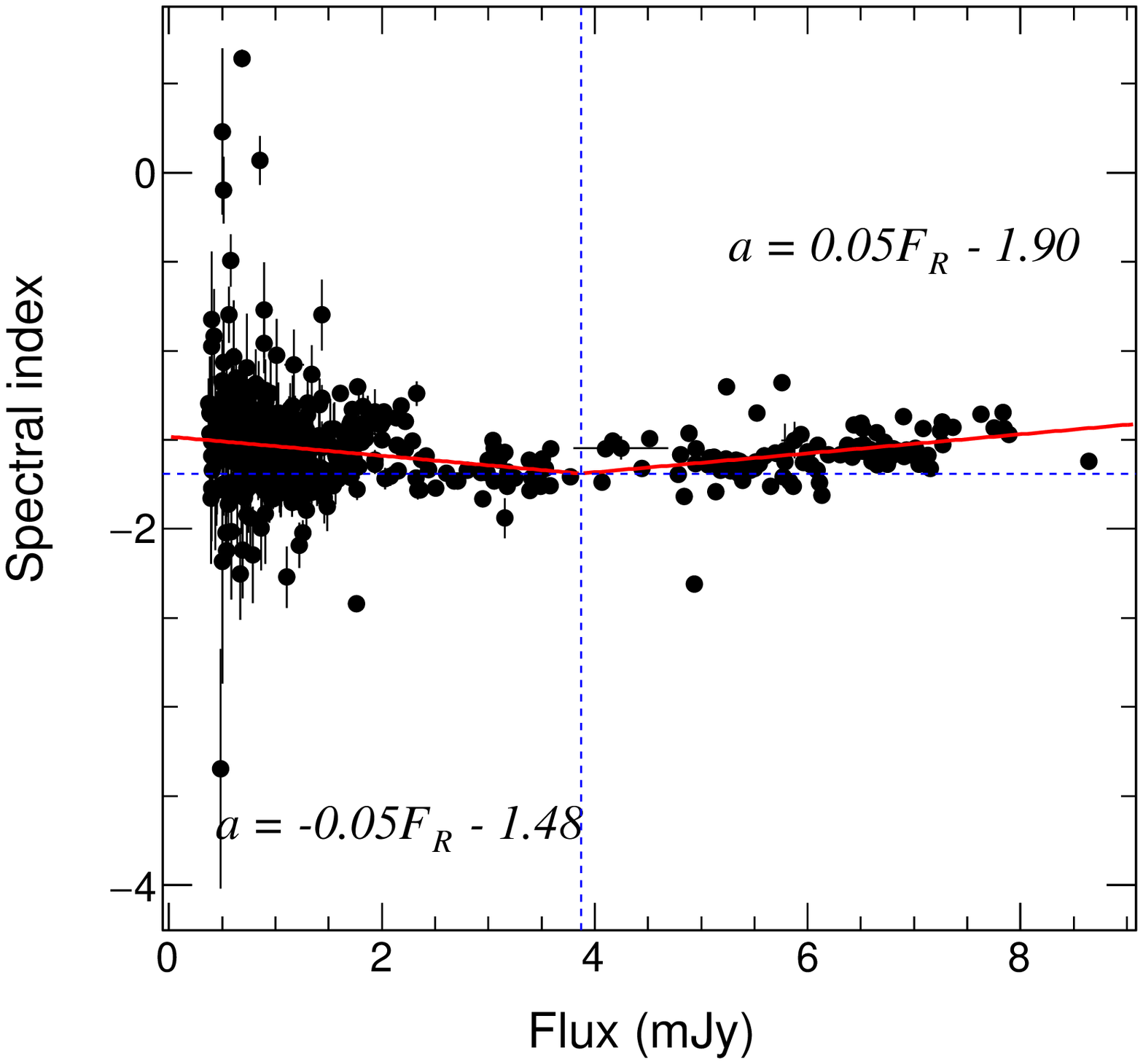}{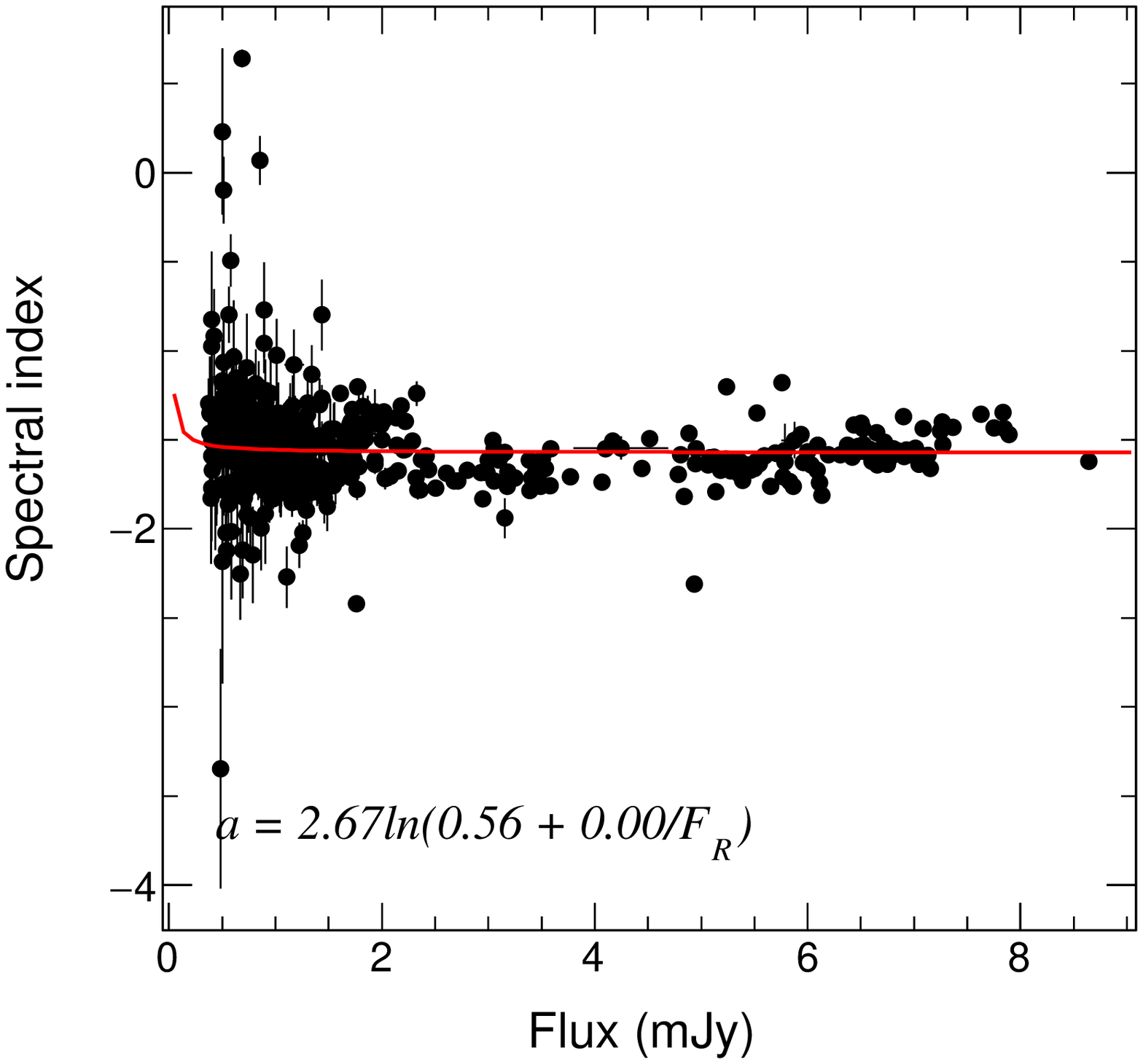}
\caption{Same as Figure~\ref{3C454alphaflux} in the main text, for PKS 1424-41. \label{1424-41}}
\end{figure}

\begin{figure}
\epsscale{0.7}
\plottwo{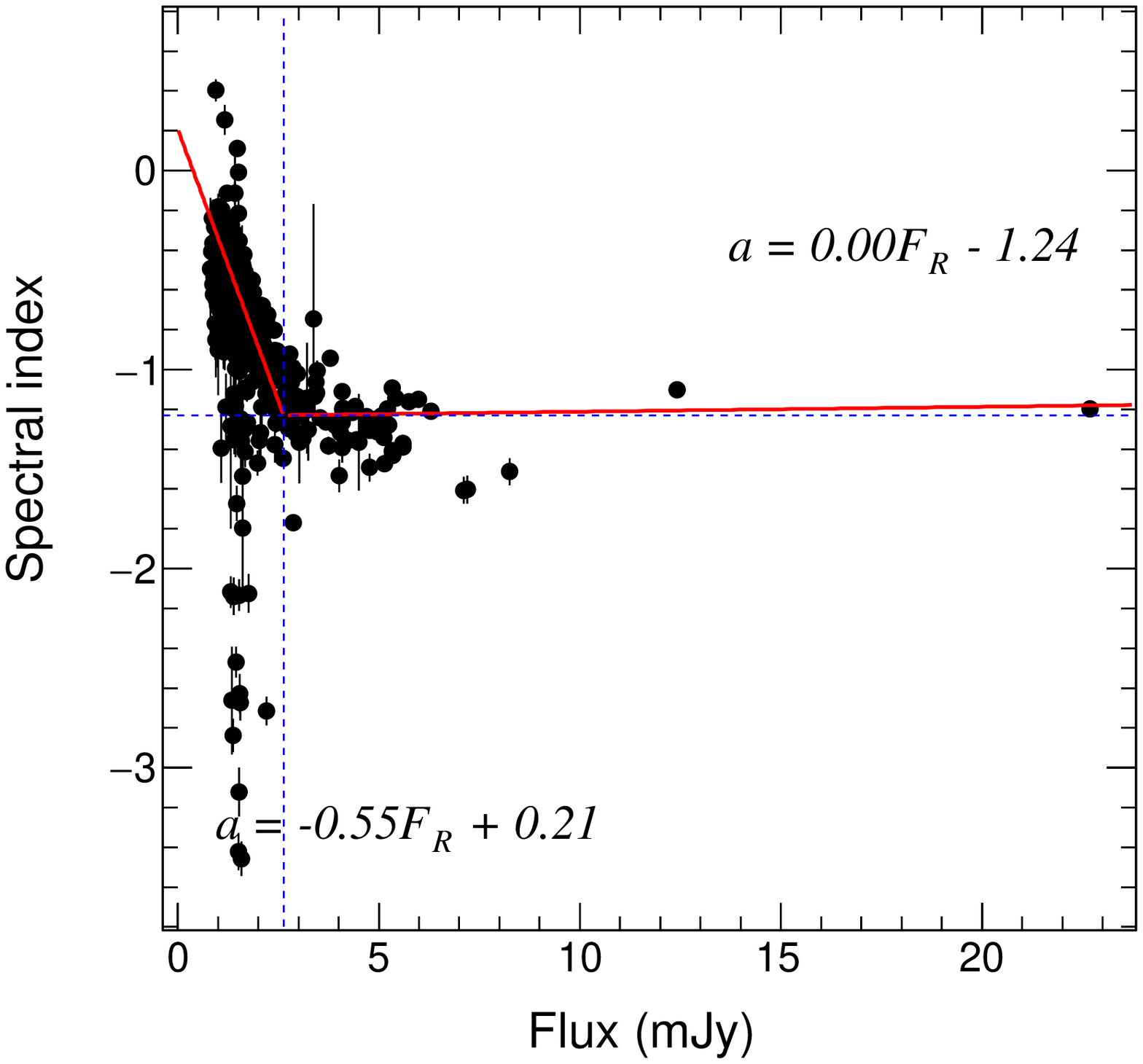}{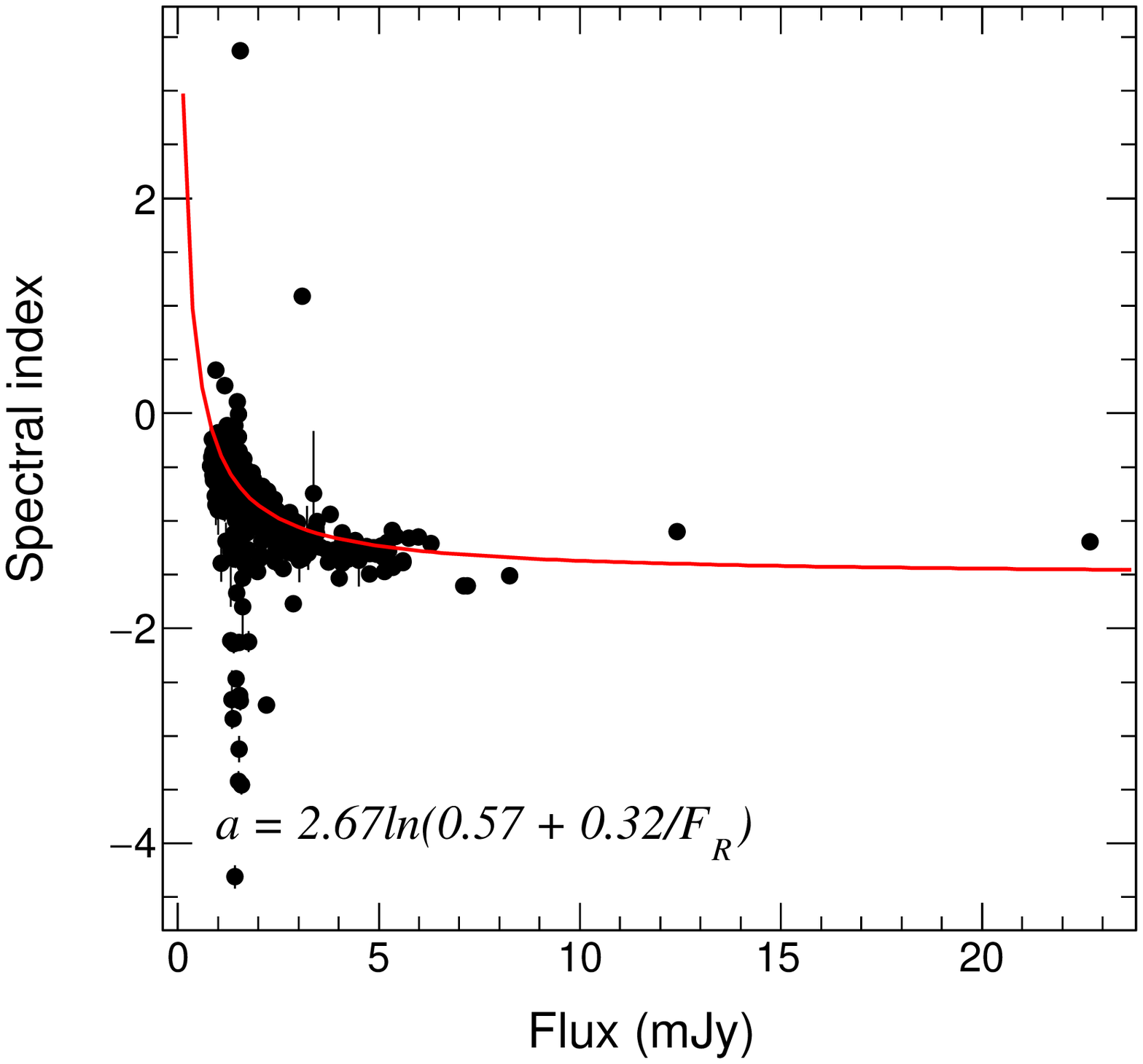}
\caption{Same as Figure~\ref{3C454alphaflux} in the main text, for PKS 1510-089. \label{1510-089}}
\end{figure}

\begin{figure}
\epsscale{0.7}
\plottwo{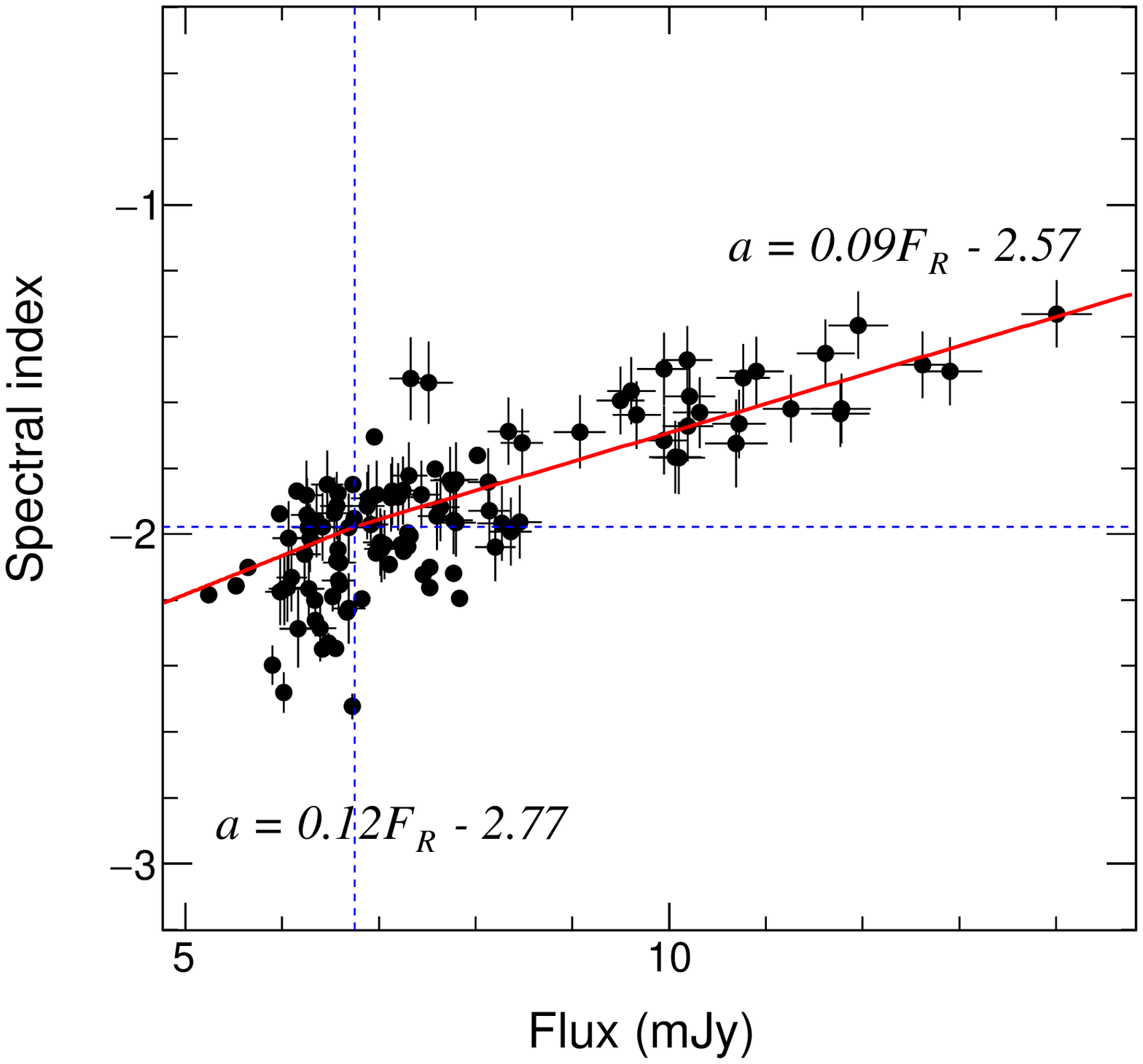}{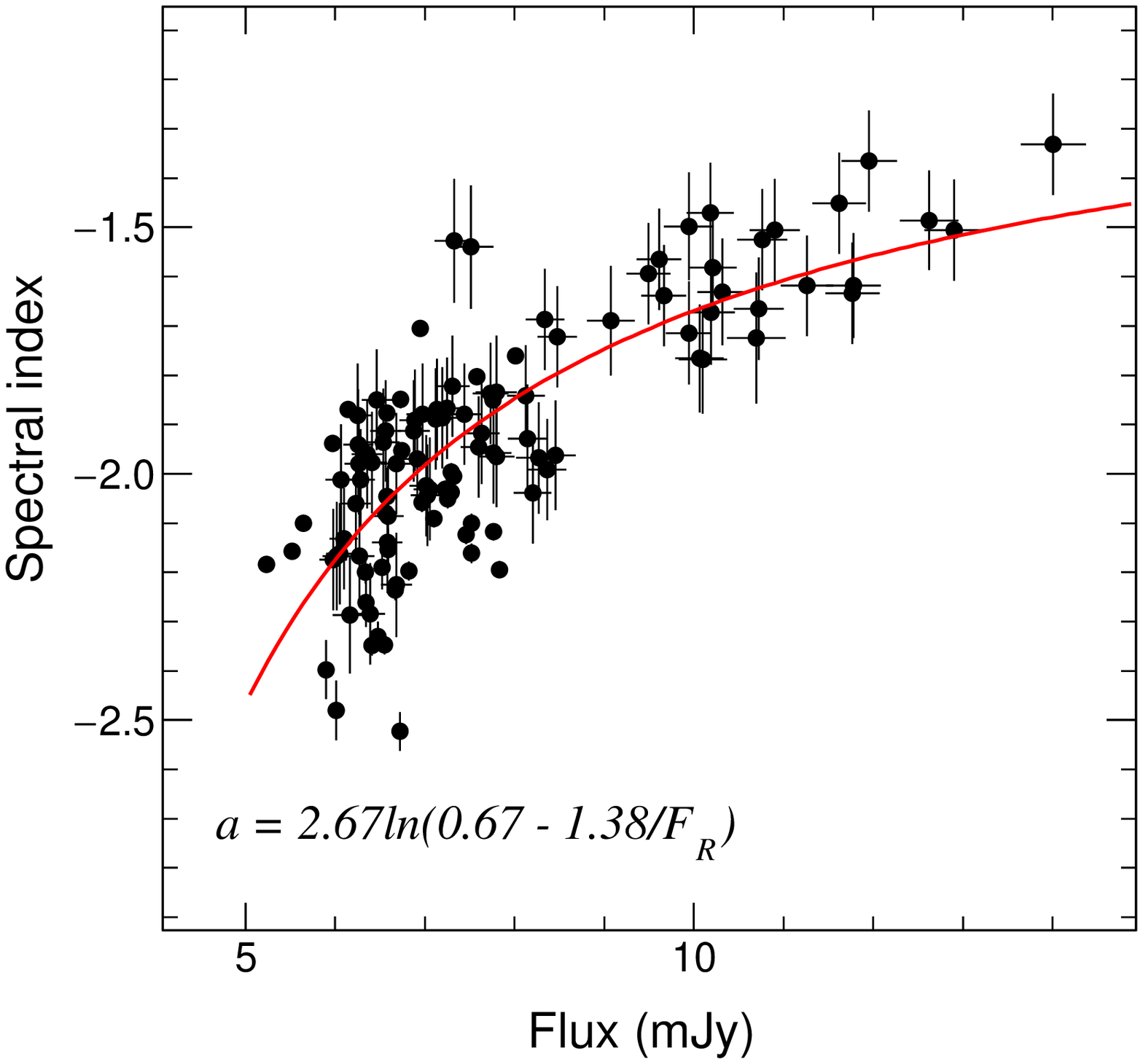}
\caption{Same as Figure~\ref{3C454alphaflux} in the main text, for AP Lib. \label{1514-241}}
\end{figure}

\begin{figure}
\epsscale{0.7}
\plottwo{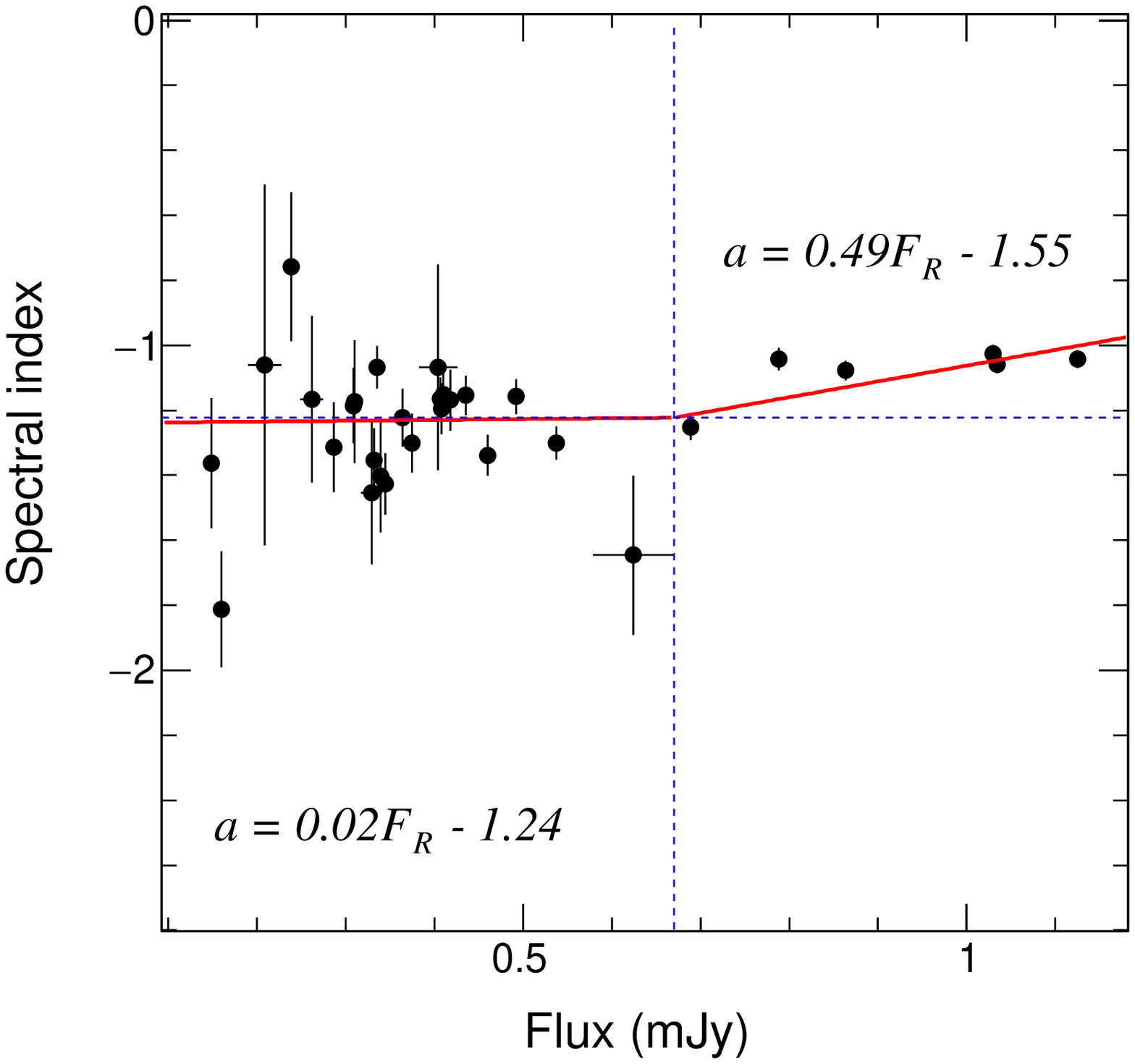}{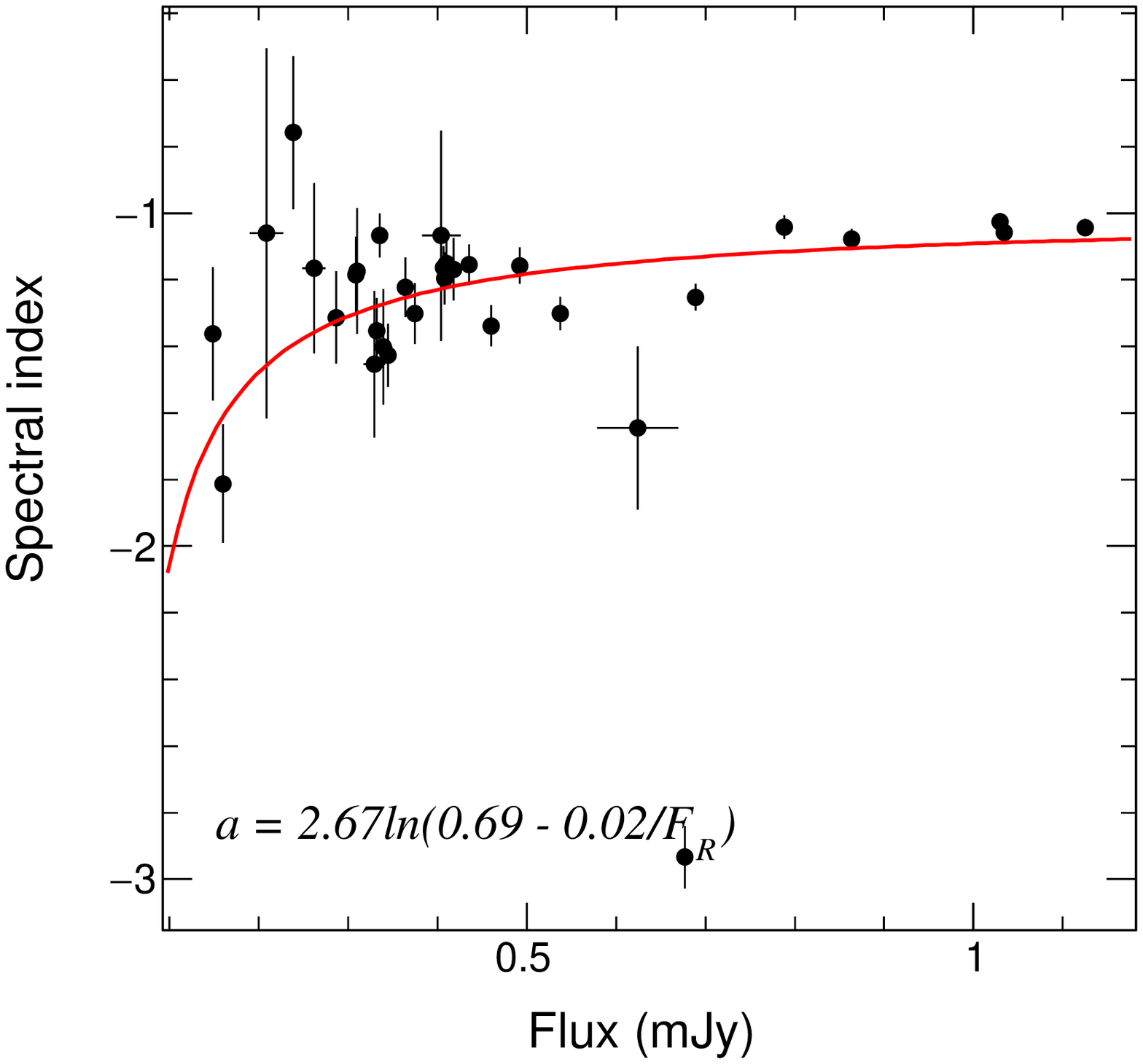}
\caption{Same as Figure~\ref{3C454alphaflux} in the main text, for PKS 1550-242. \label{1550-242}}
\end{figure}

\clearpage

\begin{figure}
\epsscale{0.7}
\plottwo{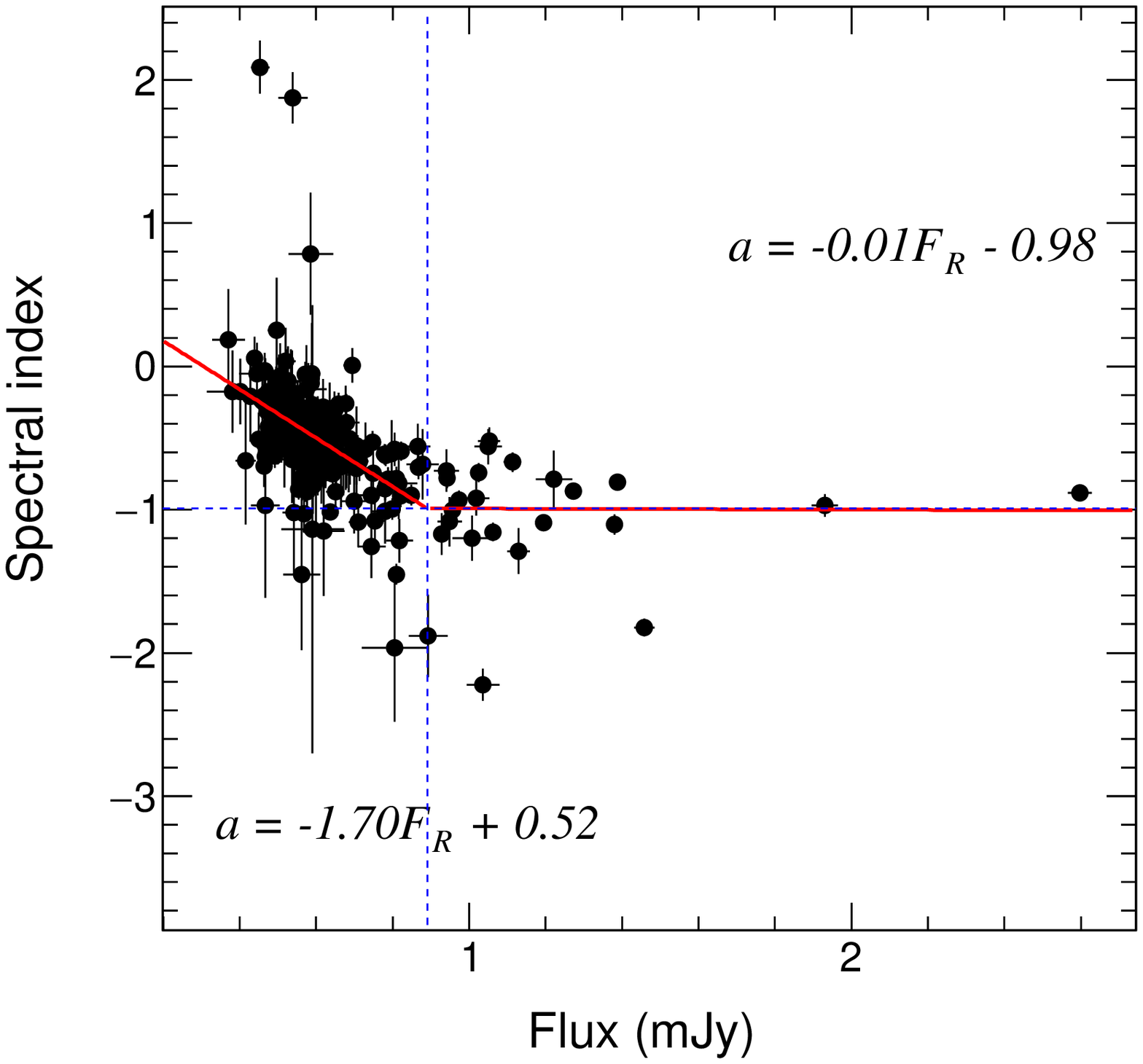}{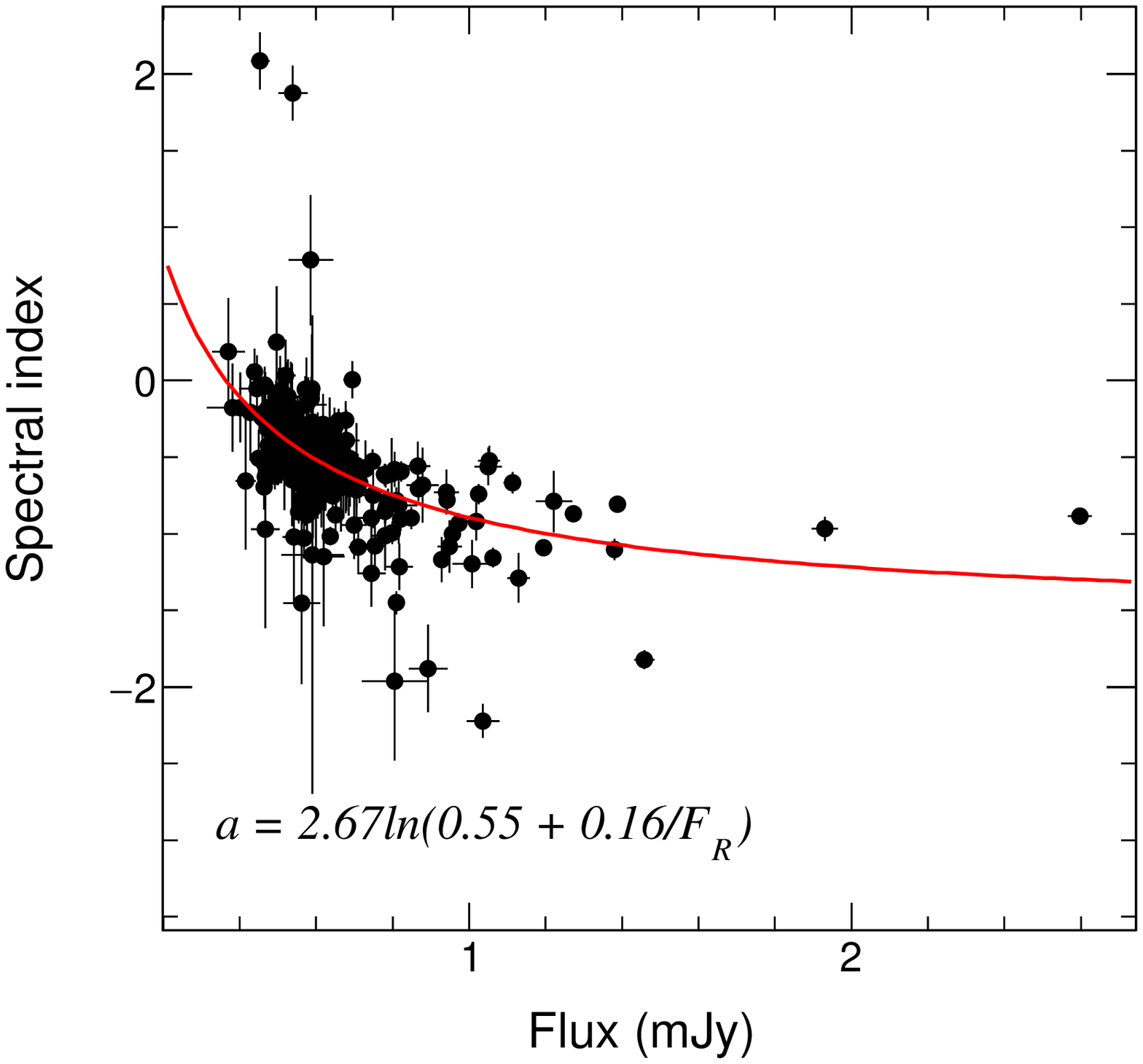}
\caption{Same as Figure~\ref{3C454alphaflux} in the main text, for PKS B1622-297. \label{1622-297}}
\end{figure}

\begin{figure}
\epsscale{0.7}
\plottwo{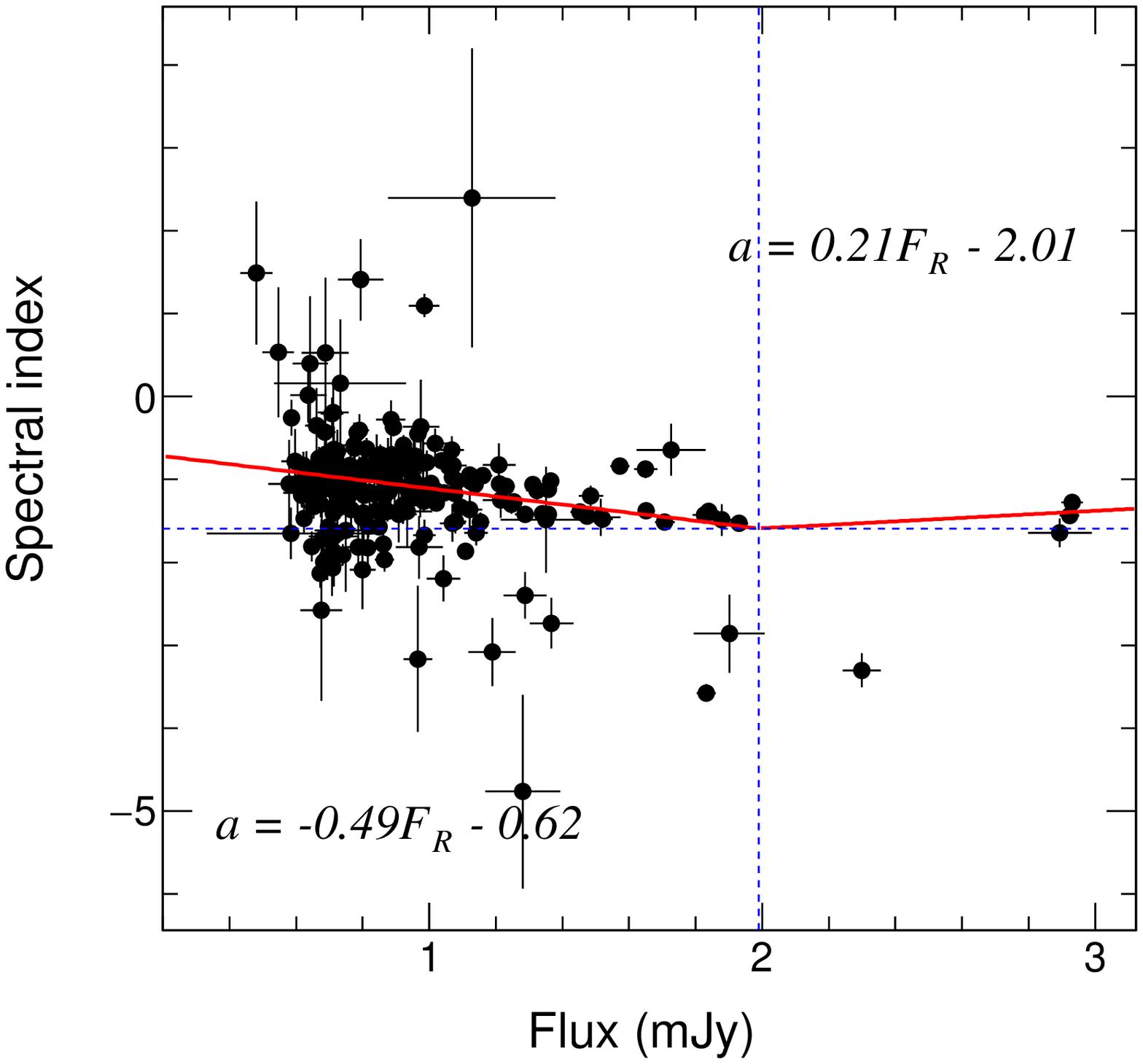}{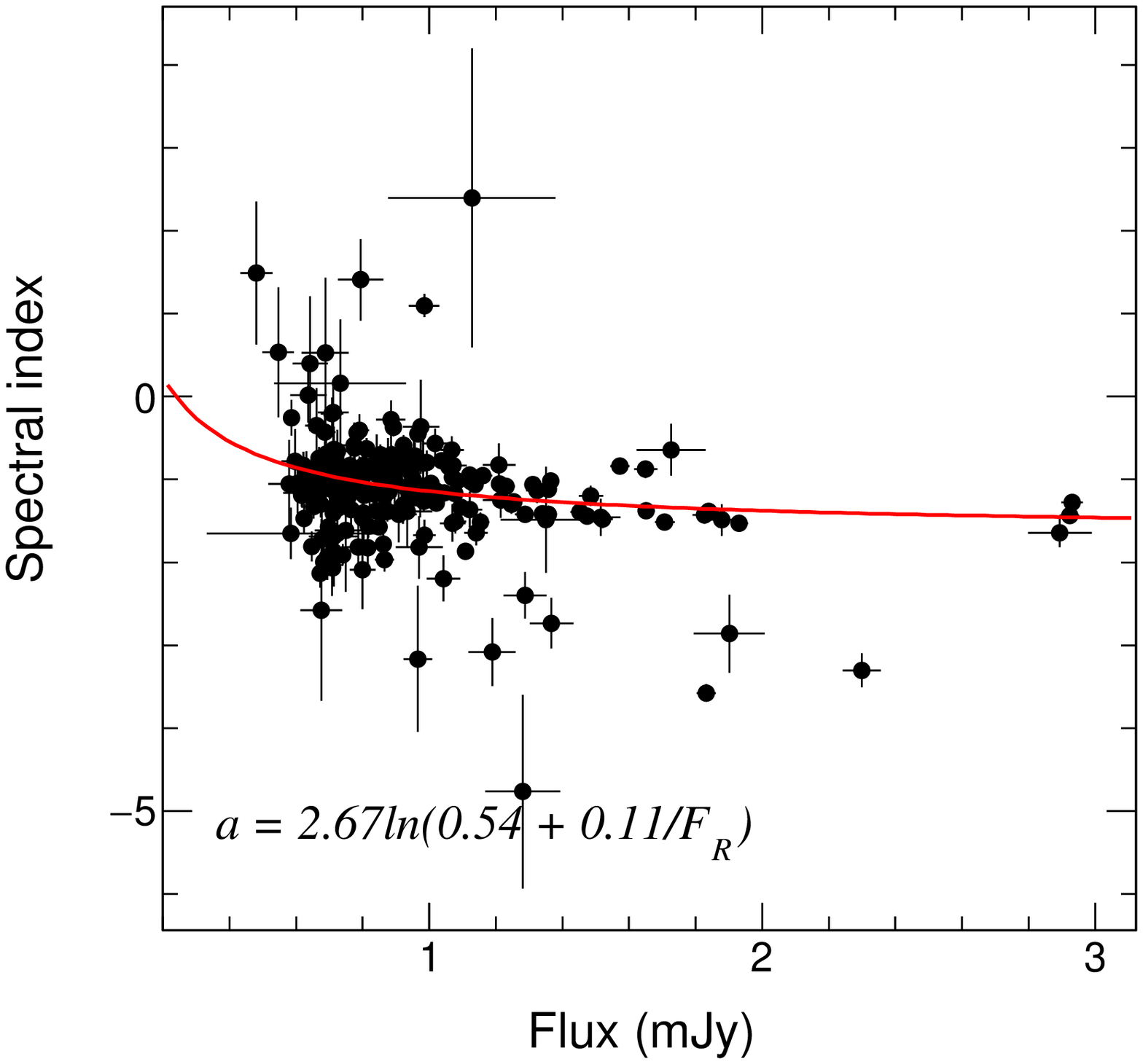}
\caption{Same as Figure~\ref{3C454alphaflux} in the main text, for PKS 1730-13. \label{1730-130}}
\end{figure}

\begin{figure}
\epsscale{0.7}
\plottwo{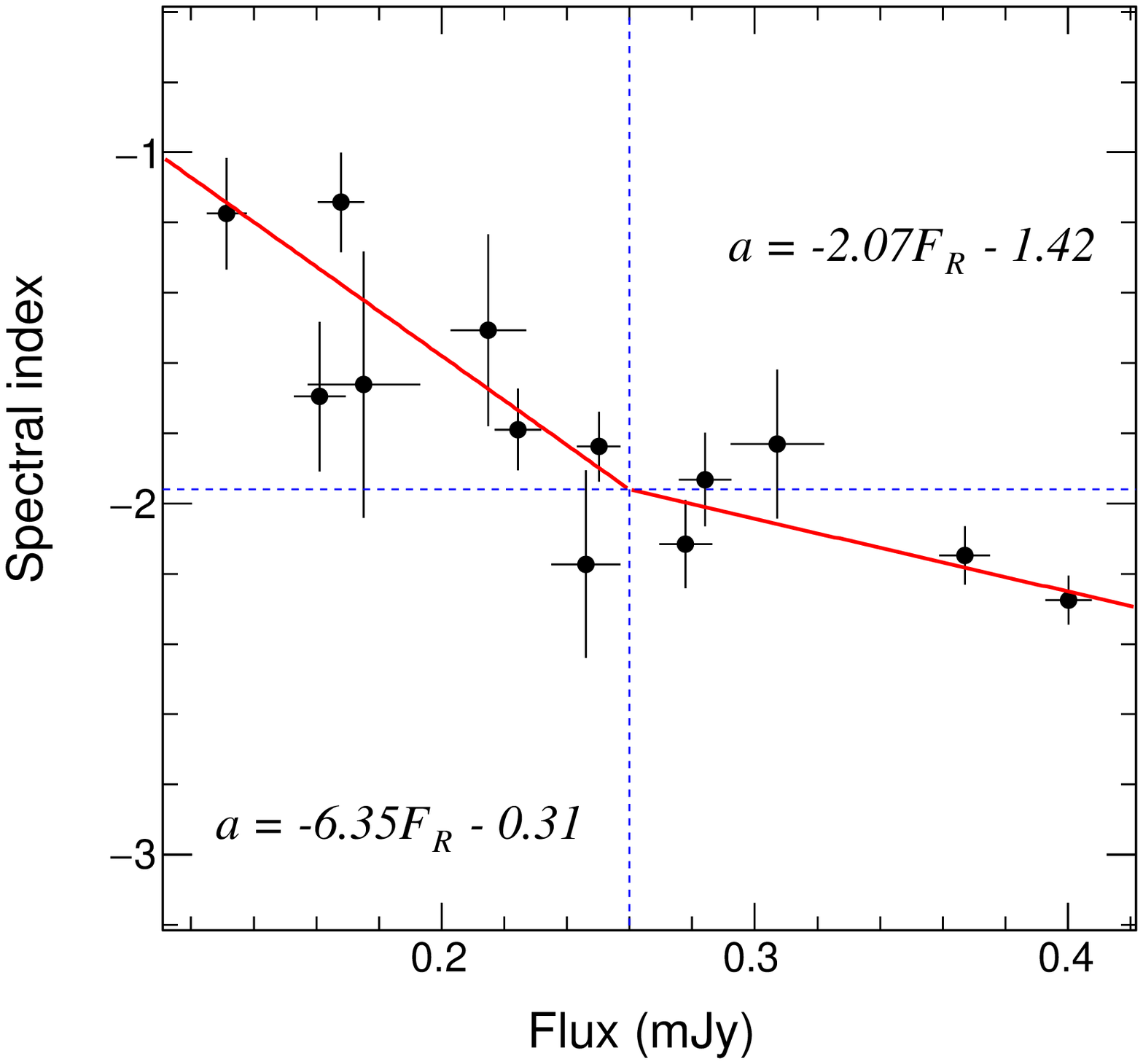}{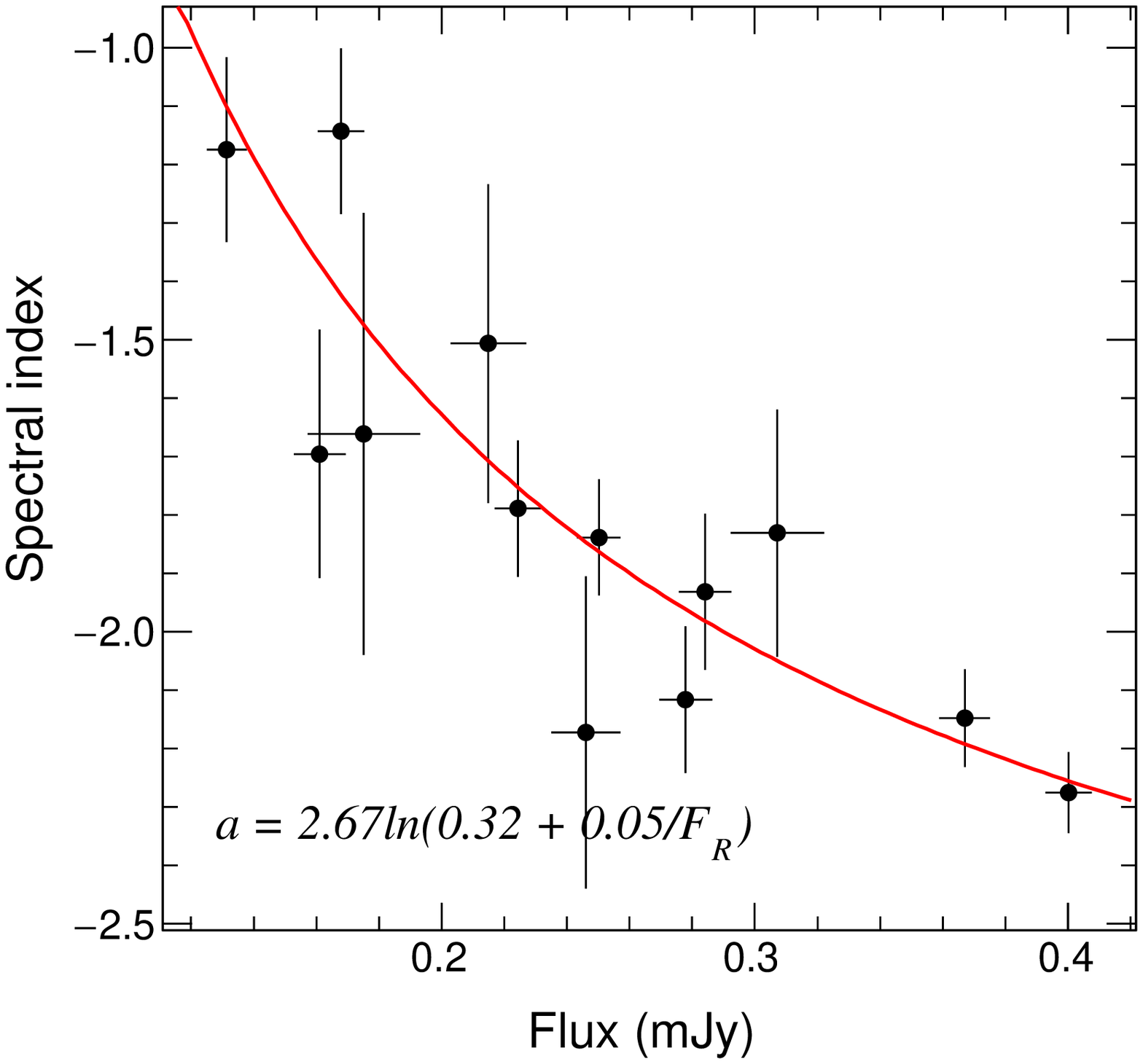}
\caption{Same as Figure~\ref{3C454alphaflux} in the main text, for PKS 1824-582. \label{1824-582}}
\end{figure}

\begin{figure}
\epsscale{0.7}
\plottwo{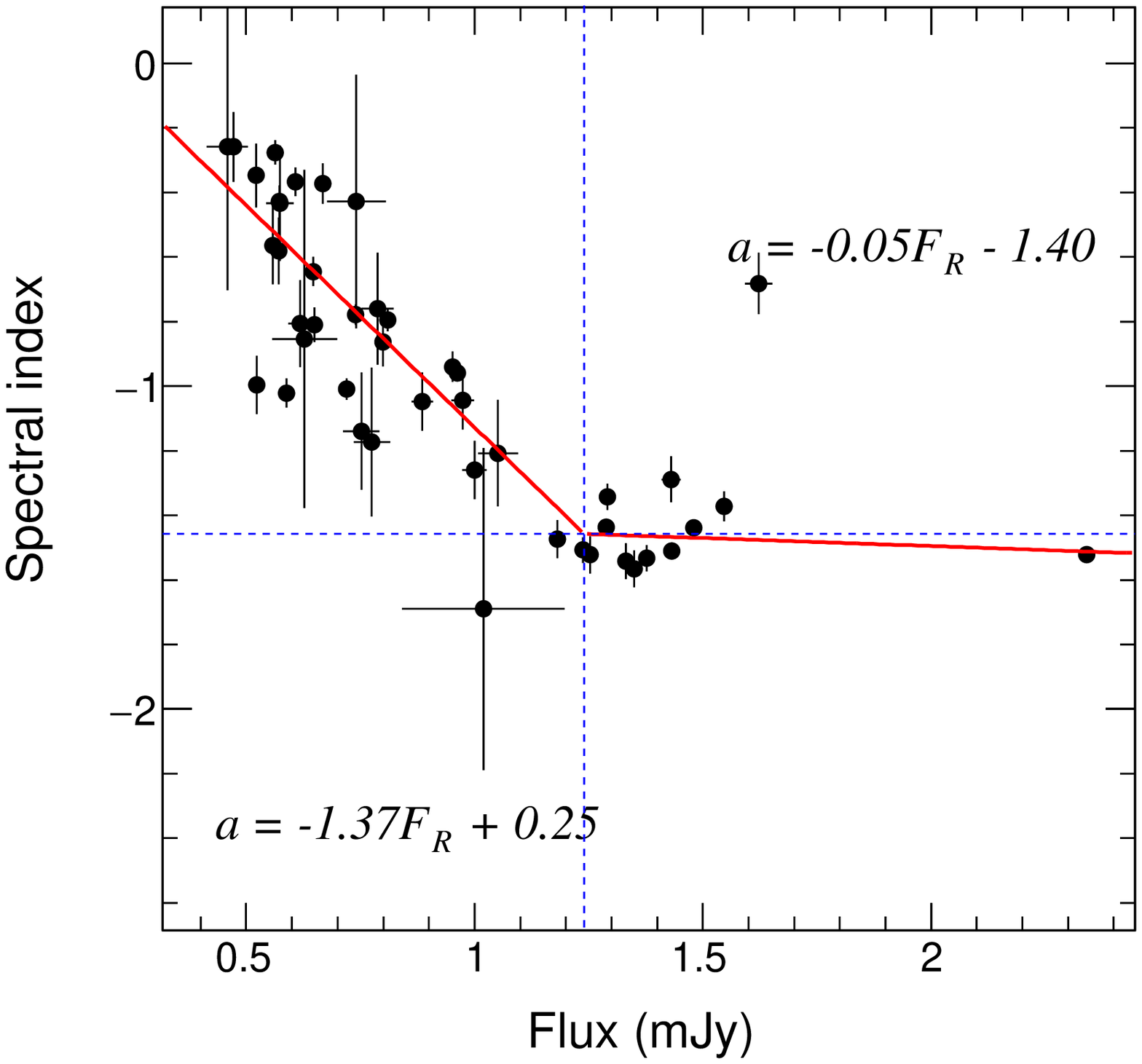}{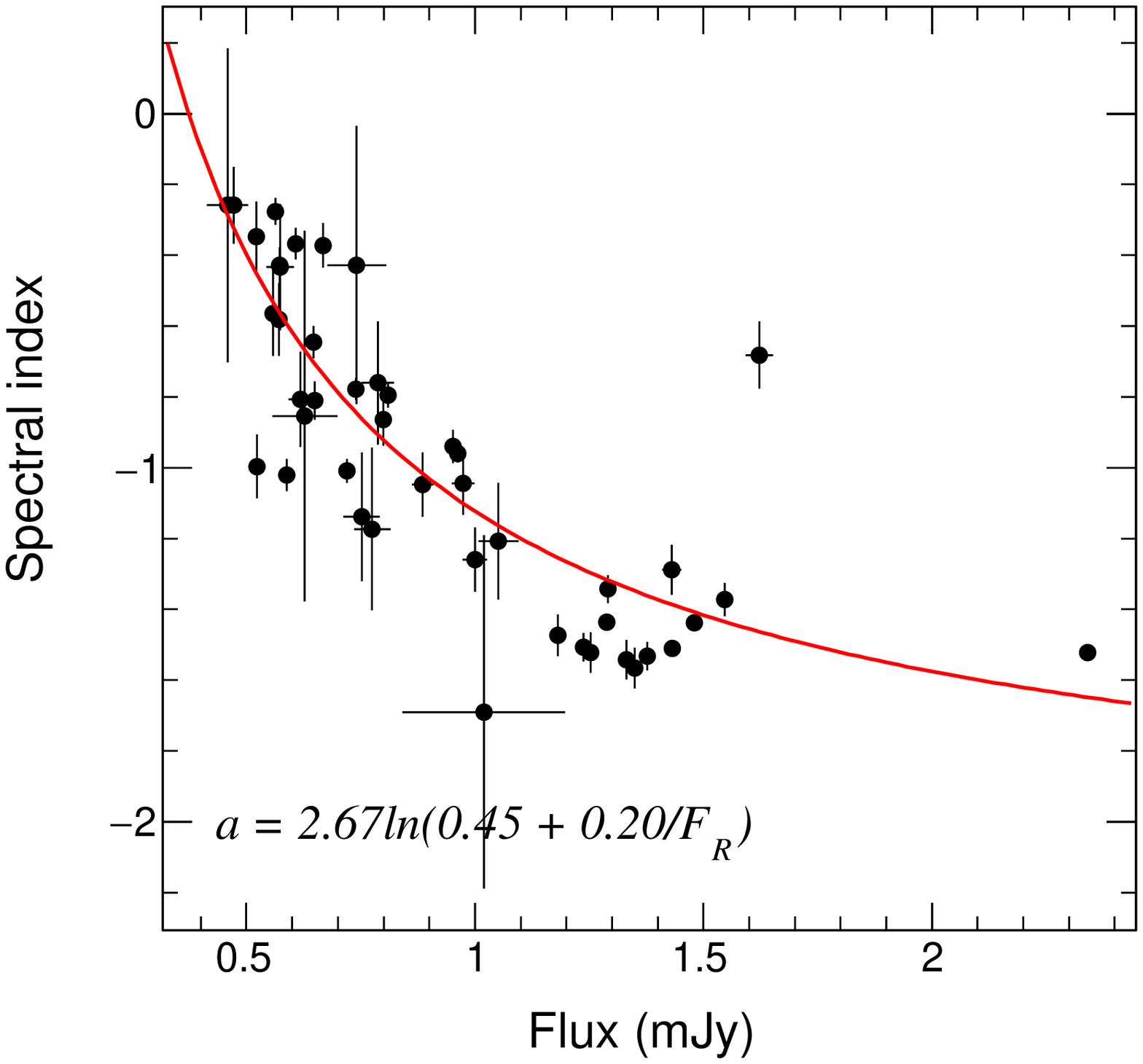}
\caption{Same as Figure~\ref{3C454alphaflux} in the main text, for PKS 1954-388. \label{1954-388}}
\end{figure}

\begin{figure}
\epsscale{0.7}
\plottwo{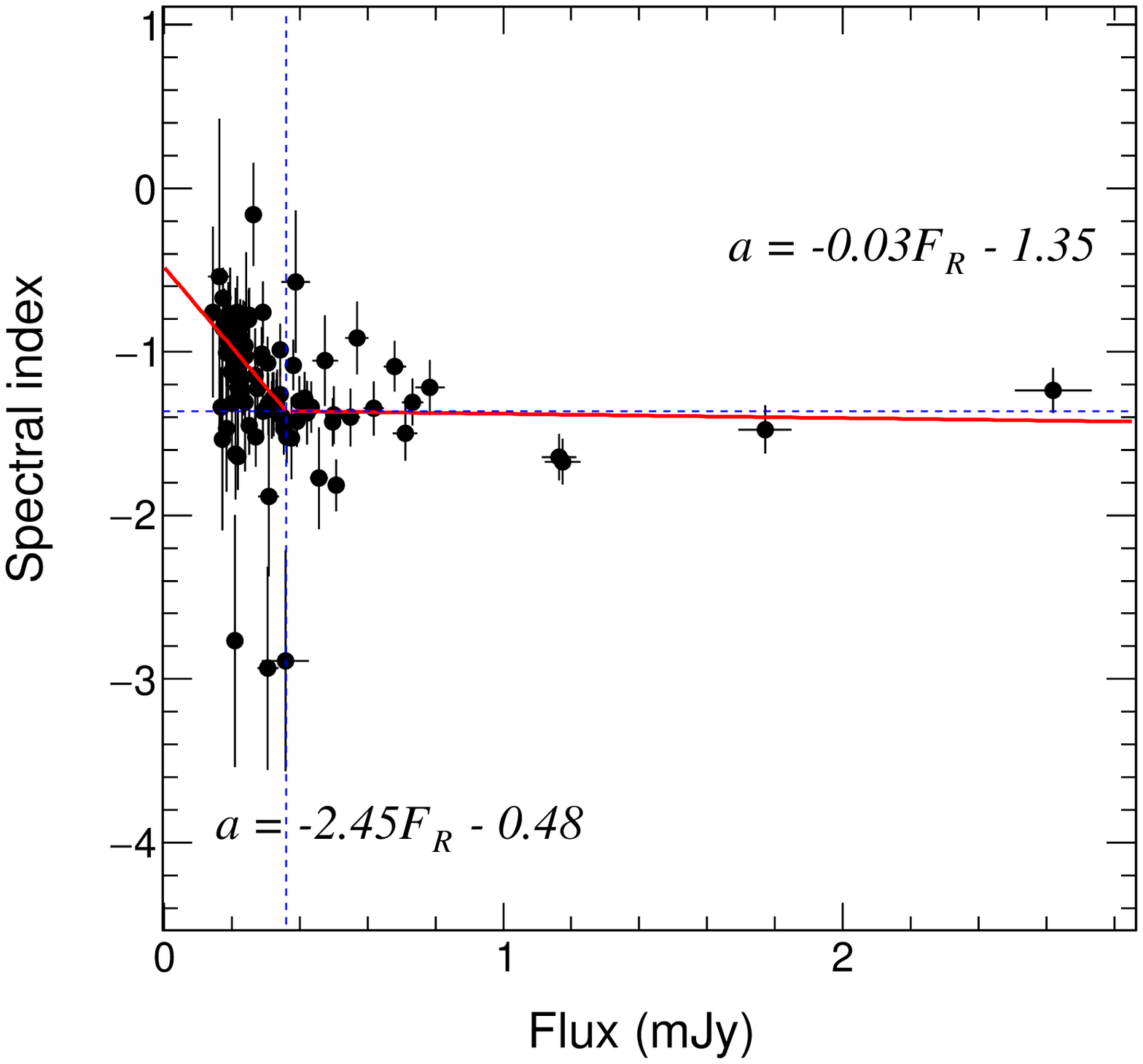}{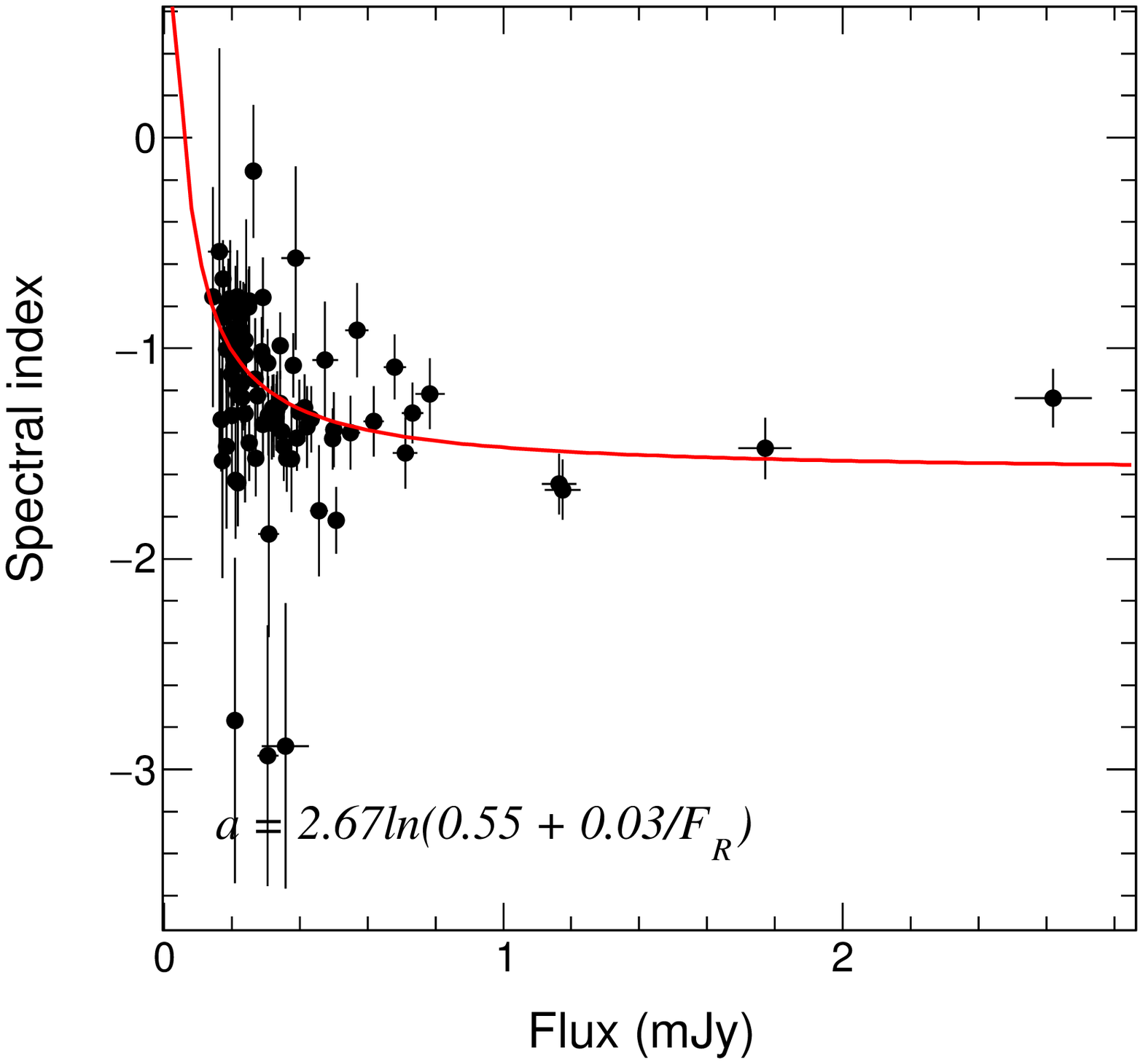}
\caption{Same as Figure~\ref{3C454alphaflux} in the main text, for PKS 2023-07. \label{2023-07}}
\end{figure}

\begin{figure}
\epsscale{0.7}
\plottwo{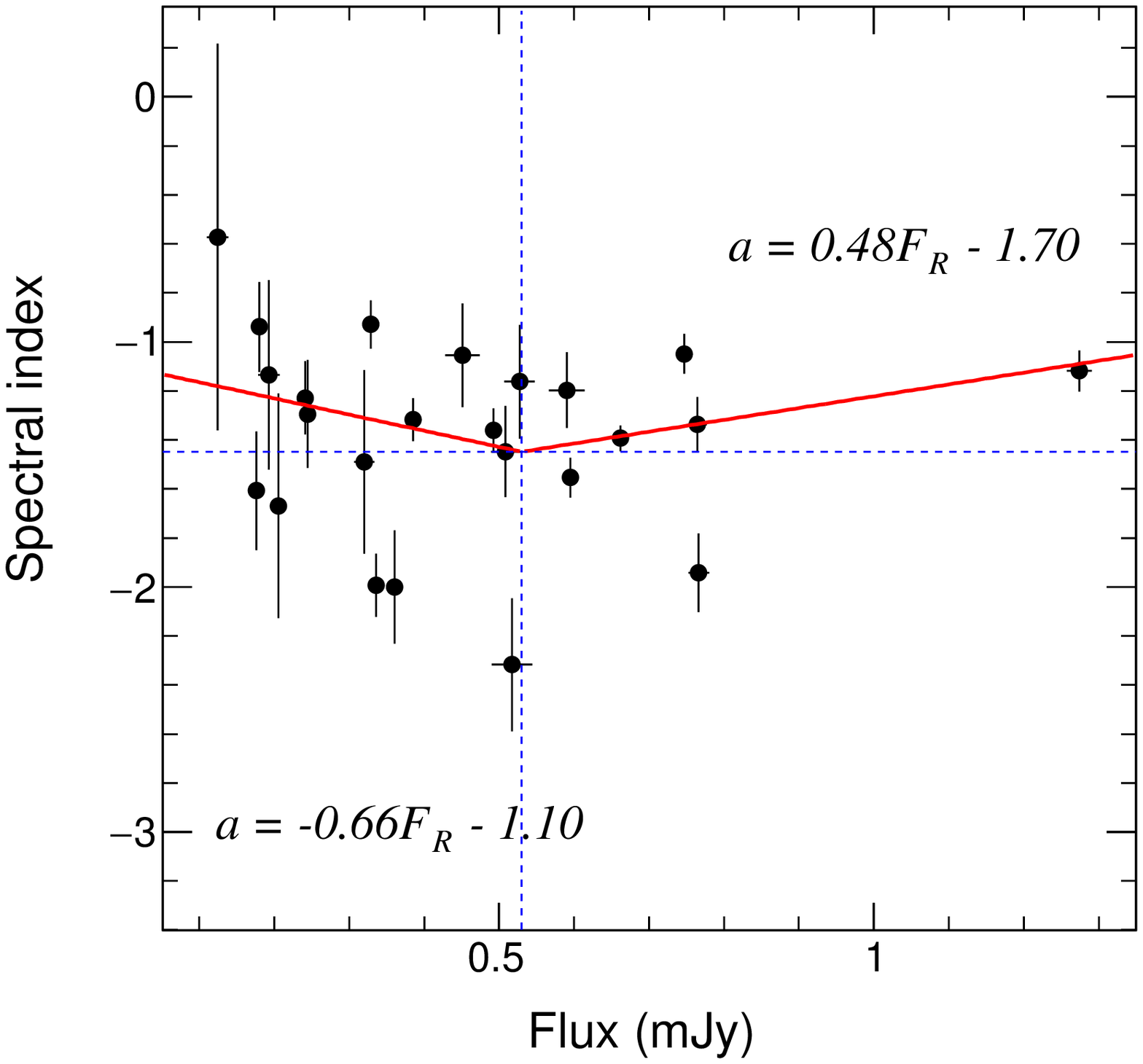}{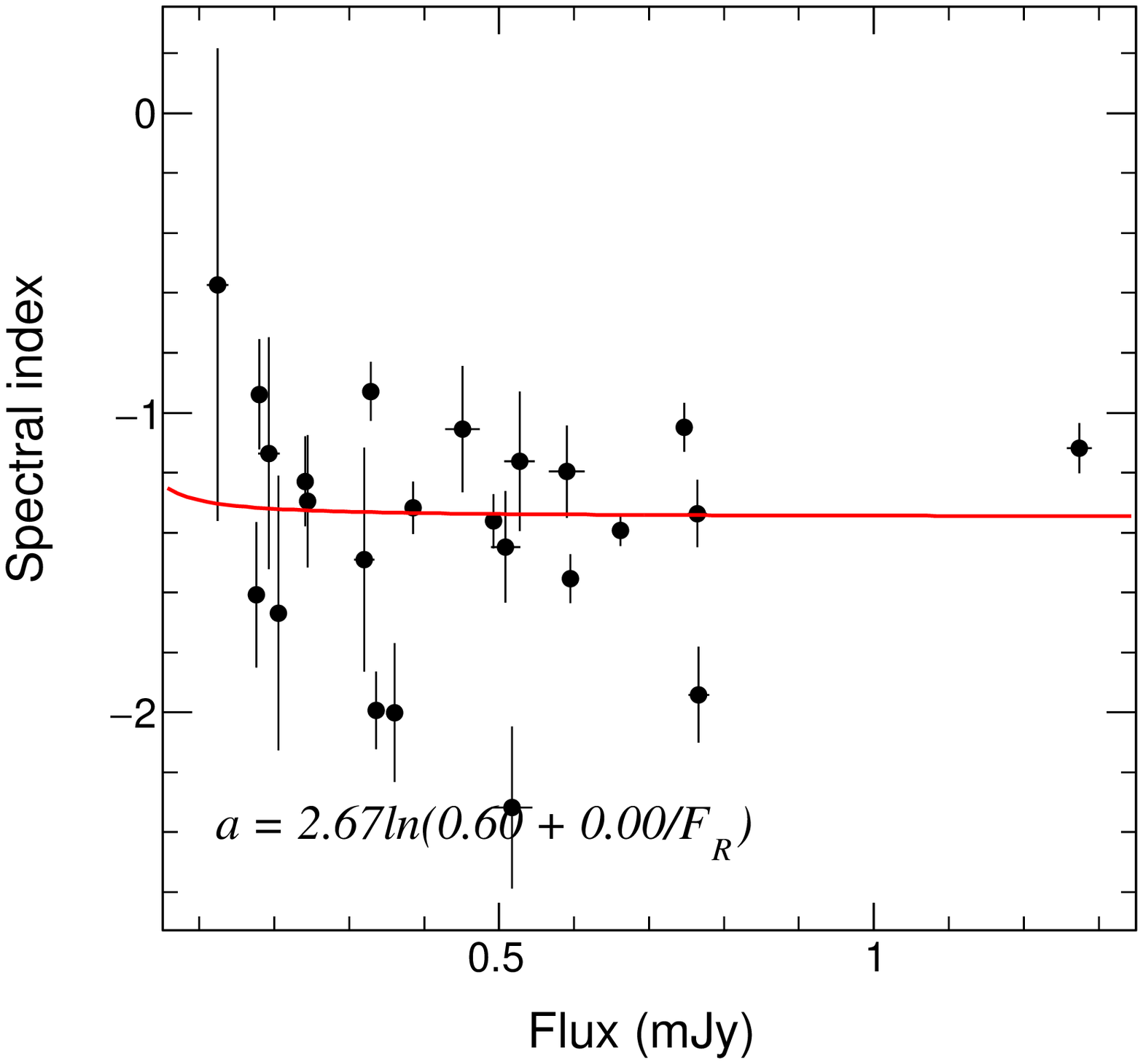}
\caption{Same as Figure~\ref{3C454alphaflux} in the main text, for PKS 2032+107. \label{2032+107}}
\end{figure}

\begin{figure}
\epsscale{0.7}
\plottwo{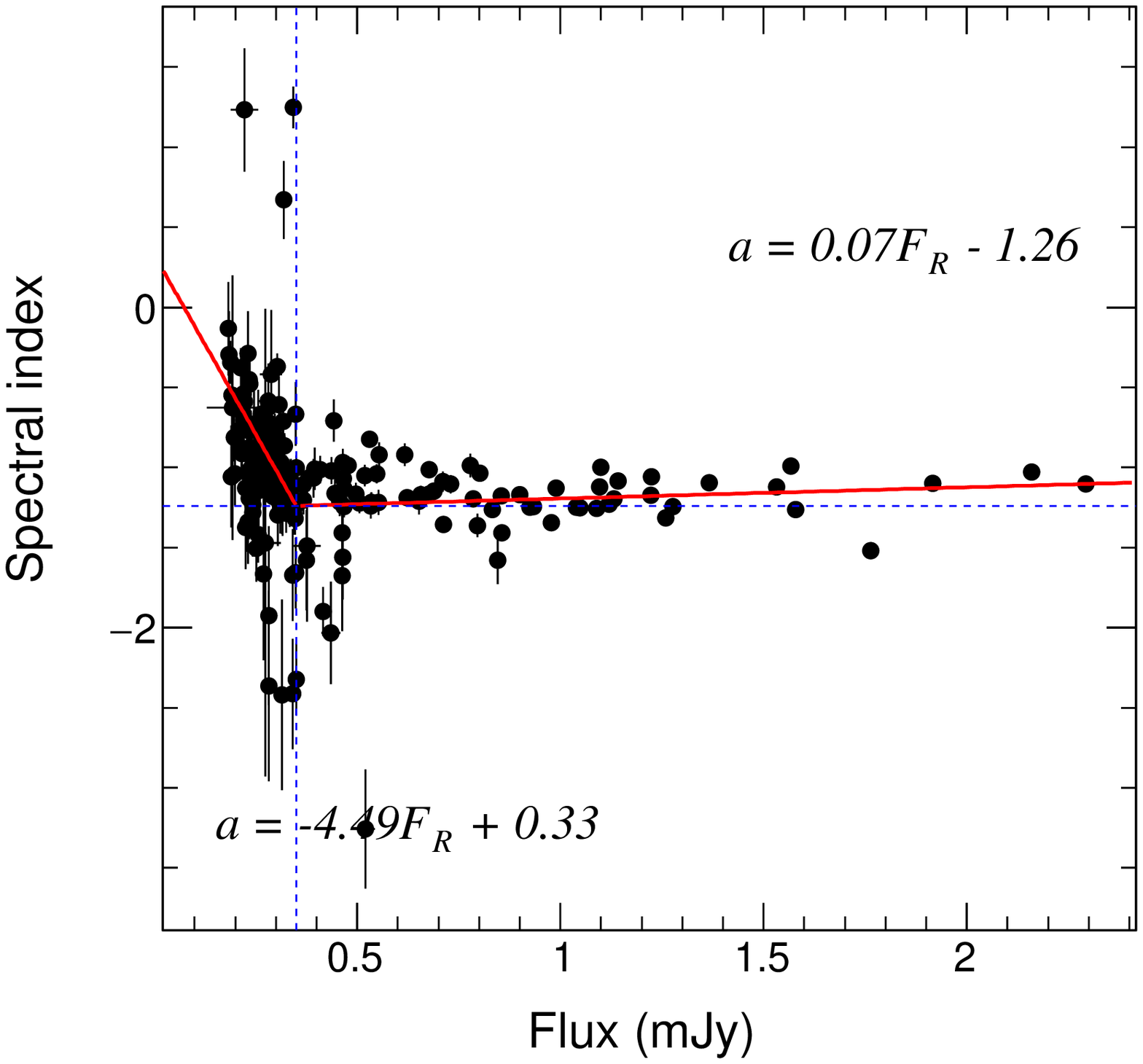}{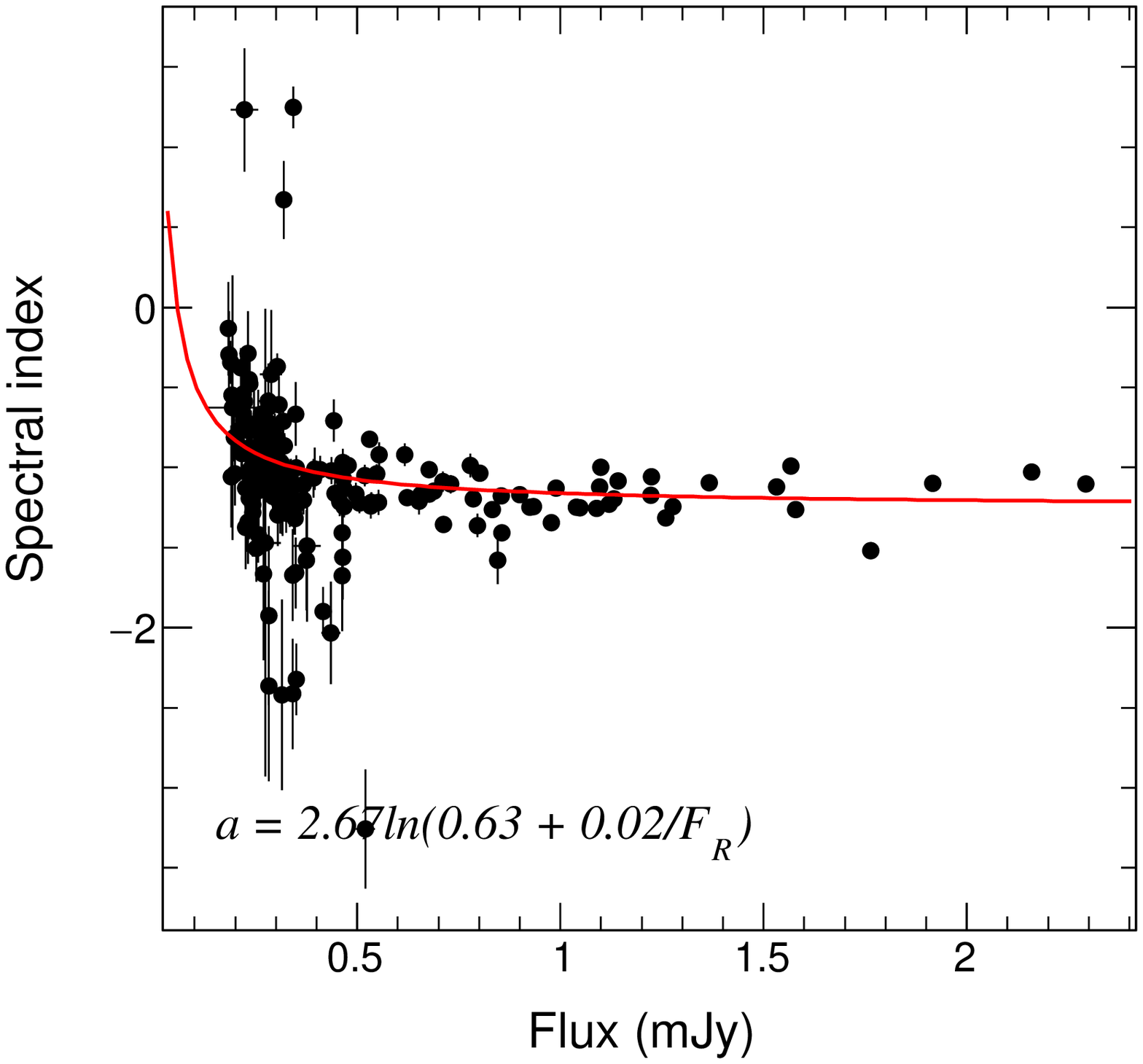}
\caption{Same as Figure~\ref{3C454alphaflux} in the main text, for PKS 2052-47. \label{2052-474}}
\end{figure}

\begin{figure}
\epsscale{0.7}
\plottwo{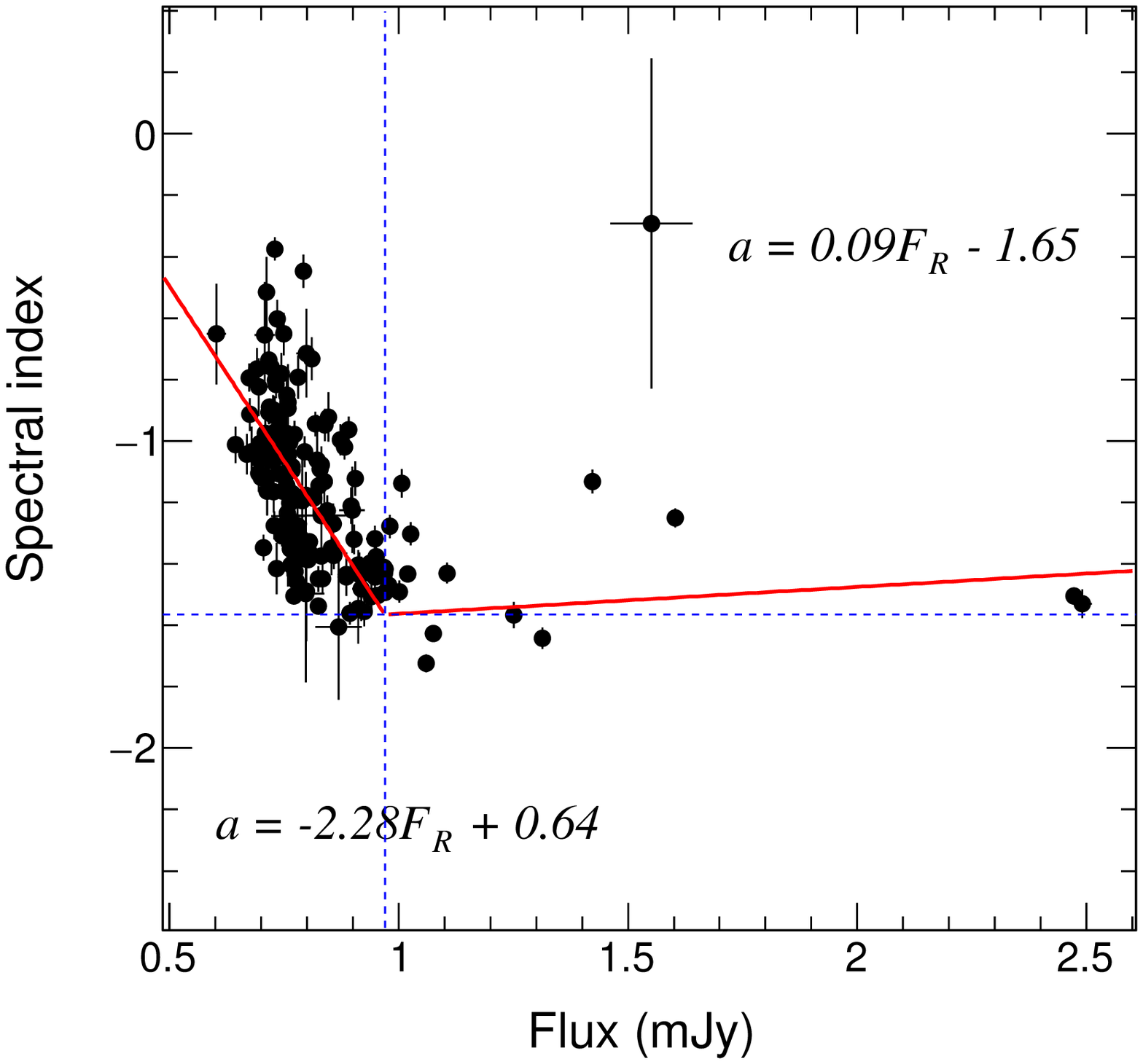}{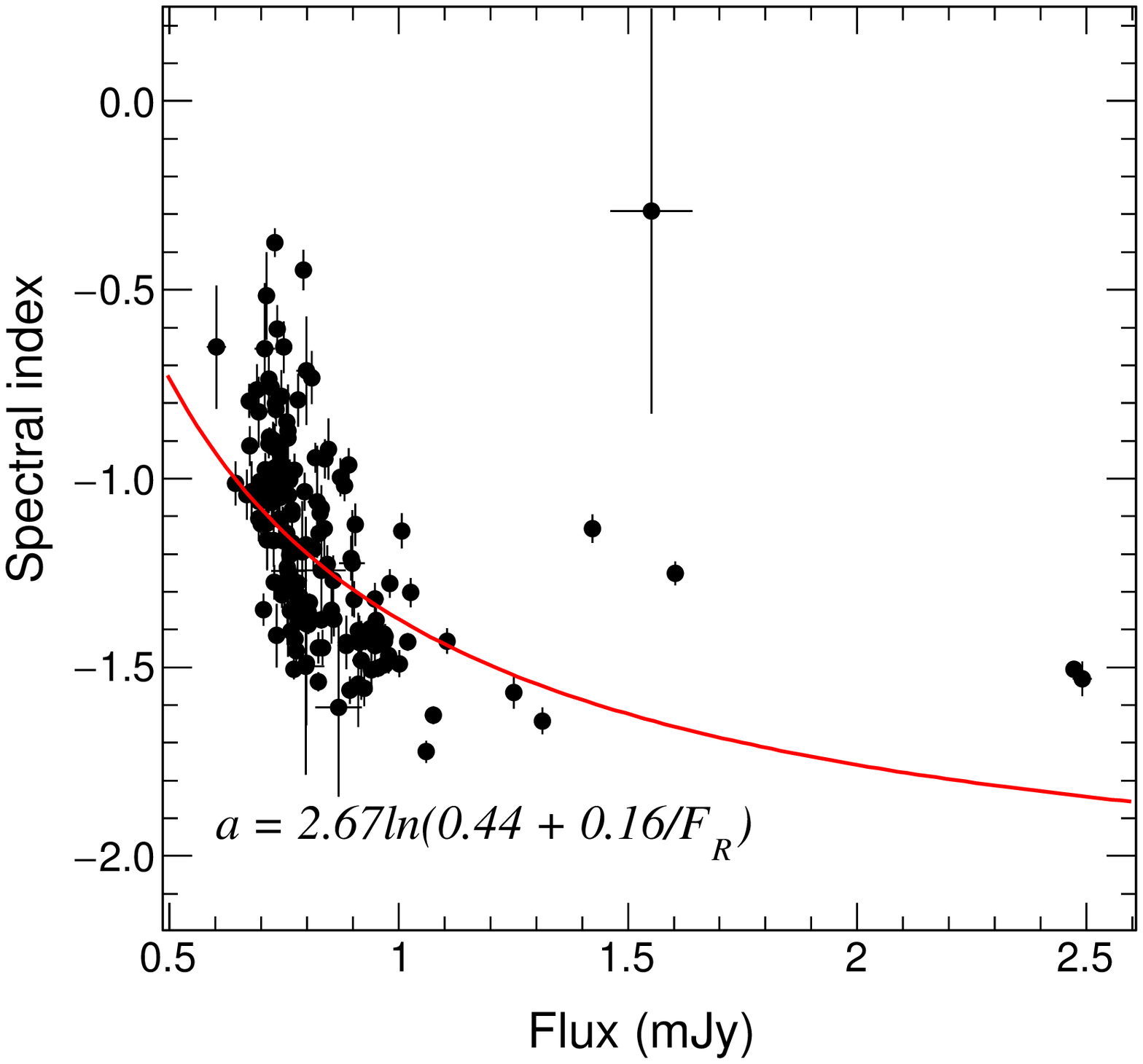}
\caption{Same as Figure~\ref{3C454alphaflux} in the main text, for PKS 2142-75. \label{2142-75}}
\end{figure}

\begin{figure}
\epsscale{0.7}
\plottwo{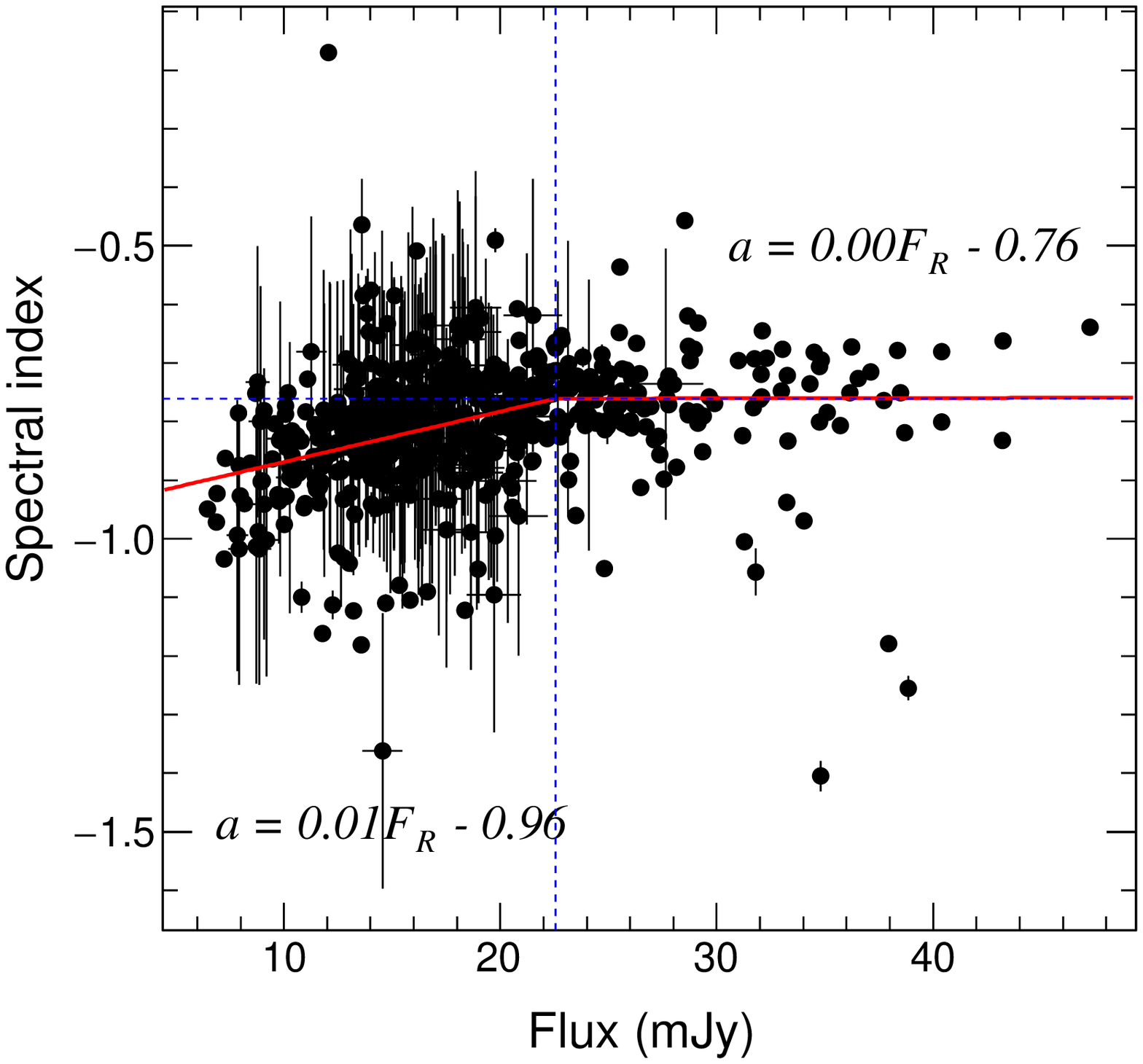}{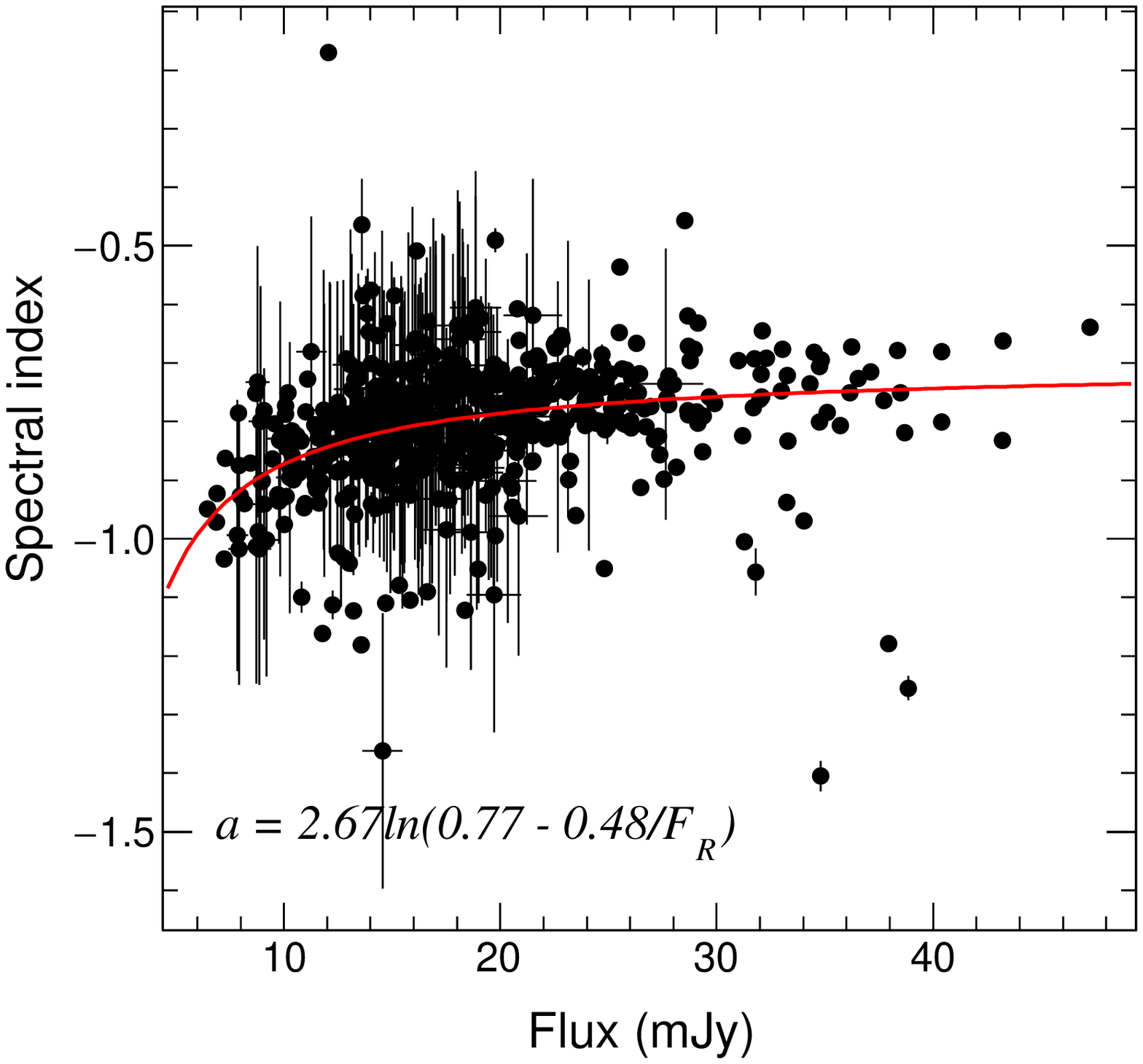}
\caption{Same as Figure~\ref{3C454alphaflux} in the main text, for PKS 2155-304. \label{2155-304}}
\end{figure}

\clearpage

\begin{figure}
\epsscale{0.7}
\plottwo{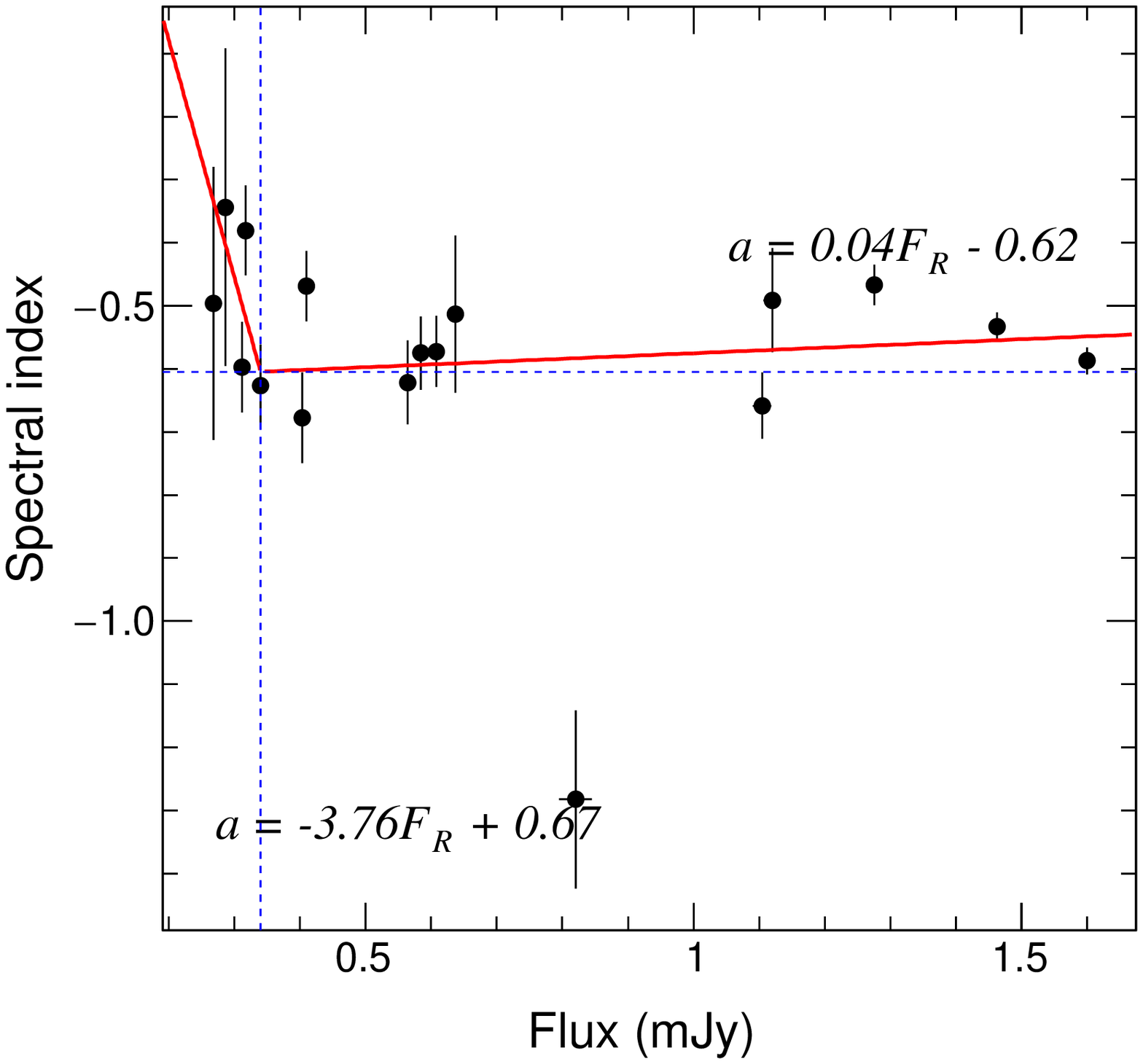}{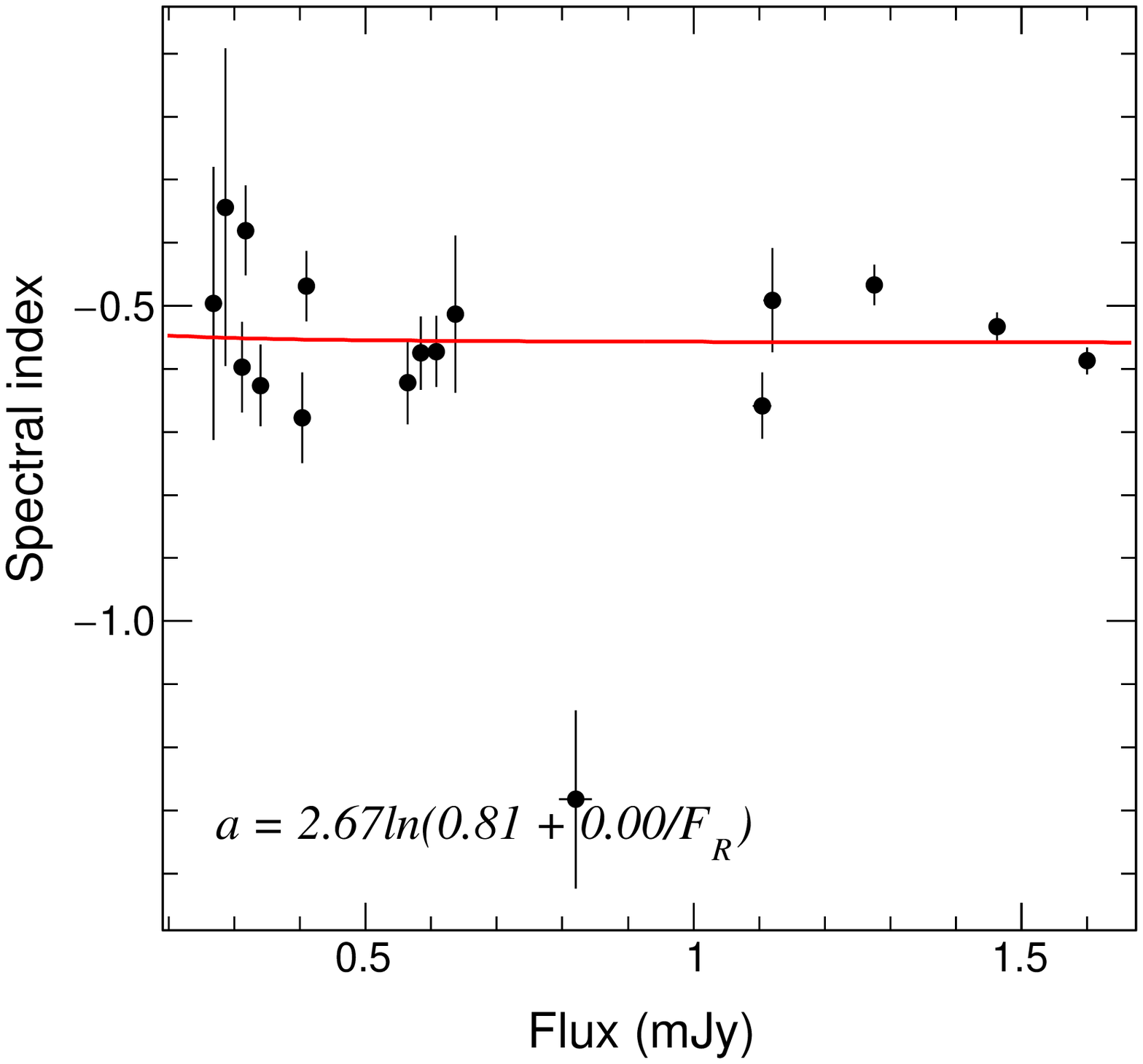}
\caption{Same as Figure~\ref{3C454alphaflux} in the main text, for PKS 2233-148. \label{2233-148}}
\end{figure}

\begin{figure}
\epsscale{0.7}
\plottwo{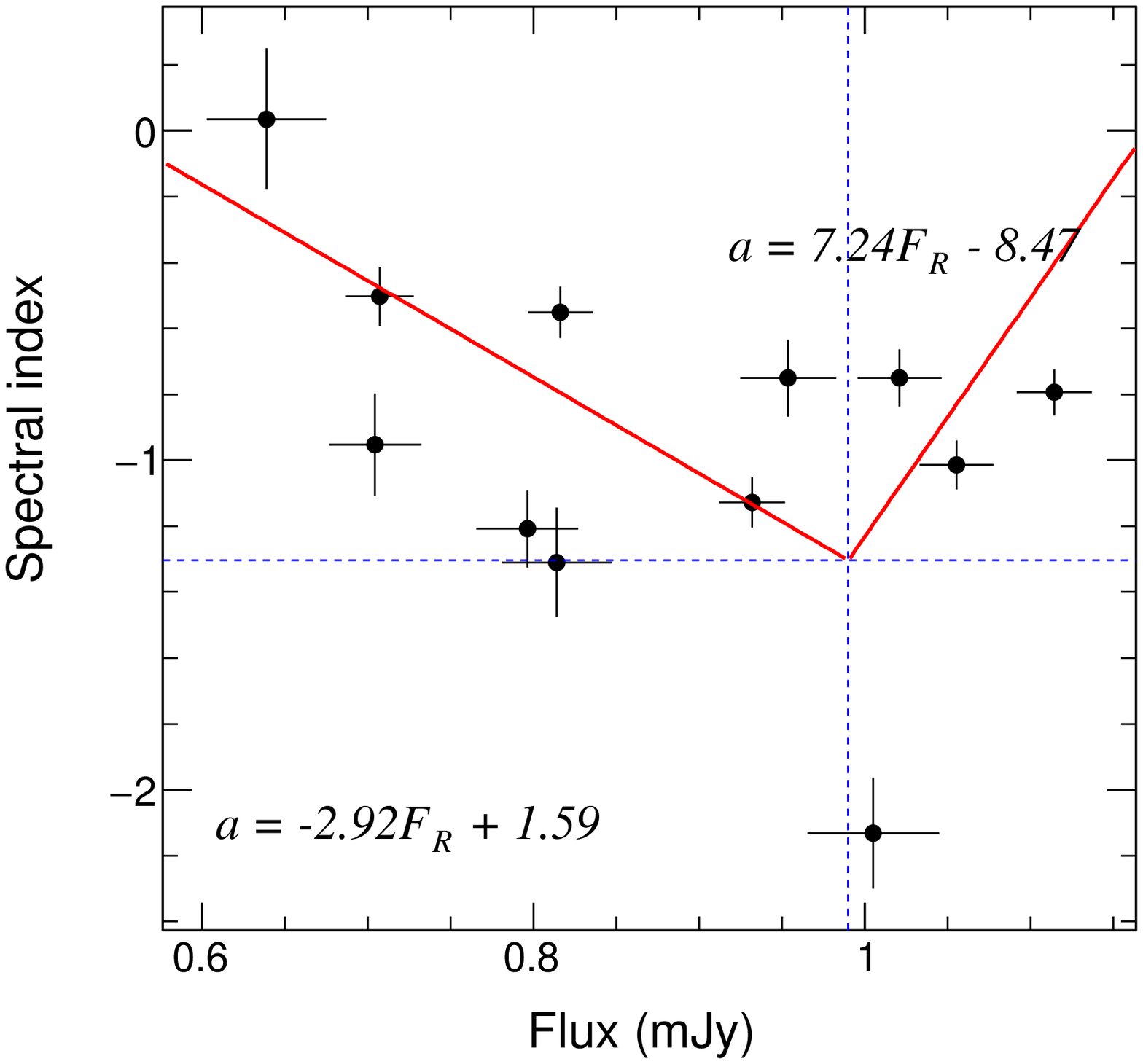}{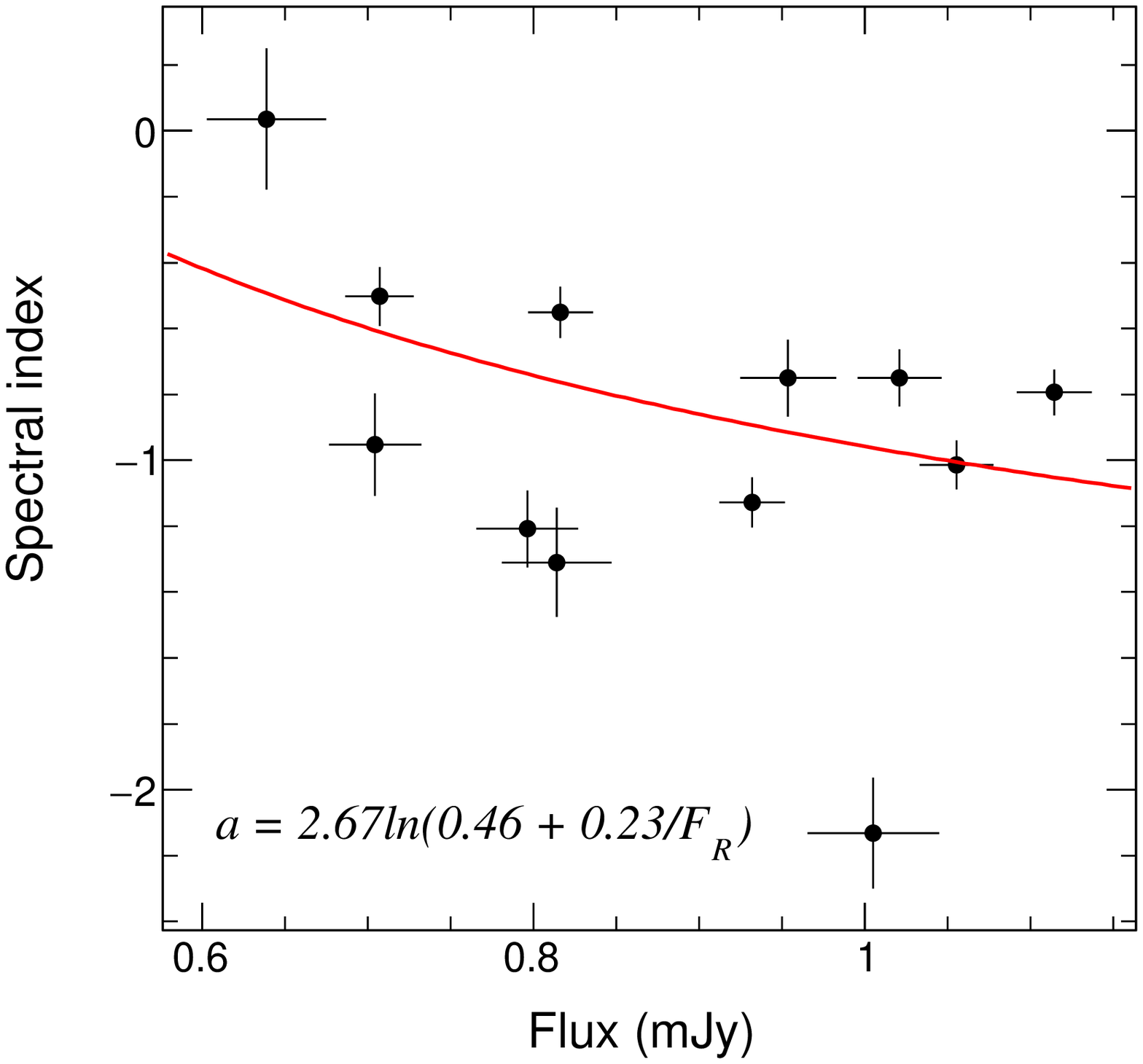}
\caption{Same as Figure~\ref{3C454alphaflux} in the main text, for PKS 2255-282. \label{2255-282}}
\end{figure}

\begin{figure}
\epsscale{0.7}
\plottwo{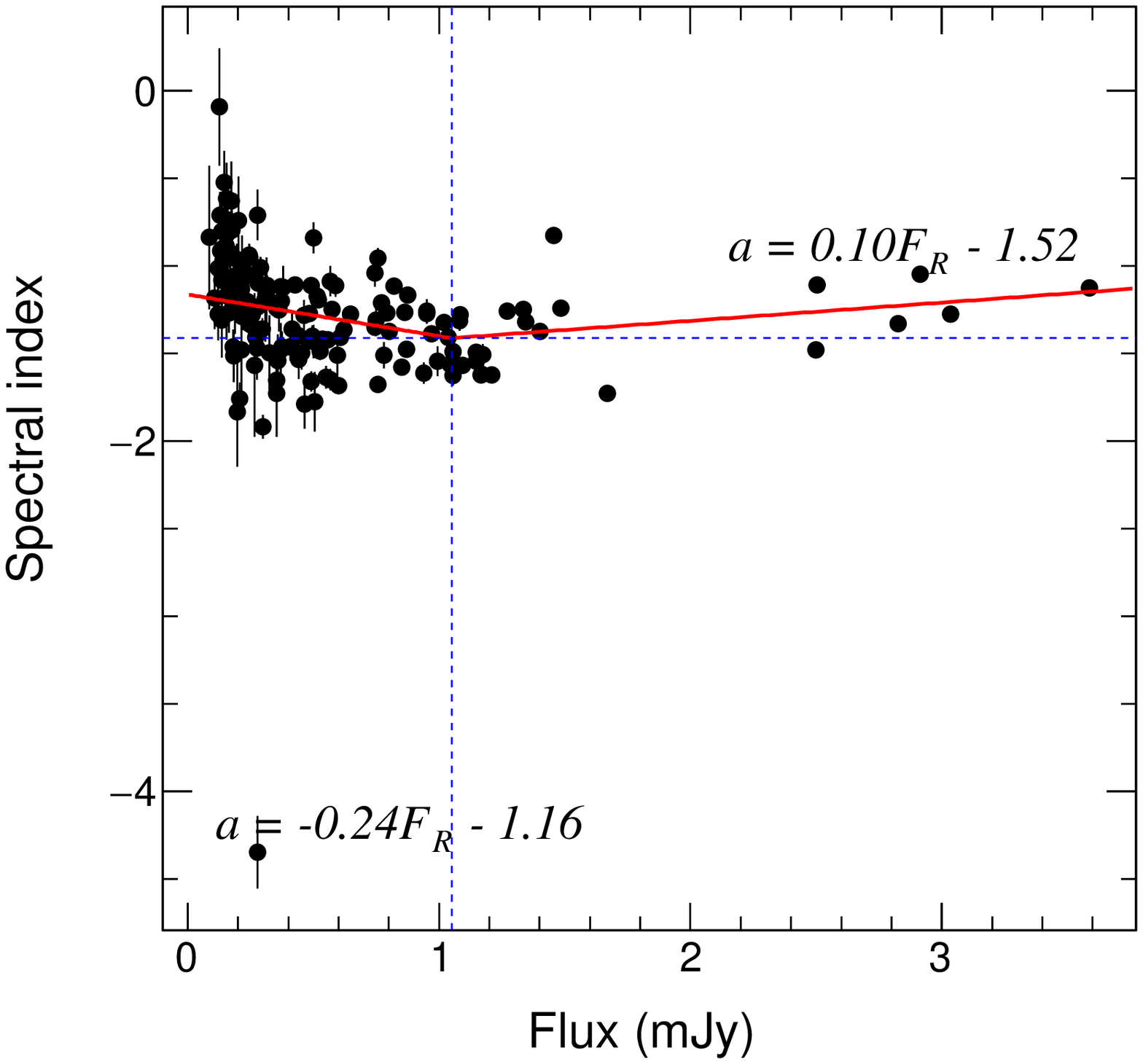}{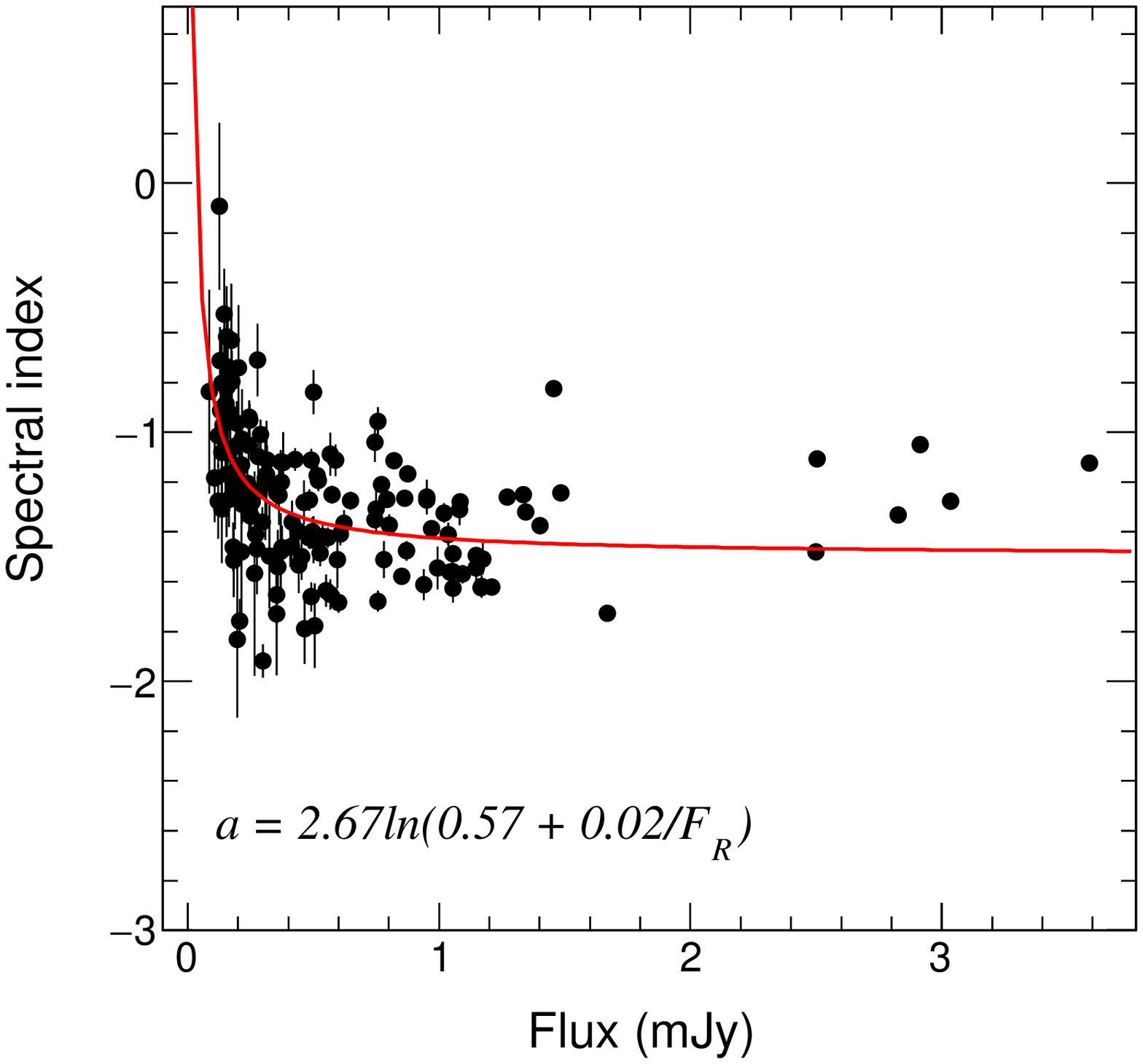}
\caption{Same as Figure~\ref{3C454alphaflux} in the main text, for PKS 2326-502. \label{2326-502}}
\end{figure}

\begin{figure}
\epsscale{0.7}
\plottwo{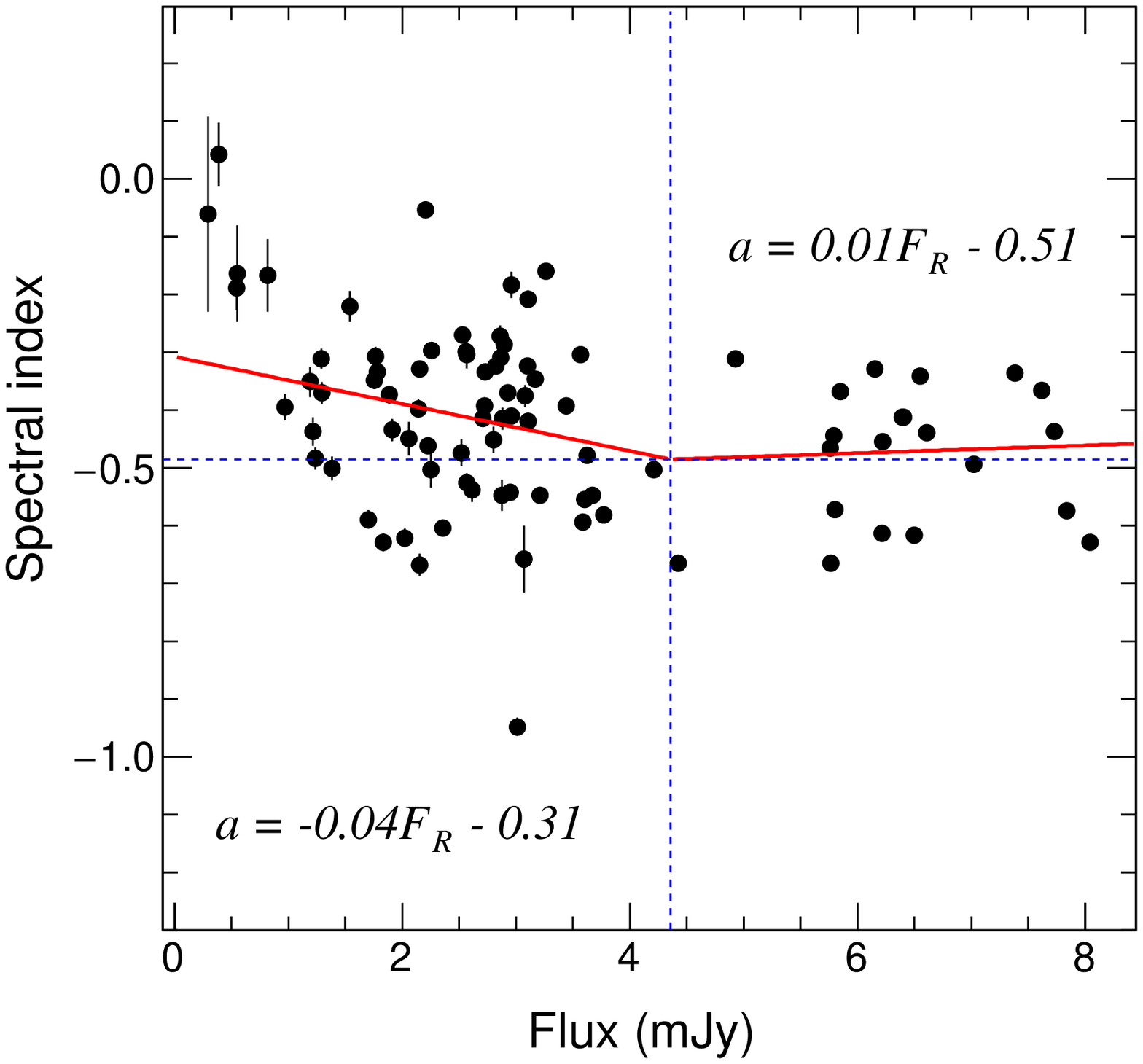}{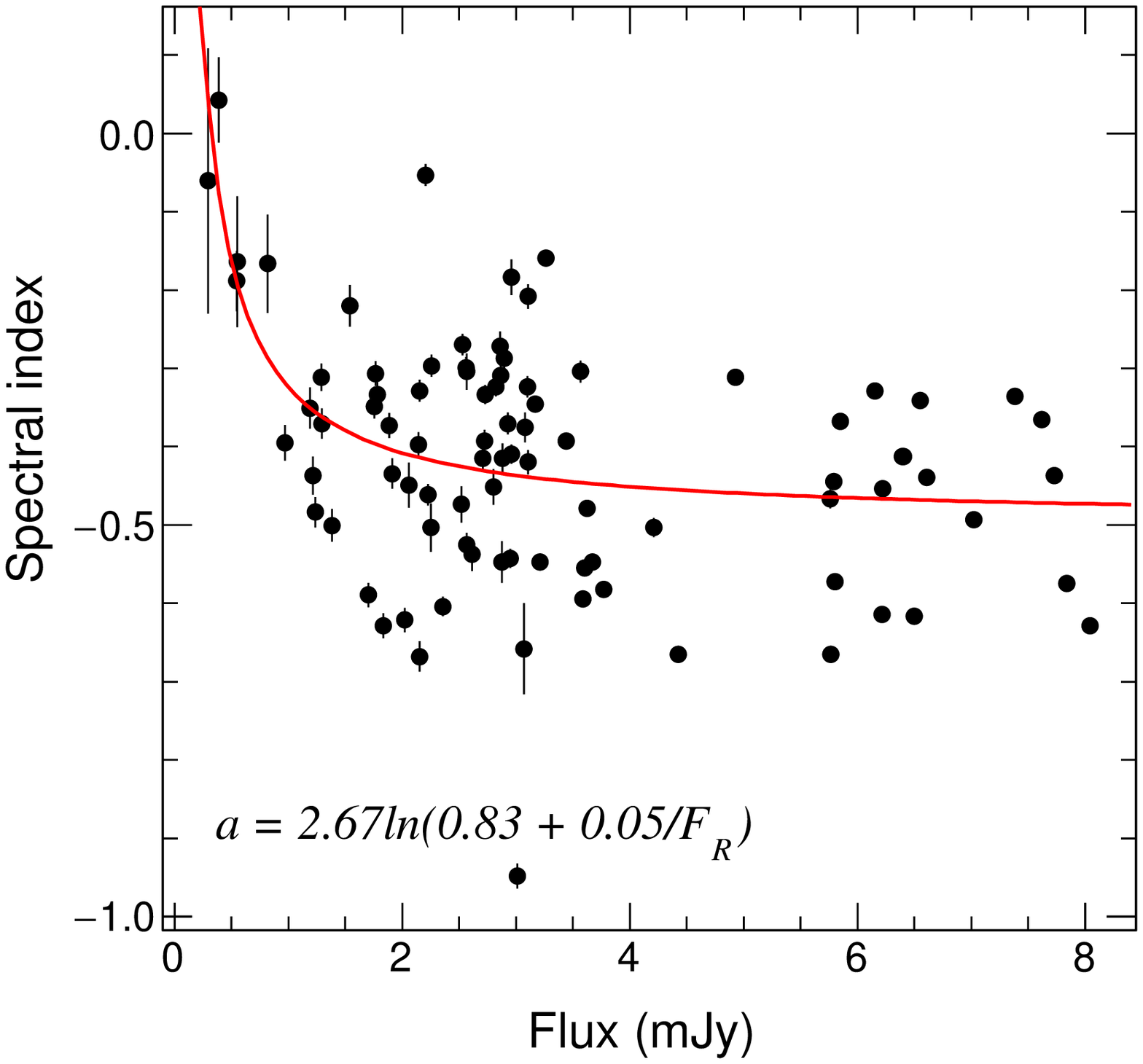}
\caption{Same as Figure~\ref{3C454alphaflux} in the main text, for PMN J2345-1555. \label{2345-1555}}
\end{figure}

\begin{figure}
\epsscale{0.7}
\plottwo{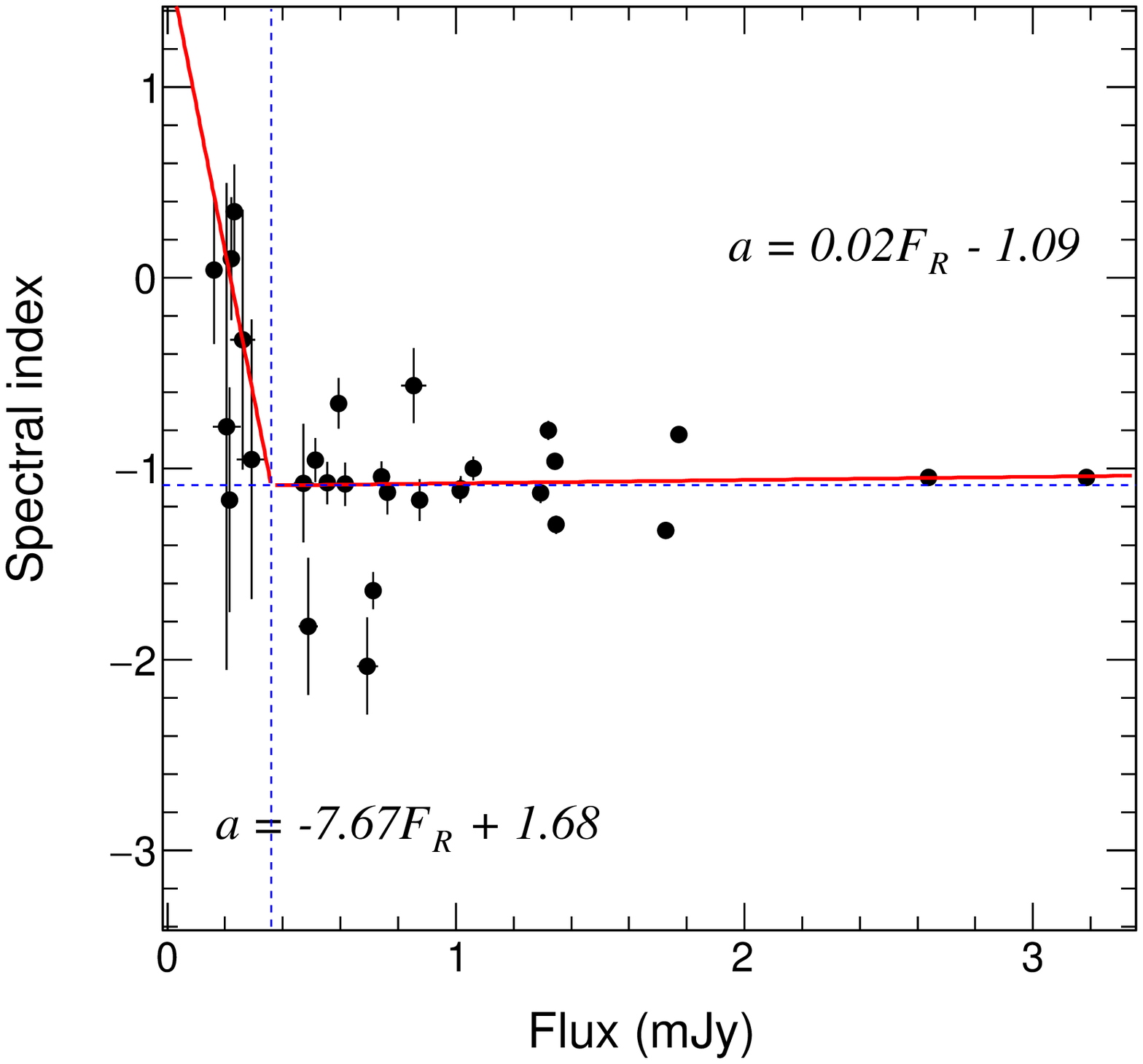}{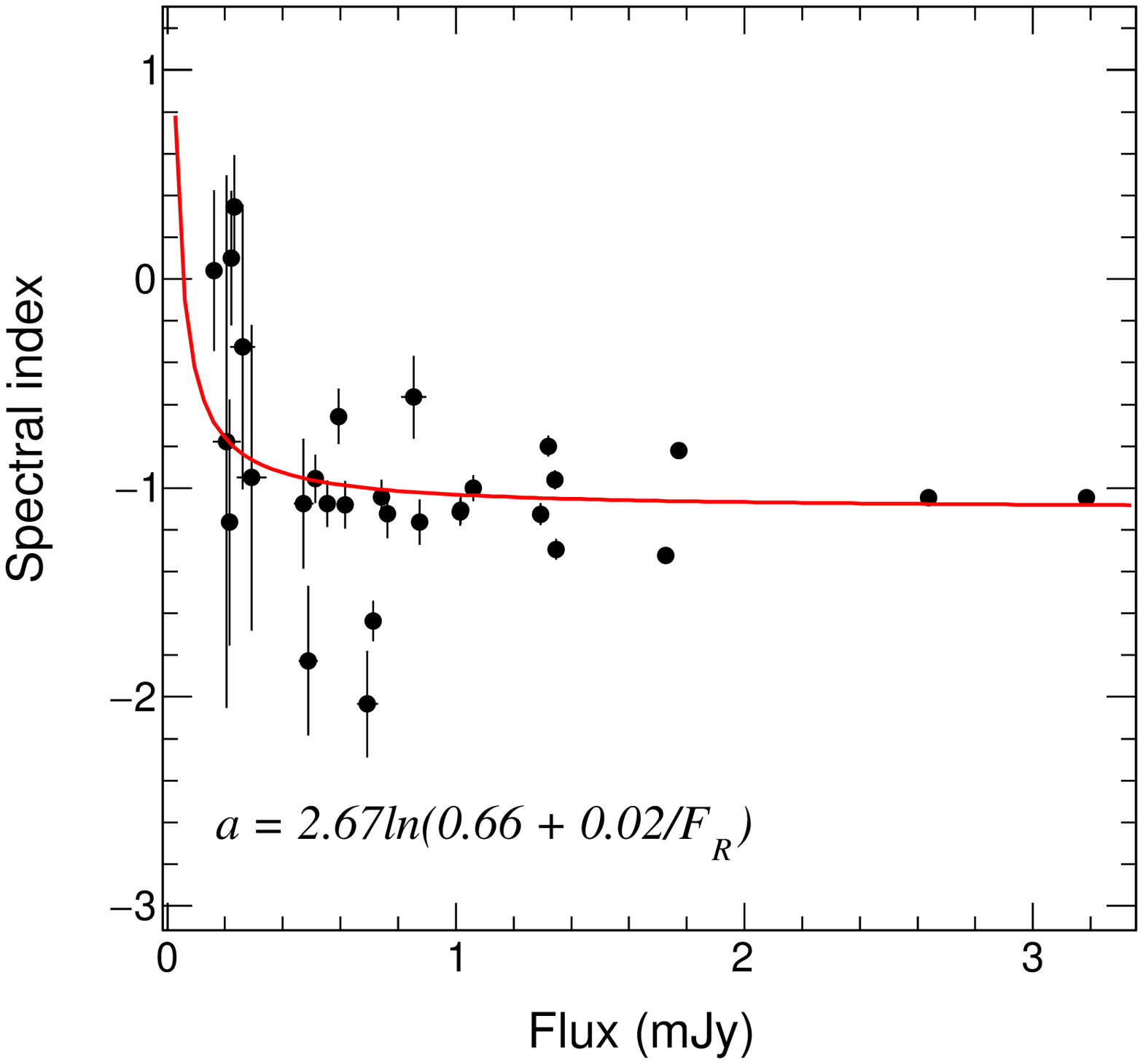}
\caption{Same as Figure~\ref{3C454alphaflux} in the main text, for PKS 2345-16. \label{2345-16}}
\end{figure}

\begin{figure}
\epsscale{0.7}
\plottwo{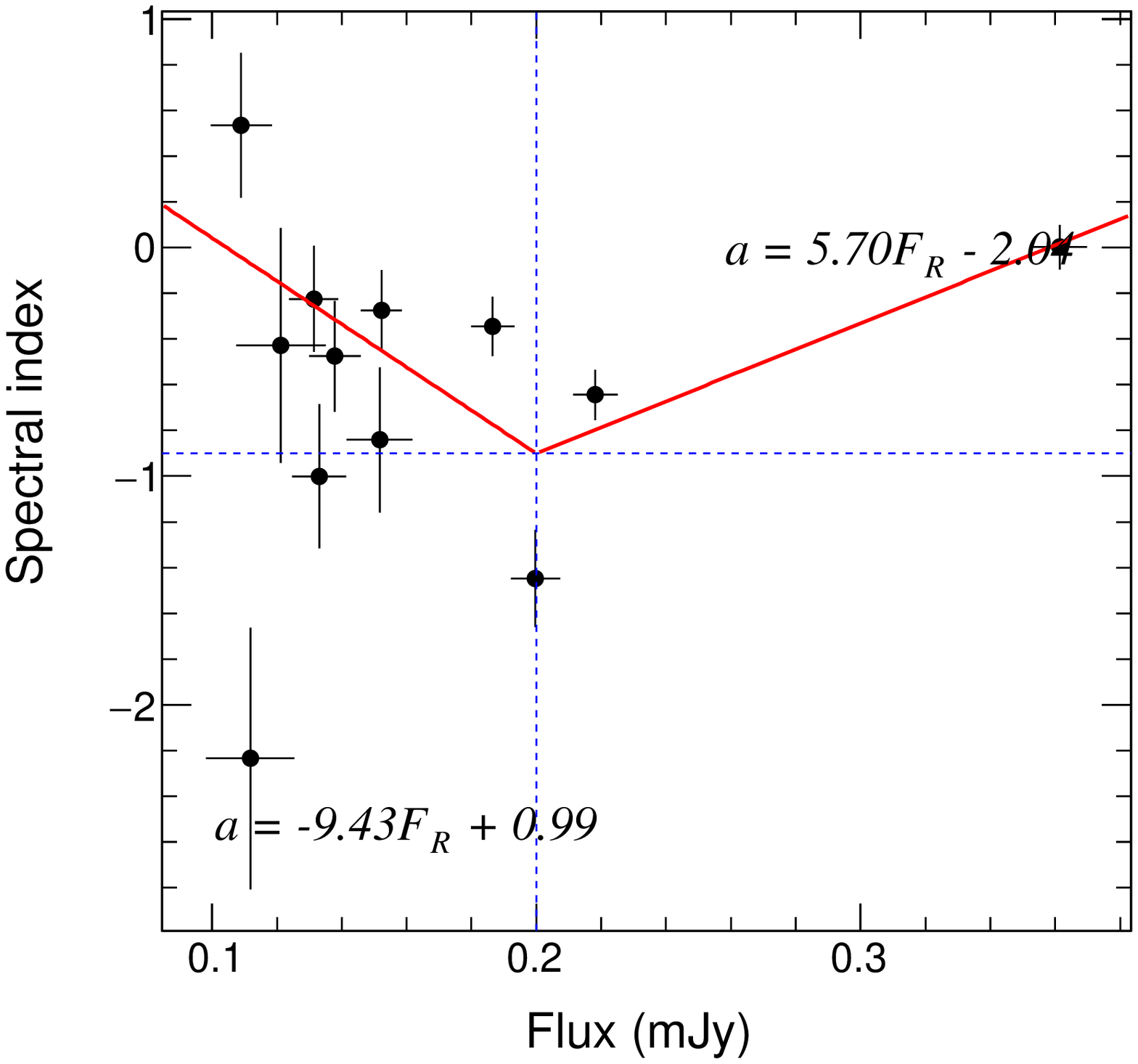}{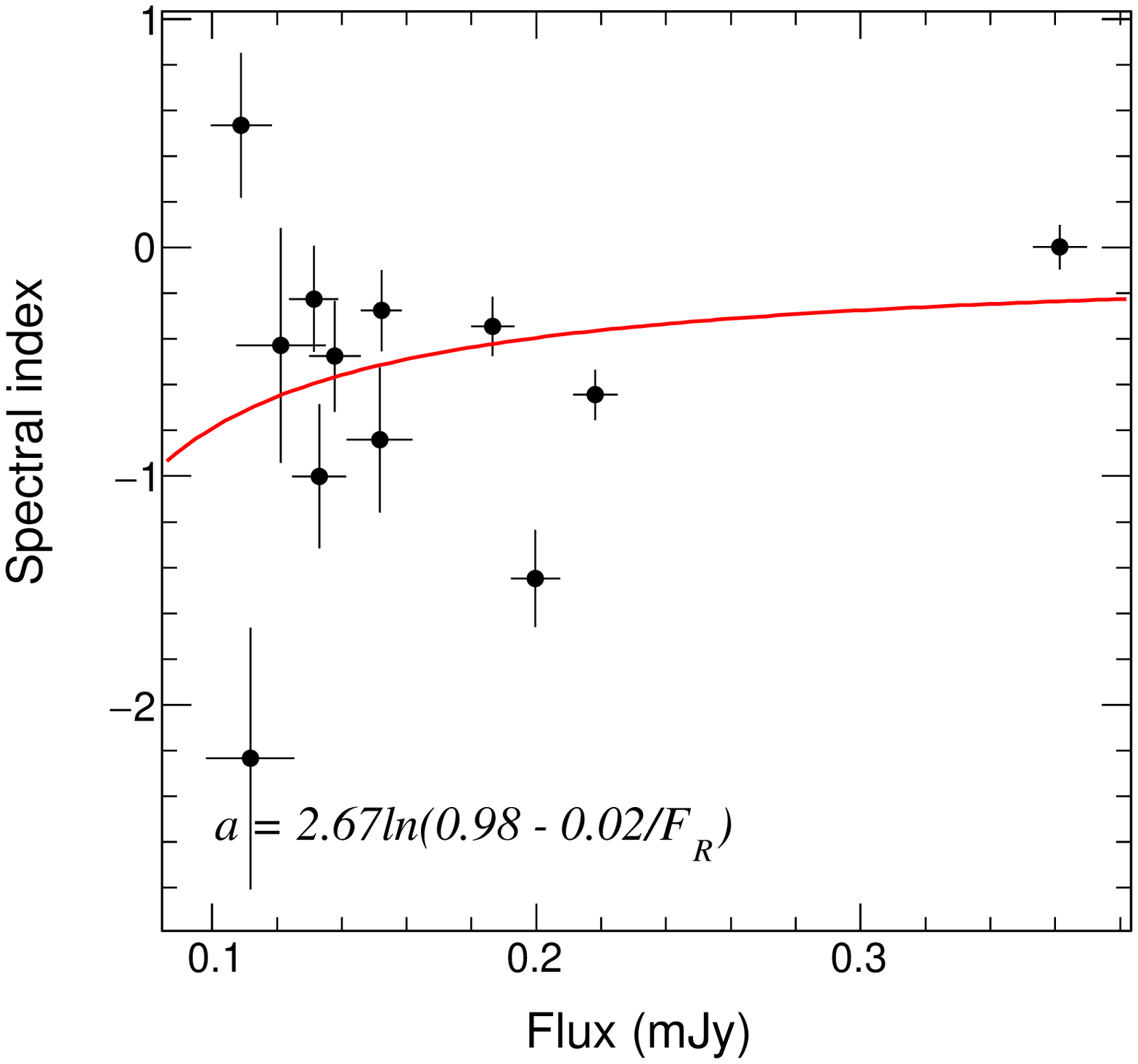}
\caption{Same as Figure~\ref{3C454alphaflux} in the main text, for PKS 2354-021. \label{2354-021}}
\end{figure}

\begin{figure}
\epsscale{0.7}
\plottwo{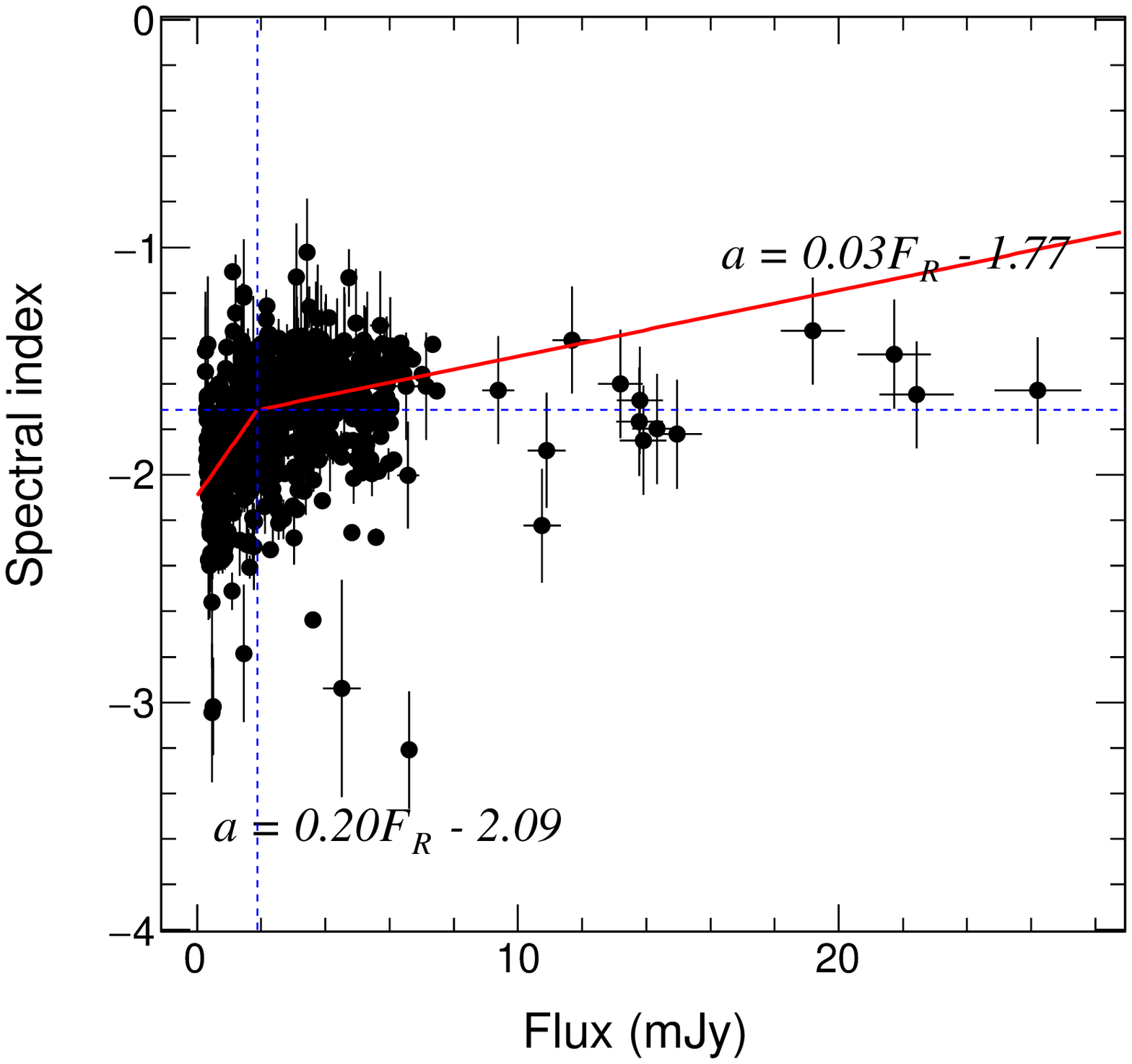}{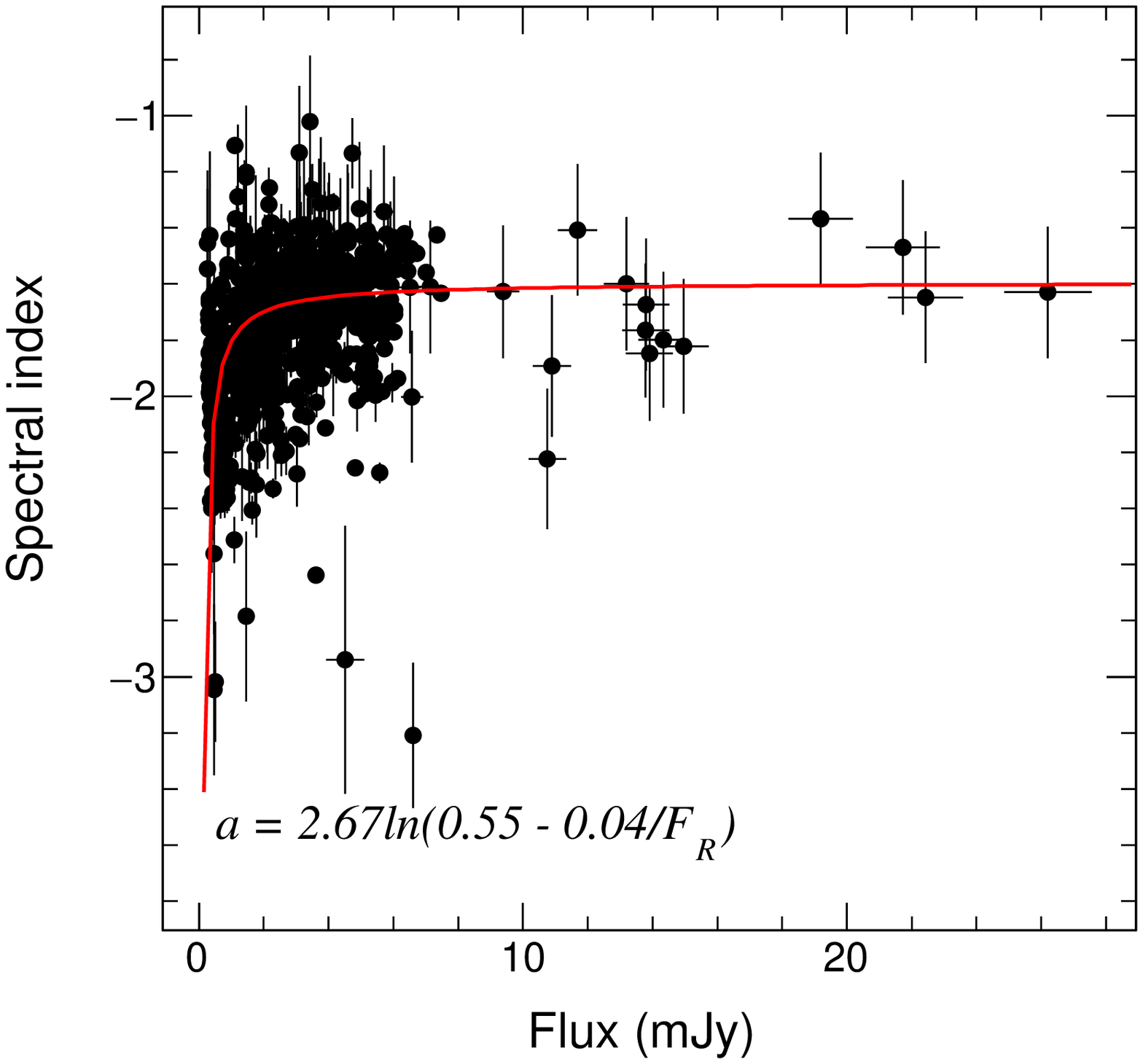}
\caption{Same as Figure~\ref{3C454alphaflux} in the main text, for 3C 279. \label{3C279}}
\end{figure}

\begin{figure}
\epsscale{0.7}
\plottwo{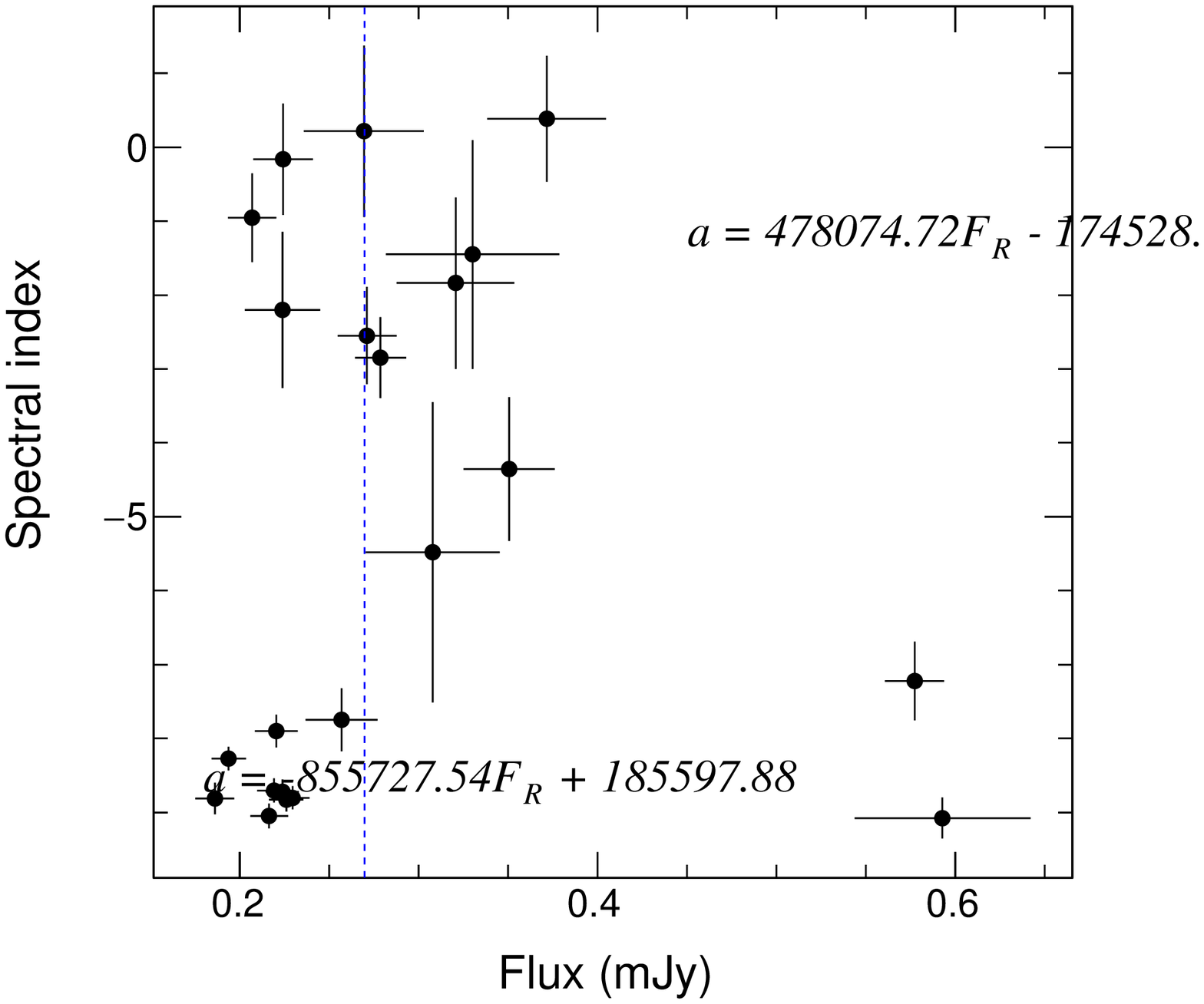}{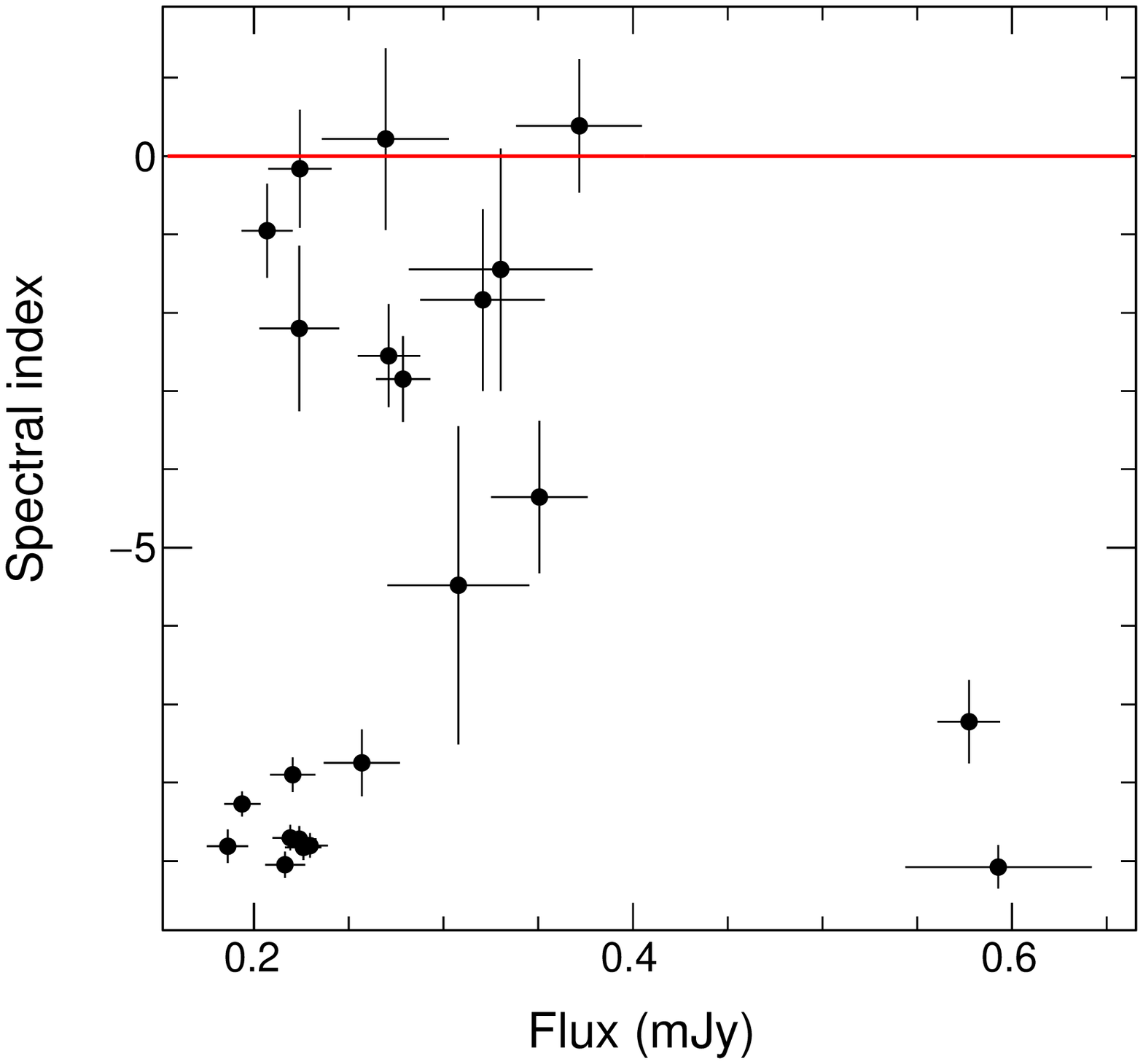}
\caption{Same as Figure~\ref{3C454alphaflux} in the main text, for 3C 446. \label{3C446}}
\end{figure}


\begin{figure}
\epsscale{0.7}
\plottwo{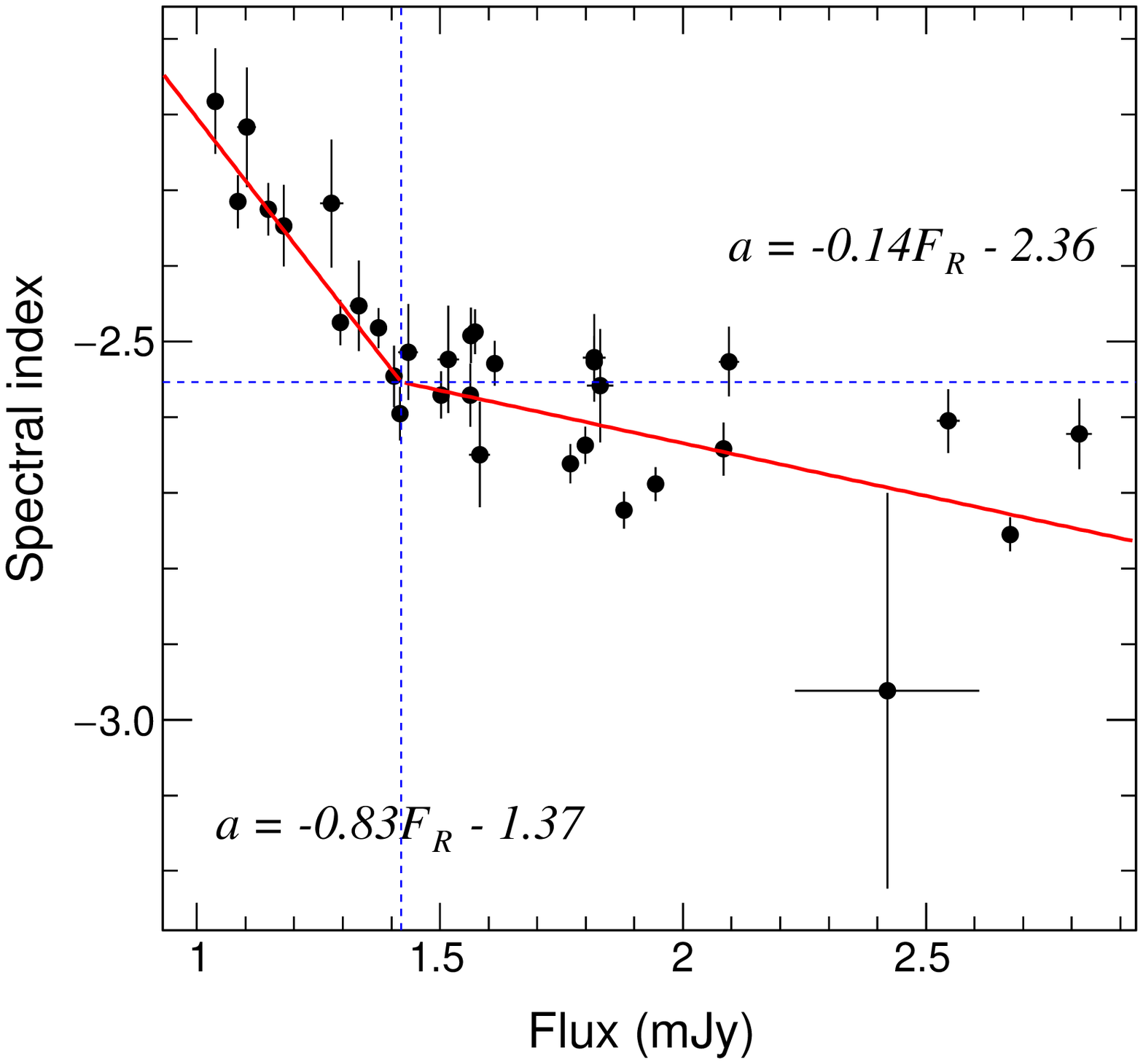}{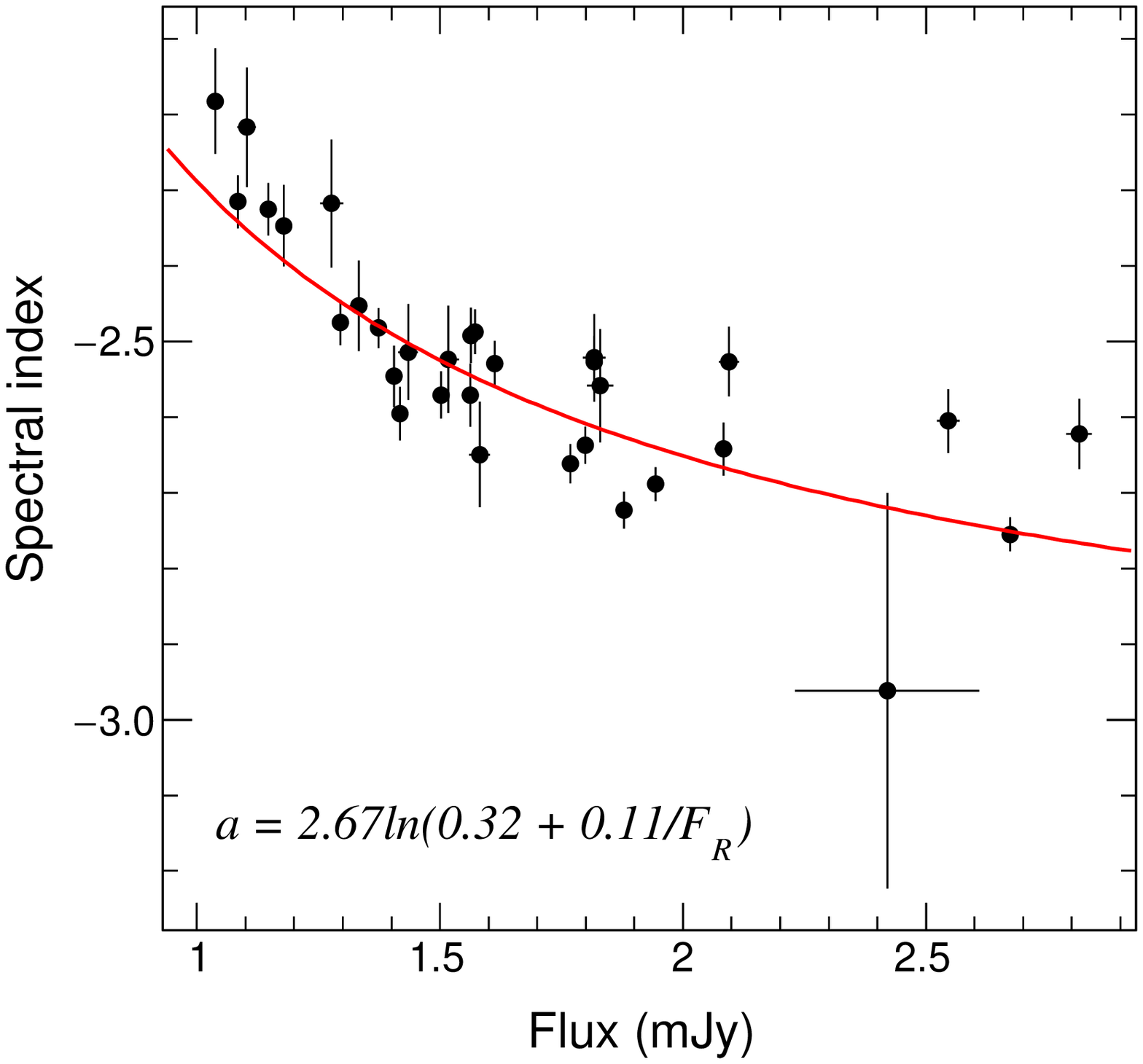}
\caption{Same as Figure~\ref{3C454alphaflux} in the main text, for CTA 102. \label{CTA102}}
\end{figure}

\begin{figure}
\epsscale{0.7}
\plottwo{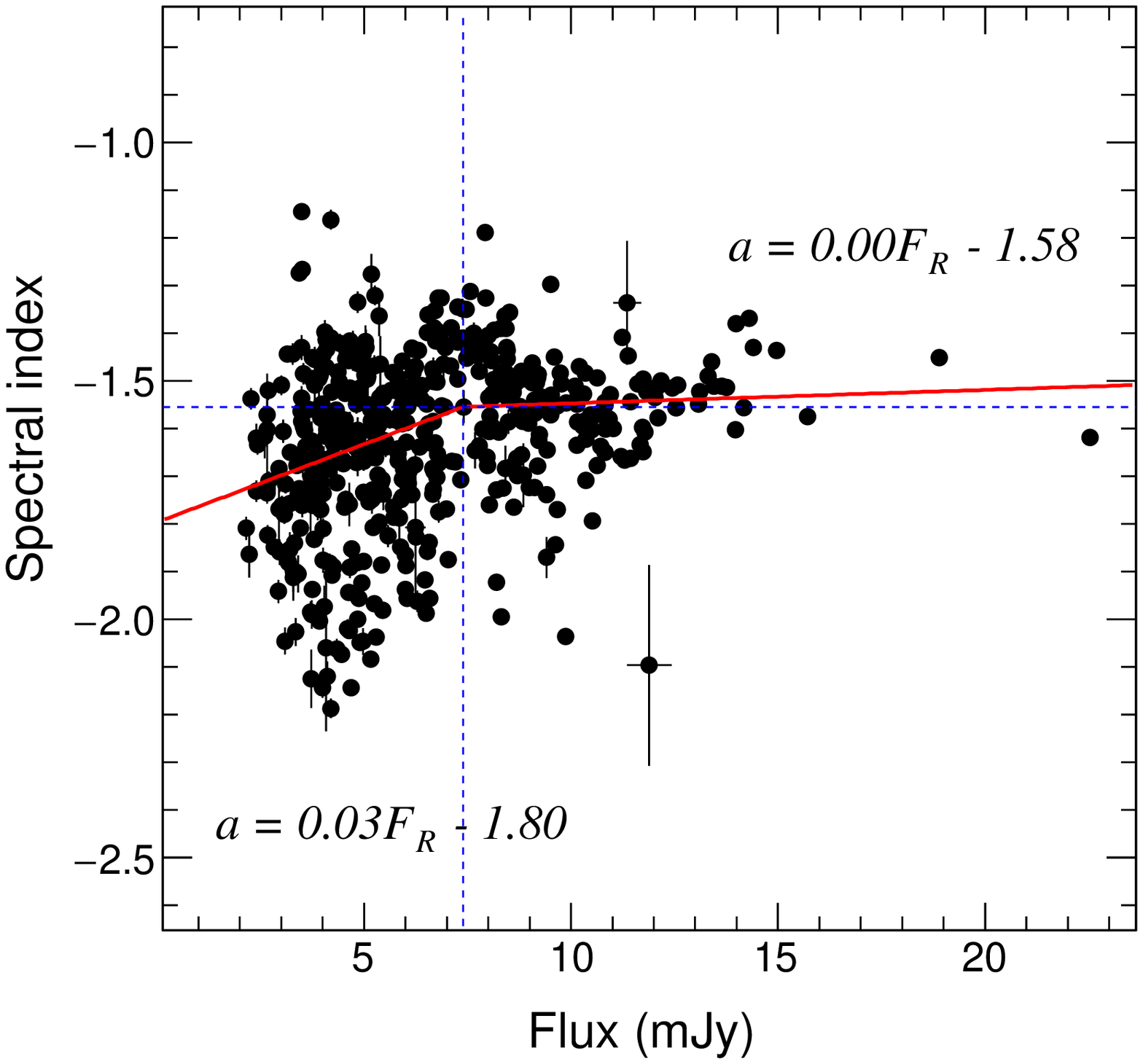}{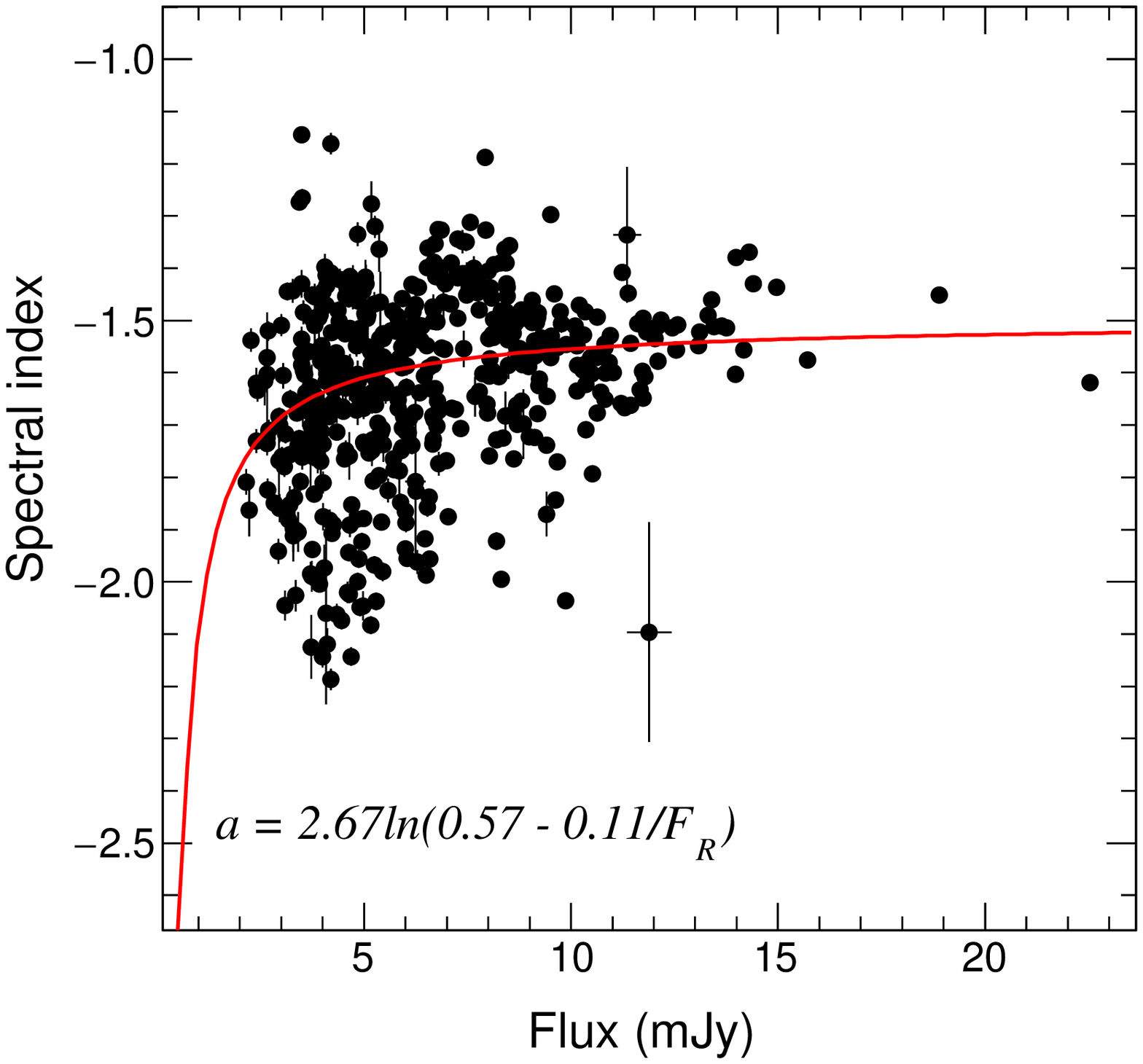}
\caption{Same as Figure~\ref{3C454alphaflux} in the main text, for OJ 287. \label{OJ287}}
\end{figure}






\clearpage
\begin{thebibliography}{}

\bibitem[Abdo et al.(2010)]{abdo10} Abdo, A. A., et al. 2010, \apj, 716, 30
\bibitem[Agarwal et al.(2015)]{agarwal15} Agarwal, A., et al. 2015, \mnras, 451, 3882
\bibitem[B\"{o}ttcher et al.(2007)]{bottcher07} B\"{o}ttcher, M. 2007, \apss, 309, 95
\bibitem[B\"{o}ttcher et al.(2010)]{bottcher10} B\"{o}ttcher, M., et al. 2010, \apj, 725, 2344
\bibitem[B\"{o}ttcher et al.(2013)]{bottcher13} B\"{o}ttcher, M., Reimer, A., Sweeney, K., \& Prakash, A. 2013, \apj, 768, 54
\bibitem[Bonning et al.(2012)]{bonning12} Bonning, E., et al. 2012, \apj, 756
\bibitem[Chatterjee, Nalewajko \& Myers(2013)]{chatterjee13} Chatterjee, R., Nalewajko, K., \& Myers, A. D. 2013, \apj, 771, L25
\bibitem[Chiang \& B\"{o}ttcher (2002)]{chiang02} Chiang, J., \& B\"{o}ttcher, M. 2002, \apj, 564, 92
\bibitem[Dai et al.(2009)]{dai09} Dai, B. Z., et al. 2009, \mnras, 392, 1181
\bibitem[Dai et al.(2011)]{dai11} Dai, Y., Wu, J., Zhu, Z.-H., Zhou, X., \& Ma, J. 2011, \aj, 141, 65
\bibitem[Dermer \& Schlickeiser (1993)]{dermer93} Dermer, C. D., \& Schlickeiser, R. 1993, \apj, 416, 458
\bibitem[Falomo, Pian \& Treves(2014)]{falomo14} Falomo, R., Pian, E., \& Treves, A. 2014, \aapr, 22, 73
\bibitem[Fan et al.(2016)]{fan16} Fan, J. H., et al. 2016, \apjs, 226, 20
\bibitem[Fan et al.(2018)]{fan18} Fan, X.-L., et al. 2018, \apj, 856, 80
\bibitem[Fiorucci, Ciprini \& Tosti(2004)]{fiorucci04} Fiorucci, M., Ciprini, S., \& Tosti, G. 2004, \aap, 419, 25
\bibitem[Gaur et al.(2012)]{gaur12} Gaur, H., et al. 2012, \mnras, 425, 3002
\bibitem[Gaur et al.(2019)]{gaur19} Gaur, H., et al. 2019, \mnras, 484, 5633
\bibitem[Ghisellini et al.(1997)]{ghisellini97} Ghisellini, G., et al. 1997, \aap, 327, 61
\bibitem[Ghisellini et al.(2019)]{ghisellini19} Ghisellini, G., et al. 2019, \aap, 627, A72
\bibitem[Ghisellini et al.(2017)]{ghisellini17} Ghisellini, G., Righi, C., Costamante, L., \& Tavecchio, F. 2017, \mnras, 469, 255
\bibitem[Ghisellini et al.(2010)]{ghisellini10} Ghisellini, G., Tavecchio, F., Foschini, L., Ghirlanda, G., Maraschi, L., \& Celotti, A. 2010, \mnras, 402, 497
\bibitem[Gu \& Ai (2011)]{gu11} Gu, M. F., \& Ai, Y. L. 2011, \aap, 528, A95
\bibitem[Gu et al.(2006)]{gu06} Gu, M. F., Lee, C.-U., Pak, S., Yim, H. S., \& Fletcher, A. B. 2006, \aap, 450, 39
\bibitem[Gupta et al.(2016)]{gupta16} Gupta, A. C., et al. 2016, \mnras, 458, 1127
\bibitem[Gupta et al.(2019)]{gupta19} Gupta, A. C., et al. 2019, \aj, 157, 95
\bibitem[Ikejiri et al.(2011)]{ikejiri11} Ikejiri, Y., et al. 2011, \pasj, 63, 639
\bibitem[Isler et al.(2017)]{isler17} Isler, J. C., et al. 2017, \apj, 844, 107
\bibitem[Kirk, Rieger \& Mastichiadis(1998)]{kirk98} Kirk, J. G., Rieger, F. M., \& Mastichiadis, A. 1998, \aap, 333, 452
\bibitem[Li et al.(2017)]{li17} Li, Y. T., et al. 2017, \pasp, 129, 014101
\bibitem[Li et al.(2018)]{li18} Li, X.-P., Luo, Y.-H., Yang, H.-T., Yang, H.-Y., Yang, C., \& Cai, Y. 2018, RAA, 18, 150
\bibitem[M\"{u}cke et al.(2003)]{mucke03} M\"{u}cke, A., Protheroe, R. J., Engel, R., Rachen, J. P., \& Stanev, T. 2003, APh, 18, 593
\bibitem[Man et al.(2016)]{man16} Man, Z., Zhang, X., Wu, J., \& Yuan, Q. 2016, \mnras, 456, 3168
\bibitem[Mao \& Zhang(2016)]{mao16} Mao, L., \& Zhang, X. 2016, \apss, 361, 345
\bibitem[Meng et al.(2018)]{meng18} Meng, N., Zhang, X., Wu, J., Ma, J., \& Zhou, X. 2018, \apjs, 237, 30
\bibitem[Murase et al.(2012)]{murase12} Murase, K., Dermer, C. D., Takami, H., \& Migliori, G. 2012, \apj, 749, 63
\bibitem[Nieppola, Tornikoski \& Valtaojaet(2006)]{nieppola06} Nieppola, E., Tornikoski, M., \& Valtaoja, E. 2006, \aap, 445, 441
\bibitem[Pandey et al.(2019)]{pandey19} Pandey, A., Gupta, A. C., Wiita, P. J., \& Tiwari, S. N. 2019, \apj, 871, 192
\bibitem[Pandey et al.(2020)]{pandey20} Pandey, A., et al. 2020, \apj, 890, 72
\bibitem[Papadakis, Villata \& Raiteri(2007)]{papadakis07} Papadakis, I. E., Villata, M., \& Raiteri, C. M. 2007, \aap, 470, 857
\bibitem[Raiteri et al.(2001)]{raiteri01} Raiteri, C. M., et al. 2001, \aap, 377, 396
\bibitem[Raiteri et al.(2008)]{raiteri08} Raiteri, C. M., et al. 2008, \aap, 491, 755
\bibitem[Raiteri et al.(2015)]{raiteri15} Raiteri, C. M., et al. 2015, \mnras, 454, 353
\bibitem[Raiteri et al.(2017)]{raiteri17} Raiteri, C. M., et al. 2017, Nature, 552, 374
\bibitem[Rajput et al.(2019)]{rajput19} Rajput, B., Stalin, C. S., Sahayanathan, S., Rakshit, S., \& Mandal, A. K. 2019, \mnras, 486, 1781
\bibitem[Ram\'{\i}rez et al.(2004)]{ramirez04} Ram\'{\i}rez, A., de Diego, J. A., Dultzin-Hacyan, D., \& Gonz\'{a}lez-P\'{e}rez, J. N. 2004, \aap, 421, 83
\bibitem[Rani et al.(2010)]{rani10} Rani, B., et al. 2010, \mnras, 404, 1992
\bibitem[Safna et al.(2020)]{safna20} Safna, P. Z., Stalin, C. S., Rakshit, S., \& Mathew, B. 2020, \mnras, 498, 3578
\bibitem[Sandrinelli, Covino \& Treves (2014)]{sandrinelli14} Sandrinelli, A., Covino, S., \& Treves, A. 2014, \aap, 562, 79
\bibitem[Sarkar et al.(2019)]{sarkar19} Sarkar, A., et al. 2019, \apj, 887, 185
\bibitem[Sokolov et al.(2004)]{sokolov04} Sokolov, A., Marscher, A. P., \& McHardy, I. M. 2004, \apj, 613, 725
\bibitem[Takalo \& Sillanp\"{a}\"{a}(1989)]{takalo89} Takalo, L. O., \& Sillanp\"{a}\"{a}, A. 1989, \aap, 218, 45
\bibitem[Urry \& Padovani(1995)]{urry95} Urry, C. M., \& Padovani, P. 1995, \pasp, 107, 803
\bibitem[Vagnetti, Trevese \& Nesci(2003)]{vagnetti03} Vagnetti, F., Trevese, D., \& Nesci, R. 2003, \apj, 590, 123
\bibitem[Villata et al.(2002)]{villata02} Villata, M., et al. 2002, \aap, 390, 407
\bibitem[Villata et al.(2004)]{villata04} Villata, M., et al. 2004, \aap, 421, 103
\bibitem[Villata et al.(2006)]{villata06} Villata, M., et al. 2006, \aap, 453, 817
\bibitem[Weaver et al.(2020)]{weaver20} Weaver, Z. R., et al. 2020, \apj, 900, 137
\bibitem[Wierzcholska et al.(2015)]{wierzcholska15} Wierzcholska, A., Ostrowski, M., Stawarz, {\L}., Wagner, S., \& Hauser, M. 2015, \aap, 573, A69
\bibitem[Wu et al.(2011)]{wu11} Wu, J., Zhou, X., Ma, J., \& Jiang, Z. 2011, \mnras, 418, 1640
\bibitem[Wu et al.(2007)]{wu07} Wu, J., Zhou, X., Ma, J., Wu, Z., Jiang, Z., \& Chen, J. 2007, \aj, 133, 1599
\bibitem[Xiong et al.(2020)]{xiong20} Xiong, D., et al. 2020, \apjs, 247, 49
\bibitem[Yuan et al.(2017)]{yuan17} Yuan, Y.-H., et al. 2017, \aap, 605, A43
\bibitem[Zhang et al.(2013)]{zhang13}Zhang, B. K., Wang, S., Zhao, X. Y., Dai, B. Z., \& Zha, M. 2013, \mnras, 428, 3630
\bibitem[Zhang et al.(2014)]{zhang14} Zhang, B.-K., Zhao, X.-Y., Wang, C.-X., \& Dai, B.-Z. 2014, RAA, 14, 933
\bibitem[Zhang et al.(2015)]{zhang15} Zhang, B.-K., Zhou, X.-S., Zhao, X.-Y., \& Dai, B.-Z. 2015, RAA, 15, 1784
\bibitem[Zhang et al.(2021)]{zhang21} Zhang, B.-K., Jin, M., Zhao, X.-Y., Zhang, L., \& Dai, B.-Z., 2021, RAA, 21, 186

\end{thebibliography}
\end{document}